\documentclass[conference]{IEEEtran}
\IEEEoverridecommandlockouts

\usepackage{cite}
\usepackage{amsmath,amssymb,amsfonts}
\usepackage{algorithmic}
\usepackage{graphicx}
\usepackage{textcomp}
\usepackage{xcolor}
\usepackage{float}
\usepackage{overpic}
\usepackage{multirow}
\usepackage{subfigure}
\usepackage[colorlinks,
linkcolor=red,
anchorcolor=red,
citecolor=green]{hyperref}
\def\BibTeX{{\rm B\kern-.05em{\sc i\kern-.025em b}\kern-.08em
    T\kern-.1667em\lower.7ex\hbox{E}\kern-.125emX}}

\begin{document}

	\title{TSN-CA: A Two-Stage Network with Channel Attention for 
		Low-Light Image Enhancement
	}

	\author{Xinxu~Wei, Xianshi Zhang\IEEEauthorrefmark{1}, Shisen~Wang, Yanlin Huang, Kaifu Yang,
	and Yongjie~Li,~\IEEEmembership{Senior Member,~IEEE}}

	\maketitle


\maketitle

\begin{abstract}
Low-light image enhancement is a challenging low-level computer vision task because after we enhance the brightness of the image, we have to deal with amplified noise, color distortion, detail loss, blurred edges, shadow blocks and halo artifacts. In this paper, we propose a Two-Stage Network with Channel Attention (denoted as TSN-CA) to enhance the brightness of the low-light image and restore the enhanced images from various kinds of degradation. In the first stage, we enhance the brightness of the low-light image in HSV space and use the information of H and S channels to help the recovery of details in V channel. In the second stage, we integrate Channel Attention (CA) mechanism into the skip connection of U-Net in order to restore the brightness-enhanced image from severe kinds of degradation in RGB space. We train and evaluate the performance of our proposed model on the LOL real-world and synthetic datasets. In addition, we test our model on several other commonly used datasets without Ground-Truth. We conduct extensive experiments to demonstrate that our method achieves excellent effect on brightness enhancement as well as denoising, details preservation and halo artifacts elimination. Our method outperforms many other state-of-the-art methods qualitatively and quantitatively. 
\end{abstract}

\begin{IEEEkeywords}
Low-light image enhancement, Image denoising, Detail preservation, Channel attention, Artifacts elimination
\end{IEEEkeywords}

\section{Introduction}
\IEEEPARstart{I}mages captured in the low-light conditions suffer from poor visibility, low contrast and severe noise. After enhancing the brightness, the noise hidden in the darkness will be amplified. In addition, color distortion, shadow blocks and halo artifacts will appear in the brightness-enhanced image. When we remove the noise from the noise-polluted image, the details are erased along with the noise, resulting in blurred edges. Therefore, it is very meaningful to propose a method which can enhance the image brightness, remove noise, correct color distortion, preserve details, restore blurred edges and eliminate artifacts simultaneously and effectively.\\ 
In this paper, inspired by\cite{yue2017contrast}, we propose a Two-Stage Network with Channel Attention (TSN-CA) to enhance the brightness of the low-light image and restore the enhanced images from various kinds of degradation.\\
In the first stage, we first transform the image from RGB space to HSV space and decompose the input image into H (Hue), S (Saturation) and V (Value). A U-Net is then trained to learn the mapping of low/normal-light V channel to enhance the brightness of V and we also fuse the information of H and S with V to help restore the details information in V channel when enhancing the brightness of V. In the second stage, we combined the enhanced and detail-restored V with the degraded H and S together and convert them from HSV space back to RGB space. We can obtain the intermediate results that are brightness-enhanced but degraded. And then we train a U-Net to recover degraded images from various kinds of degradation, such as amplified noise, color distortion, blurred edges, shadow blocks and halo artifacts. Inspired by\cite{zhao2019deep}, we introduce channel attention mechanism into U-Net to integrate residual features, guide the network to ignore useless degraded features, learn non-degraded features better, and help the network restore the degraded images from high degree of degradation as well as eliminate shadow and halo artifacts. \\
We highlight the contributions of this paper as follows:
\begin{itemize}
	\item Inspired by\cite{yue2017contrast}, we propose a novel Two-Stage Network to enhance the brightness of low-light image in HSV space and restore the enhanced image from severe kinds of degradation in RGB space.\\ 
	\item Following the enhancement network of DA-DRN\cite{wei2021dadrn}, we train an enhancer which is able  to enhance the V channel in HSV directly which can restore the details information of V channel with the help of H and S.\\
	\item We introduce channel attention mechanism into U-Net to restore highly degraded images and eliminate shadow blocks and halo artifacts effectively. \\
	\item Extensive experiments are conducted to demonstrate that our method outperforms many other state-of-the-art methods qualitatively and quantitatively.
\end{itemize}

\section{Related Works}
Many effective methods have been developed in low-light image enhancement. These methods can be divided into two categories: traditional enhancement methods and deep learning-based enhancement methods.

\subsubsection{Traditional Enhancement Methods}
NPE \cite{wang2013naturalness} improve the contrast while keeping the naturalness of the image. 
LIME \cite{guo2016lime} uses structure prior to estimate a structure-aware illumination map and then enhance it.
CRM \cite{ying2017new} propose a novel enhancement method using the response characteristics of cameras. EFF\cite{ying2017new02} designs a weight matrix for image fusion. JED \cite{ren2018joint} is a joint low-light enhancement and denoising strategy. 

\subsubsection{Deep Learning-based Enhancement Methods}
GLADNet \cite{wang2018gladnet} propose an enhancement method by estimating global illumination and achieve good effect in terms of details preservation. 
MBLLEN \cite{lv2018mbllen} fuses the different enhanced results generated by multiple subnets. RetinexNet \cite{wei2018deep} decomposes the low-light image into reflectance and illumination, and then denoise on the reflectance and increase brightness on illumination. EnlightenGan (EnGan) \cite{jiang2021enlightengan} proposes an unsupervised generative adversarial network (GAN). KinD \cite{zhang2019kindling} first decomposes low-light images into a noisy reflectance and a smooth illumination and then uses a deep U-Net to recover reflectance and a CNN to enhance the brightness of illumination. RDGAN \cite{wang2019rdgan} combines Retinex decomposition with GAN. 
Zero-DCE \cite{guo2020zero} estimates the optimal brightness curve of the input image by a lightweight network. 
KinD++\cite{zhang2021beyond} proposes MSIA module to deal with color distortion and noise in its previous work.
DA-DRN\cite{wei2021dadrn} proposes a Degradation-Aware Deep Retinex Network to directly restore the degraded reflectance and preserve details information during the decomposition stage by leveraging the dependency between reflectance and illumination map.

\begin{figure}
	\centering
	
	
	\subfigure[H]{
		\begin{minipage}[b]{0.15\textwidth}
			\includegraphics[width=2.8cm]{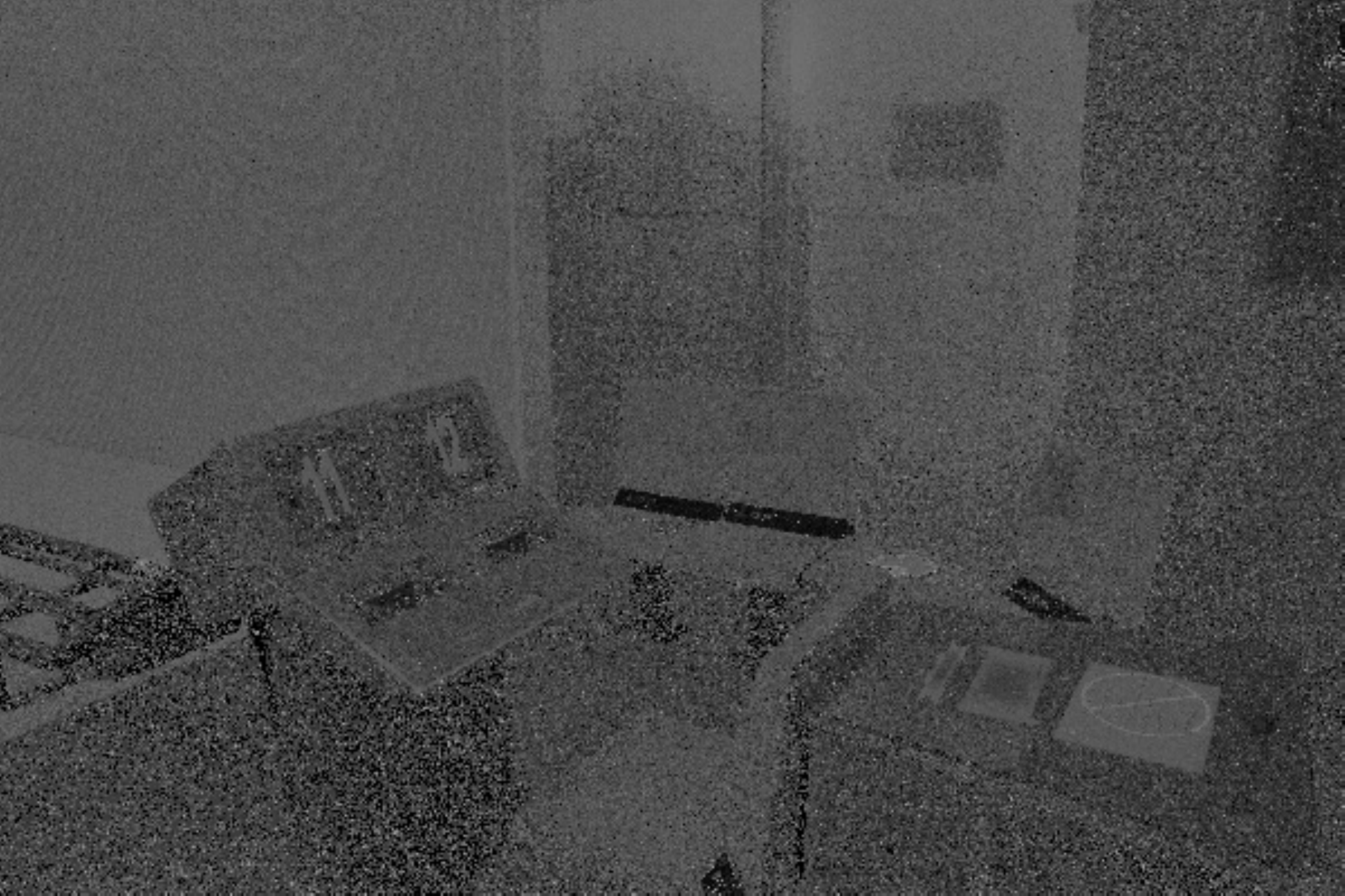}\vspace{2pt} \\
			\includegraphics[width=2.8cm]{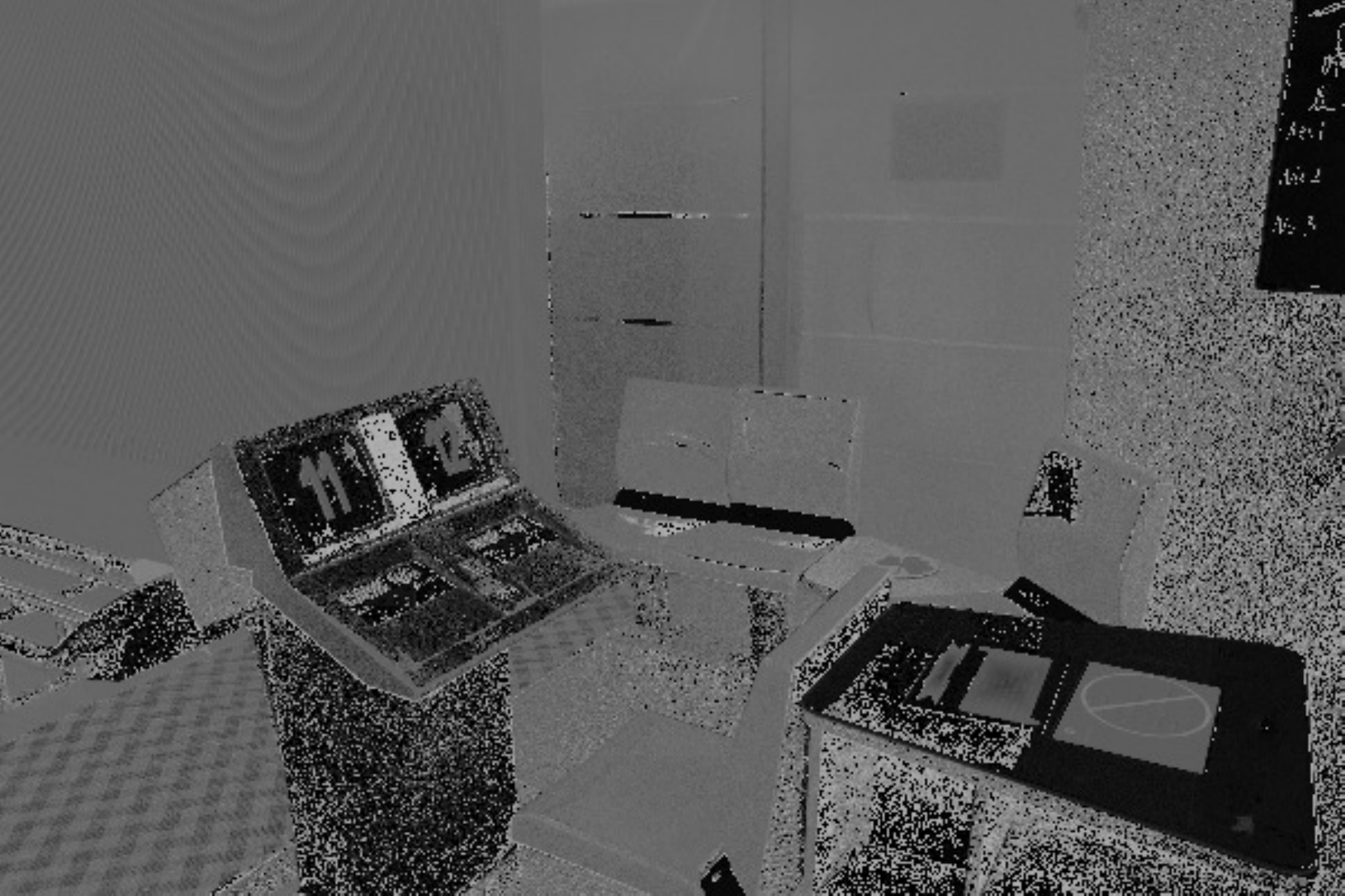}\vspace{0pt}
		\end{minipage}
	}\hspace{-5pt}
	\subfigure[S]{
		\begin{minipage}[b]{0.15\textwidth}
			\includegraphics[width=2.8cm]{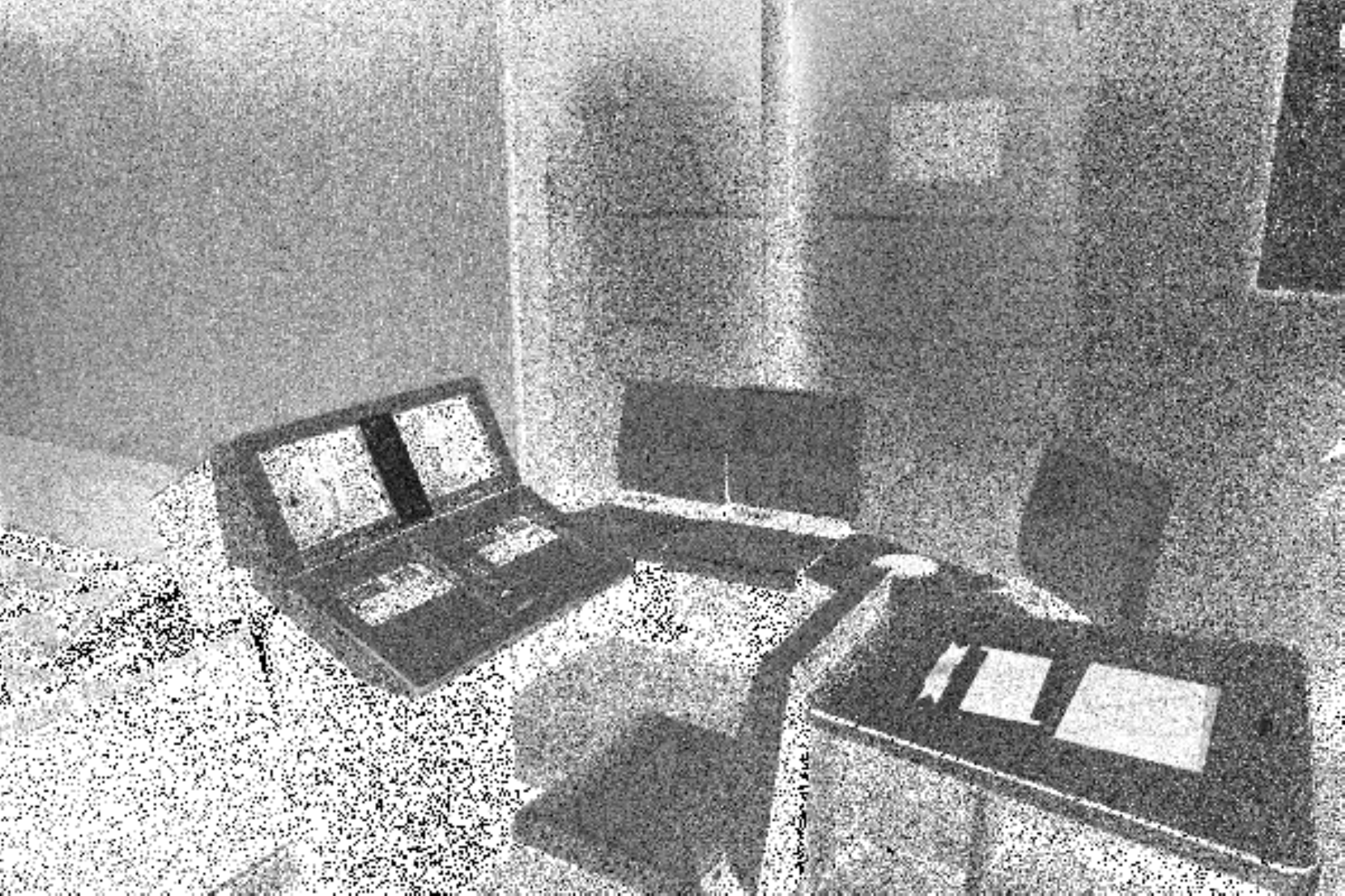}\vspace{2pt} \\
			\includegraphics[width=2.8cm]{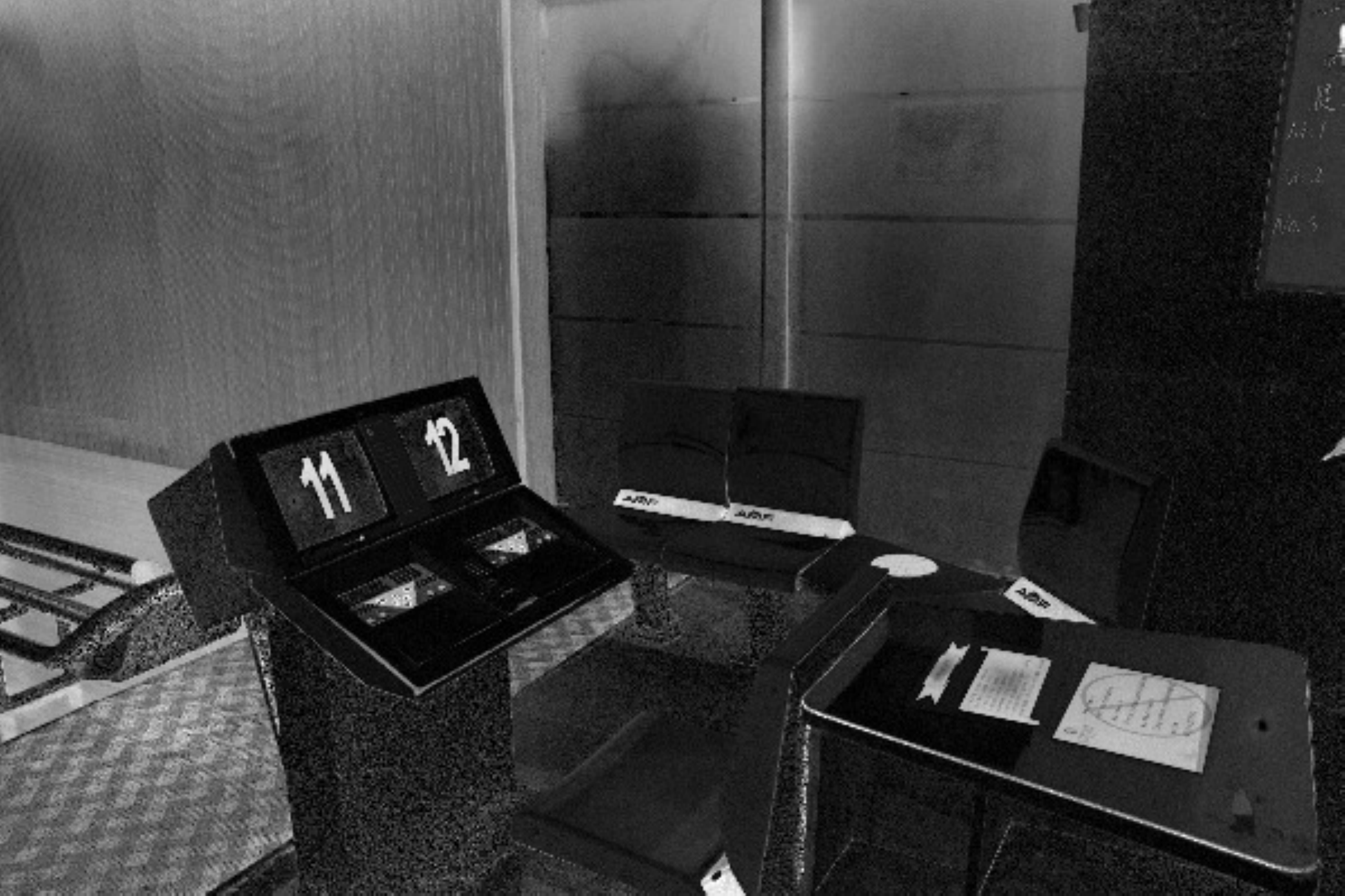}\vspace{0pt}
		\end{minipage}
	}\hspace{-5pt}
	\subfigure[V]{
		\begin{minipage}[b]{0.15\textwidth}
			\includegraphics[width=2.8cm]{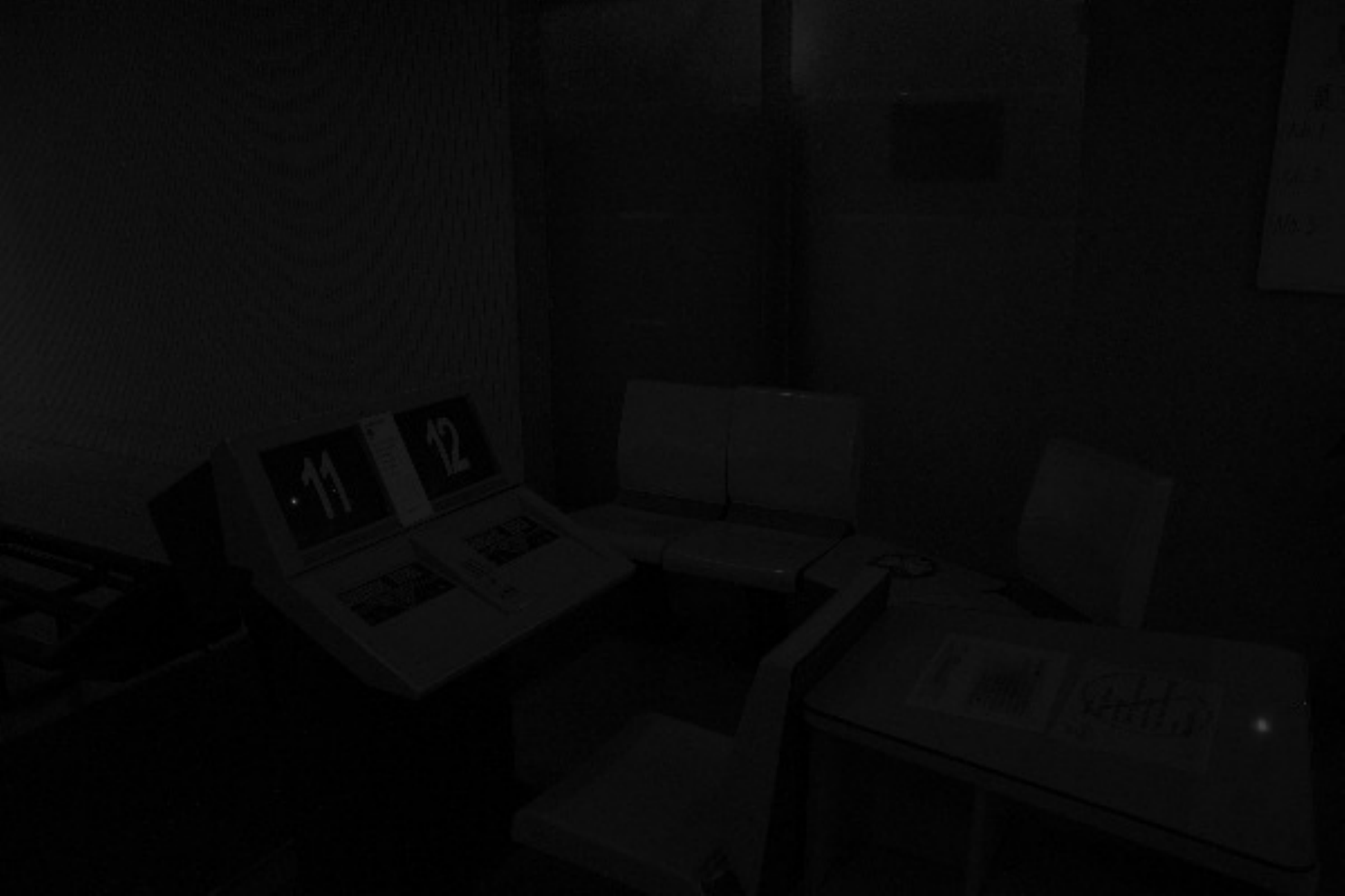}\vspace{2pt} \\
			\includegraphics[width=2.8cm]{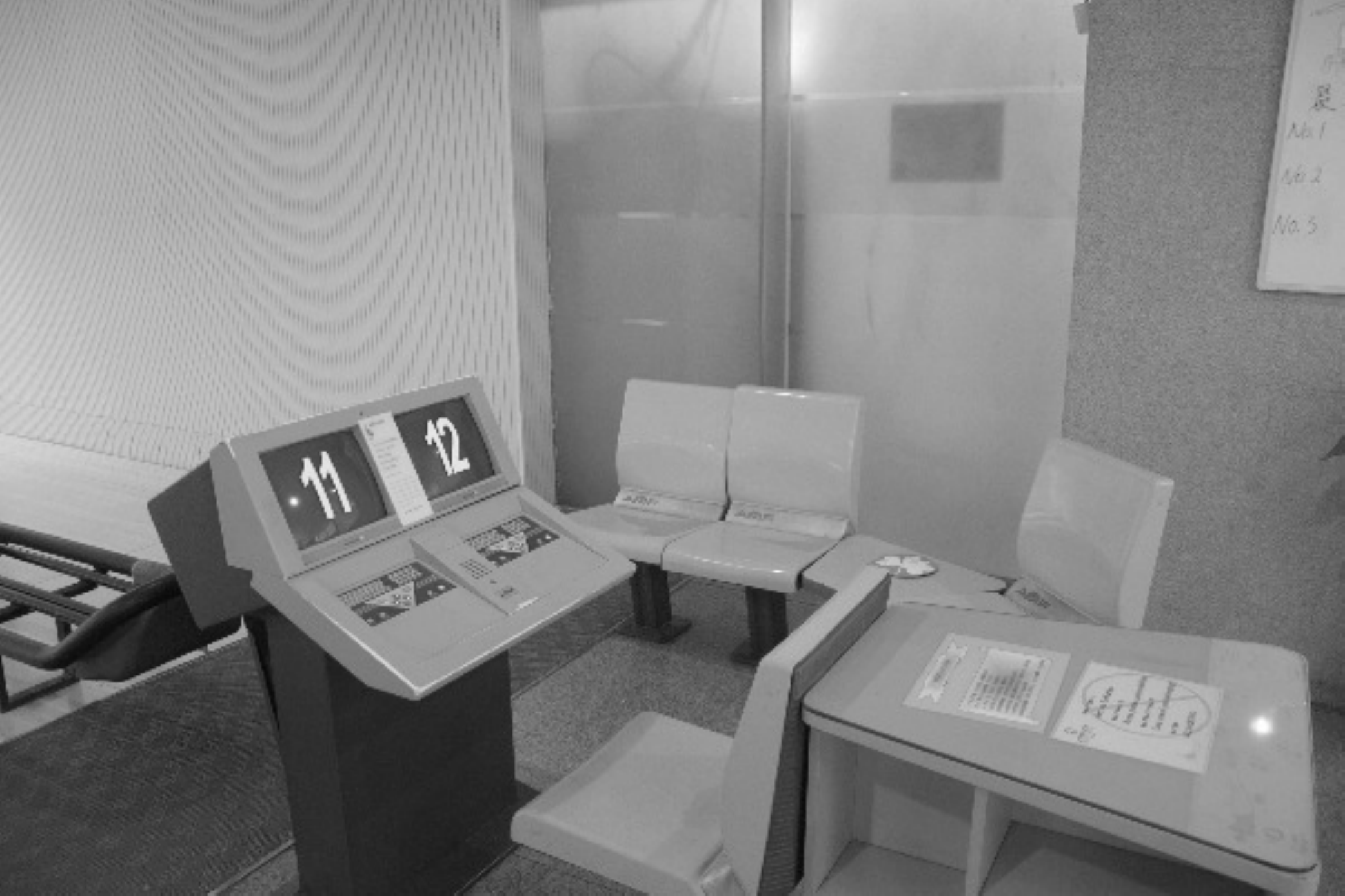}\vspace{0pt}
		\end{minipage}
	}\hspace{-5pt}
	\caption{Visual comparison of H, S and V channels decoupled from low-light image and normal-light Ground-Truth. The first row is the H, S, V channels of the low-light image. The second row is the H, S, V channels of the corresponding normal-light Ground-Truth.}
	\label{665_hsv}
\end{figure}

\begin{figure*}[htbp]
	\centering
	\includegraphics[width=18cm]{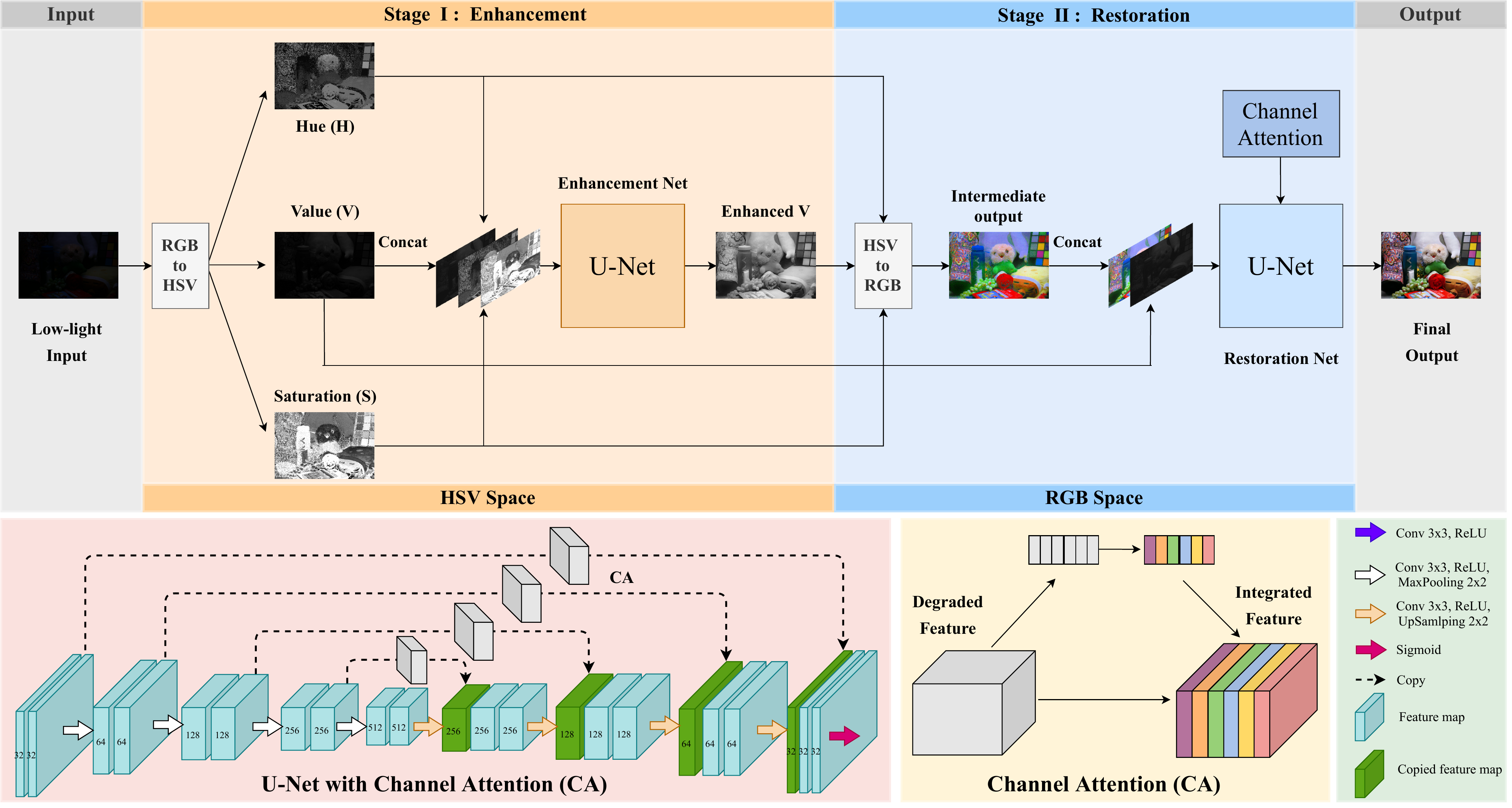}
	\caption{The network architecture of TSN-CA.}
	\label{model}
\end{figure*}

\section{Methodology}
Inspired by\cite{yue2017contrast}, we develop a Two-Stage Network with Channel Attention (TSN-CA) for low-light image enhancement and restoration after enhancement. As shown in Fig.\ref{model}, in stage one, we enhance the brightness of V channel while preserving details with the help of H and S channels in HSV space. In stage two, we combine the enhanced V channel with the noisy and degraded H and S channels to convert the enahcned but degraded image from HSV space back to RGB space, and then introduce channel attention mechanism into U-Net to restore the degraded image from noise and color distortion, especially eliminate shadow blocks and halo artifacts.

\subsection{Stage $\uppercase\expandafter{\romannumeral1}$: Enhancement in HSV Space}
Images captures from real-world low-light conditions suffer from various kinds of degradation, such as severe and complex noise, low contrast and poor visibility.
As shown in Fig.\ref{665_hsv}, when a noise-free image captured in normal-light conditions is decoupled into H, S and V, the V channel is normal in brightness and contains a lot of high-frequency details information without noise. In addition, there are little noise and degradation in H and S channel. However, when a low-light image is decoupled into H, S and V, the V channel suffers from very low brightness and invisible noise and details information are hidden in the darkness. The H and S channels decoupled from low-light image also suffer from severe noise and many other types of degradation.\\
As shown in Fig.\ref{model}, in stage one, we firstly converted low-light images from RGB space to HSV space and then the three channels H (Hue), S (Saturation) and V (Value) are separated out from the HSV image. Following the enhancer of DA-DRN\cite{wei2021dadrn}, we train a deep U-Net to learn the mapping of normal/low-light images. According to DA-DRN\cite{wei2021dadrn}, plain CNN without up-and-down sampling structrue may amplify noise. The V channel differs from an illumination map, according to Retinex Theory\cite{land1977retinex}, illumination map is smooth enough to have no high-frequency noise, however, after the image is converted to HSV and the three channels are separated out, the noise is distributed into H, S and V channels separately. \cite{yue2017contrast} only focuses on enhancing the brightness and contrast of V channel, but ignores the noise and other kinds of degradation in H and S channels. We train a deep U-Net for directly learn the mapping of normal/low-light image pairs of V channel which can enhance the brightness, suppress noise and restore details information of V channel with the help of H and S channels.

\begin{figure*}
	
	
	\subfigure[Low]{
		\begin{minipage}[b]{0.19\textwidth}
			\includegraphics[width=3.5cm]{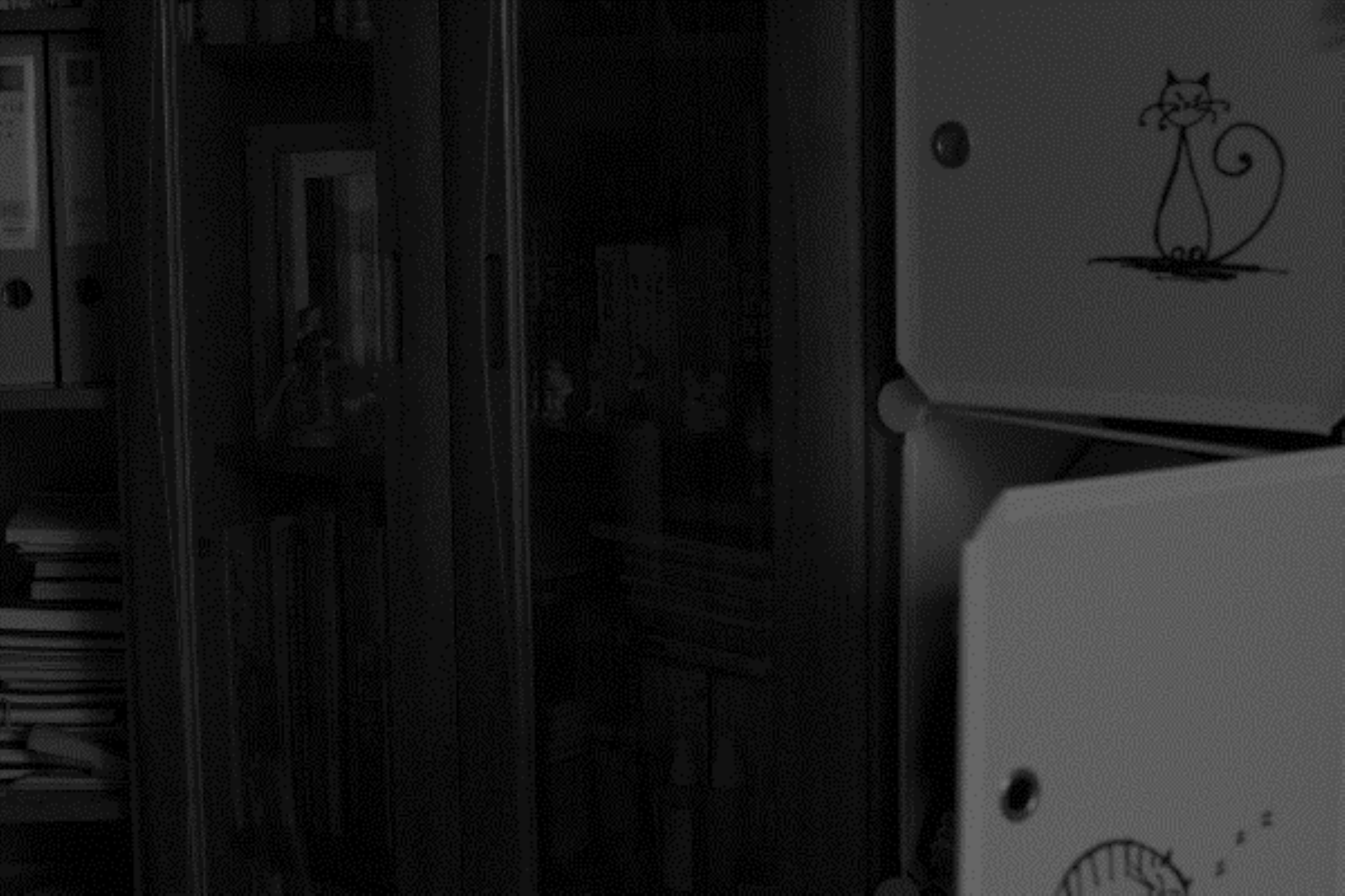}\vspace{-10pt} \\
		\end{minipage}
	}\hspace{-5pt}
	\subfigure[w/o H,S w/o SSIM Loss]{
		\begin{minipage}[b]{0.19\textwidth}
			\includegraphics[width=3.5cm]{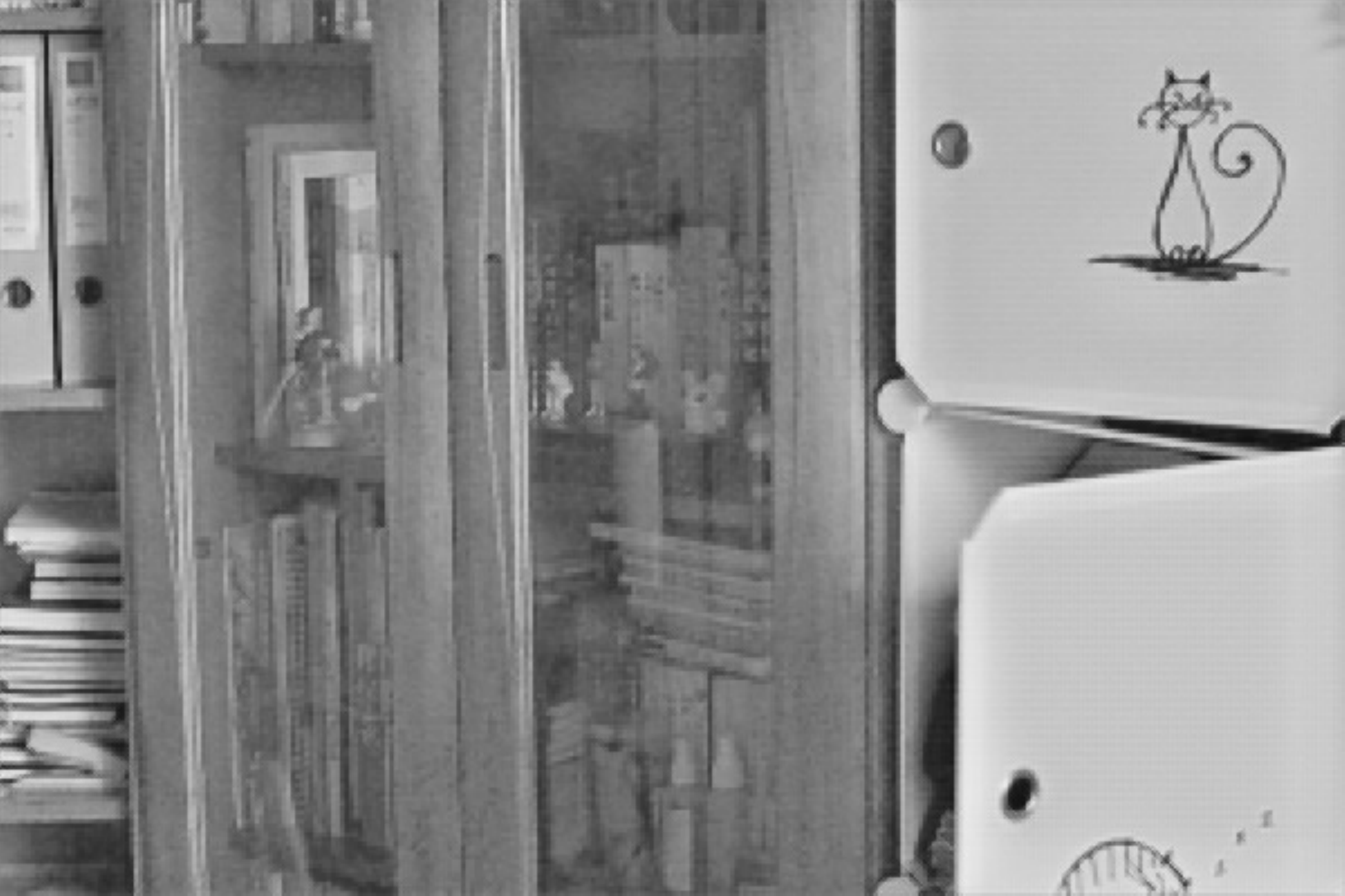}\vspace{-10pt} \\
		\end{minipage}
	}\hspace{-5pt}
	\subfigure[with H,S w/o SSIM Loss]{
		\begin{minipage}[b]{0.19\textwidth}
			\includegraphics[width=3.5cm]{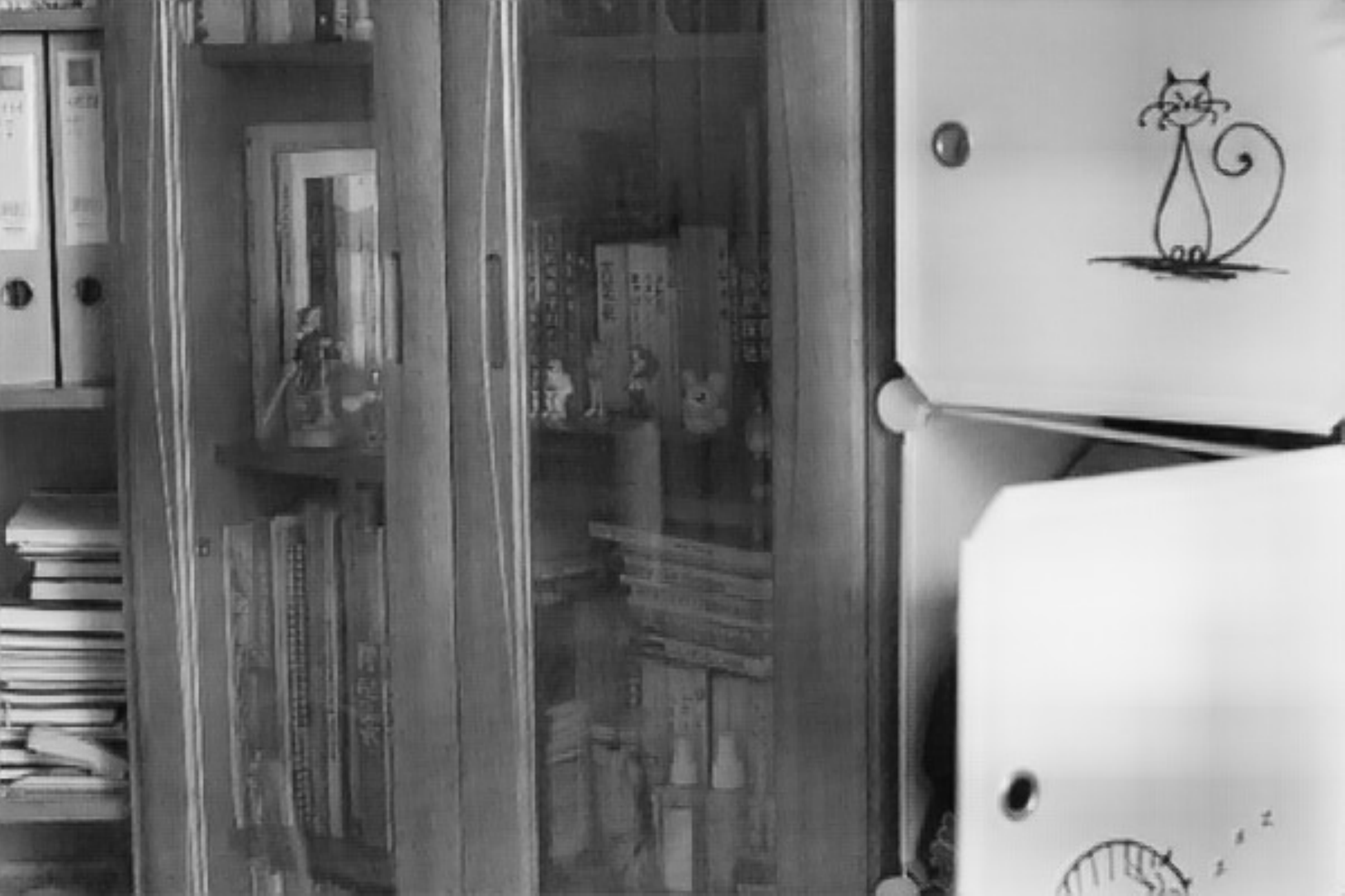}\vspace{-10pt} \\
		\end{minipage}
	}\hspace{-5pt}
	\subfigure[with H,S with SSIM Loss]{
		\begin{minipage}[b]{0.19\textwidth}
			\includegraphics[width=3.5cm]{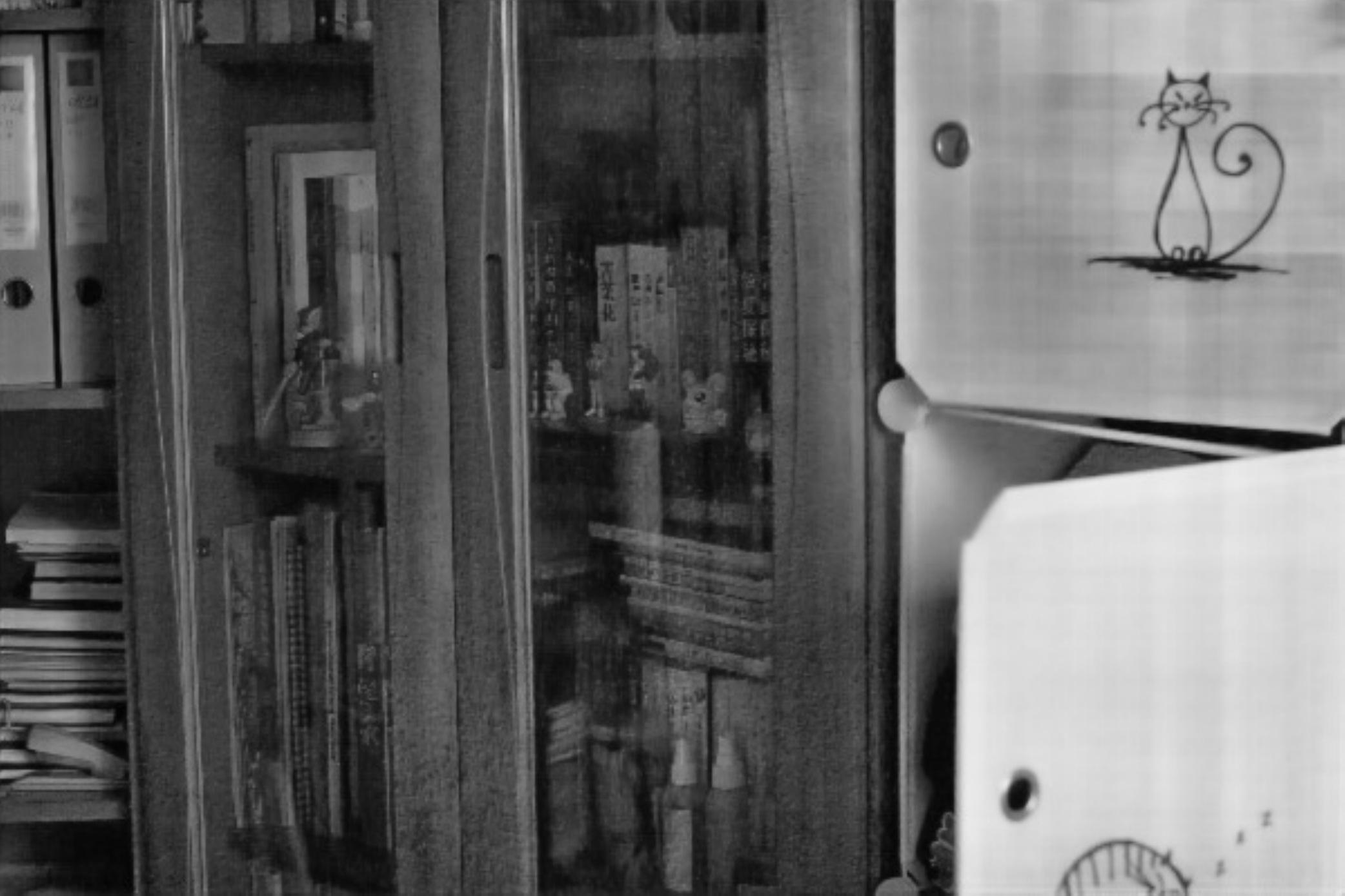}\vspace{-10pt} \\
		\end{minipage}
	}\hspace{-5pt}
	\subfigure[Ground-Truth]{
		\begin{minipage}[b]{0.19\textwidth}
			\includegraphics[width=3.5cm]{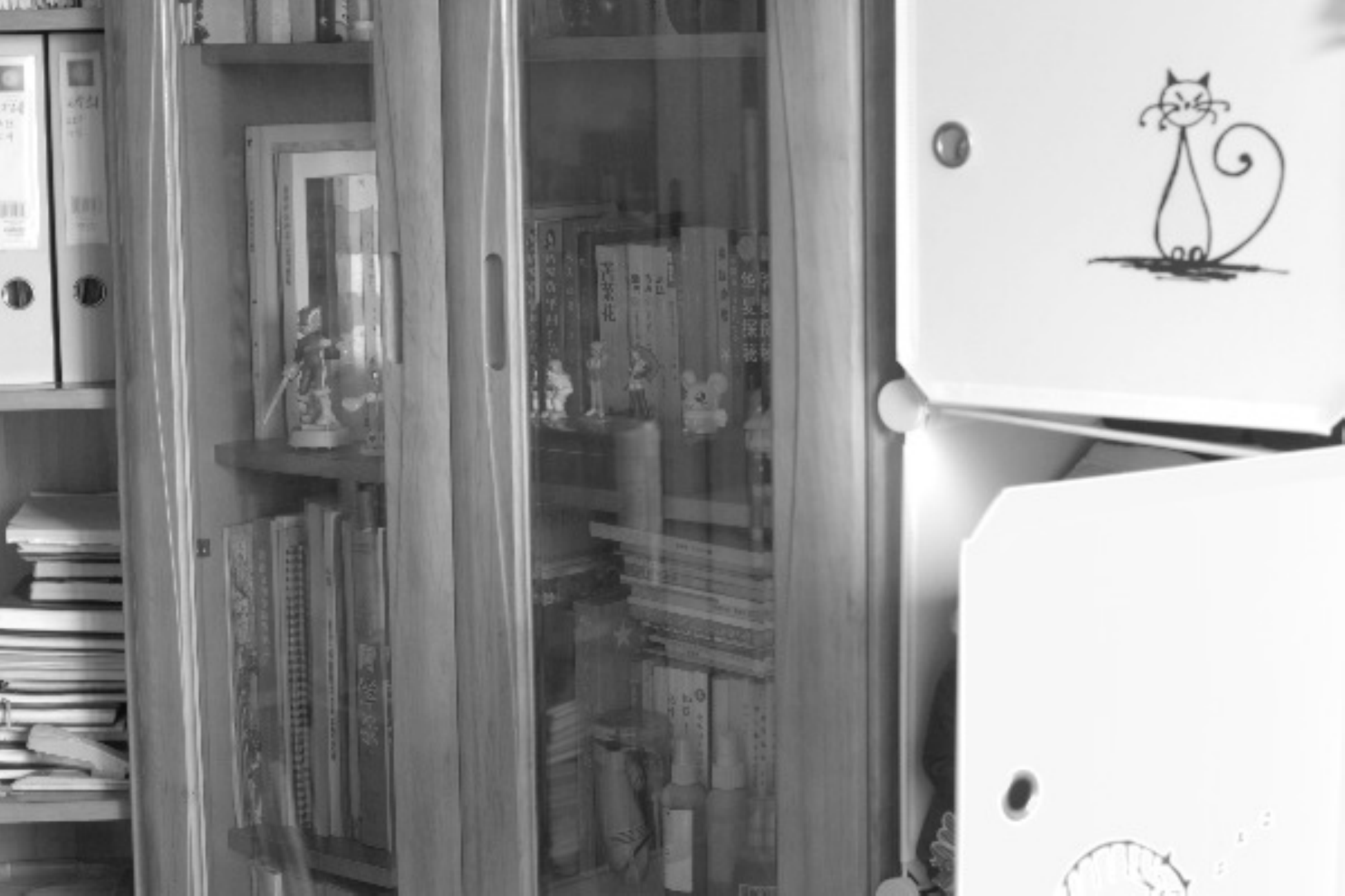}\vspace{-10pt} \\
		\end{minipage}
	}
	\caption{Visual comparison of the output of stage one and ablation study of different loss functions in stage one.}
	\label{abla_v}
\end{figure*}

As shown in Fig.\ref{abla_v}, we notice that if the image of V channel is directly used as the input to train the enhancement network, the noise in the brightness-enhanced output results is removed, but the high-frequency details are also lost, resulting in blurred edges in the enhanced V channel. SSIM loss can effectively help network to recover and reconstruct high-frequency details, but as shown in Fig.\ref{abla_v}, it also allows noise to be retained, and also introduces significant large shadow blocks and halo artifacts. So instead of using SSIM loss, we fuse the V, H and S channels together to recover the details information. Although H and S decoupled from low-light contain a lot of noise, and they differ greatly from their counterparts decoupled from normal-light Ground-Truth, they contain relatively complete details and texture information, which is benefit to the restoration of details in V channel. Therefore, we concatenate H and S together with the V channel as a three-channel input tensor of the enhancement network in stage one, the output is the corresponding single-channel denoised V with enhanced brightness and restored details.

The total enhancement loss $L_{total}^{\uppercase\expandafter{\romannumeral1}}$ in stage one is as follow:
\begin{align}
	L_{total}^{\uppercase\expandafter{\romannumeral1}}&=\left \| V_{output} - V_{high} \right \|_{1} + \left \| \triangledown(V_{output})-\triangledown(V_{high})  \right \|_{1} 
	\notag
	\\&+ \frac{1}{CHW}\left \| F (V_{output})-F (V_{high}) \right \|_{2}^{2}
\end{align}
where $V_{output}$ and $V_{high}$ denotes the V channel generated by our enhancement network and decoupled from normal-light Ground-Truth. $\triangledown$ denotes the gradients in the horizontal and vertical directions. $F$ is the 31st feature map obtained by the VGG16 \cite{simonyan2014very} network pre-trained on ImageNet database. C,H,W represents the number of channels, the height and the width of the input image, respectively.


\subsection{Stage $\uppercase\expandafter{\romannumeral2}$: Restoration in RGB Space}
\begin{figure}
	\subfigure[Input]{
		\begin{minipage}[b]{0.48\linewidth}
			\includegraphics[width=4.25cm]{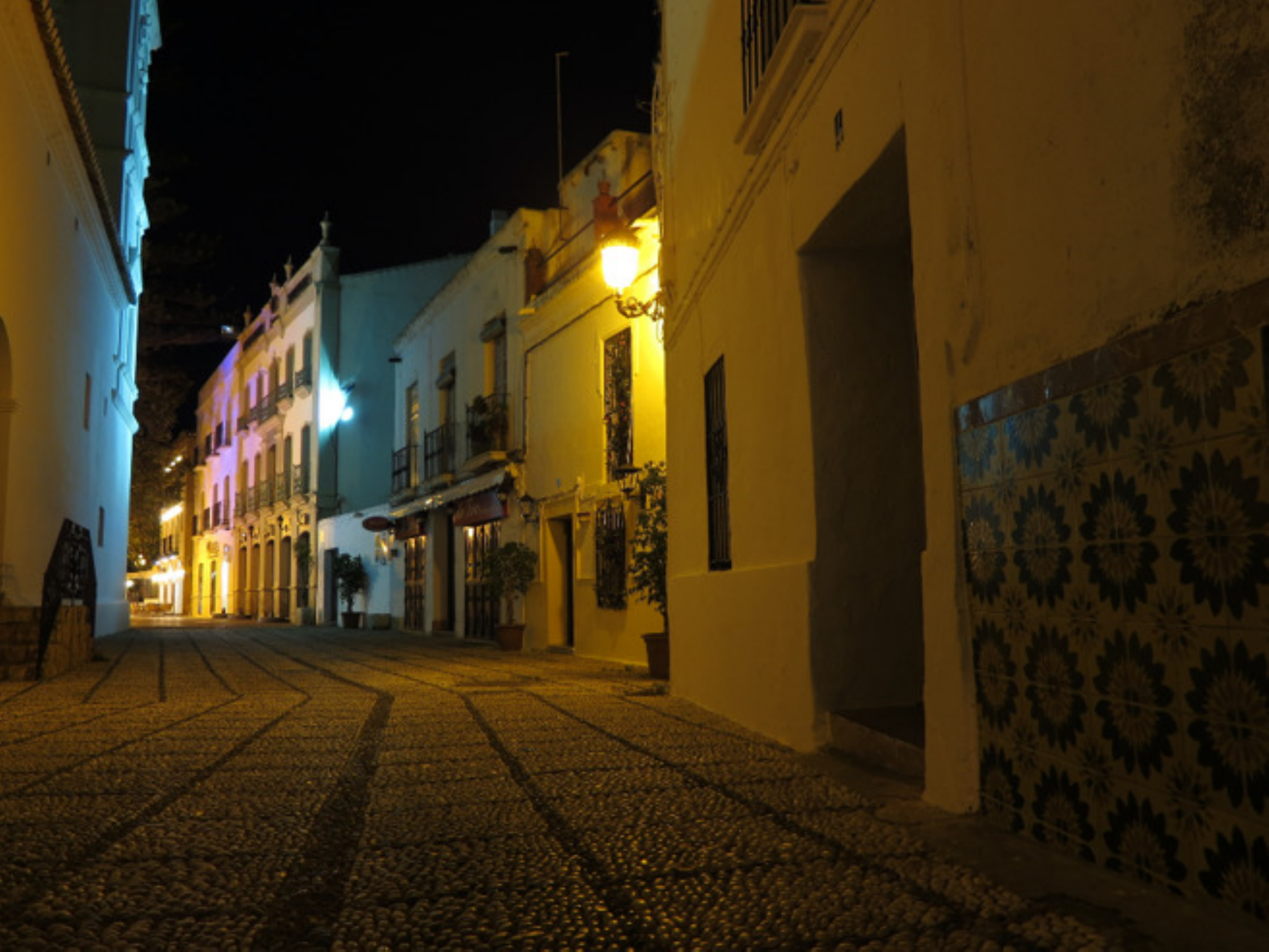}\vspace{-3pt}
	\end{minipage}}
	\vspace{-6pt}
	\subfigure[Input of stage two]{
		\begin{minipage}[b]{0.48\linewidth}
			\includegraphics[width=4.25cm]{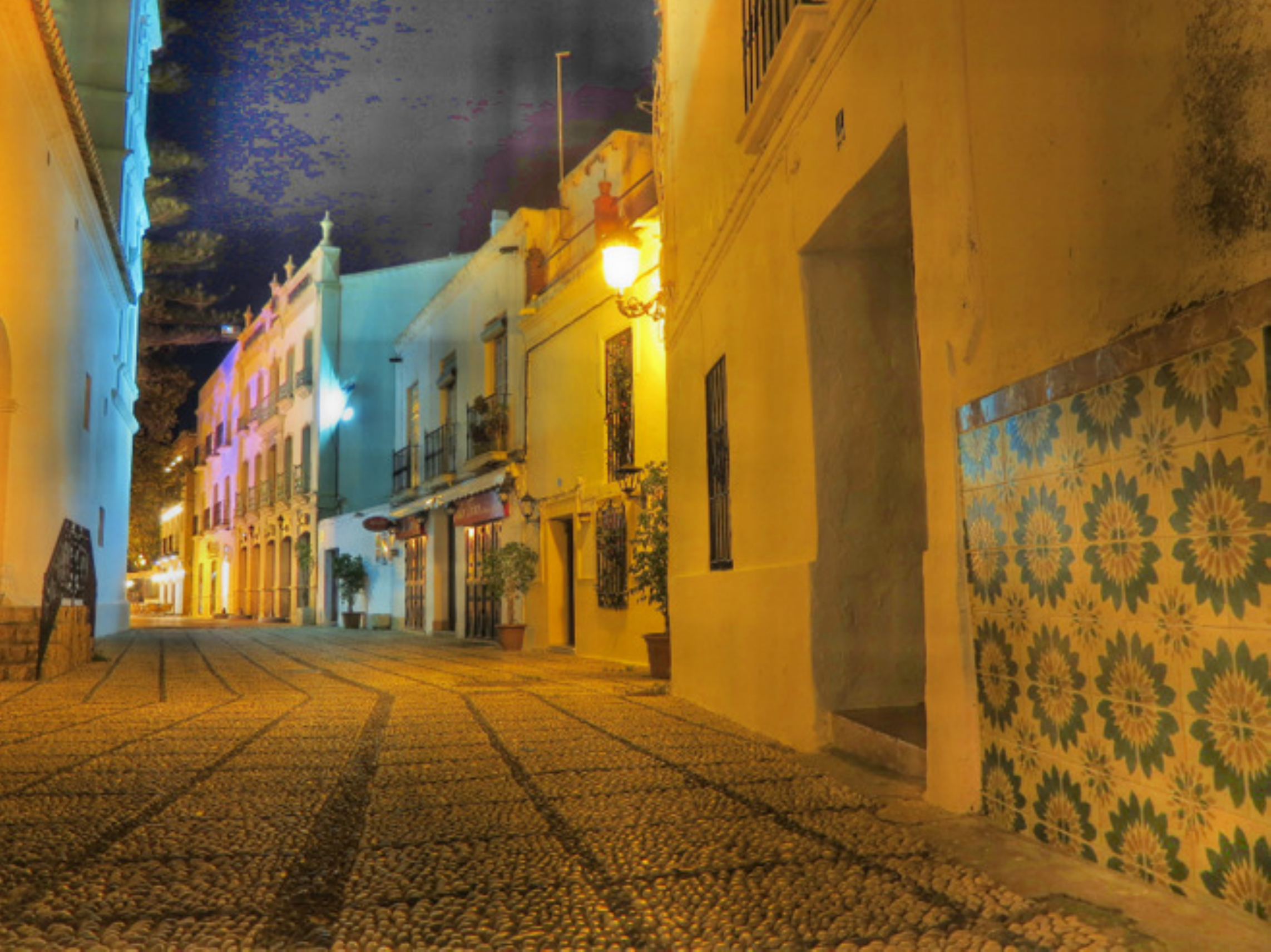}\vspace{-3pt}
	\end{minipage}}
	\vspace{-6pt}
	\subfigure[Output of stage two w/o CA]{
		\begin{minipage}[b]{0.48\linewidth}
			\includegraphics[width=4.25cm]{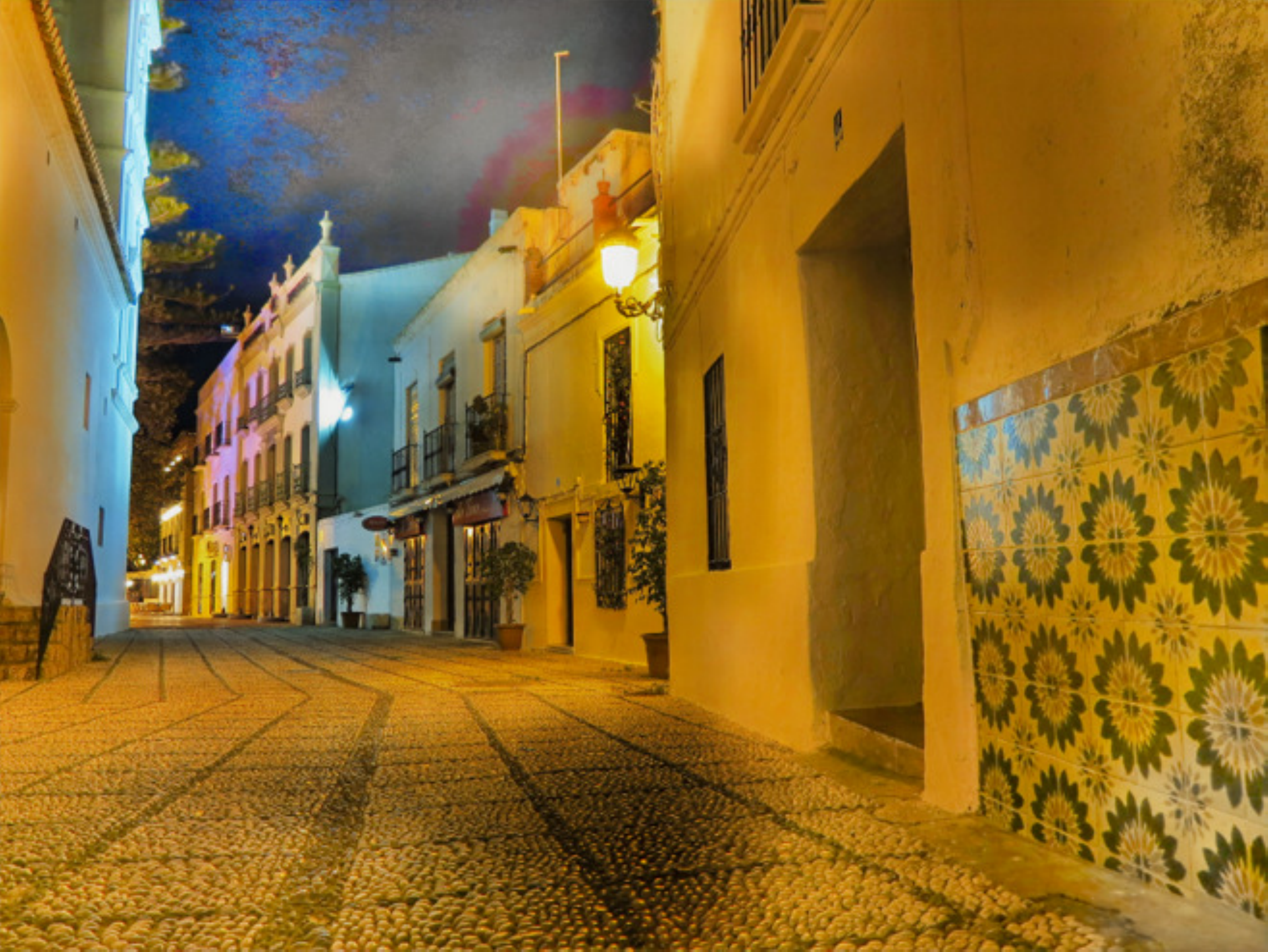}\vspace{-3pt}
	\end{minipage}}
	\vspace{-4pt}
	\subfigure[Output of stage two with CA]{
		\begin{minipage}[b]{0.48\linewidth}
			\includegraphics[width=4.25cm]{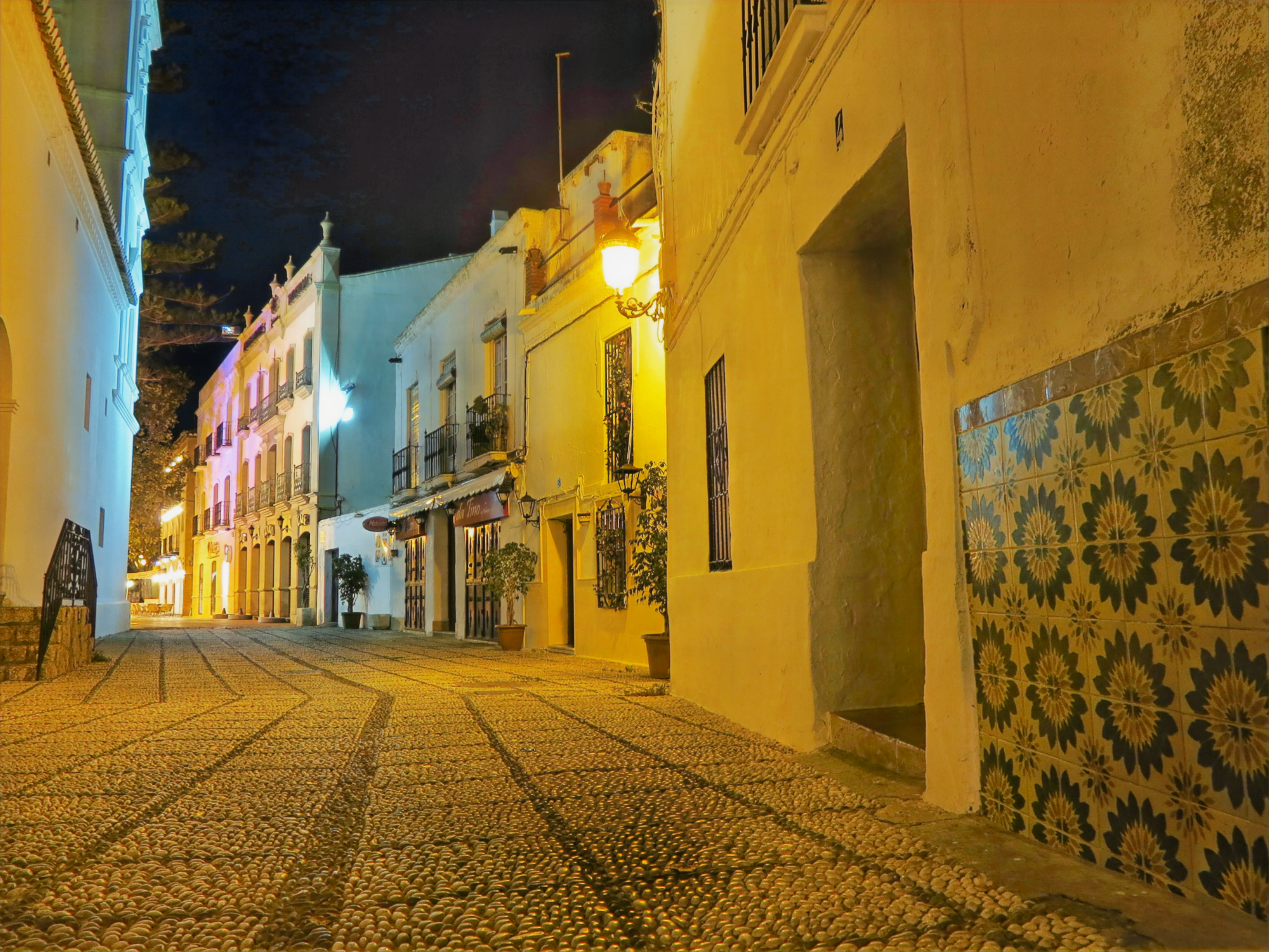}\vspace{-3pt}
	\end{minipage}}
	\vspace{-4pt}
	\caption{Ablation study of the output in different stages and the function of Channel Attention Module.}
	\label{abla_st}
\end{figure}

\begin{figure*}
	\centering
	
	
	\subfigure[Input]{
		\begin{minipage}[b]{0.155\textwidth}
			\includegraphics[width=2.8cm]{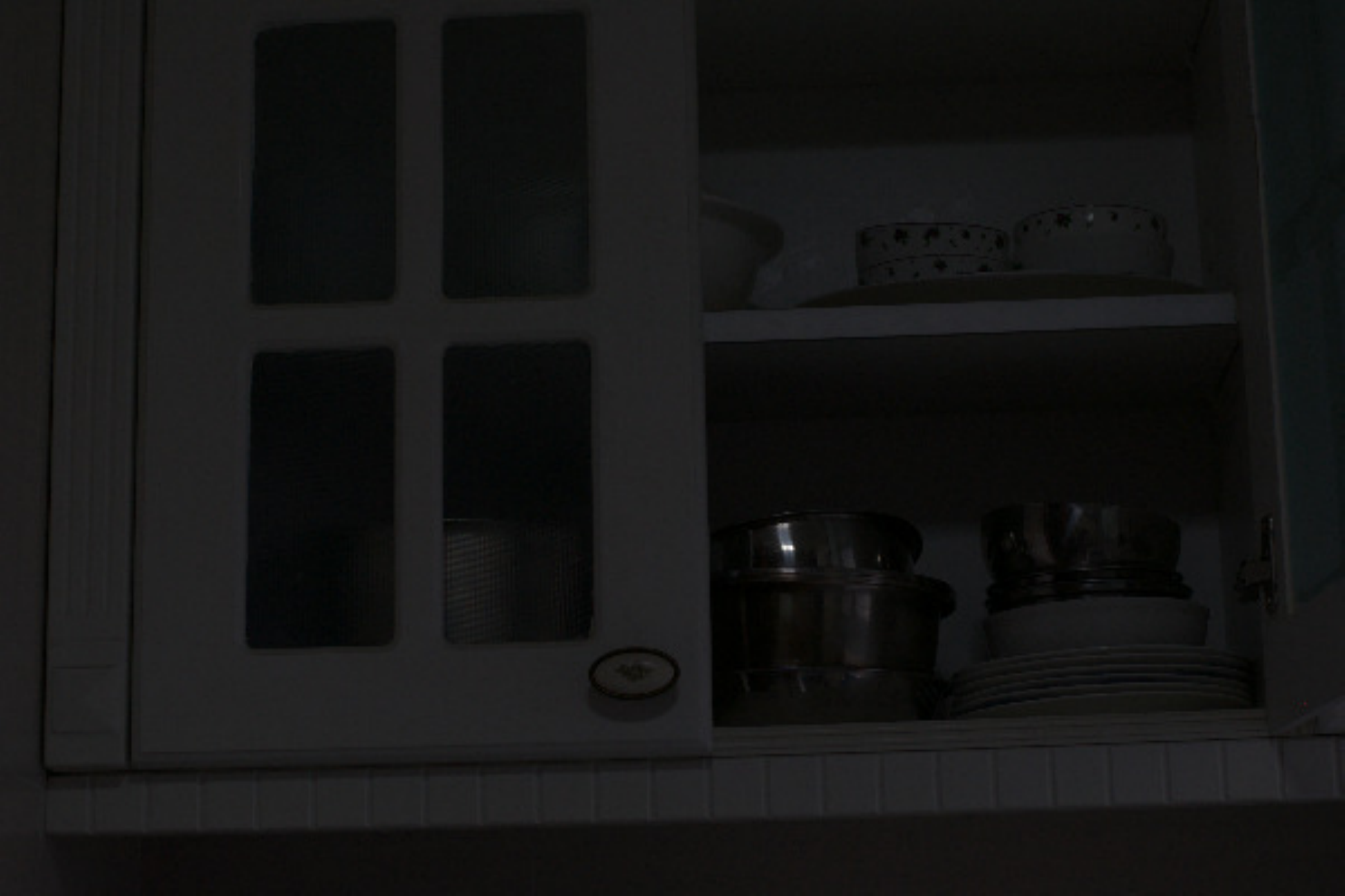}\vspace{3pt} \\
			\includegraphics[width=2.8cm]{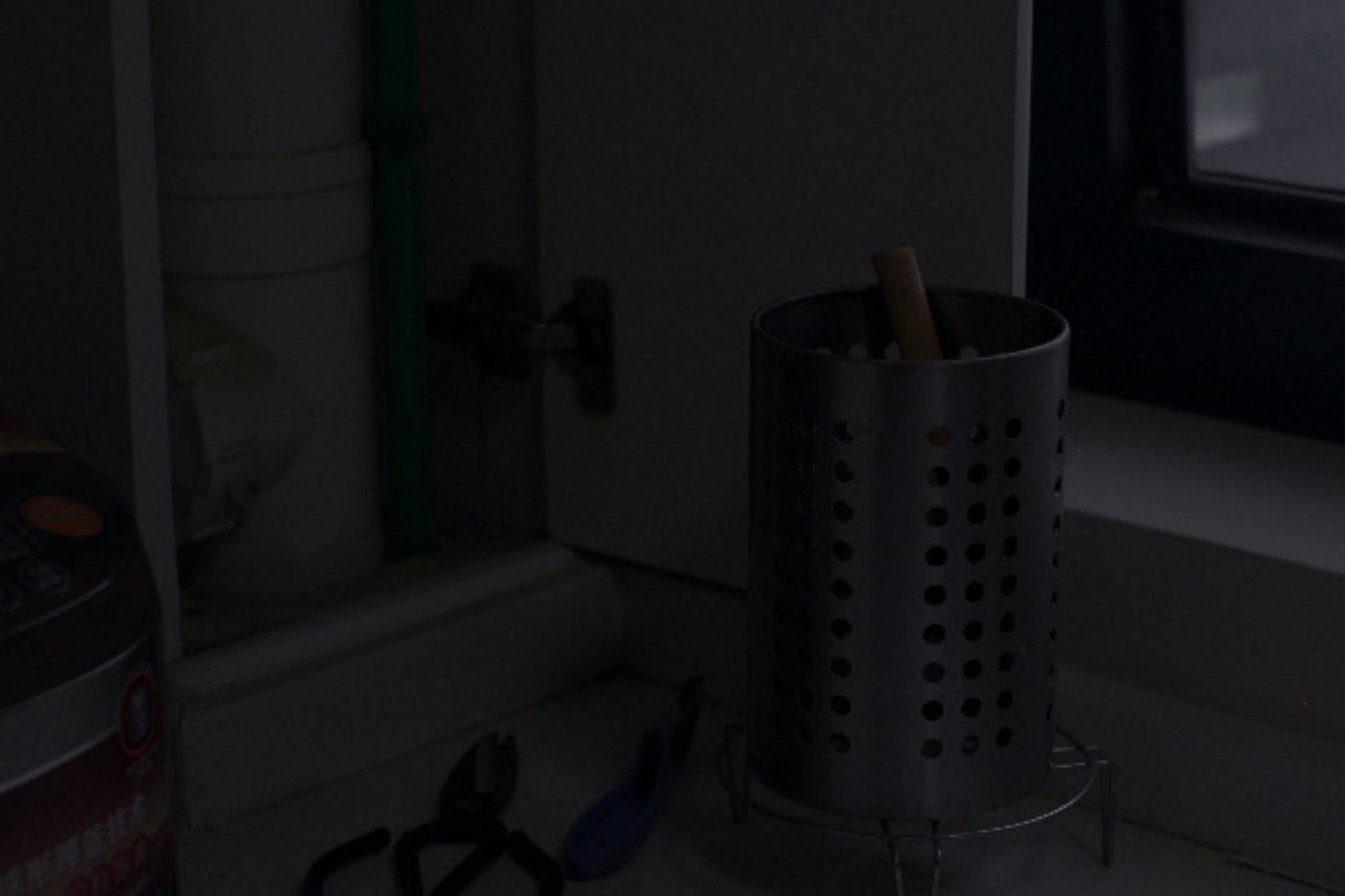}\vspace{3pt}
		\end{minipage}
	}\hspace{-5pt}
	\subfigure[Enhanced V Channel]{
		\begin{minipage}[b]{0.155\textwidth}
			\includegraphics[width=2.8cm]{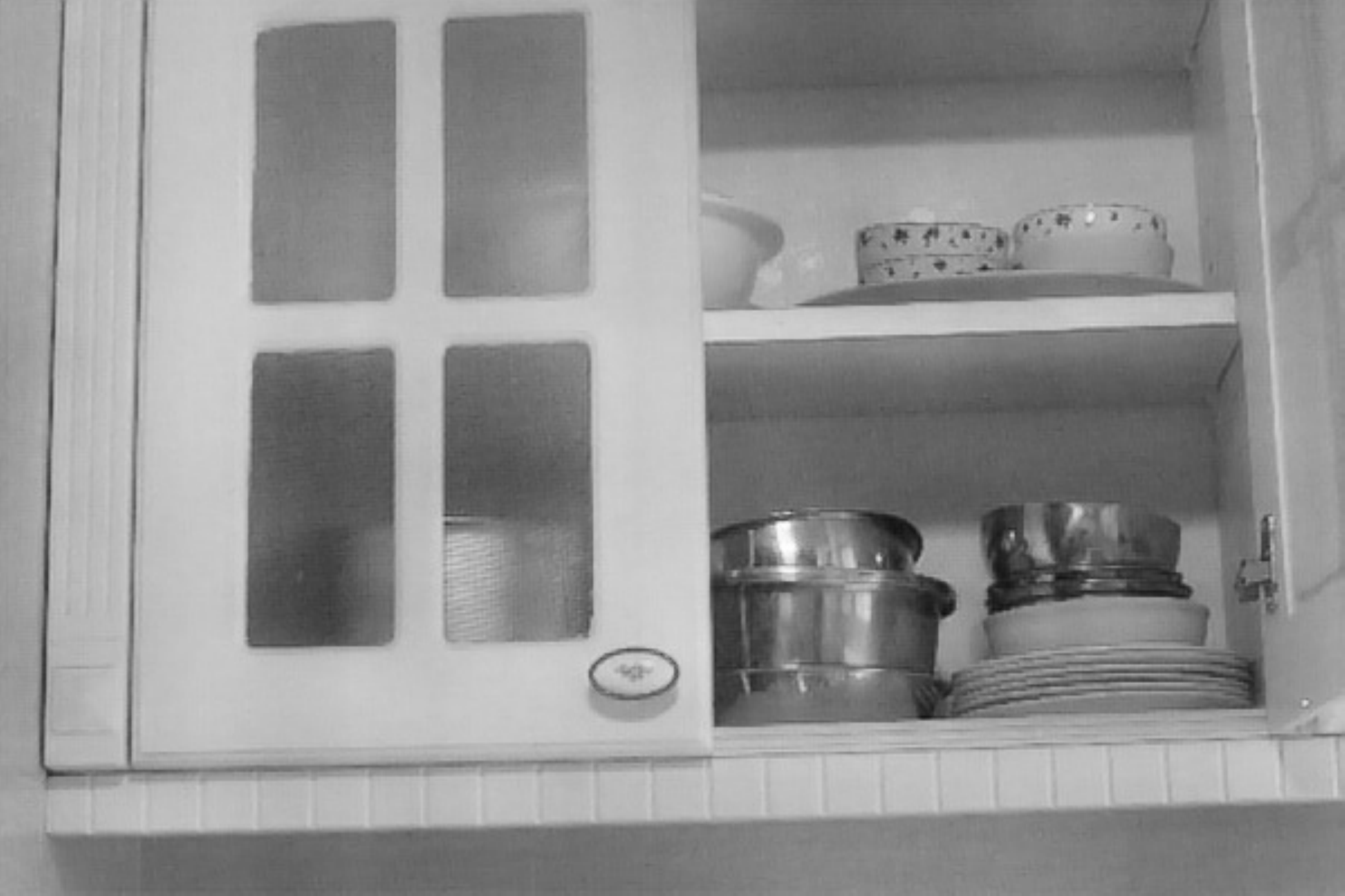}\vspace{3pt} \\
			\includegraphics[width=2.8cm]{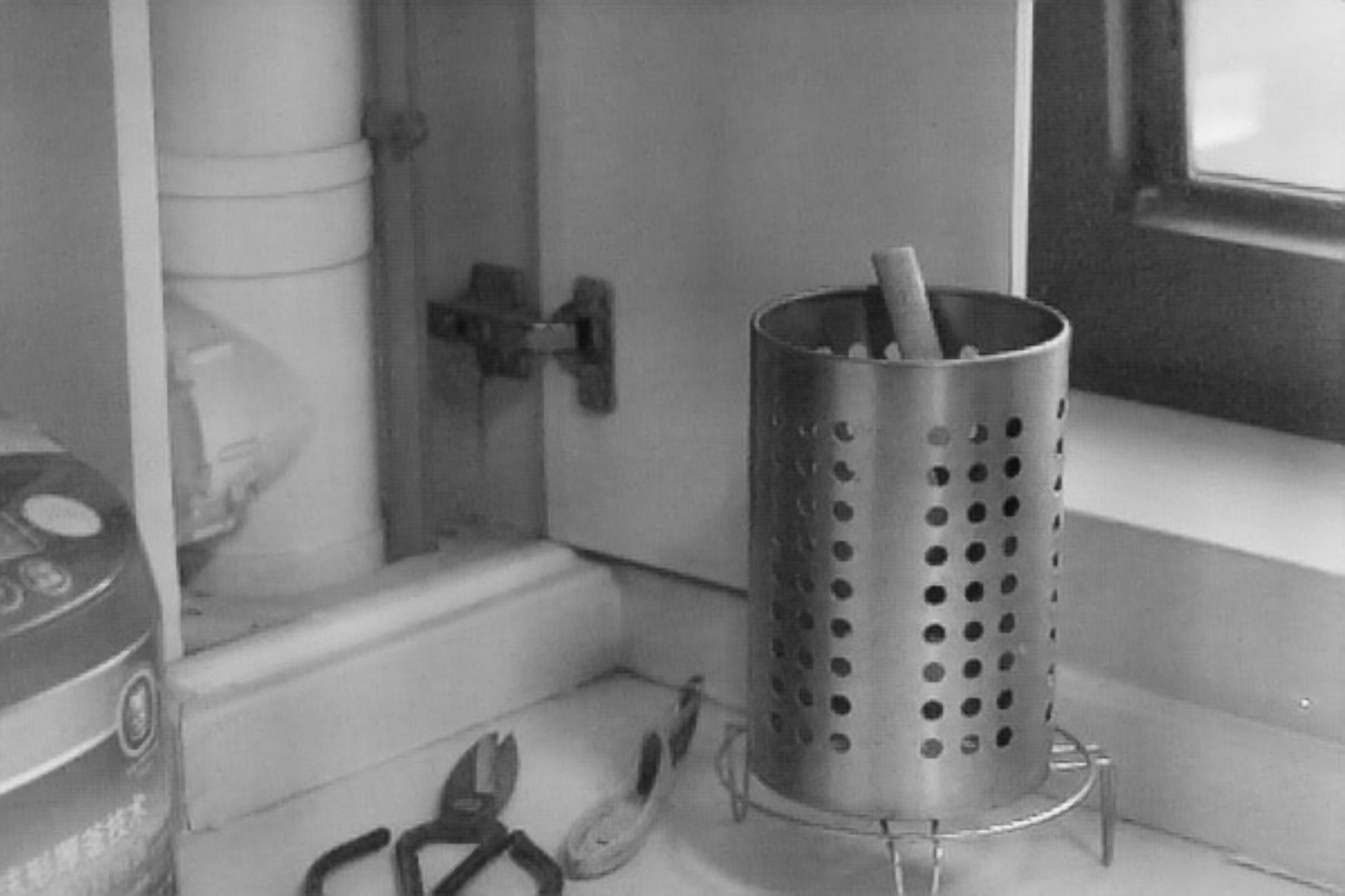}\vspace{3pt}
		\end{minipage}
	}\hspace{-5pt}
	\subfigure[Input of stage two]{
		\begin{minipage}[b]{0.155\textwidth}
			\includegraphics[width=2.8cm]{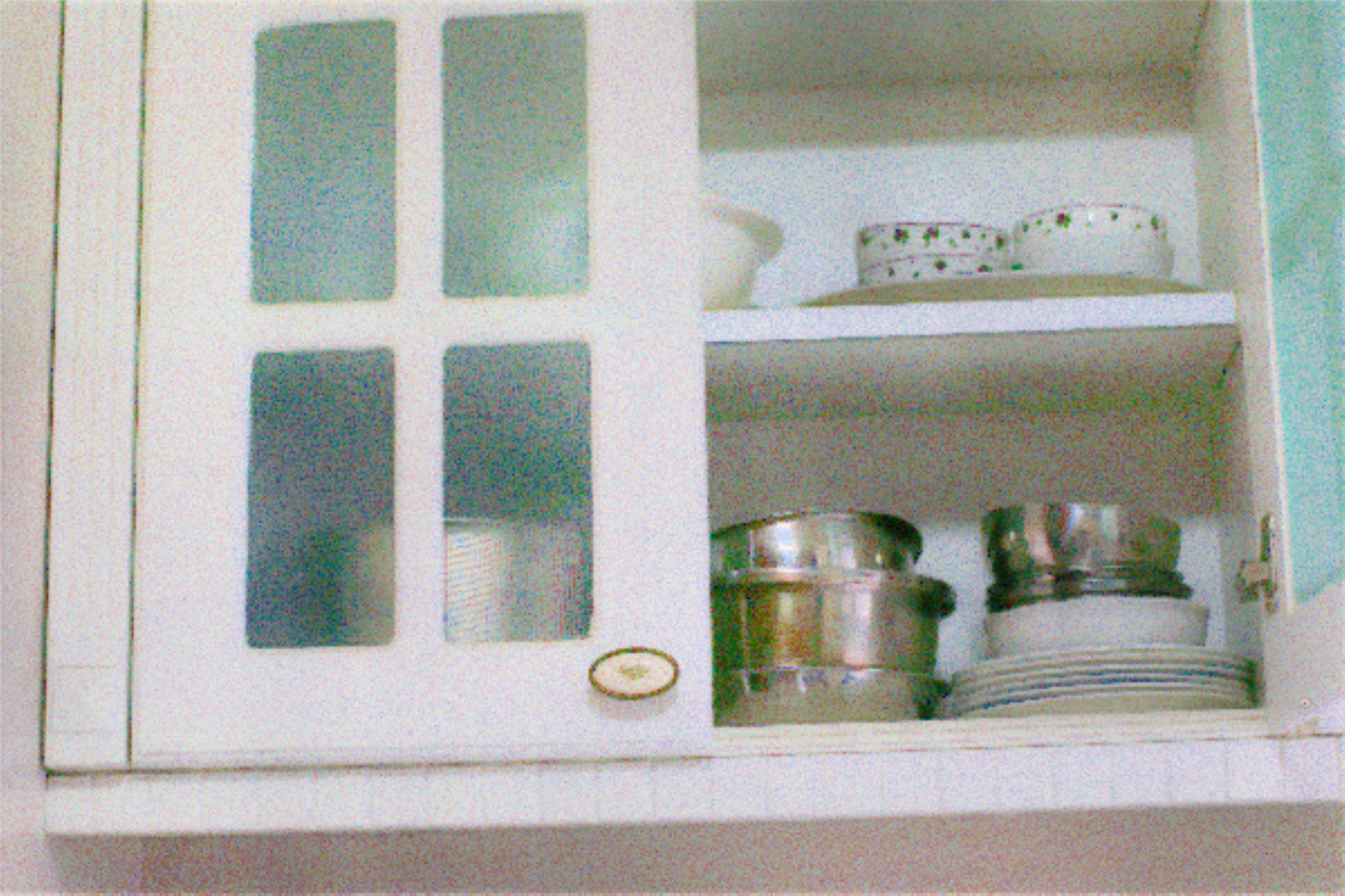}\vspace{3pt} \\
			\includegraphics[width=2.8cm]{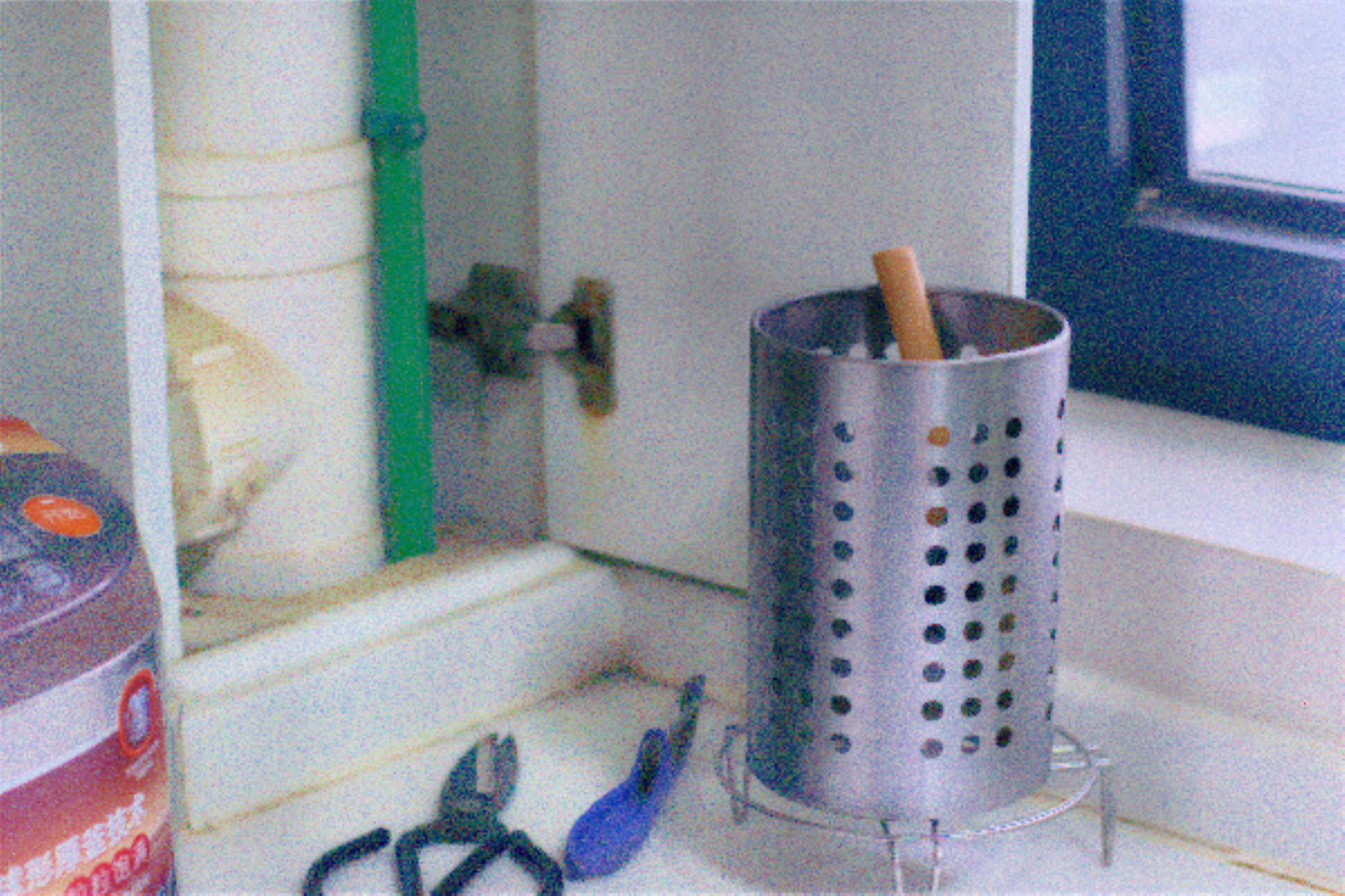}\vspace{3pt}
		\end{minipage}
	}\hspace{-5pt}
	\subfigure[Final output w/o CA]{
		\begin{minipage}[b]{0.155\textwidth}
			\includegraphics[width=2.8cm]{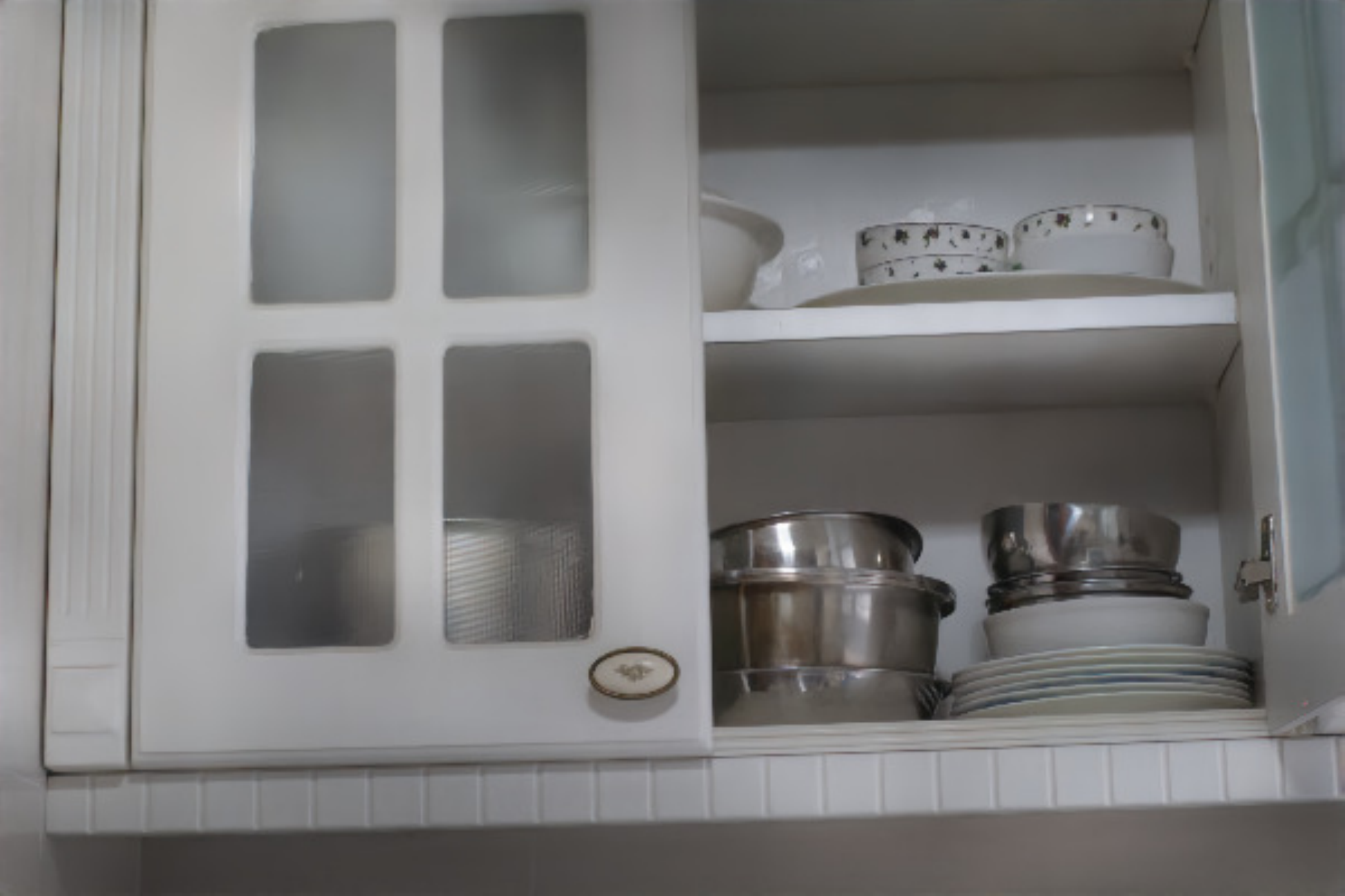}\vspace{3pt} \\
			\includegraphics[width=2.8cm]{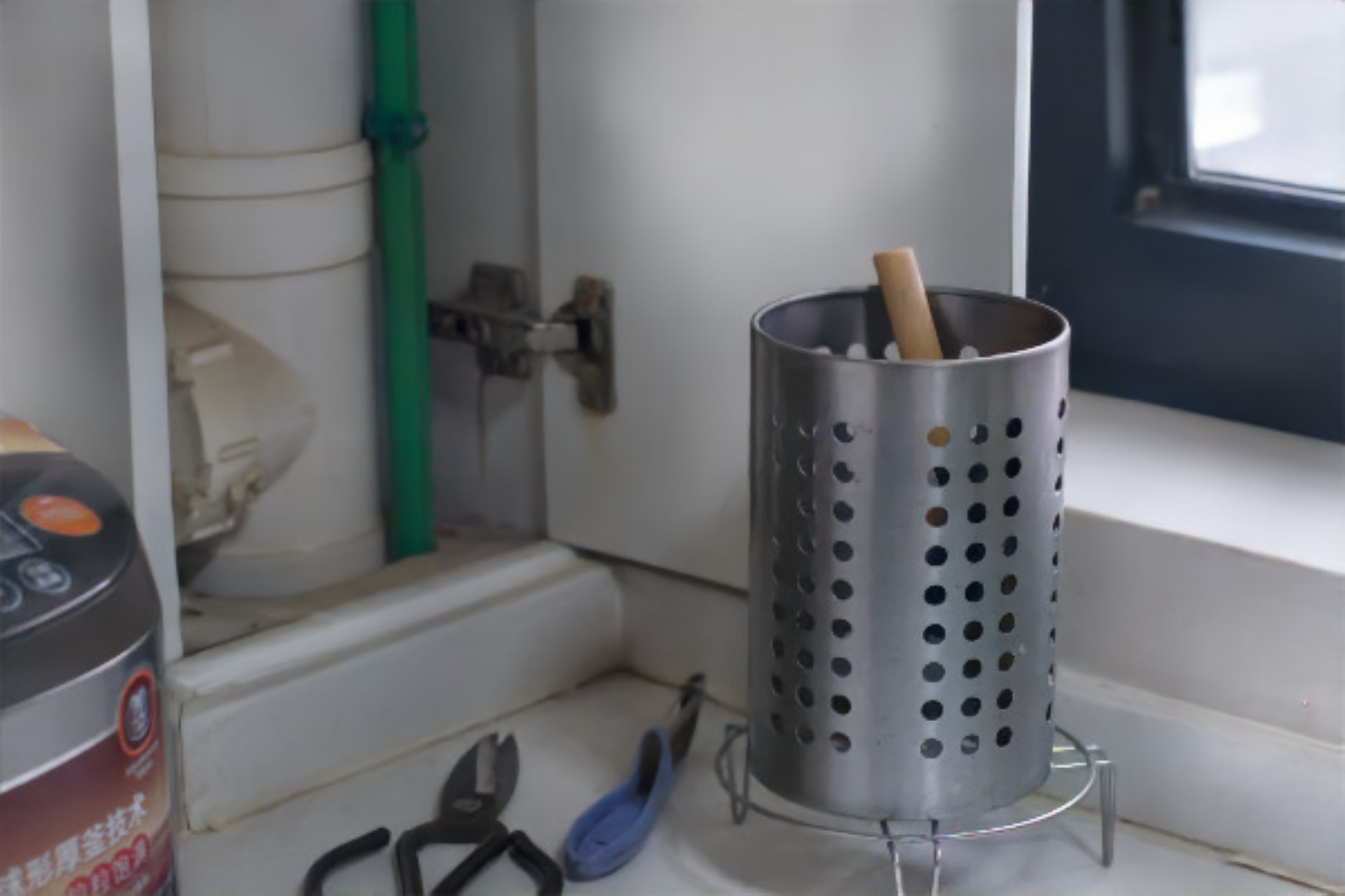}\vspace{3pt}
		\end{minipage}
	}\hspace{-5pt}
	\subfigure[Final output with CA]{
		\begin{minipage}[b]{0.155\textwidth}
			\includegraphics[width=2.8cm]{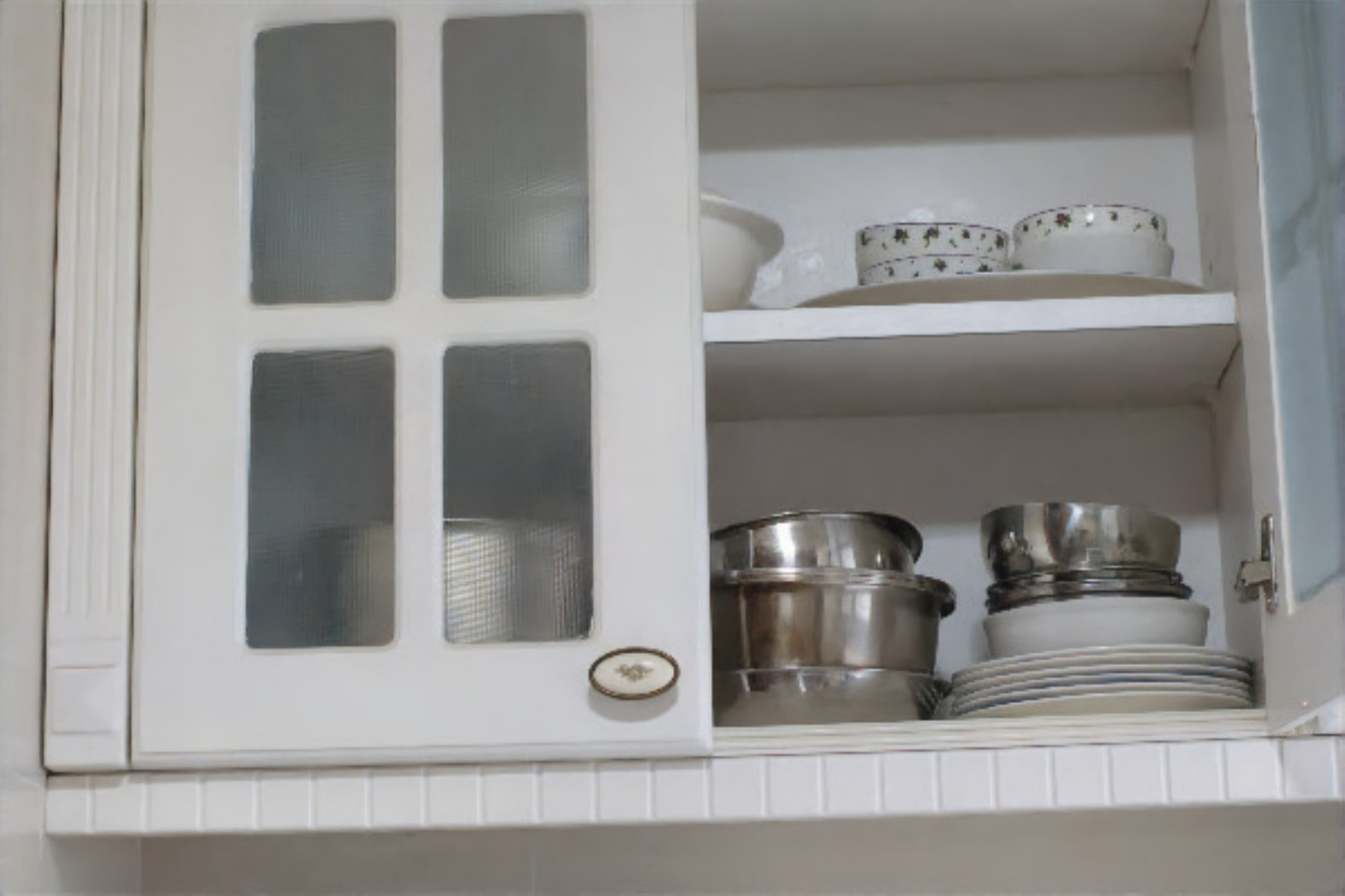}\vspace{3pt} \\
			\includegraphics[width=2.8cm]{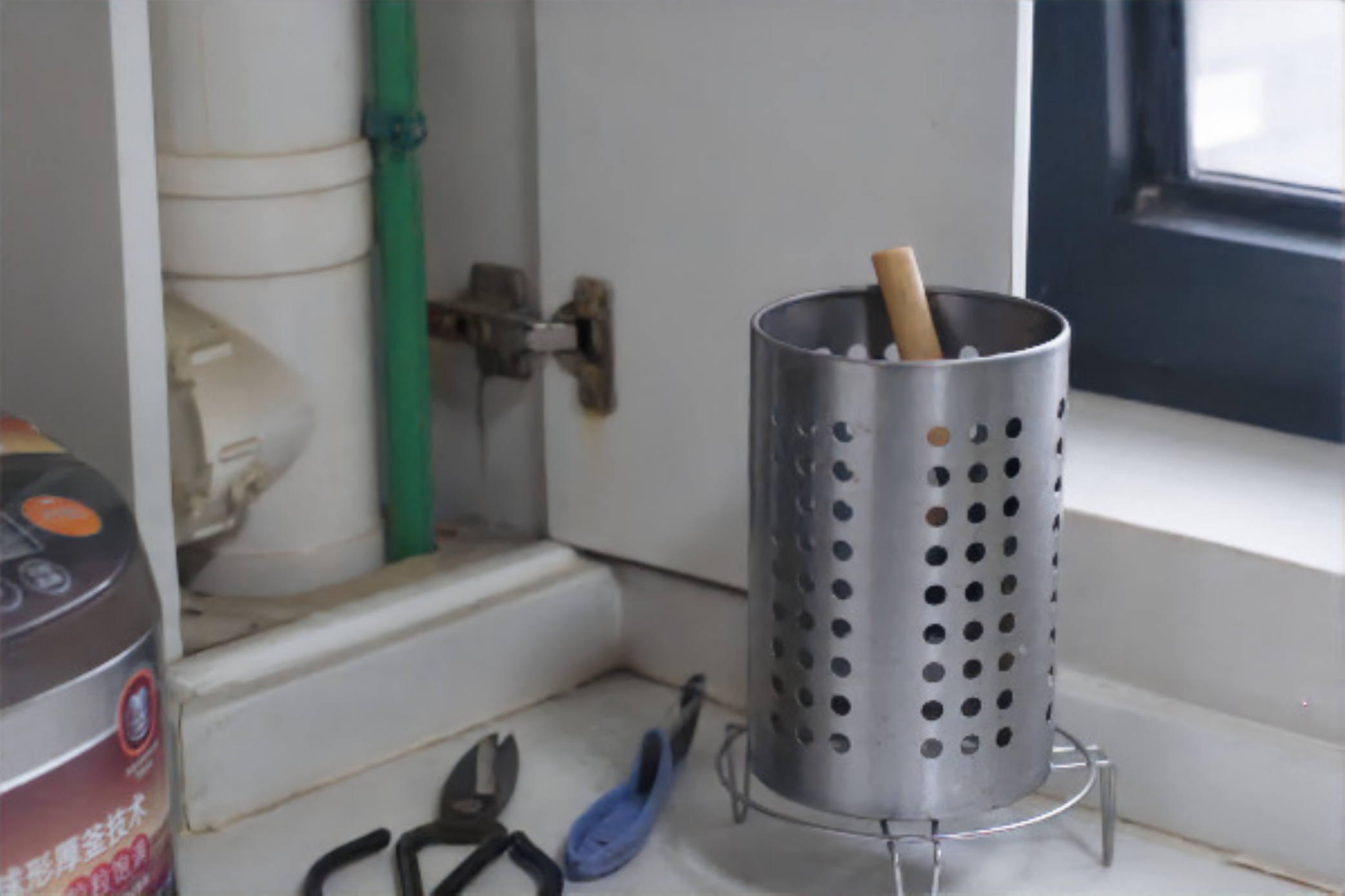}\vspace{3pt}
		\end{minipage}
	}\hspace{-5pt}
	\subfigure[Ground-Truth]{
		\begin{minipage}[b]{0.155\textwidth}
			\includegraphics[width=2.8cm]{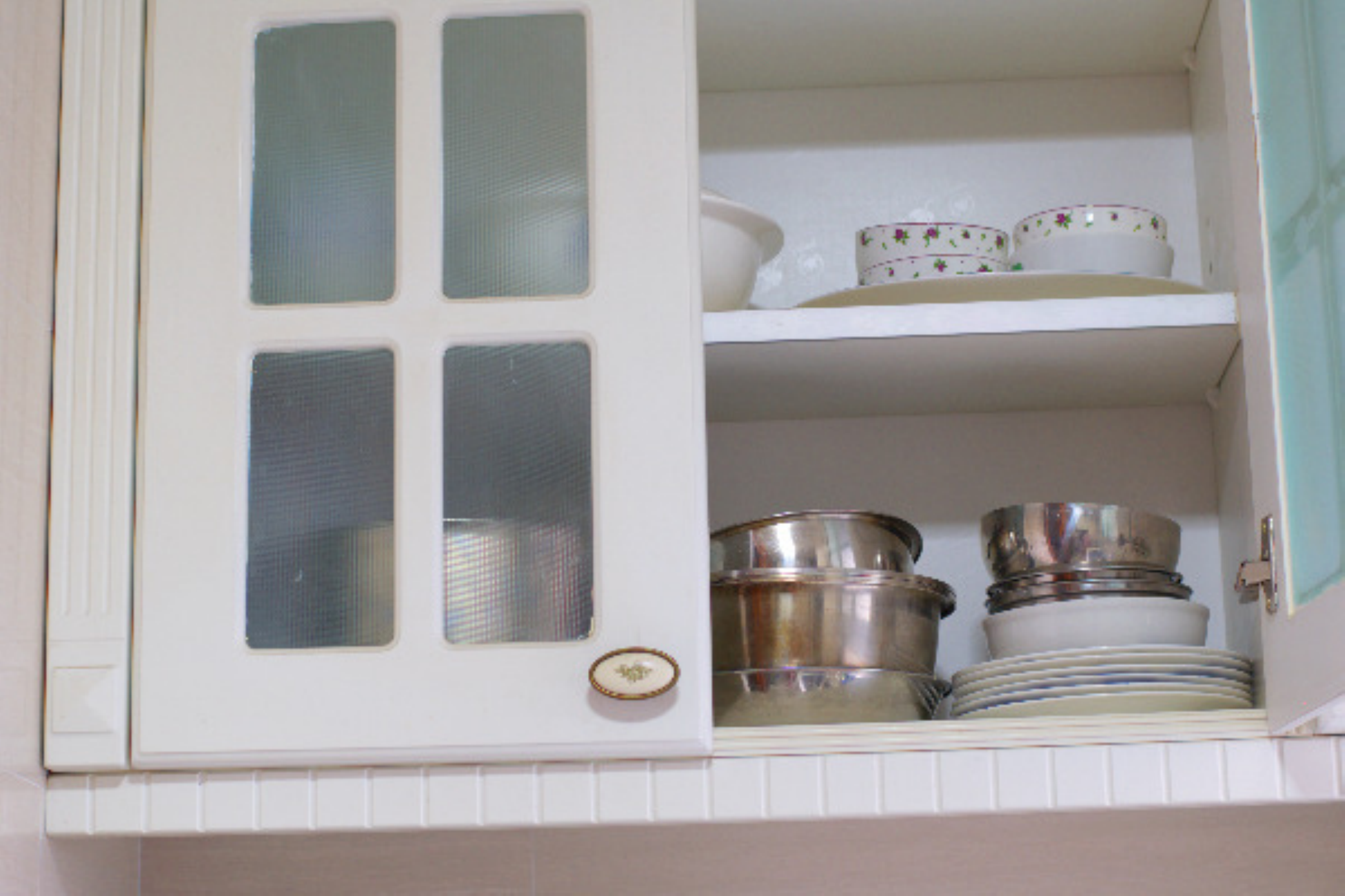}\vspace{3pt} \\
			\includegraphics[width=2.8cm]{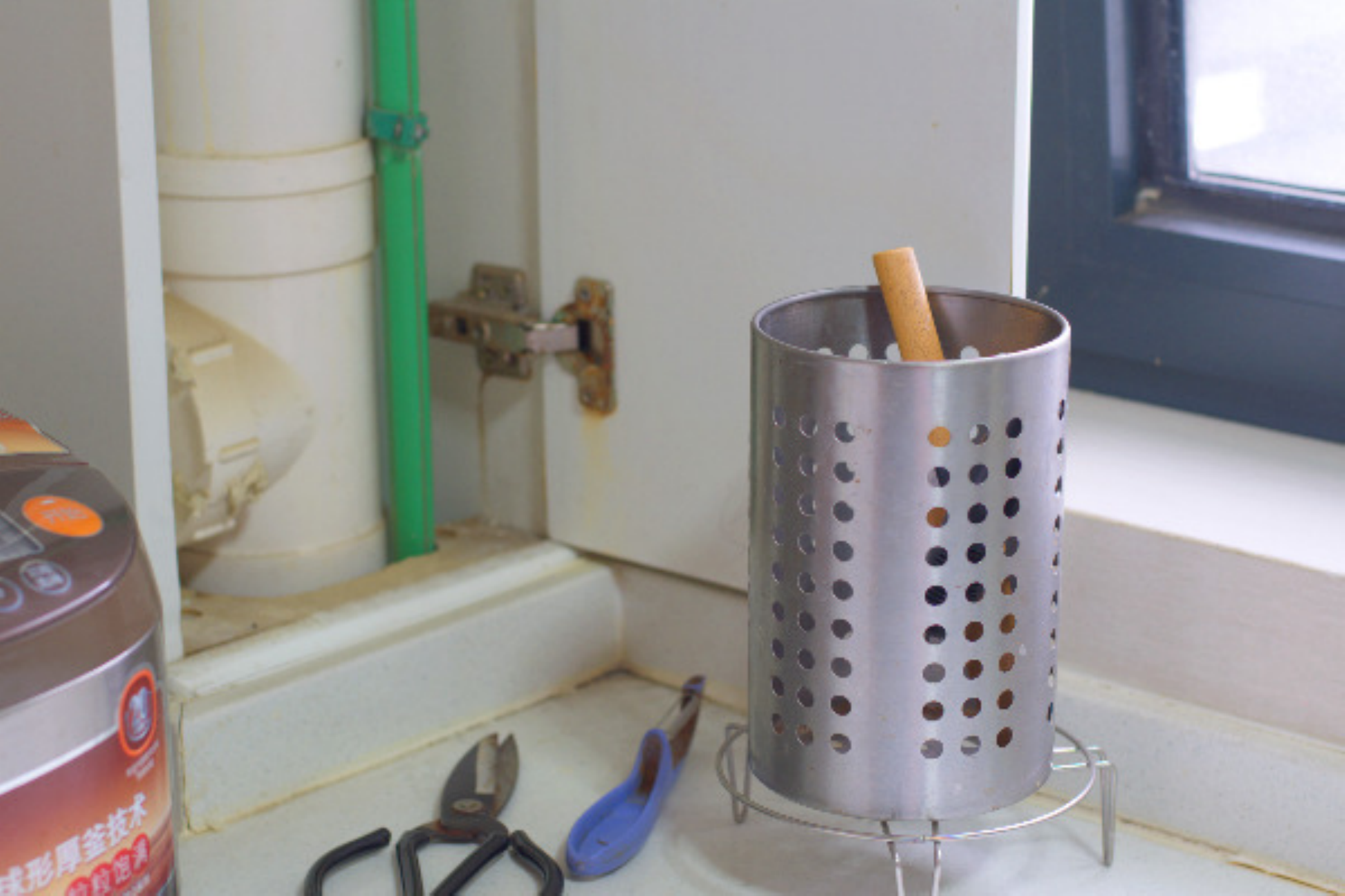}\vspace{3pt}
		\end{minipage}
	}
	\caption{Visual comparison of the output of different stages and ablation study of channel attention (CA).}
	\label{abla_se_st}
\end{figure*}

As shown in Fig.\ref{model}, after stage one, we combine the enhanced V channel with the original degraded H and S channels then convert them back to the RGB space. In this way, we can obtain the intermediate enhanced results. However, although the brightness is enhanced, there is serious noise in the brightness-enhanced images, and the detail information is drowned in the severe noise and other kinds of degradation, which makes the image edges blurred. \\
In addition, as shown in Fig.\ref{abla_se_st}, we notice that color over-saturation emerges in the enhanced image because the V channel in the image is only responsible for the brightness, and the color is determined by H and S channels. The H and S channels that make up these intermediate brightness-enhanced images are decoupled directly from the input low-light images, the degree of degradation in these H and S channels is high. 
What's worse, due to the under-fitting phenomenon after directly estimating the V channel, when we converted the three channels from HSV space back to RGB space, unpleasant shadow blocks and halo artifacts appear. So it is essential for us to restore these highly degraded images from varoius kinds of degradation.\\
As shown in Fig.\ref{model}, inspired by the \cite{zhang2019kindling}, we adopt U-Net for the restoration of enhanced but degraded images. The total restoration loss $L_{total}^{\uppercase\expandafter{\romannumeral2}}$ in stage two is as follow:
\begin{align}
L_{total}^{\uppercase\expandafter{\romannumeral2}}&=\left \| I_{low} - I_{high}\right \|_{2}^{2}-SSIM(I_{low}, I_{high})
\notag
\\&+\left \| \triangledown(I_{output})-\triangledown(I_{high})  \right \|_{2}^{2}
\end{align}
where $I_{low}$ and $I_{high}$ represent the output of stage one which is normal-light but degraded and the normal-light Ground-Truth, respectively. SSIM means the SSIM Loss. $\triangledown$ denotes the gradients in the horizontal and vertical directions.\\
Although this can achieve good denoising and restoration effects, we found that the PSNR and SSIM indexes of the restored images are still not very high, which means that there is still some noise left in the restored results, and there is still unreasonable in the structure and details. To make matters worse, as shown in Fig.\ref{abla_st}, there are obvious shadow blocks and halo artifacts in the restored results after stage two. Because before restoration, in the results obtained through stage one, details, structure and other useful information are covered by severe noise, and all useful features are hidden under the degraded useless features. It is because of these severe kinds of degradation that the training process of network learning and restoration of useful features such as details, structure and corrected color information, becomes difficult, resulting in the degradation still exists in the generated restoration results after stage two. In addition, there are some shadow blocks and halo artifacts in the enhanced results after stage one because the estimation of V channel is not so accurate. As shown in Fig.\ref{abla_se_st}, instead of being eliminated, these shadow blocks and halo artifacts are amplified after stage two. All of these problems resulted in the final results of the restoration being not very good in terms of quantitative metrics and visual effects. To solve these problems, we introduce channel attention mechanism (SE Module) into the skip connection of U-Net restoration network in stage two.

\subsection{Channel Attention for Image Restoration}
Inspired by SENet\cite{hu2018squeeze} in image recognition, we introduce channel attention mechanism into image enhancement and restoration. We embedded the channel attention module (SE Module) into the skip connection of U-Net to better remove noise, restore details and especially eliminate shadow blocks and halo artifacts.\\
We found that the reason why residual noise, shadow blocks and halo artifacts appeared in the generated restoration results is that the skip connections of U-Net directly pass the severely degraded features to the upsampling stage by conatenating the upsampled features with the previous downsampled features which still contain severe degradation. This resulted in the preservation of degraded features. \\
As shown in Fig.\ref{model}, we embedded the channel attention module (SE Module) in the skip connection of U-Net, by integrating the downsampled image features in the channel dimension before pass them to the unsampling stage. The channel attention mechanism is similar to the ventral 'What' pathway in the human brain. Because human attention resources are limited, the human brain and visual system usually pay more attention to useful things that interest them, and ignore unimportant things that can not interest them. Channel attention mechanism can simulate this selective attention mechanism of human, it can integrate image features in channel dimensions by assigning more weights to useful features, such as correct color, detail and texture features, allows the network to learn these useful features better, while assigning less weights to the less important features, such as the feature of noise, distorted color, shadow blocks and halo artifacts, or even assigning no weights at all. \\
In this way, the network can selectively learn which features in the degraded image are useful and which features are belong to useless degradation, and then selectively suppress the useless features and focus more attention and computing resources on the learning of useful features.\\
By passing the integrated features to the upsampling stage of U-Net and then fusing them with the upsampled features, shadow blcoks and halo artifacts can be effectively eliminated, because the degraded features such as the feature of shadows and halos have been selectively ignored after the feature integration.

\section{Experiments}
\subsection{Implementation details}
We train the model on the LOL \cite{wei2018deep} real-world and synthetic training datasets individually and evaluate it on the LOL real-world and synthetic validation datasets. In addition, we test our model on four popular test datasets: LIME \cite{guo2016lime}, DICM \cite{lee2013contrast}, MEF\cite{ma2015perceptual} and NPE\cite{wang2013naturalness} datasets. We use the PyTorch to train our model on an Nvidia TITAN XP GPU.
We use the Adam\cite{kingma2014adam} optimizer for the training and set the training batch-size to four and the patch-size of random crop to 384x384. The number of inner channel and the reduction ratio of SE block were set to 64 and 4, respectively.

\subsection{Quantitative Comparison}
We adopt PSNR, SSIM\cite{wang2004image}, VIF\cite{sheikh2006image}, LPIPS \cite{zhang2018unreasonable}, FSIM \cite{zhang2011fsim}, UQI \cite{wang2002universal}, Signal to Reconstruction Error Ratio (SRER), Root-MSE (RMSE) and Spectral Angle Mapper (SAM)\cite{de2000spectral} as metrics to evaluate the quality of enhanced images. Following DA-DRN\cite{wei2021dadrn}, we also use Angular Error\cite{hordley2004re} and DeltaE\cite{sharma2005ciede2000} as the indicators of color distortion. We test our model on LOL real-world and synthetic datasets as well as four commonly used datasets without Ground-Truth: LIME \cite{guo2016lime}, DICM \cite{lee2013contrast}, MEF \cite{ma2015perceptual} and NPE\cite{wang2013naturalness} datasets. We use NIQE\cite{mittal2012making} as the non-reference metric. As shown in Table.\ref{lol_real}, \ref{lol_syn}, \ref{wogt}, our method achieves very good effects and  outperforms many other state-of-the-art methods in terms of several widely used indicators quantitatively.

\begin{table*}[htbp] 
	\centering \caption{Quantitative comparison of several metrics between our method
		and other state-of-the-art methods on \textbf{LOL REAL-WORLD} dataset. Mean, Median and Avg represent the mean and median values of the Angular Error and the average value of them, respectively. “↑” indicates the higher the better, “↓” indicates the lower the	better. \color{red}Red: the best, \color{blue}Blue: the second best.}
	\begin{tabular}{  c | c  c  c  c  c  c  c  c  c  c  c  c }
		\hline  Methods  & PSNR↑   & SSIM↑  &VIF↑  & LPIPS↓  &FSIM↑   & UQI↑  & SRER↑  & RMSE↓  & SAM↑   & Mean↓   & Median↓   &DeltaE↓ \\
		\hline
		\hline  Input & 7.7733 & 0.1914  & 0.2407  & 0.4173 &0.7190   &0.0622  & 47.5772 & 0.0264 & 76.5801  &3.8061 & 3.9728  & 76.5837\\
		LIME \cite{guo2016lime}  & 16.7586 & 0.4449 & 0.4500 & 0.4183 &0.8549   &0.8805 & 52.1989 & 0.0094 & 86.9102 
		& 3.2096 & 4.0825  & 21.1816\\
		NPE \cite{wang2013naturalness}  & 16.9697 & 0.4839 & 0.3943 & 0.4156 &0.8964   &0.8943 & 52.2944 & 0.0093 & 87.0226 & 3.5588 & 4.2505 & 22.6374\\
		JED \cite{ren2018joint}  & 13.6857 & 0.6509 & 0.3985 & 0.3549 &0.8812   &0.7143 & 50.5667 & 0.0146 & 87.3038 & 3.4064 & 3.8651 & 33.8342\\
		CRM \cite{ying2017new}  & 17.2032 & 0.6229 & 0.4114 & 0.3748 & 0.9456   &0.8441 & 52.4903 & 0.0099 & 87.0542 & 3.4396 & 3.6790  & 23.7405\\
		EFF \cite{ying2017new02}  & 13.8752 & 0.5949 & 0.3906 & 0.3673 &0.9263   &0.7088 & 50.6598 & 0.0141 & 86.6089 & 3.4004 & 3.5187  & 33.8820\\
		MBLLEN \cite{lv2018mbllen}  & 17.8583 & 0.7247 & 0.4911 & 0.3672 &0.9262   &0.8261 & 52.7664 & 0.0086 & 86.1212 & 3.2716 & 4.4620  & 21.5774\\
		RetinexNet \cite{wei2018deep}  & 16.7740 & 0.4249 & 0.2370 & 0.4670 &0.8642   &0.9110 & 52.2075 & 0.0094 & \color{blue}88.2461 & 3.7501 & 4.4975  & 21.3550\\
		GLAD \cite{wang2018gladnet}  & 19.7182 & 0.6820 & 0.4091 & 0.3994 & 0.9329   &0.9204 & 53.7990 & 0.0070 & 88.2170  & 3.3110 & 3.8021  & 16.0393\\
		RDGAN \cite{wang2019rdgan}  & 15.9363 & 0.6357 & 0.3620 & 0.3985 &0.9276   &0.8296 & 51.7681 & 0.0114 & 87.4576 & 4.3899 & 5.3027  & 26.3796\\
		Zero-DCE \cite{guo2020zero}  & 14.8671 & 0.5623 & 0.3849 & 0.3852 &0.9276   &0.7205 & 51.2269 & 0.0126 & 85.9968 & 4.1051 & 4.6860  & 31.4451\\
		EnGan \cite{jiang2021enlightengan}  & 17.4828 & 0.6515 & 0.4234 & 0.3903 &0.9226  &0.8499 & 52.5934 & 0.0095 & 87.7195 & 4.5296 & 5.2536 & 21.9113\\
		KinD \cite{zhang2019kindling}  & 20.3792 & 0.8056 & \color{blue}0.5137 & 0.2711 &0.9397  &0.9250 & 54.1233 & 0.0066 & 87.5607 & 2.2947 & 2.6376  & 13.9618\\
		KinD++ \cite{zhang2021beyond}  & \color{blue}21.8037 & \color{blue}0.8253 & 0.4954 & \color{blue}0.2592 &0.9275  &\color{red}0.9620 & \color{blue}54.8074 & \color{blue}0.0053 & 87.7490 & 2.2537 & 2.6731  & \color{red}11.0270\\
		DA-DRN\cite{wei2021dadrn}  & 20.7282 & 0.7939 & 0.4327 & 0.3126 &\color{blue}0.9458   &0.9378  &  54.1478 & 0.0061 & \color{red}88.2747 &\color{red}2.1638 & \color{red}2.3149  &12.9350\\
		\hline 
		\hline 
		TSN & 21.4727 & 0.8375 & 0.5460 & 0.2592 &0.9572   &0.9315 & 54.6206 & 0.0056 & 88.1231 & 2.4883 & 3.0704  & 13.1755\\
		TSN-CA & \color{red}22.4301 & \color{red}0.8452 & \color{red}0.5624 & \color{red}0.2433 &\color{red}0.9631   &\color{blue}0.9338  &  \color{red}55.1780 & \color{red}0.0043 & 88.1604 &\color{blue}2.2462 & \color{blue}2.5946  & \color{blue}12.4946\\
		\hline\end{tabular}\vspace{0cm}
	\label{lol_real}
\end{table*}

\begin{table*}[htbp] 
	\centering \caption{Quantitative comparison of several metrics between our method
		and other state-of-the-art methods on \textbf{LOL SYNTHETIC} dataset. Mean, Median and Avg represent the mean and median values of the Angular Error and the average value of them, respectively. “↑” indicates the higher the better, “↓” indicates the lower the	better. \color{red}Red: the best, \color{blue}Blue: the second best.}
	\begin{tabular}{  c | c  c  c  c  c  c  c  c  c  c  c  c }
		\hline  Methods  & PSNR↑   & SSIM↑  & VIF↑  & LPIPS↓  &FSIM↑   & UQI↑  & SRER↑  & RMSE↓  & SAM↑   & Mean↓   & Median↓   &DeltaE↓ \\
		\hline
		\hline  Input & 10.2533 & 0.4193 & 0.4248  & 0.2871 &0.7802   &0.3502  & 48.9243 & 0.0112 & 77.5350  &3.2315 & 3.1539  & 51.9337\\
		LIME \cite{guo2016lime}  & 17.0682 & 0.7606 & 0.6311 & 0.2040 &0.8617   &0.8804 & 52.3908 & 0.0092 & 85.8743 
		& 3.2096 & 4.0825   & 21.1816\\
		NPE \cite{wang2013naturalness}  & 14.6603 & 0.7724 & 0.5708 & 0.1866 &0.9036   &0.7921 & 51.1505 & 0.0123 & 85.1261 & 2.4371 & 2.6856  & 23.4608\\
		JED \cite{ren2018joint}  & 15.0805 & 0.7145 & 0.4397 & 0.2562 &0.8815   &0.7990 & 51.3495 & 0.0118 & 84.7191 & 3.4064 & 3.8651  & 33.8342\\
		CRM \cite{ying2017new}  & 14.9942 & 0.7689 & 0.6011 & 0.1831 &0.9115   &0.7850 & 51.3881 & 0.0122 & 85.5286 & 3.9513 & 4.5929  & 23.9757\\
		EFF \cite{ying2017new02}  & 18.7439 & 0.8519  & \color{blue}0.6342 & 0.1778 &0.9305   &0.8956 & 53.4375 & 0.0077 & 86.5873 & 4.0354 & 5.0109  & 16.5619\\
		MBLLEN \cite{lv2018mbllen}  & 14.2620 & 0.6552 & 0.4726 & 0.2903 &0.9039   &0.7013 & 50.9951 & 0.0132 & 84.1075 & 2.5991 & 3.1658  & 27.5349\\
		RetinexNet \cite{wei2018deep}  & 17.2025 & 0.7639 & 0.3512 & 0.2467 &0.8639   &0.8888 & 52.4594 & 0.0095 & \color{red}88.1026 & 1.7625 & 2.7897  & 18.2853\\
		GLAD \cite{wang2018gladnet}  & 16.2292 & 0.8007 & 0.6005 & 0.1888 & 0.9378   &0.8406 & 52.1234 & 0.0105 & 86.2192  & 3.6618 & 3.8868  & 19.4709\\
		RDGAN \cite{wang2019rdgan}  & 18.2270 & 0.8368 & 0.6006 & 0.1706  &0.9415   &0.8971 & 53.1087 & 0.0084 & 87.1588 & 3.1857 & 3.4799  & 16.6347\\
		Zero-DCE \cite{guo2020zero}  & 16.5206 & 0.8173 & 0.5809  & 0.1772 &0.9256   &0.8150 & 52.2576 & 0.0102 & 85.2074 & 4.0482 & 3.6468 & 21.8502\\
		EnGan \cite{jiang2021enlightengan}  & 15.2653 & 0.7516 & 0.5390 & 0.1754 &0.8947  &0.7953 & 51.4678 & 0.0117 & 85.9107 & 3.0516 & 3.8443  & 22.0353\\
		KinD \cite{zhang2019kindling}  & 16.2156 & 0.8173 & 0.5825 & 0.1457 &0.9306  &0.8257 & 51.9733 & 0.0102 & 85.5904 & 1.7839 &3.1954  & 18.7326\\
		KinD++ \cite{zhang2021beyond}  & 16.4247 & 0.7845 & 0.4949 & 0.2618 &0.8864  &0.8639 & 52.1112 & 0.0097 & 85.8265 & 3.4677 &4.6003  & 19.4102\\
		DA-DRN\cite{wei2021dadrn}  & \color{blue}20.5360 & \color{blue}0.8388 & 0.4627 & \color{blue}0.1691 &\color{blue}0.9549   &\color{blue}0.9359  &  \color{blue}54.1278 & \color{blue}0.0063 & \color{blue}87.2474 &\color{blue}1.5082 & \color{blue}1.7007  & \color{blue}12.3788\\
		\hline 
		\hline 
		TSN & 22.1771 & 0.9094 & 0.6632 & 0.1093 &0.9715  &0.9418 & 54.9230 & 0.0056 & 86.9985 & 1.4571 & 2.1176  & 10.6888\\
		TSN-CA  & \color{red}22.3467 & \color{red}0.9208 & \color{red}0.7195 & \color{red}0.0767 &\color{red}0.9801   &\color{red}0.9481  &  \color{red}55.1215 & \color{red}0.0056 & 87.2226 &\color{red}1.3551 & \color{red}1.6288  & \color{red}9.5615\\
		\hline\end{tabular}\vspace{0cm}
	\label{lol_syn}
\end{table*}

\begin{table}[htbp] 
	\centering \caption{Quantitative comparison in terms of \textbf{NIQE} Metric between our method
		and other state-of-the-art methods on \textbf{LIME, DICM, MEF and NPE} datasets.}
	\begin{tabular}{  c | c  c  c  c }
		\hline  Methods  & LIME   & DICM   & MEF  &NPE \\
		\hline
		\hline  Input & 4.3577 & 3.8608 & 5.1884  & 3.6784  \\
		LIME \cite{guo2016lime}  & 4.1549 & 3.0005 & 4.4466 & 3.7715 \\
		NPE \cite{wang2013naturalness}  & 3.9048 & \color{blue}2.8448 & 4.2556 & 3.3997\\
		JED \cite{ren2018joint}  & 4.1456 & 3.5704 & 4.7250 & 3.5947 \\
		CRM \cite{ying2017new}  &3.8546 &2.9908 & 4.0080 & 3.4867\\
		EFF \cite{ying2017new02}  &3.8596 &2.9142 & 4.0533 & 3.4317\\
		MBLLEN \cite{lv2018mbllen}  & 4.5138 & 3.6654 & 4.6901 & 3.9788 \\
		RetinexNet \cite{wei2018deep}  & 4.5978 & 4.5779 & 5.1747 & 4.5472 \\
		GLAD \cite{wang2018gladnet}  & 4.1282 & 3.1147 & 3.6897 & 3.5311  \\
		RDGAN \cite{wang2019rdgan}  & 4.1186  & 3.0737  & 3.6314  &3.5836  \\
		Zero-DCE \cite{guo2020zero}  & 3.7690 & 2.8348 & 4.0240 & 3.5862\\
		EnGan \cite{jiang2021enlightengan}  & \color{blue}3.6574 & 2.9172 & \color{blue}3.5373 & 3.5623 \\
		KinD \cite{zhang2019kindling}  & 4.7632 & 3.5651 & 4.7514 & 3.8605 \\
		KinD++ \cite{zhang2021beyond}  & 3.7362 & 2.9573 & 3.7818 & \color{blue}3.3596 \\
		DA-DRN \cite{wei2021dadrn}  & 4.9852 & 3.7964 & 4.3252 & 4.1270 \\
		\hline 
		\hline 
		TSN   & 3.6749 & 2.7985 & 3.5462 & 3.2560 \\
		TSN-CA   & \color{red}3.5947 & \color{red}2.7106 & \color{red}3.5233 & \color{red}3.2234 \\
		\hline\end{tabular}\vspace{0cm}
	\label{wogt}
\end{table}

\begin{figure*}
	\centering
	
	
	\subfigure[Input]{
		\begin{minipage}[b]{0.13\textwidth}
			\includegraphics[width=2.5cm]{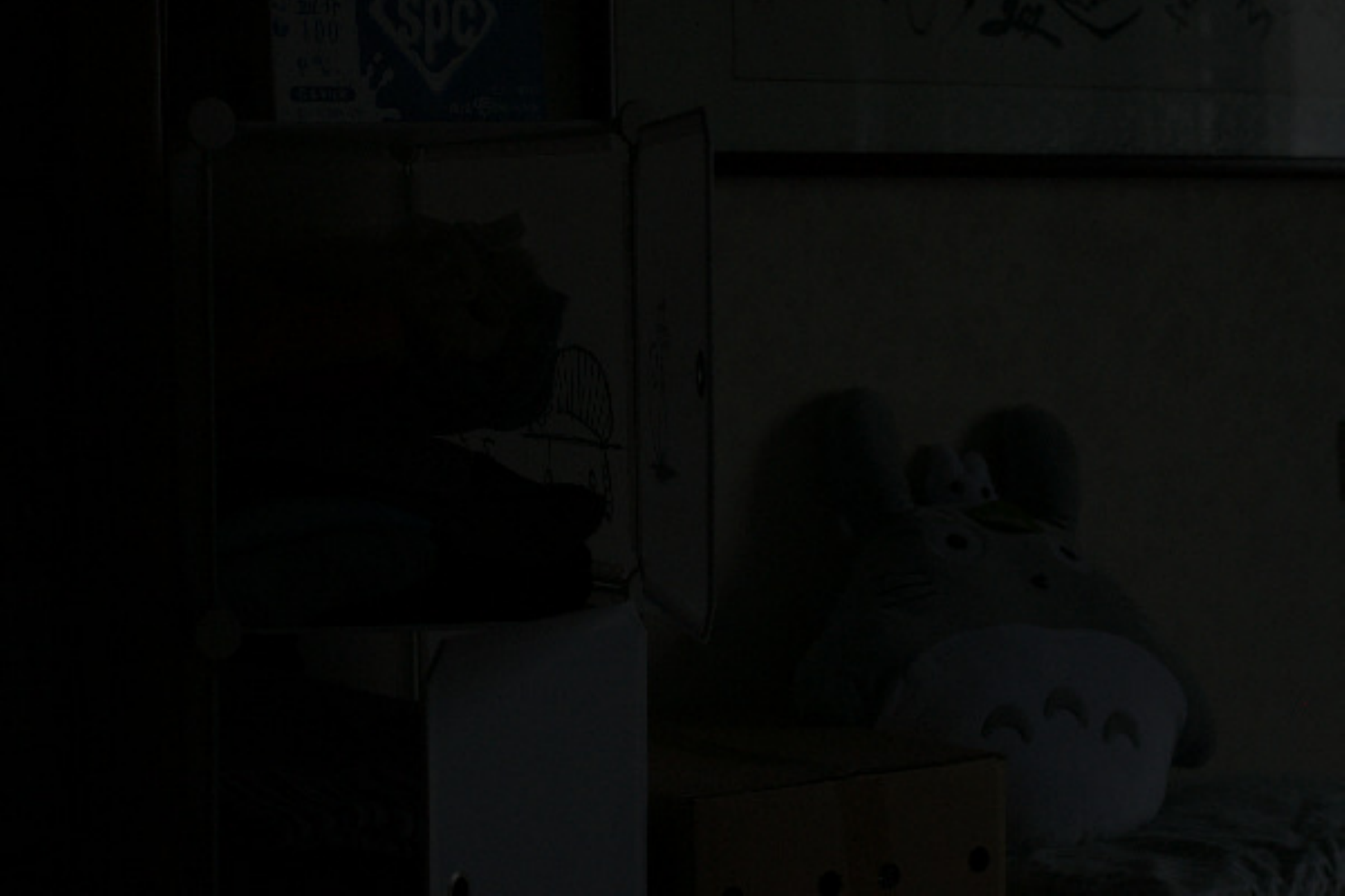}\vspace{2pt} \\
			\includegraphics[width=2.5cm]{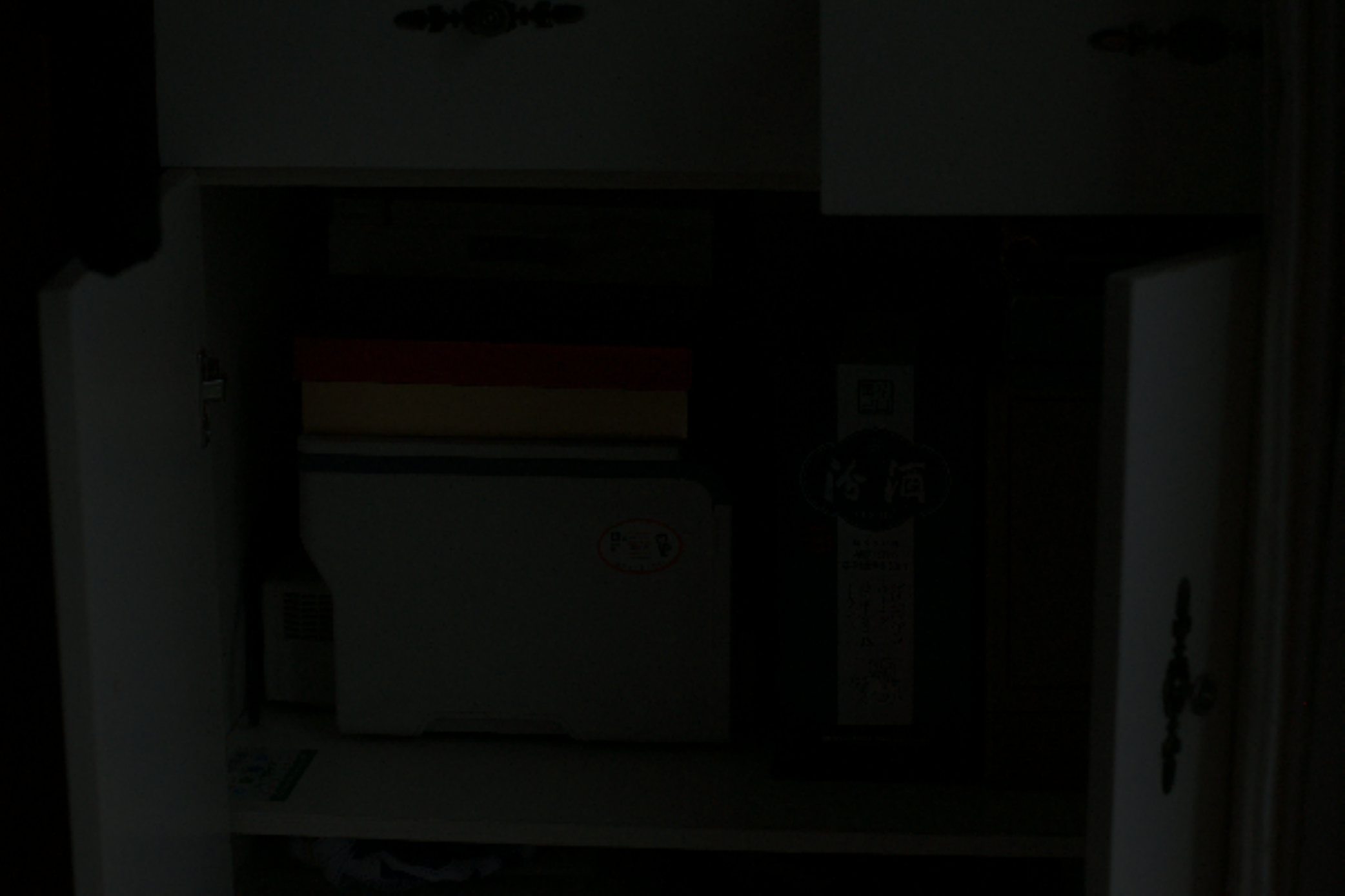}\vspace{2pt}
			\includegraphics[width=2.5cm]{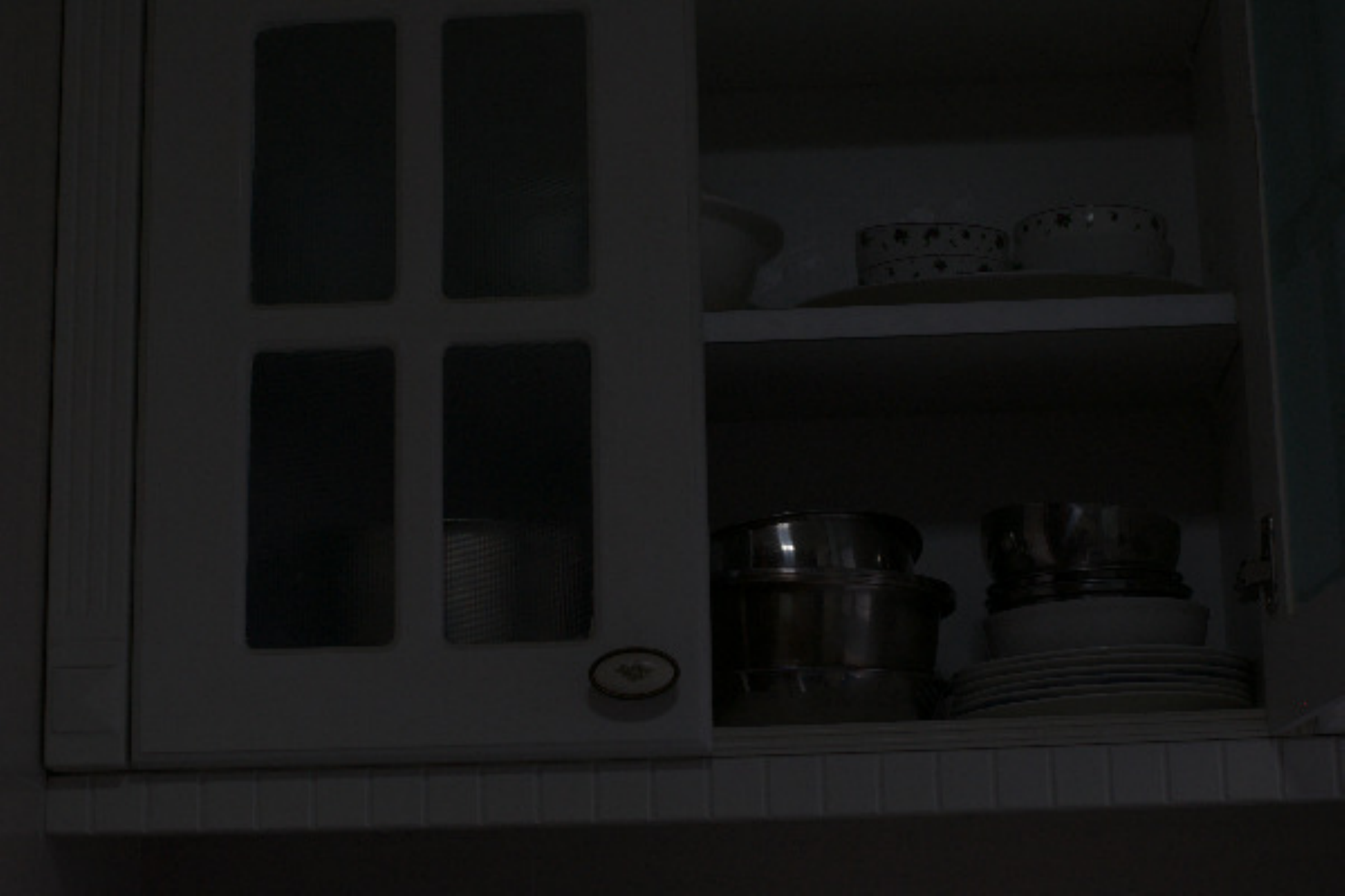}\vspace{2pt}
			\includegraphics[width=2.5cm]{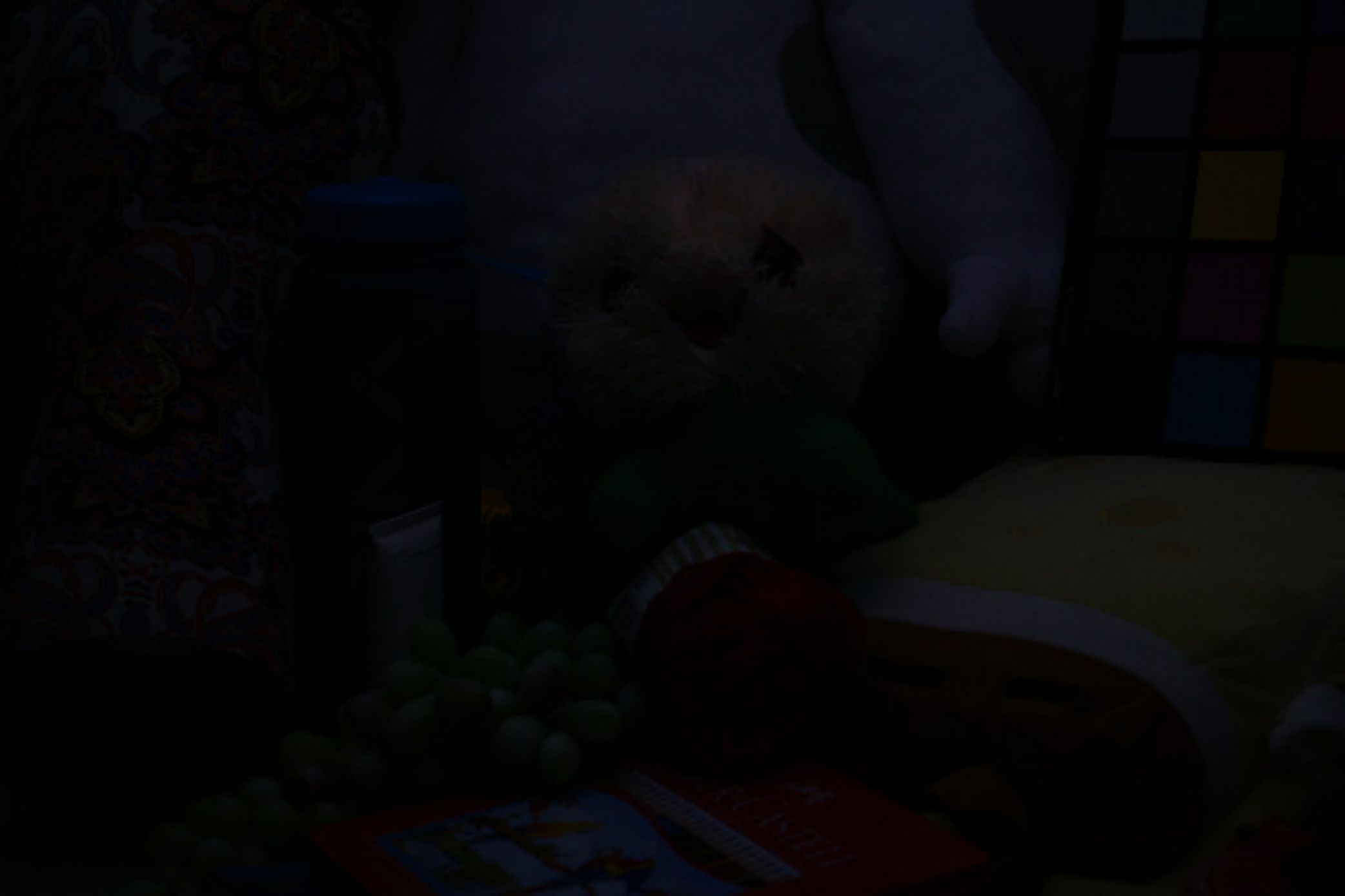}\vspace{2pt}
			\includegraphics[width=2.5cm]{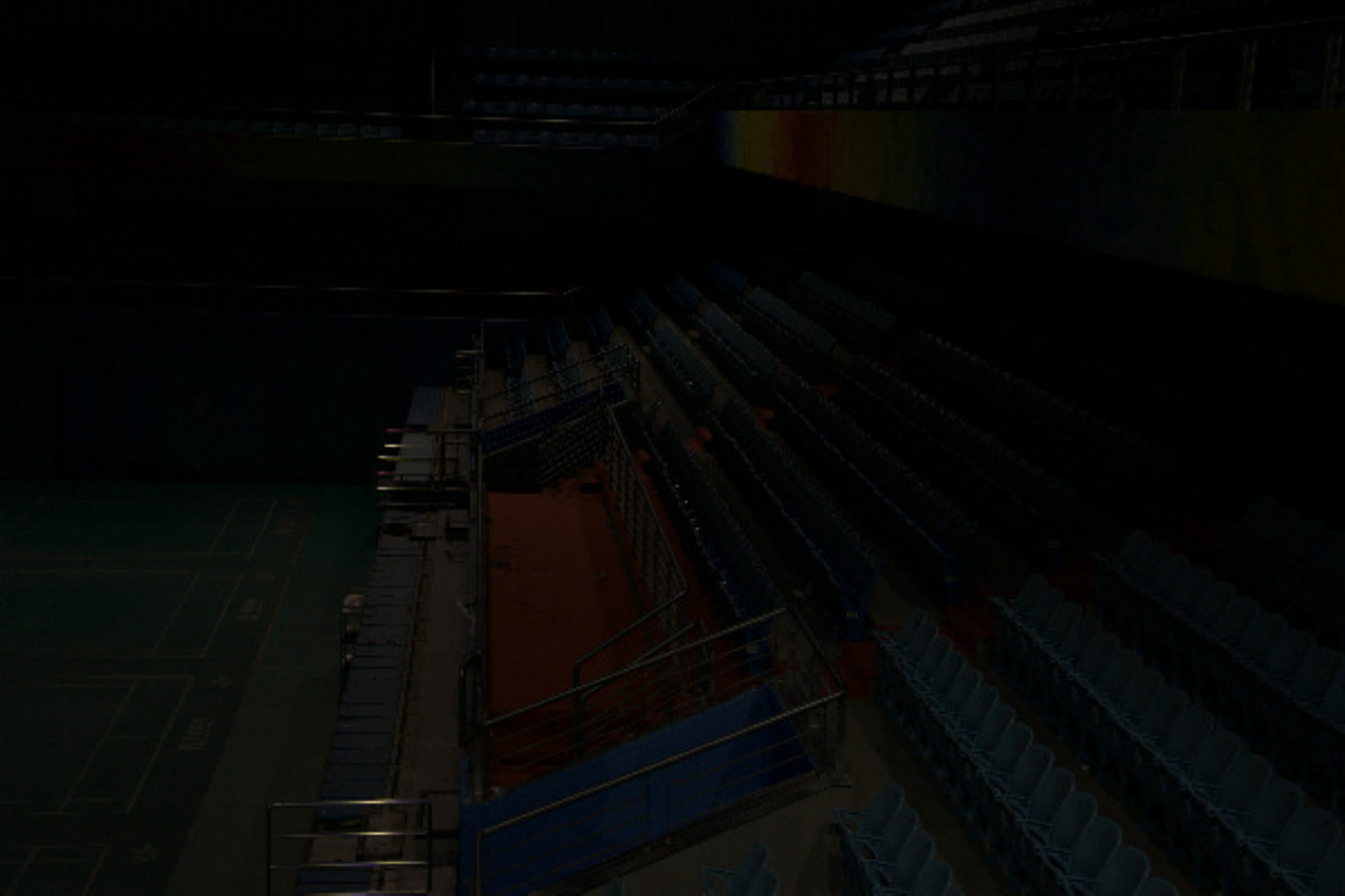}
		\end{minipage}
	}\hspace{-5pt}
	\subfigure[EnGan\cite{jiang2021enlightengan}]{
		\begin{minipage}[b]{0.13\textwidth}
			\includegraphics[width=2.5cm]{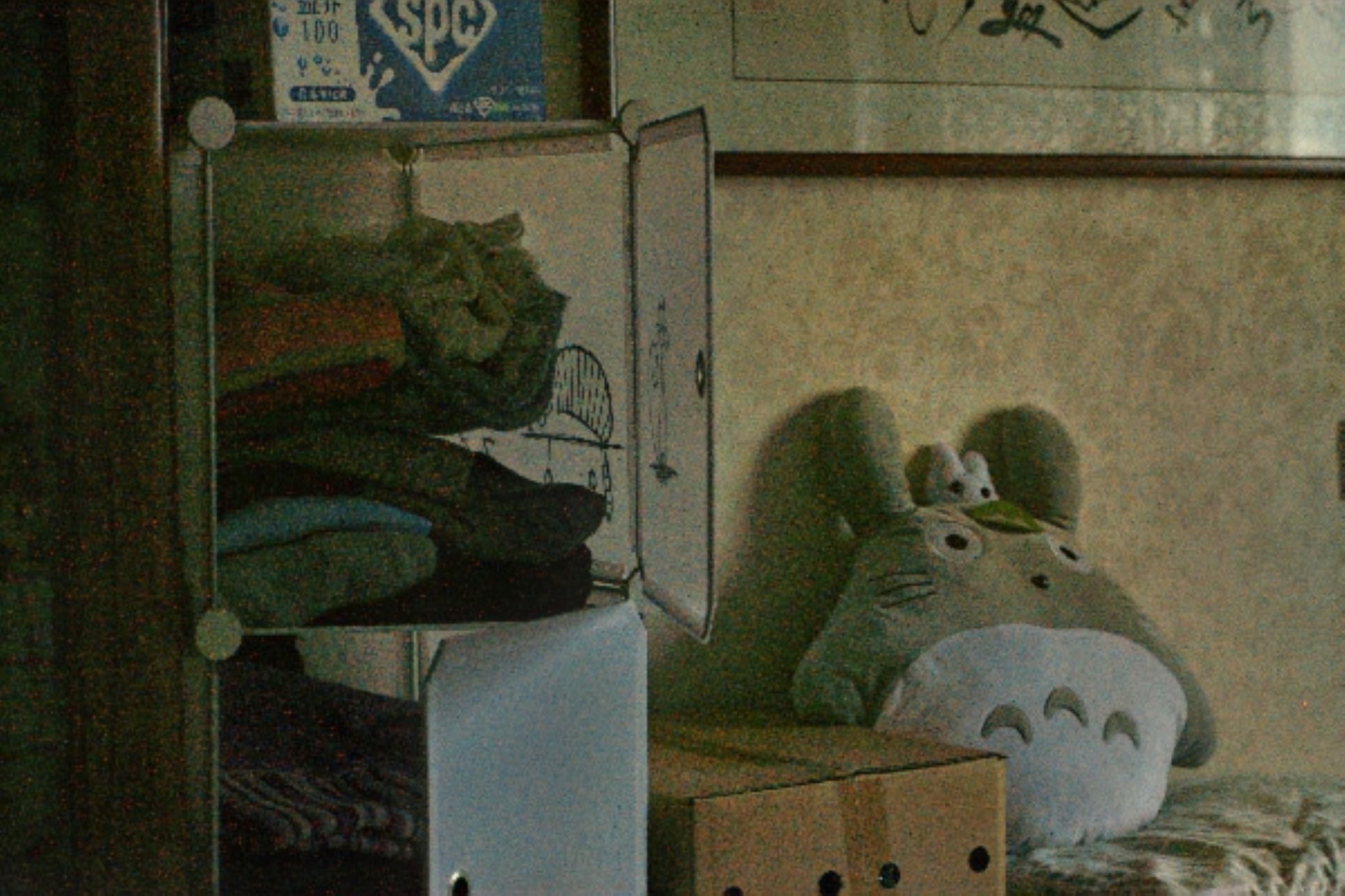}\vspace{2pt} \\
			\includegraphics[width=2.5cm]{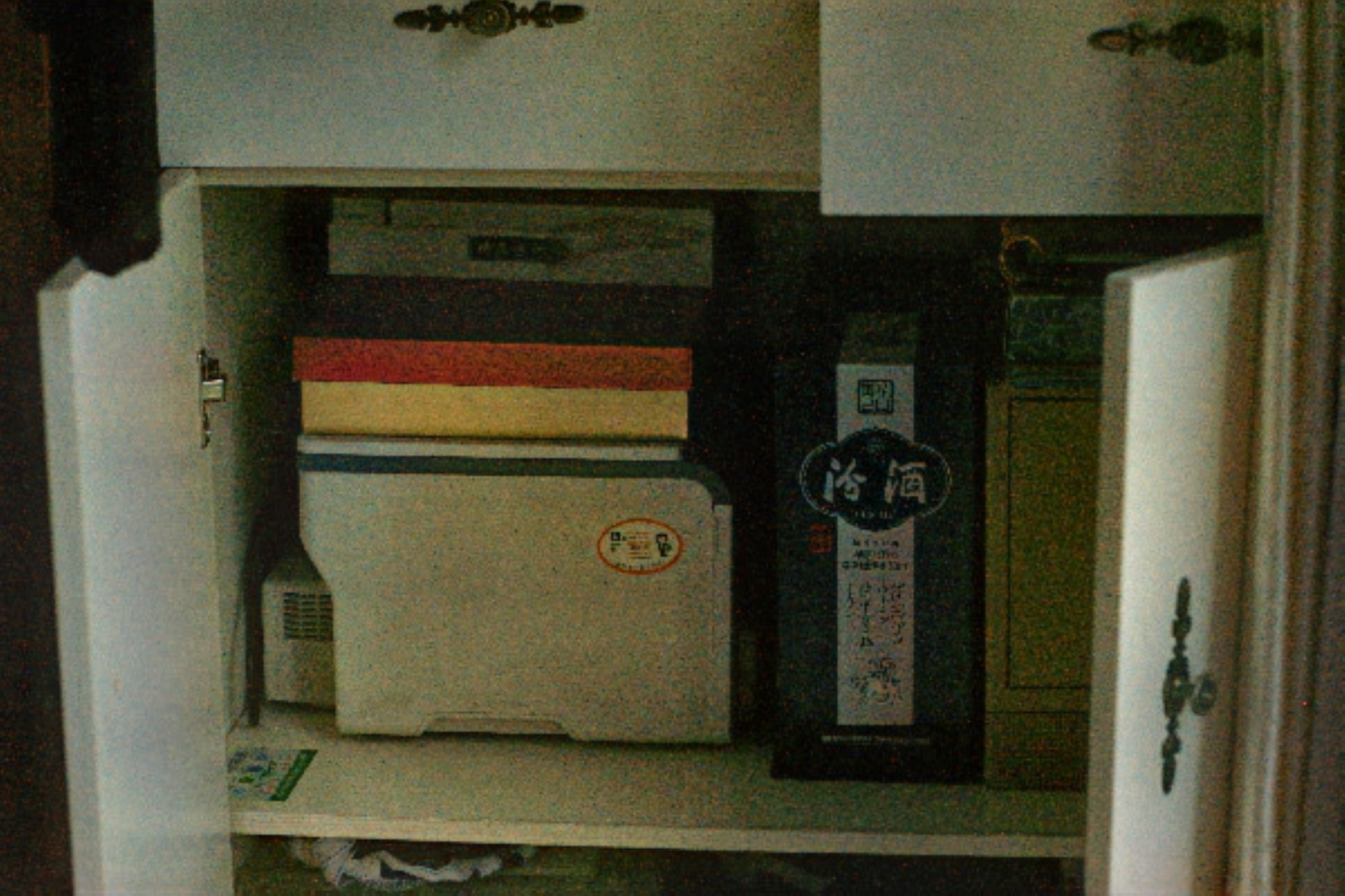}\vspace{2pt}
			\includegraphics[width=2.5cm]{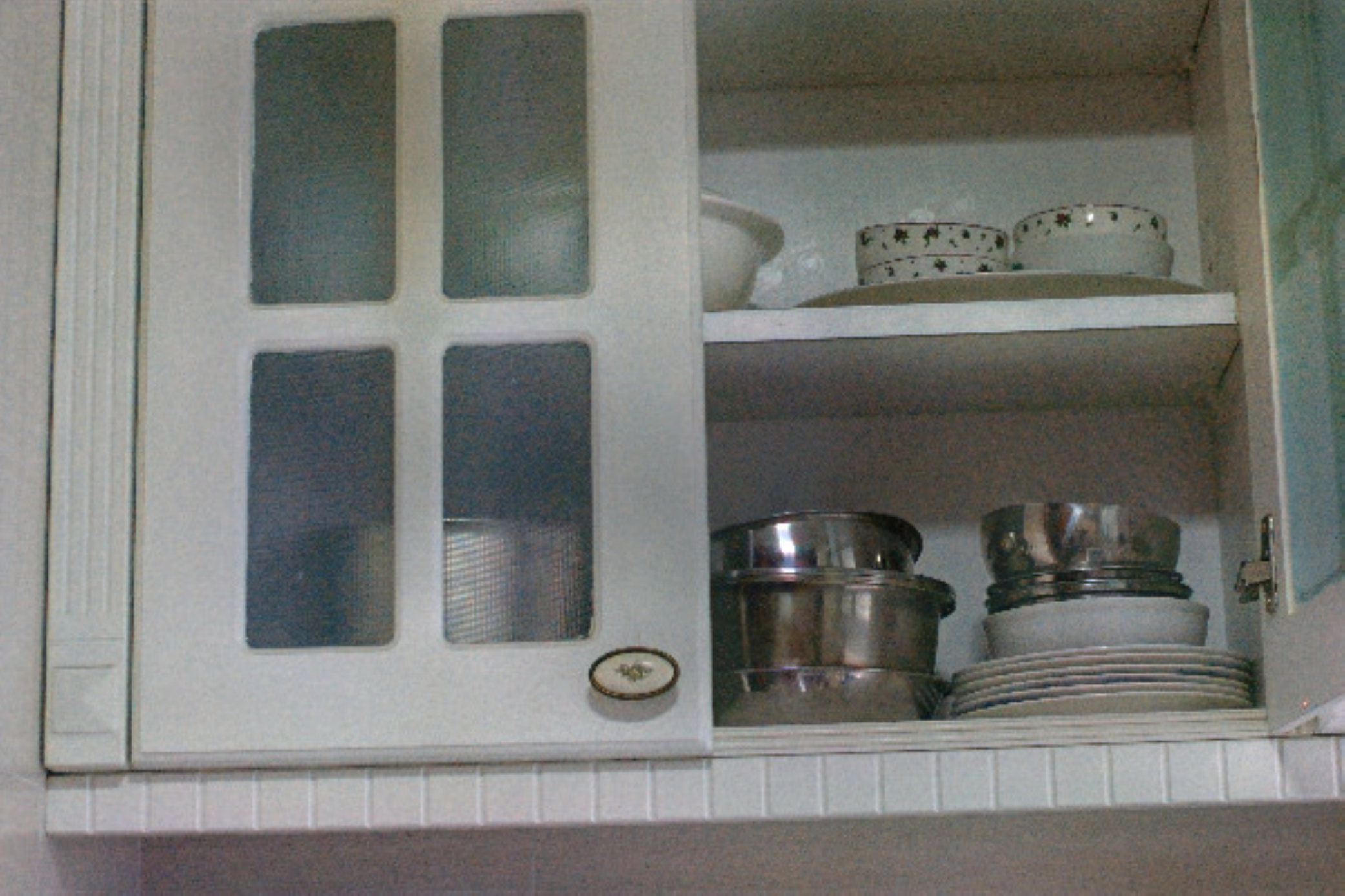}\vspace{2pt}
			\includegraphics[width=2.5cm]{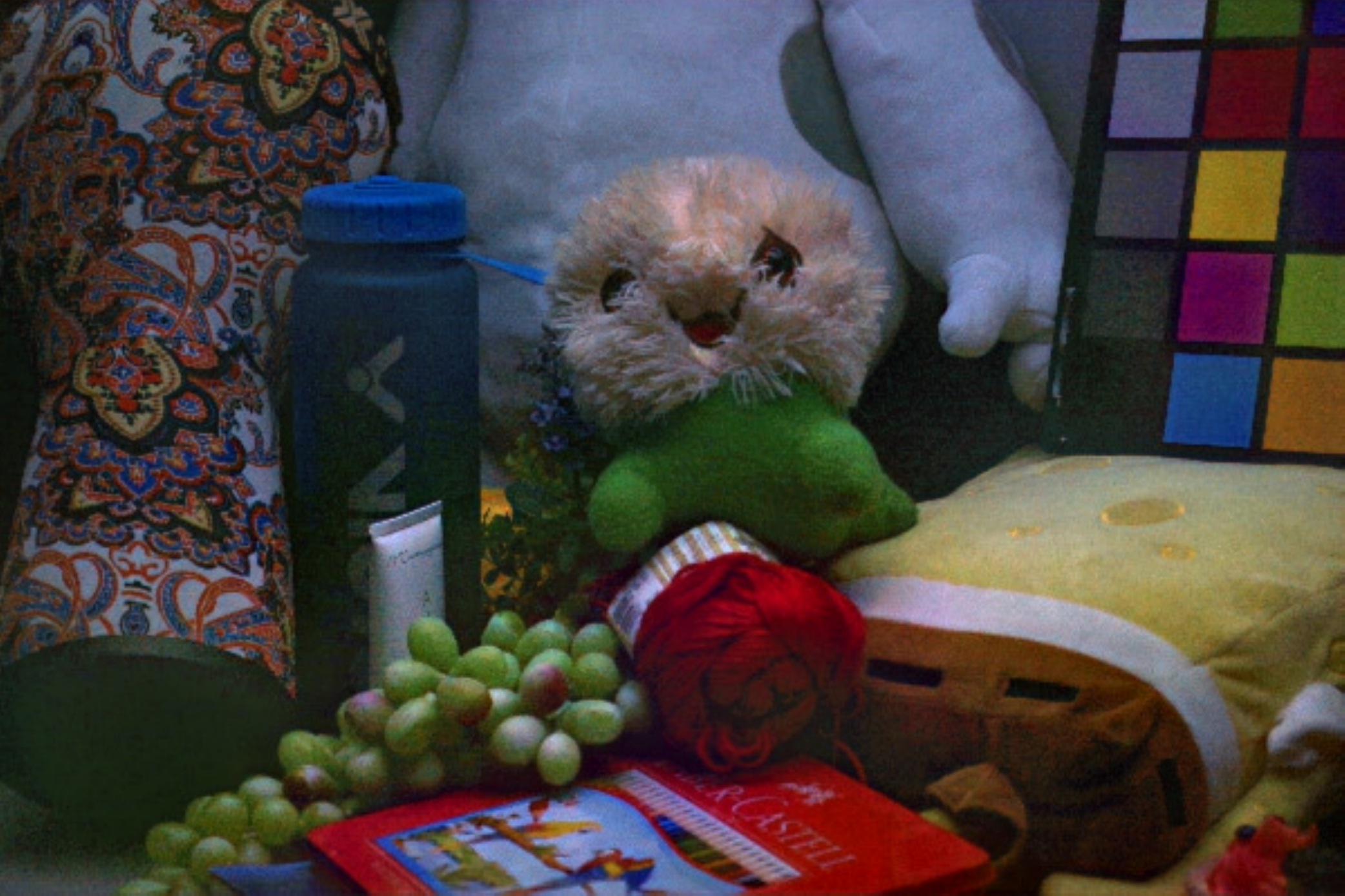}\vspace{2pt}
			\includegraphics[width=2.5cm]{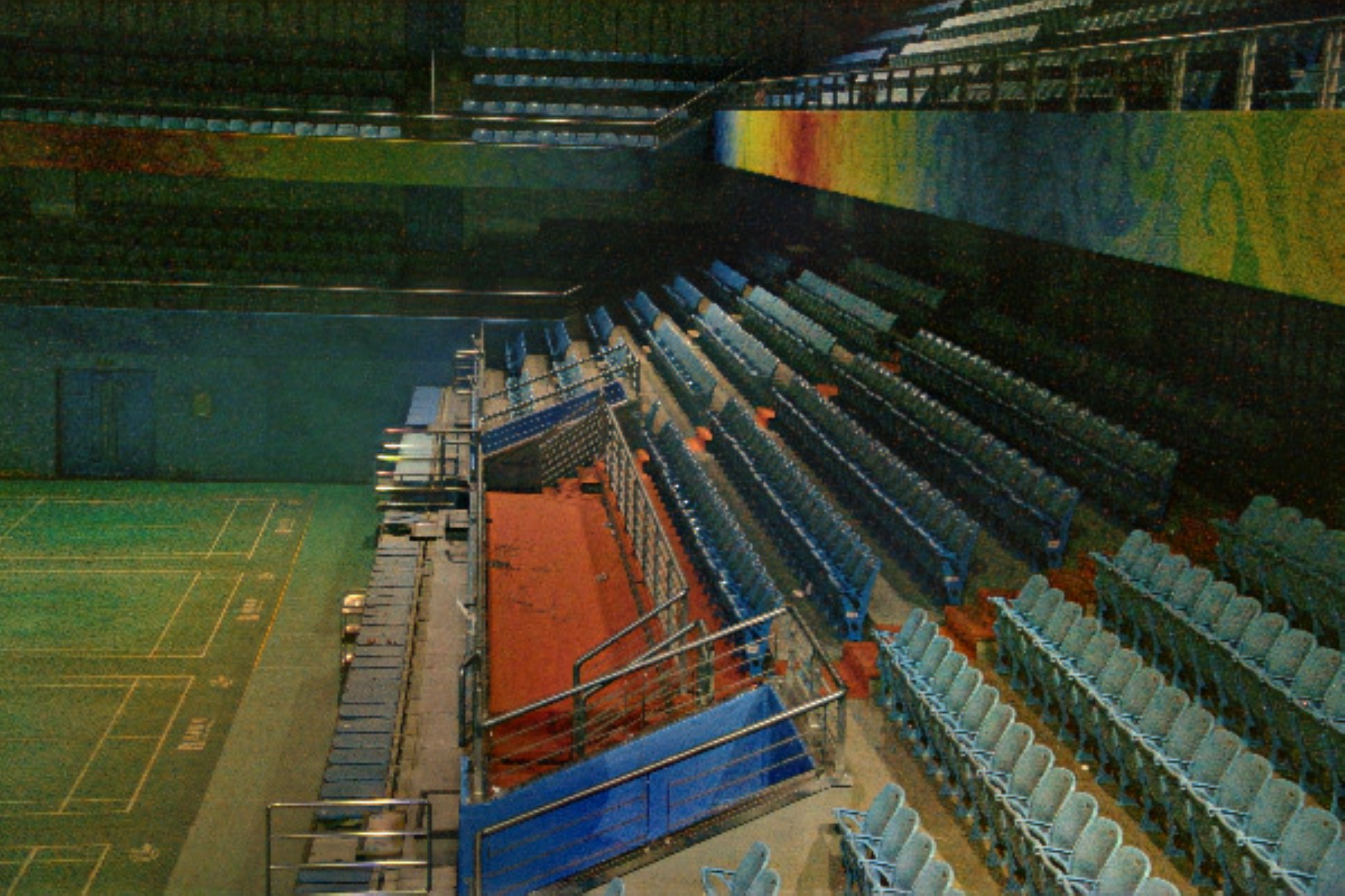}
		\end{minipage}
	}\hspace{-5pt}
	\subfigure[GLAD\cite{wang2018gladnet}]{
		\begin{minipage}[b]{0.13\textwidth}
			\includegraphics[width=2.5cm]{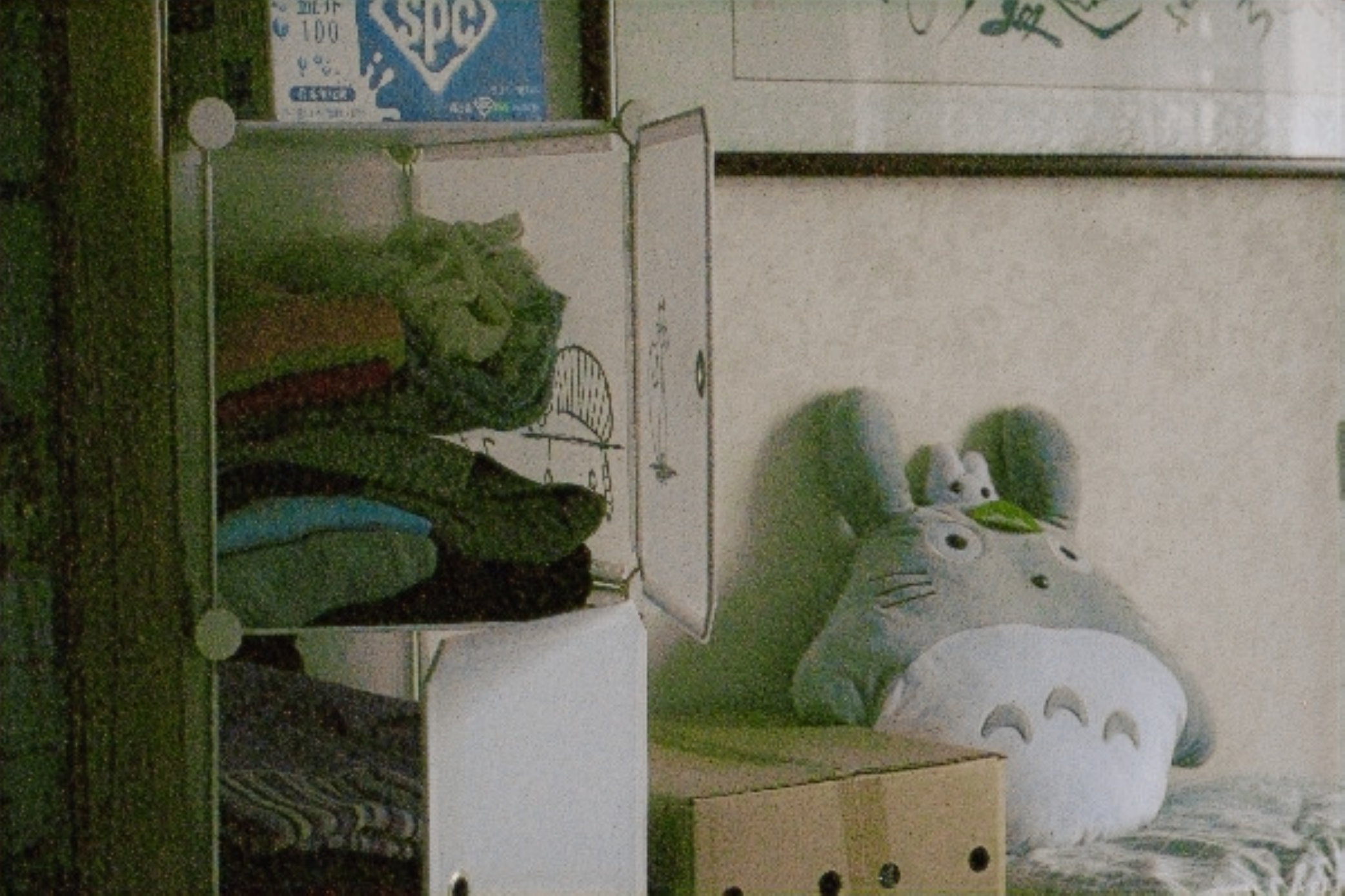}\vspace{2pt} \\
			\includegraphics[width=2.5cm]{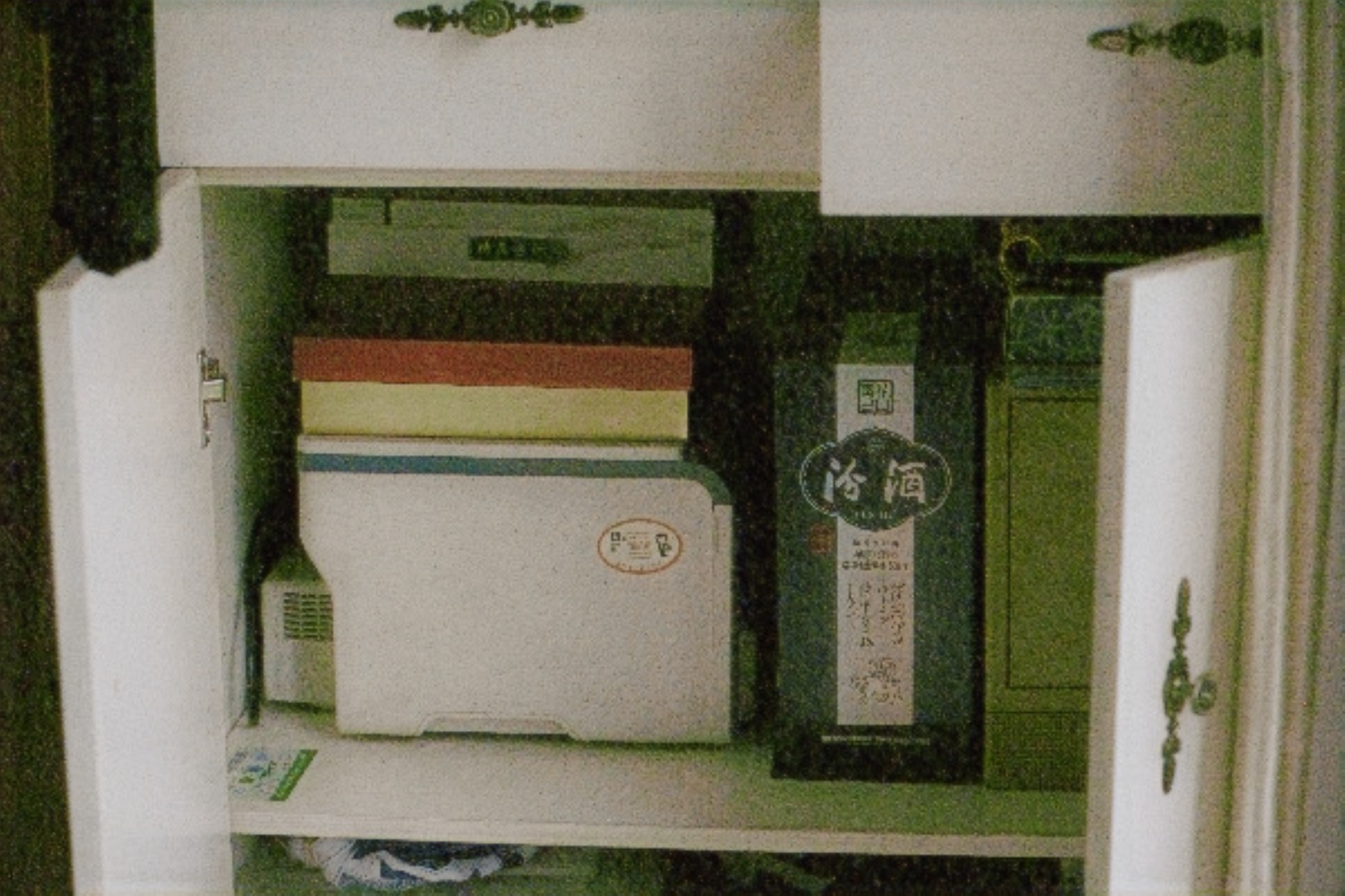}\vspace{2pt}
			\includegraphics[width=2.5cm]{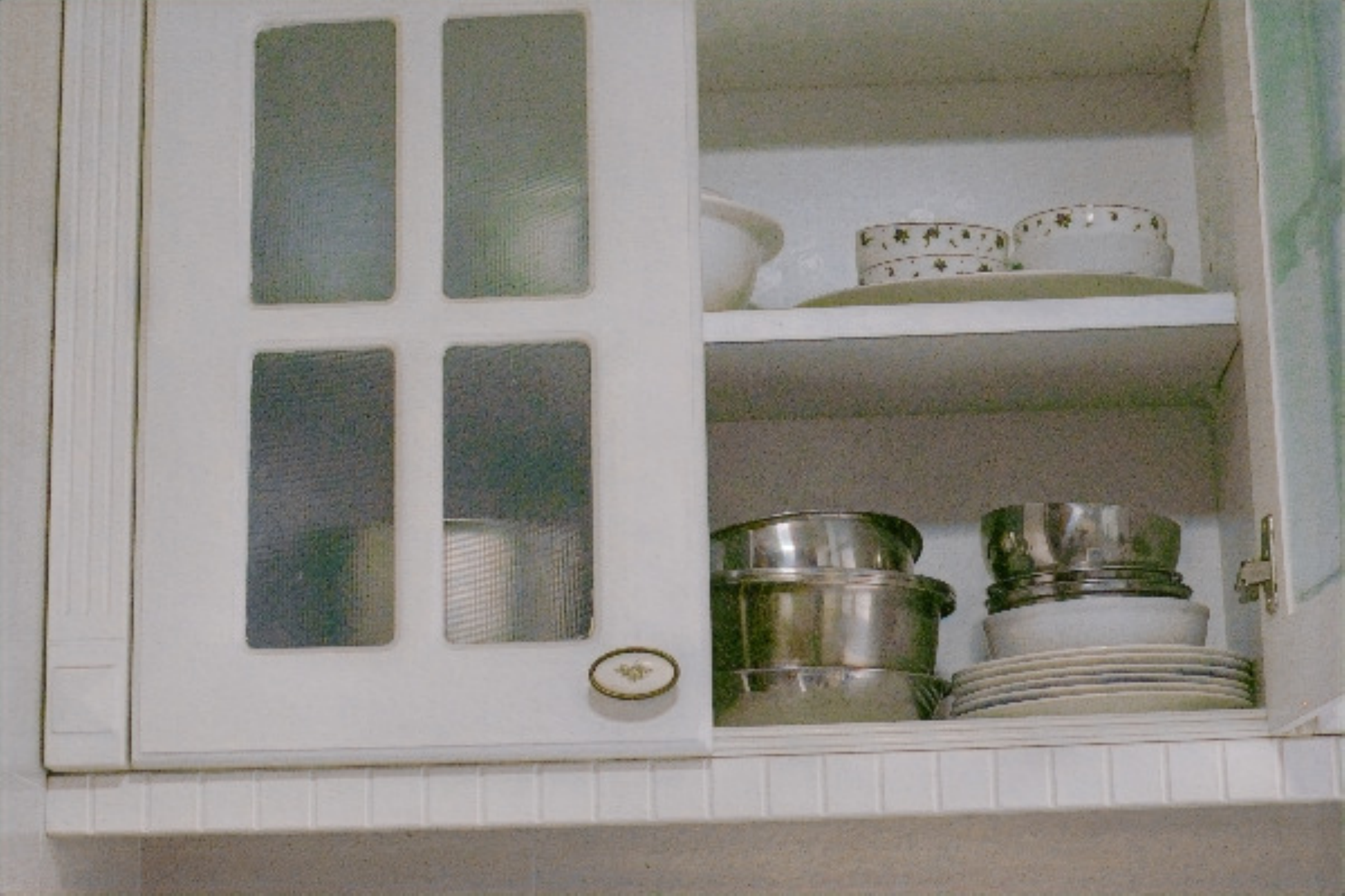}\vspace{2pt}
			\includegraphics[width=2.5cm]{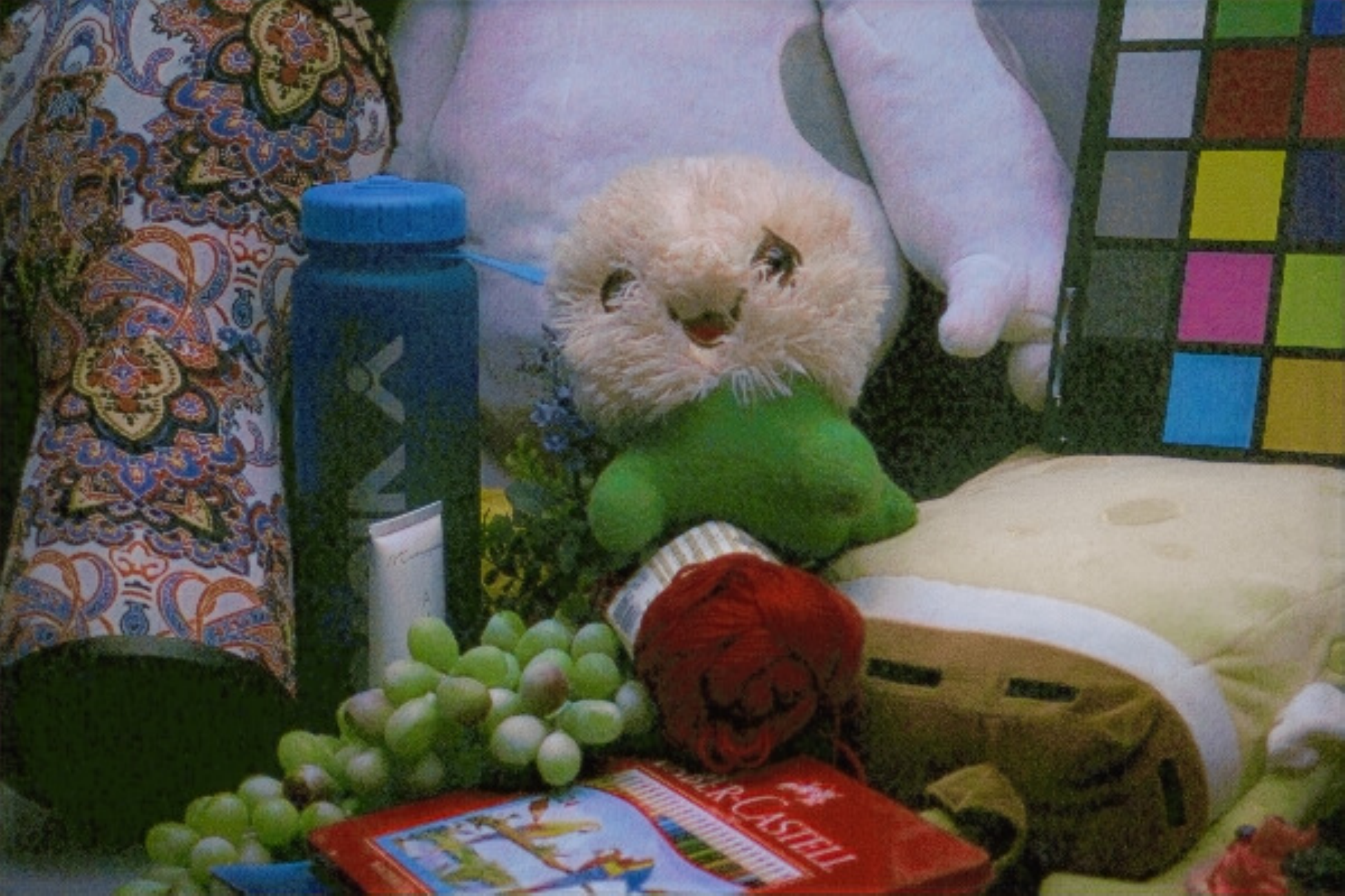}\vspace{2pt}
			\includegraphics[width=2.5cm]{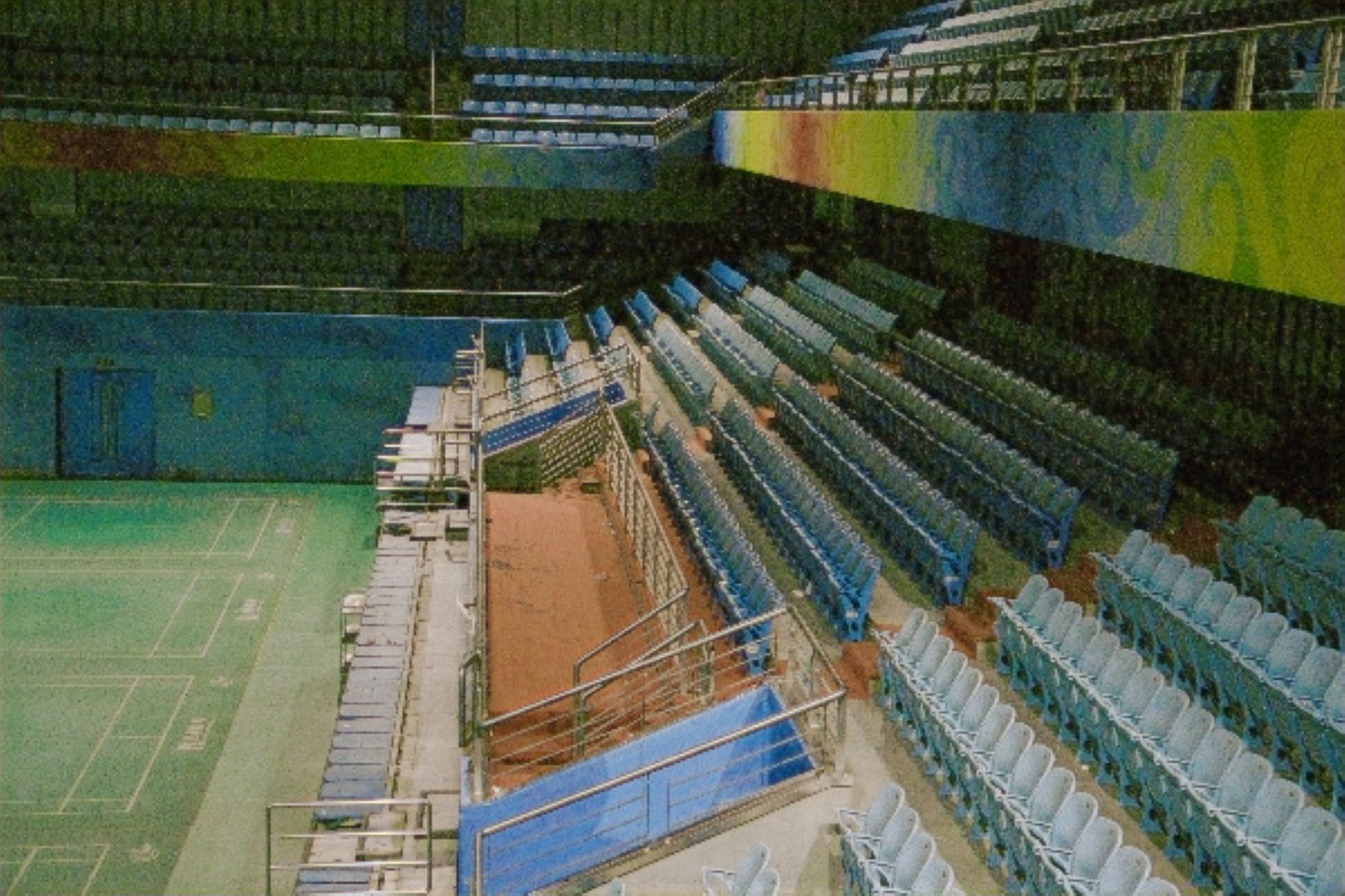}
		\end{minipage}
	}\hspace{-5pt}
	\subfigure[KinD\cite{zhang2019kindling}]{
		\begin{minipage}[b]{0.13\textwidth}
			\includegraphics[width=2.5cm]{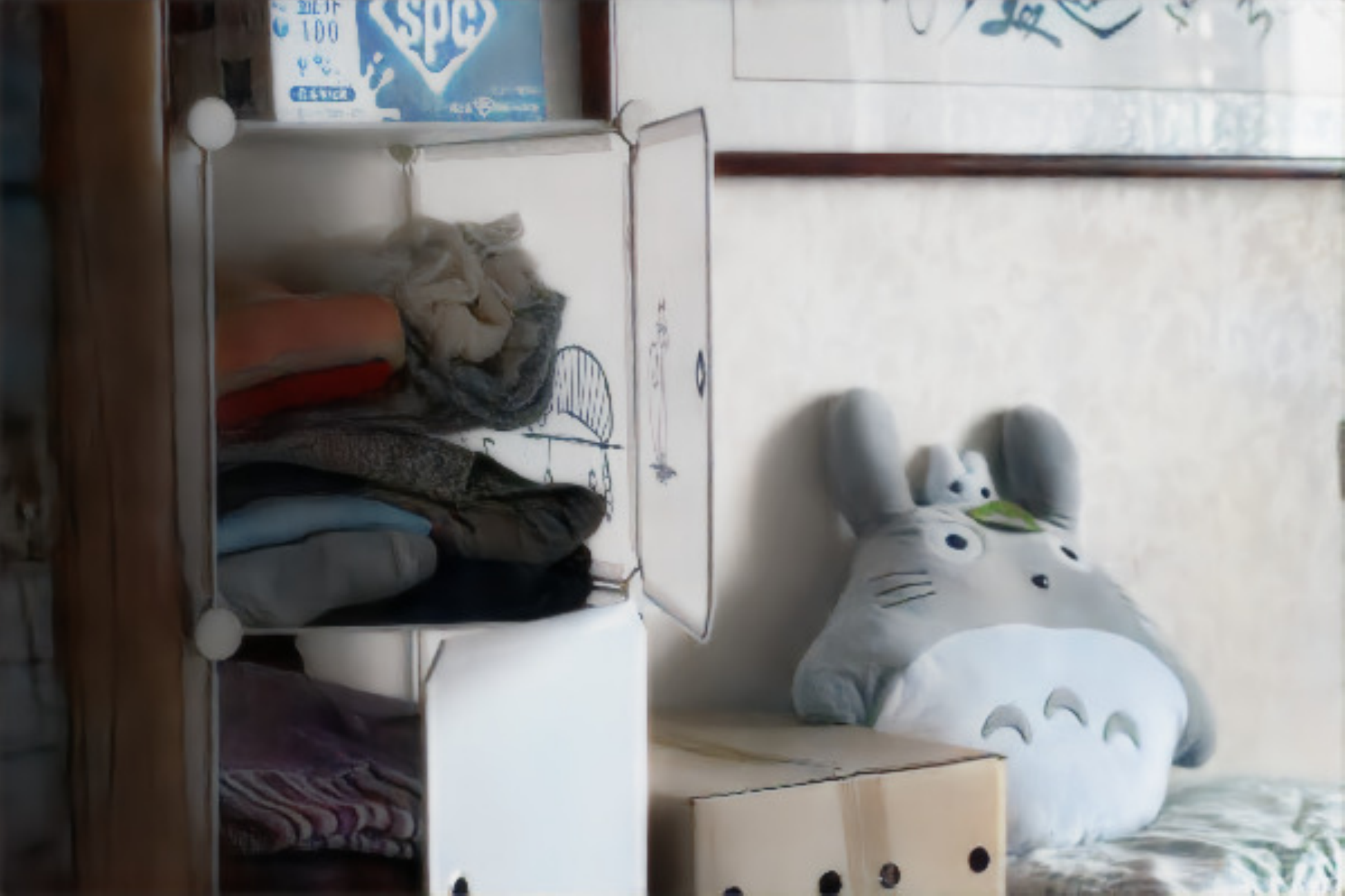}\vspace{2pt} \\
			\includegraphics[width=2.5cm]{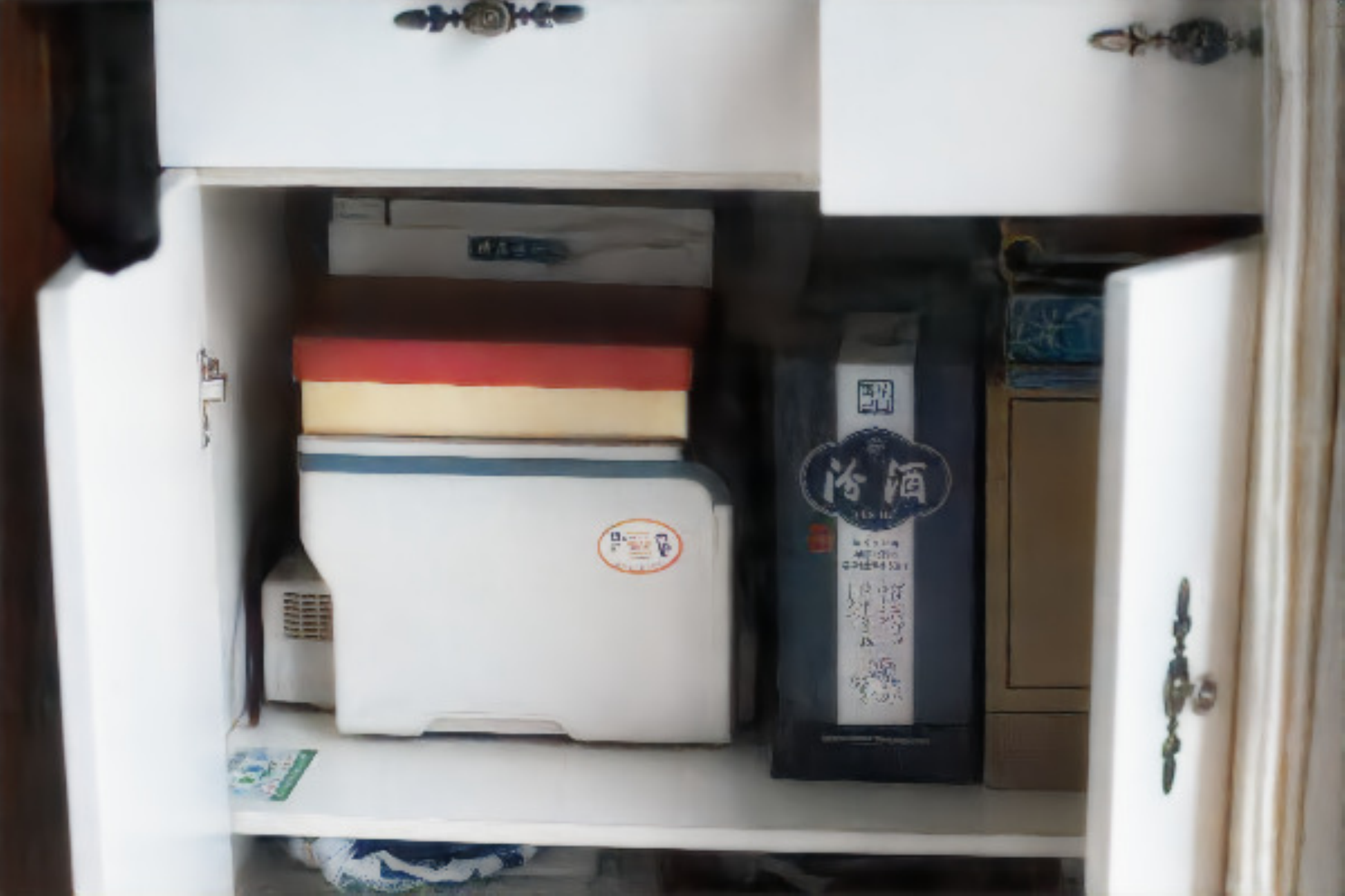}\vspace{2pt}
			\includegraphics[width=2.5cm]{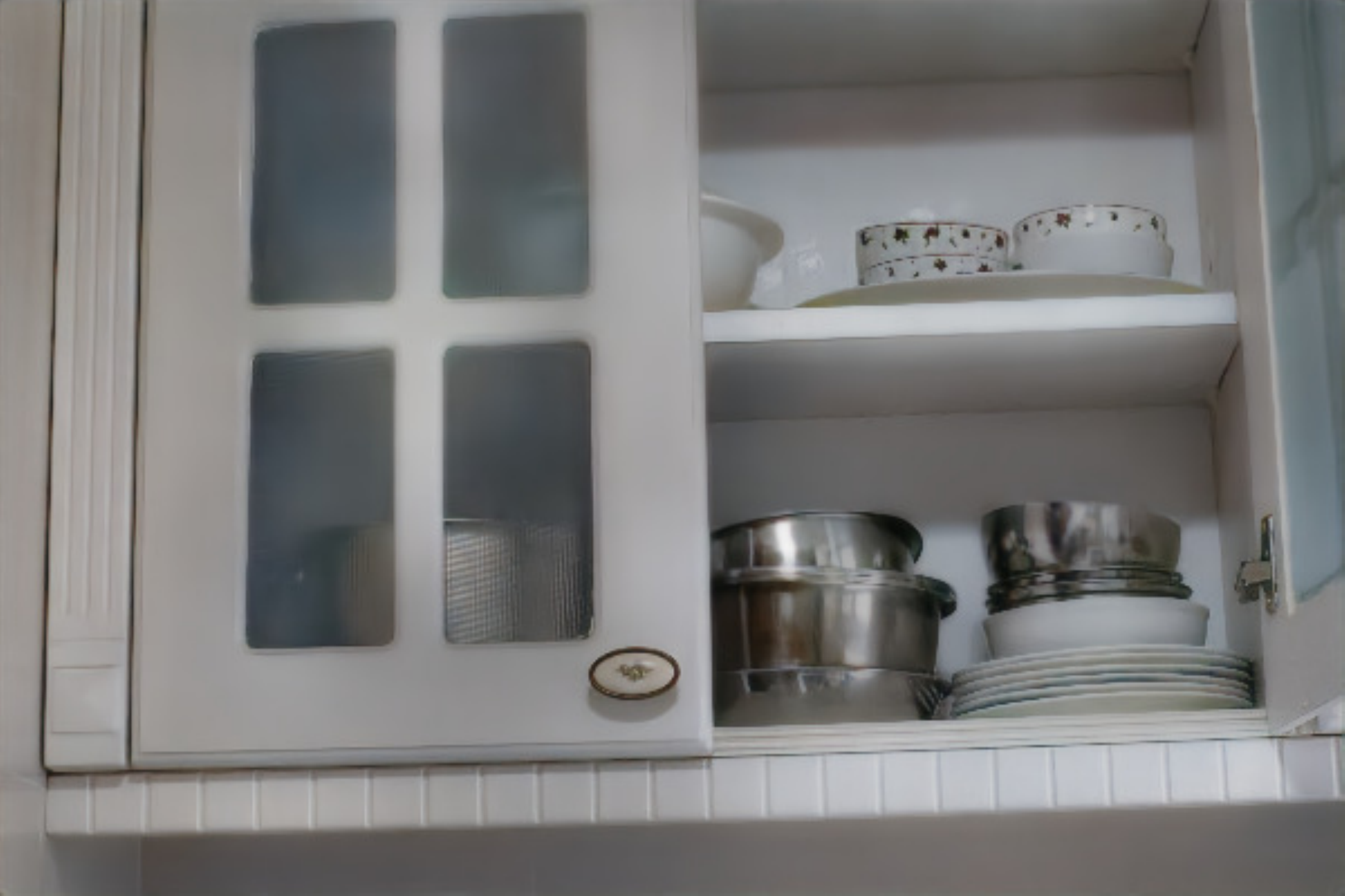}\vspace{2pt}
			\includegraphics[width=2.5cm]{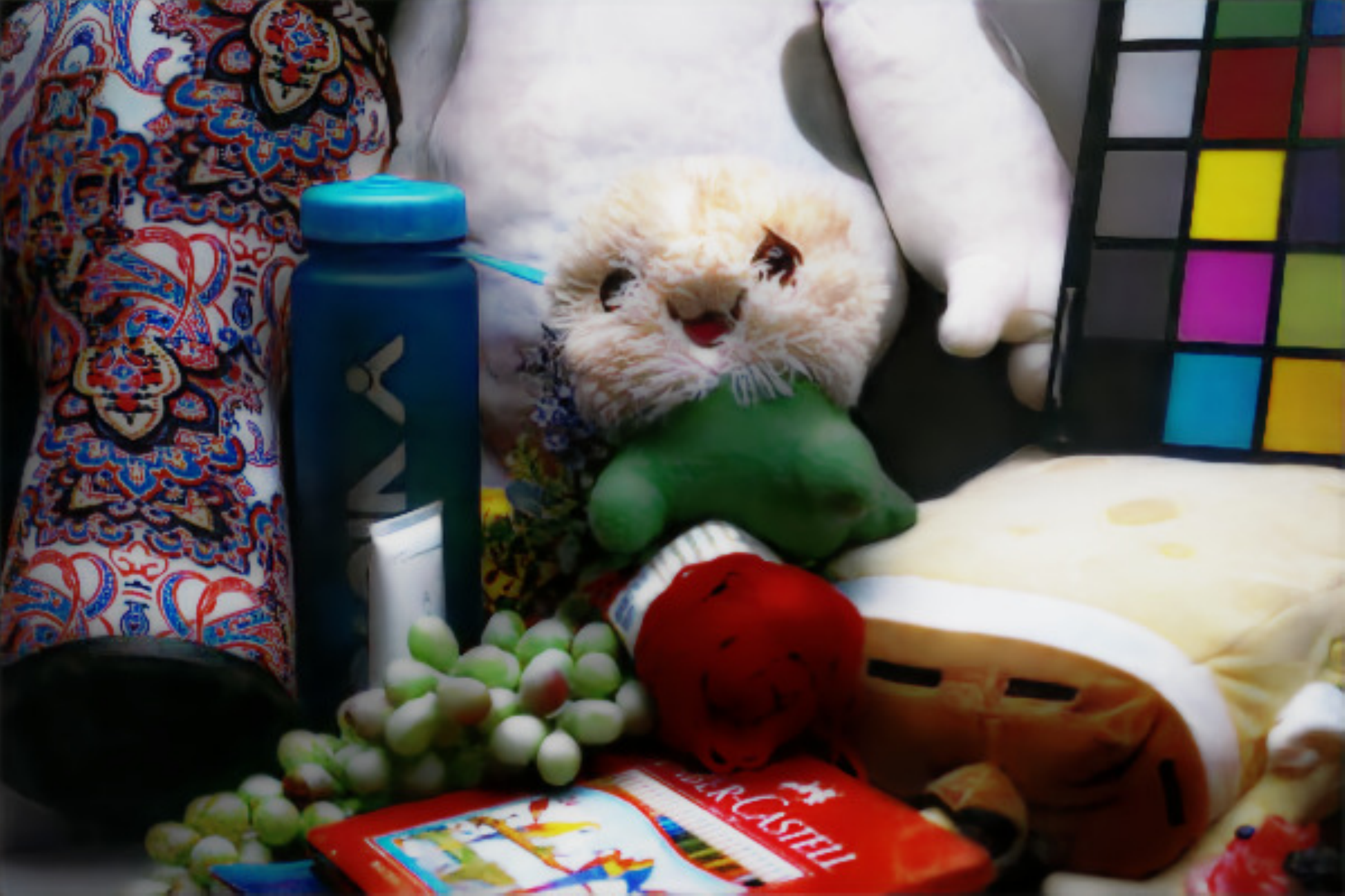}\vspace{2pt}
			\includegraphics[width=2.5cm]{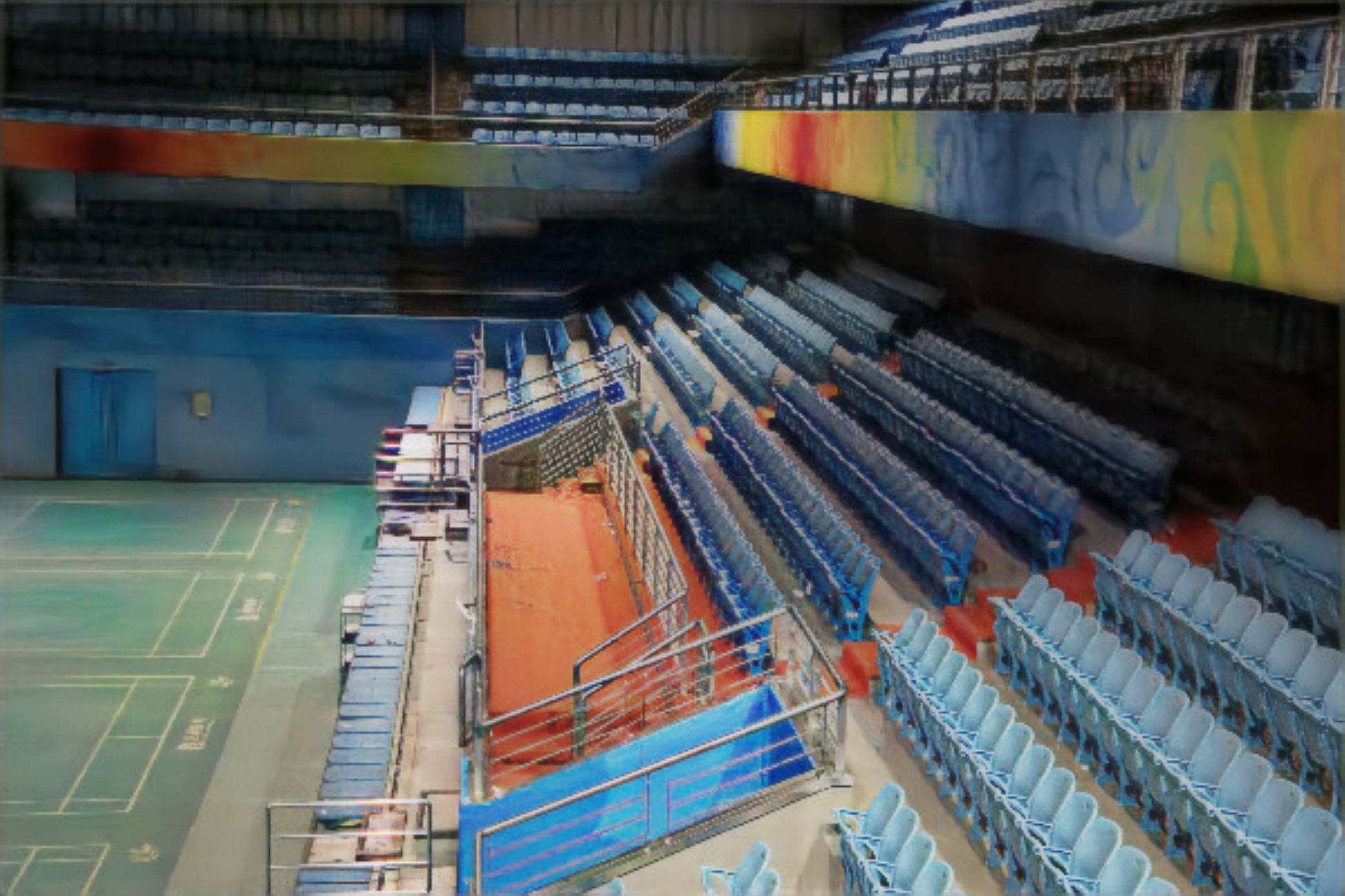}
		\end{minipage}
	}\hspace{-5pt}
	\subfigure[KinD++\cite{zhang2021beyond}]{
		\begin{minipage}[b]{0.13\textwidth}
			\includegraphics[width=2.5cm]{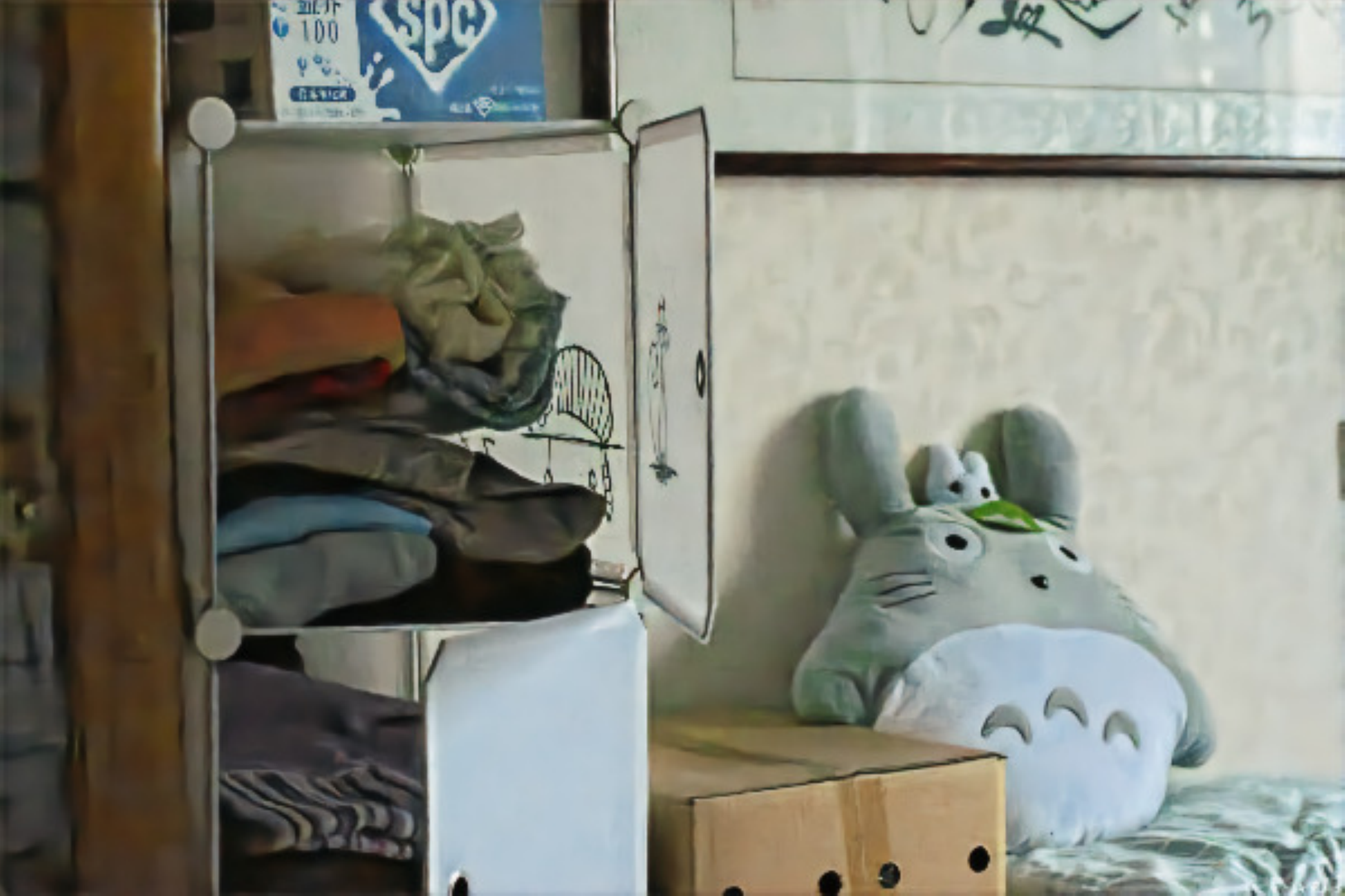}\vspace{2pt} \\
			\includegraphics[width=2.5cm]{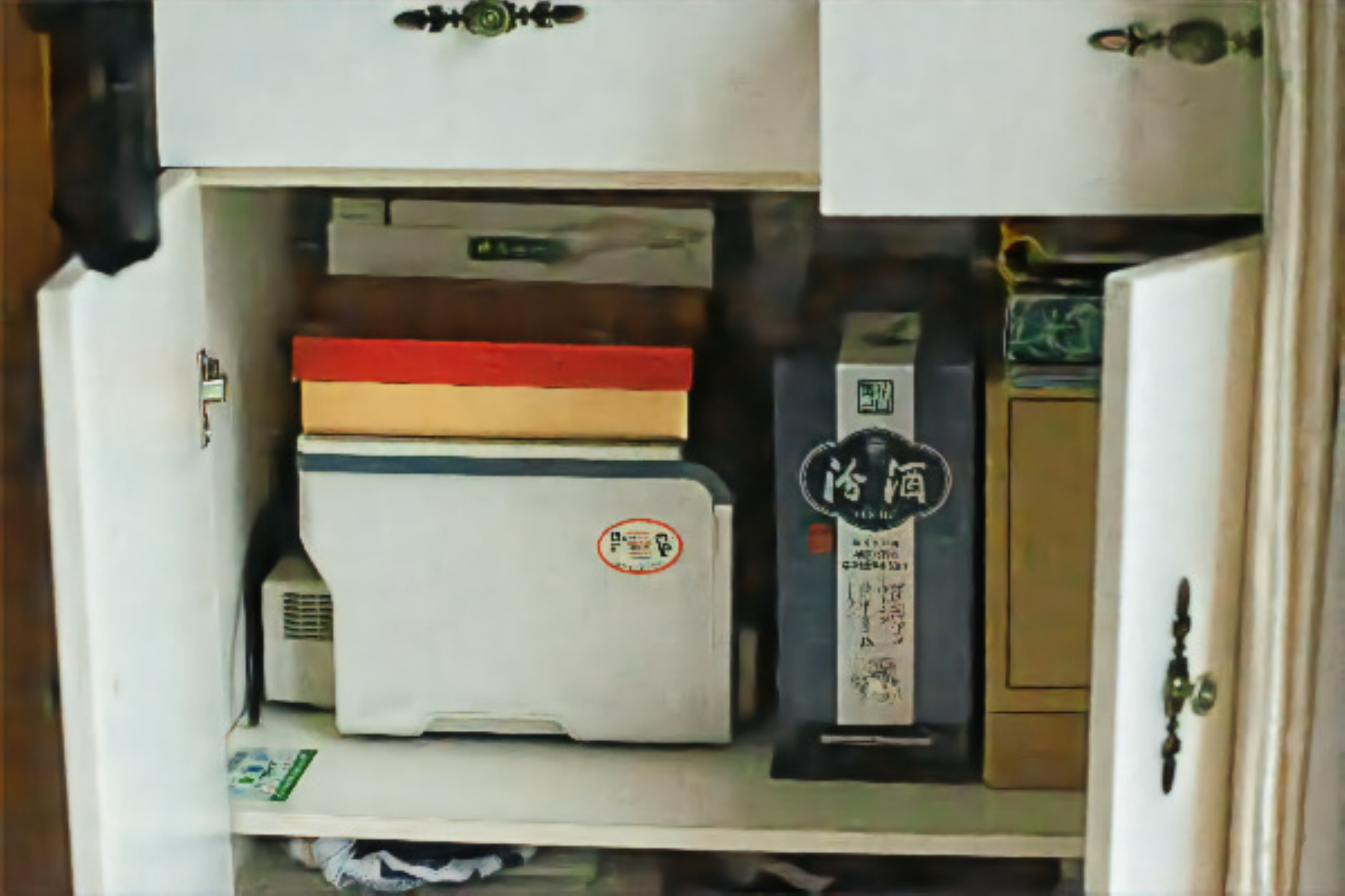}\vspace{2pt}
			\includegraphics[width=2.5cm]{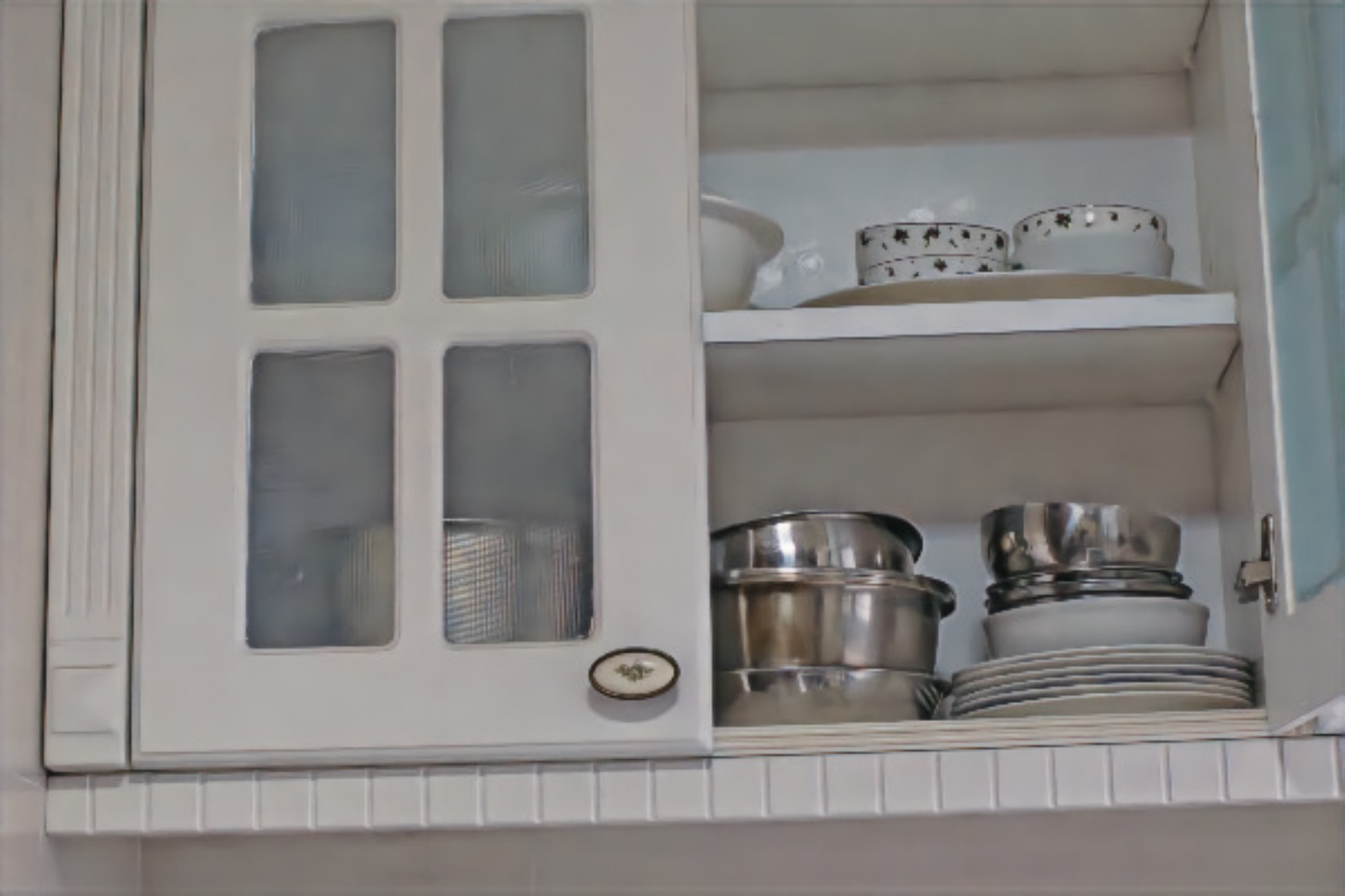}\vspace{2pt}
			\includegraphics[width=2.5cm]{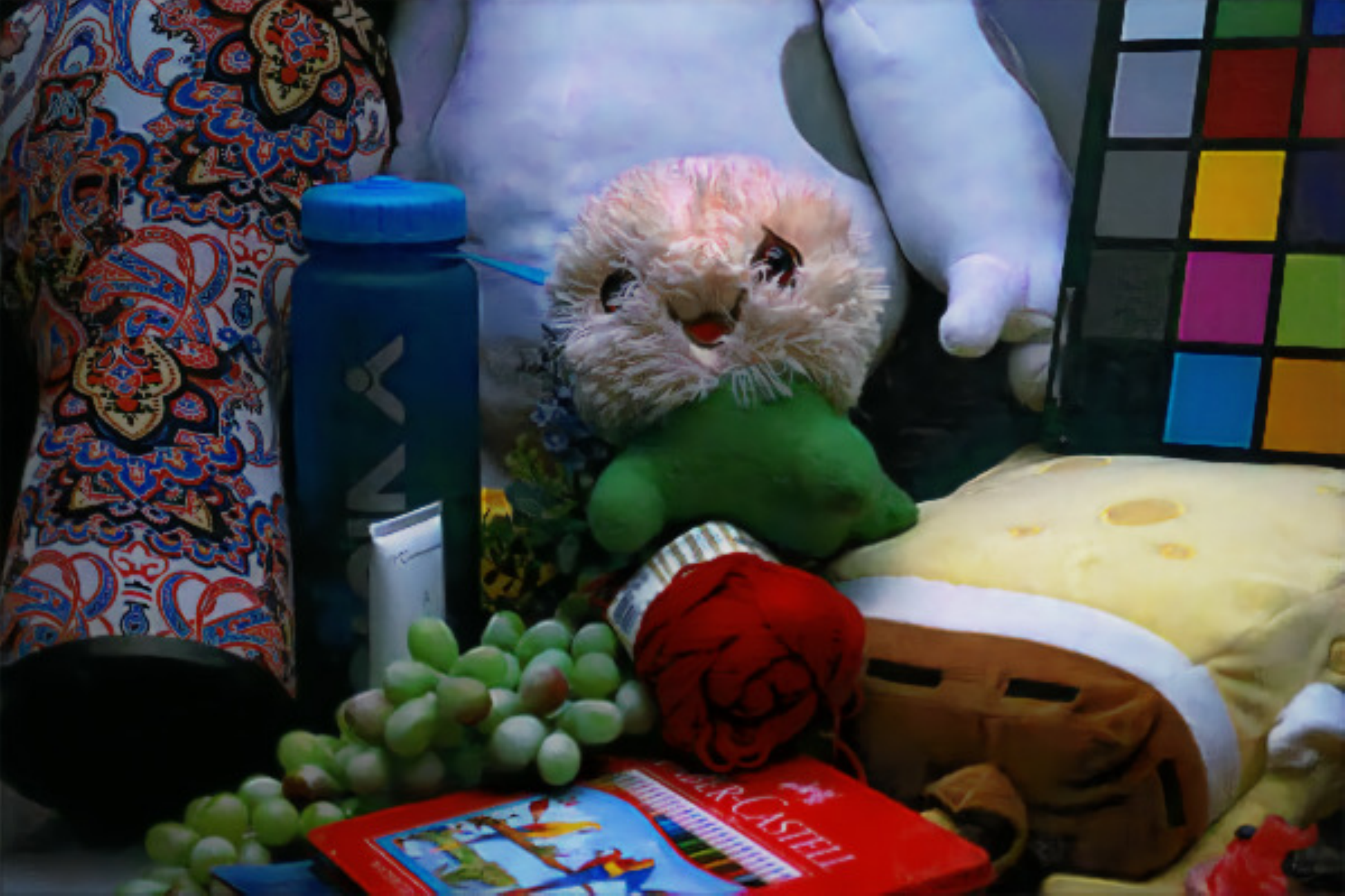}\vspace{2pt}
			\includegraphics[width=2.5cm]{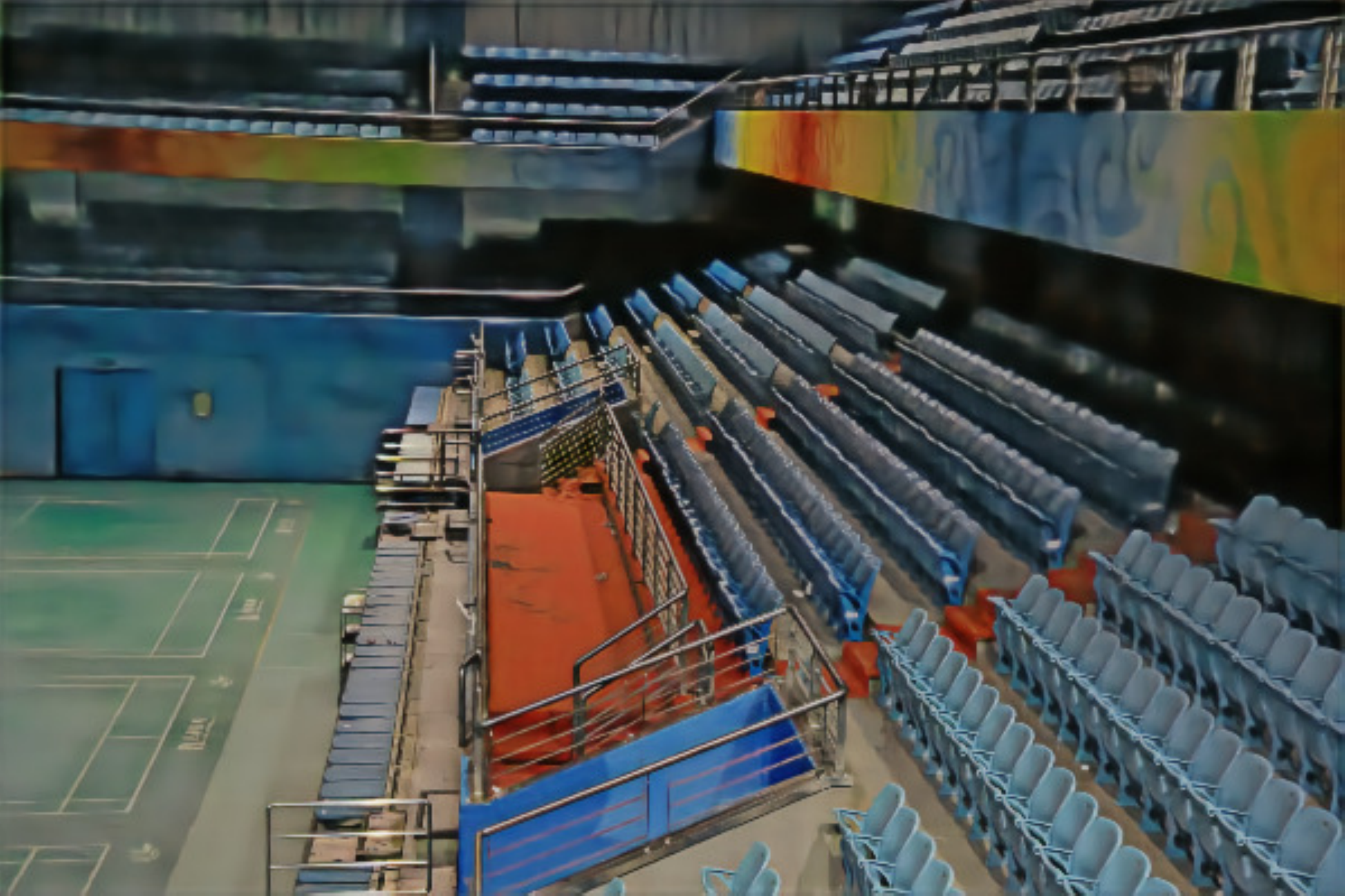}
		\end{minipage}
	}\hspace{-5pt}
	\subfigure[TSN-CA]{
		\begin{minipage}[b]{0.13\textwidth}
			\includegraphics[width=2.5cm]{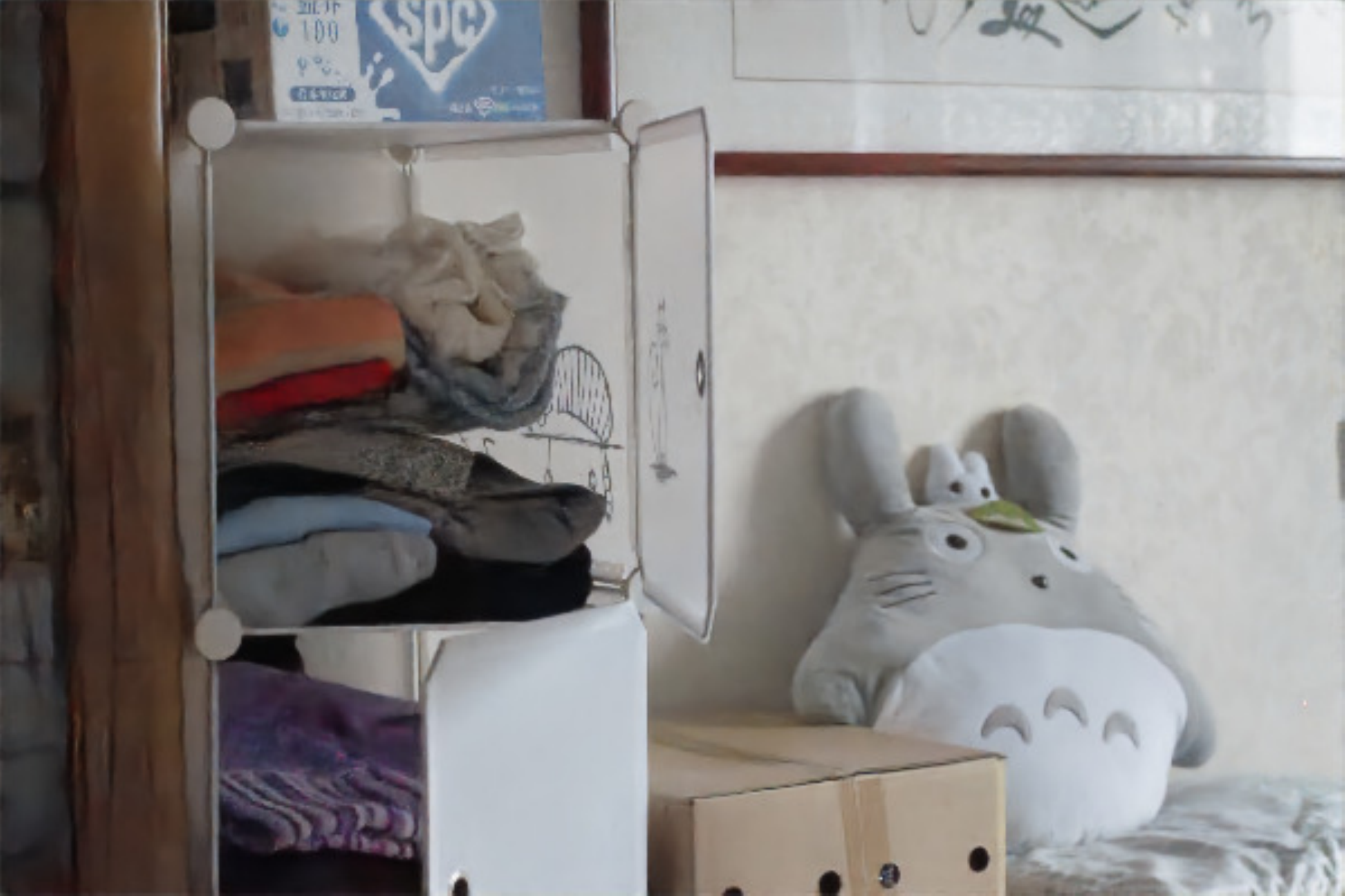}\vspace{2pt} \\
			\includegraphics[width=2.5cm]{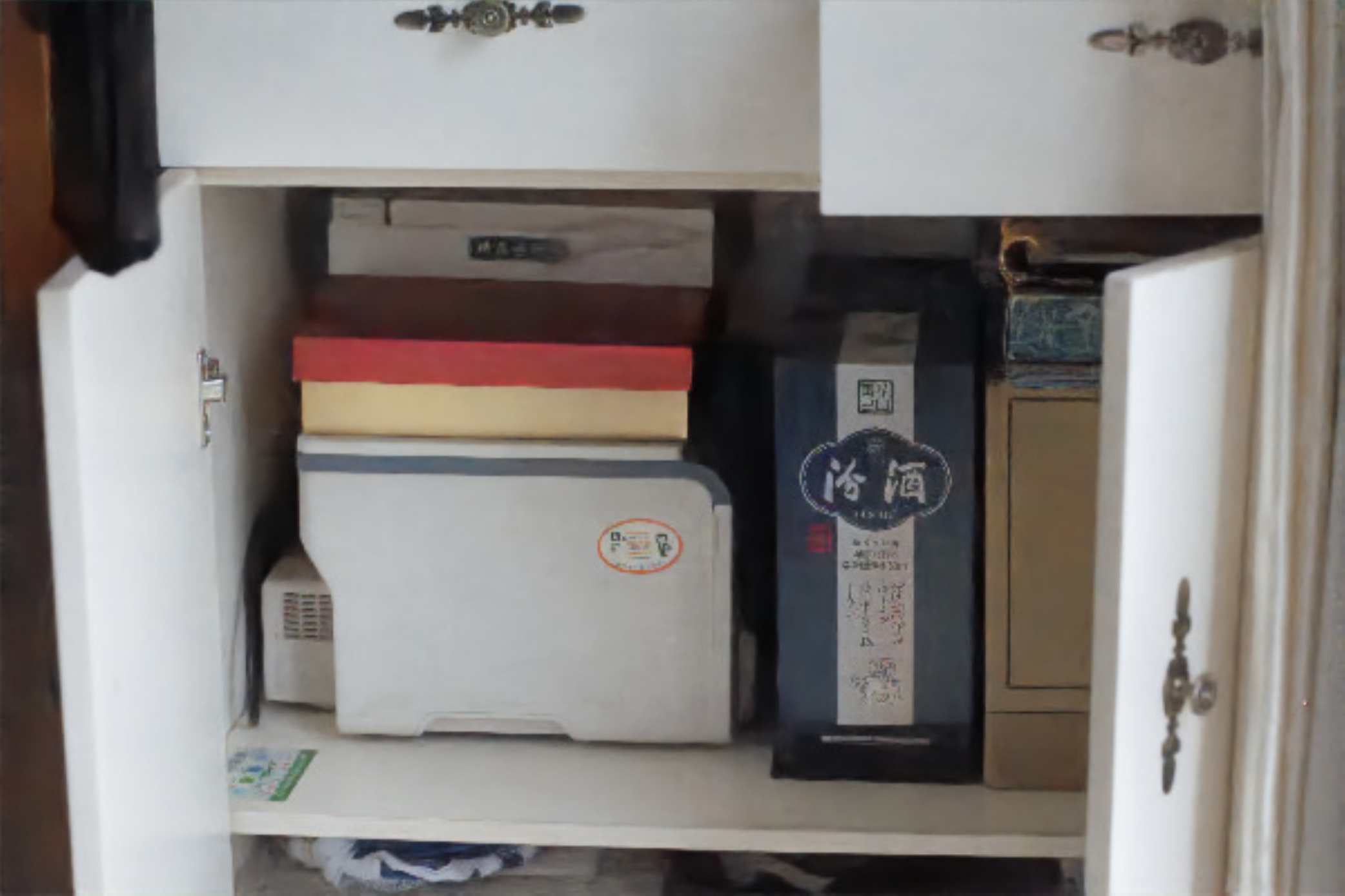}\vspace{2pt}
			\includegraphics[width=2.5cm]{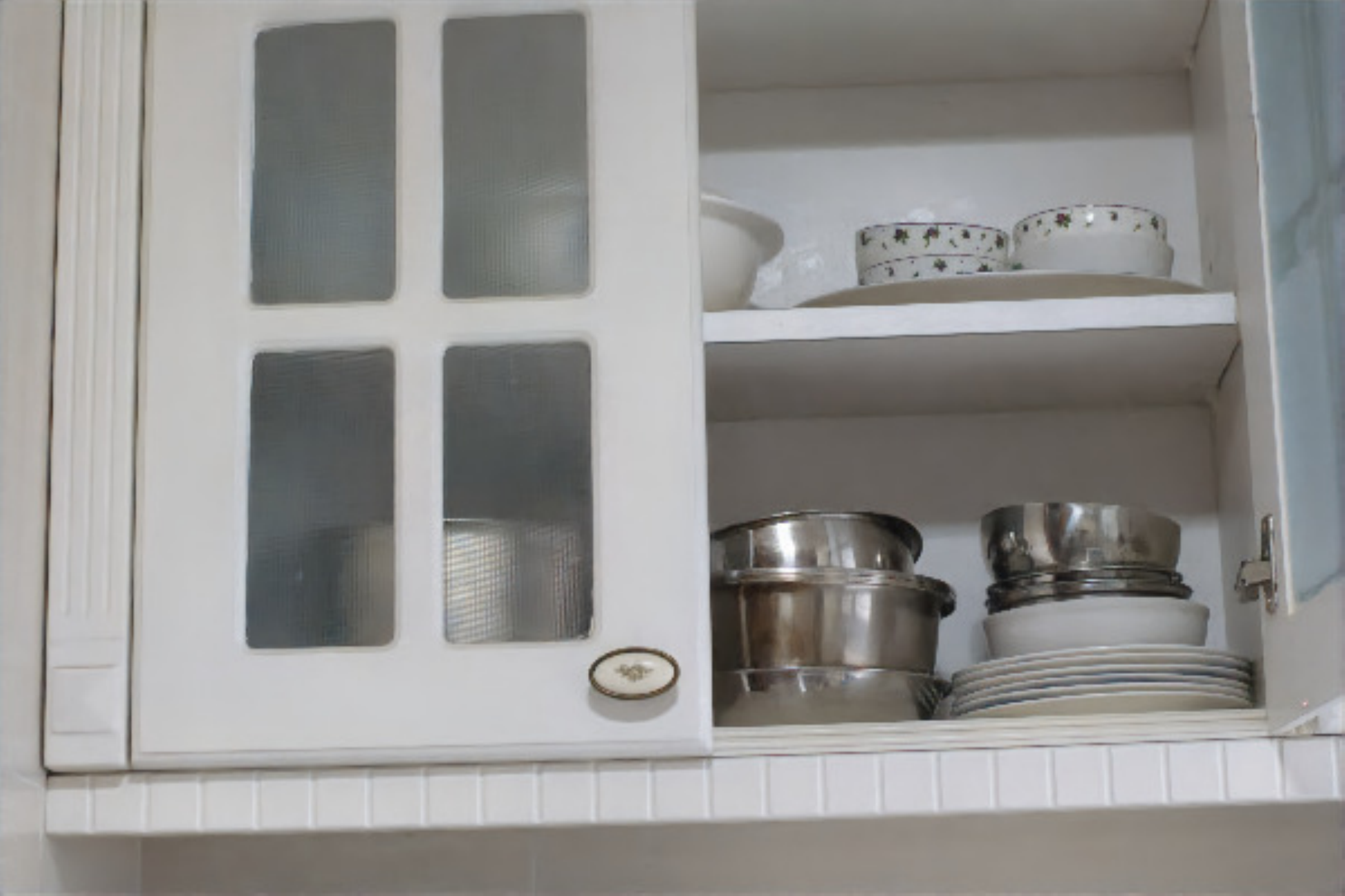}\vspace{2pt}
			\includegraphics[width=2.5cm]{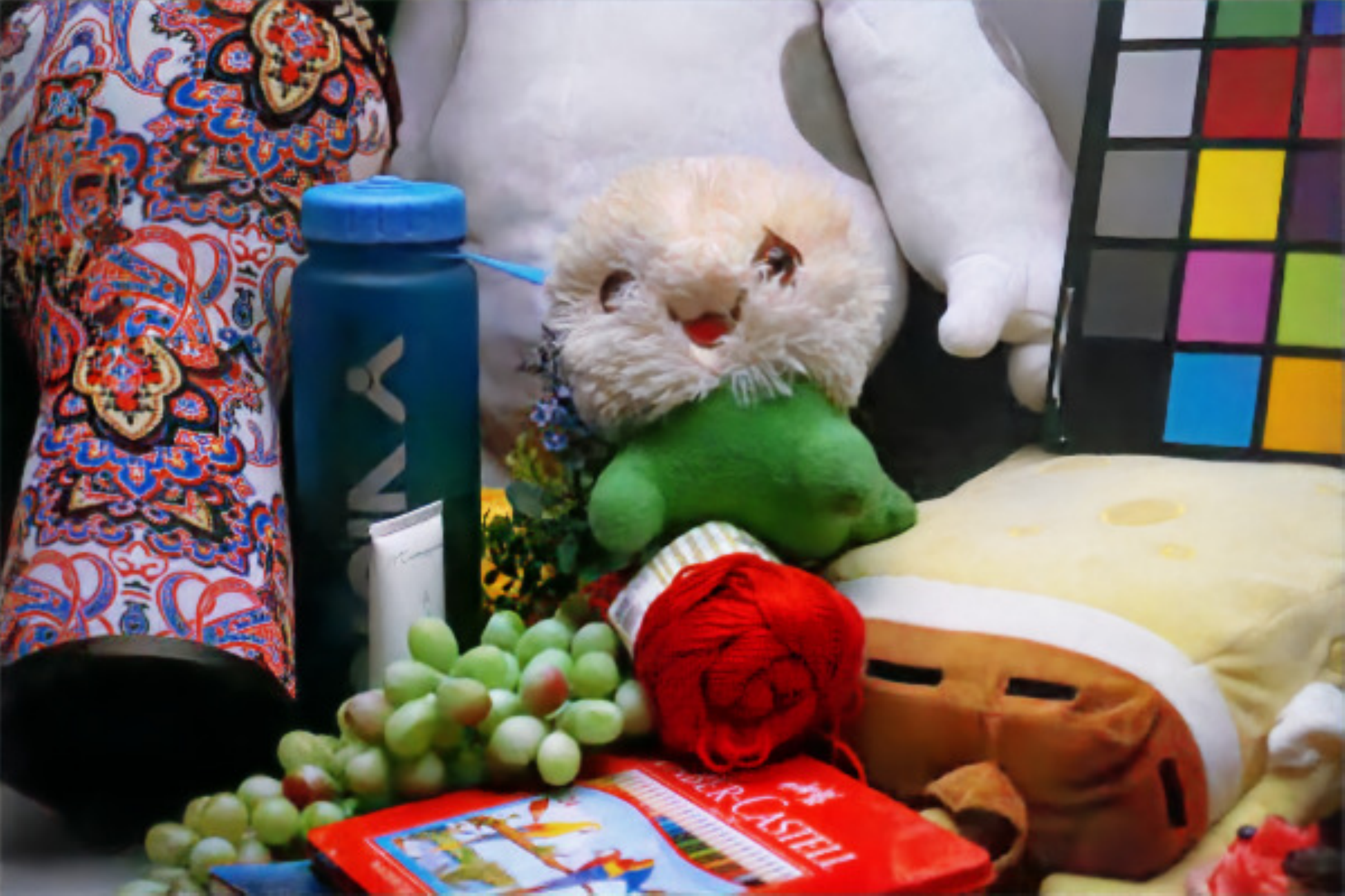}\vspace{2pt}
			\includegraphics[width=2.5cm]{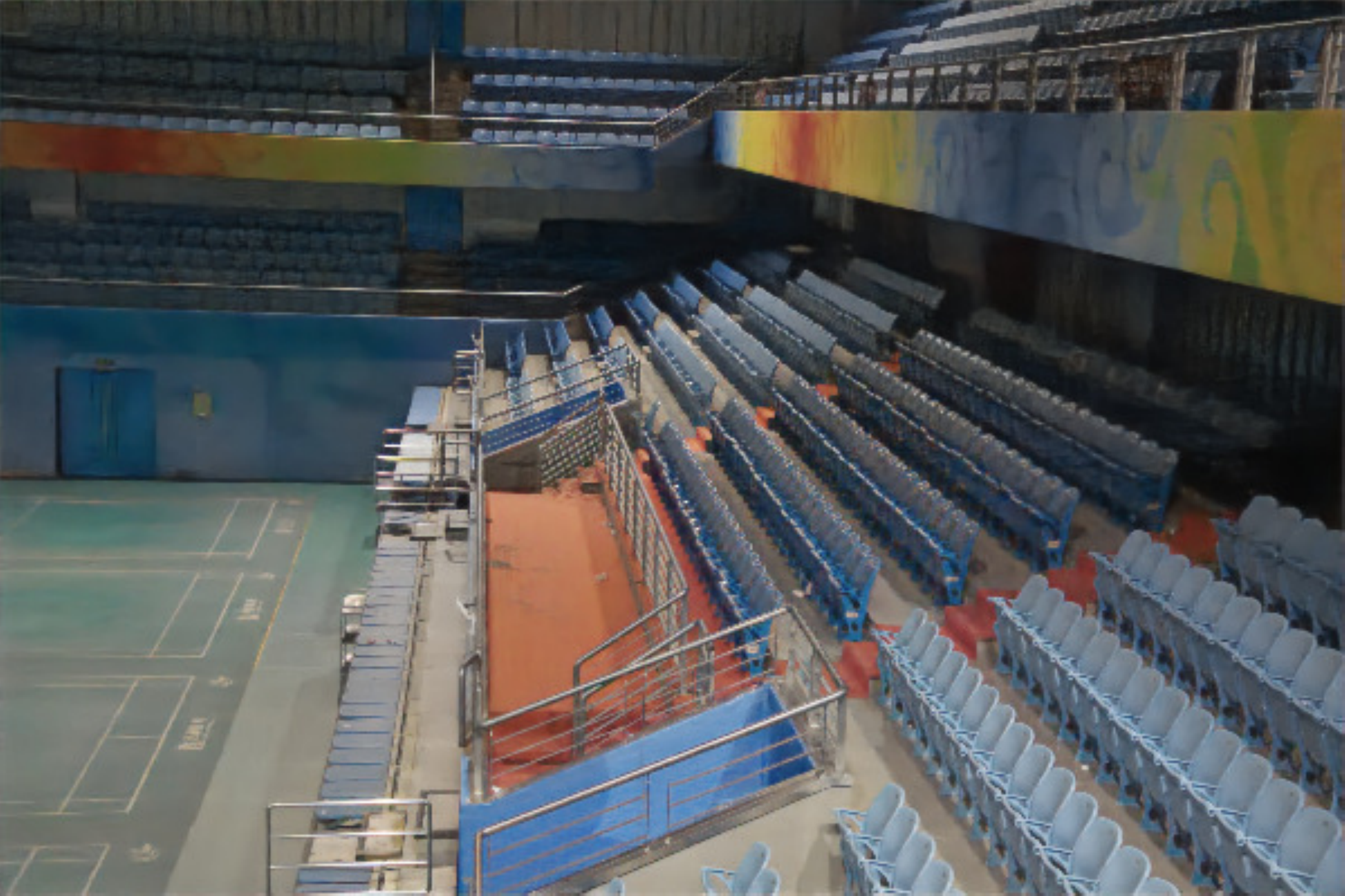}
		\end{minipage}
	}\hspace{-5pt}
	\subfigure[Ground-Truth]{
		\begin{minipage}[b]{0.13\textwidth}
			\includegraphics[width=2.5cm]{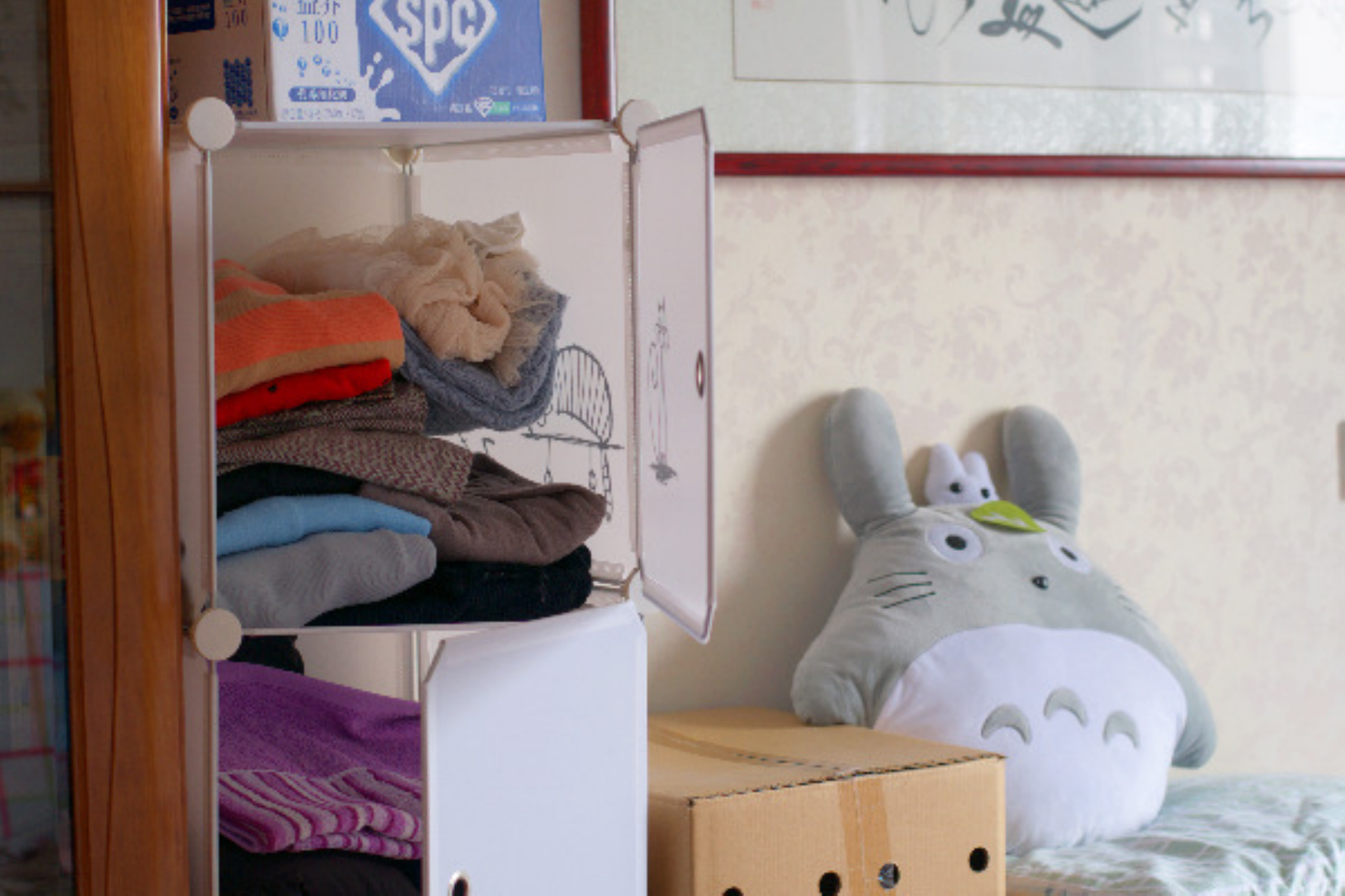}\vspace{2pt} \\
			\includegraphics[width=2.5cm]{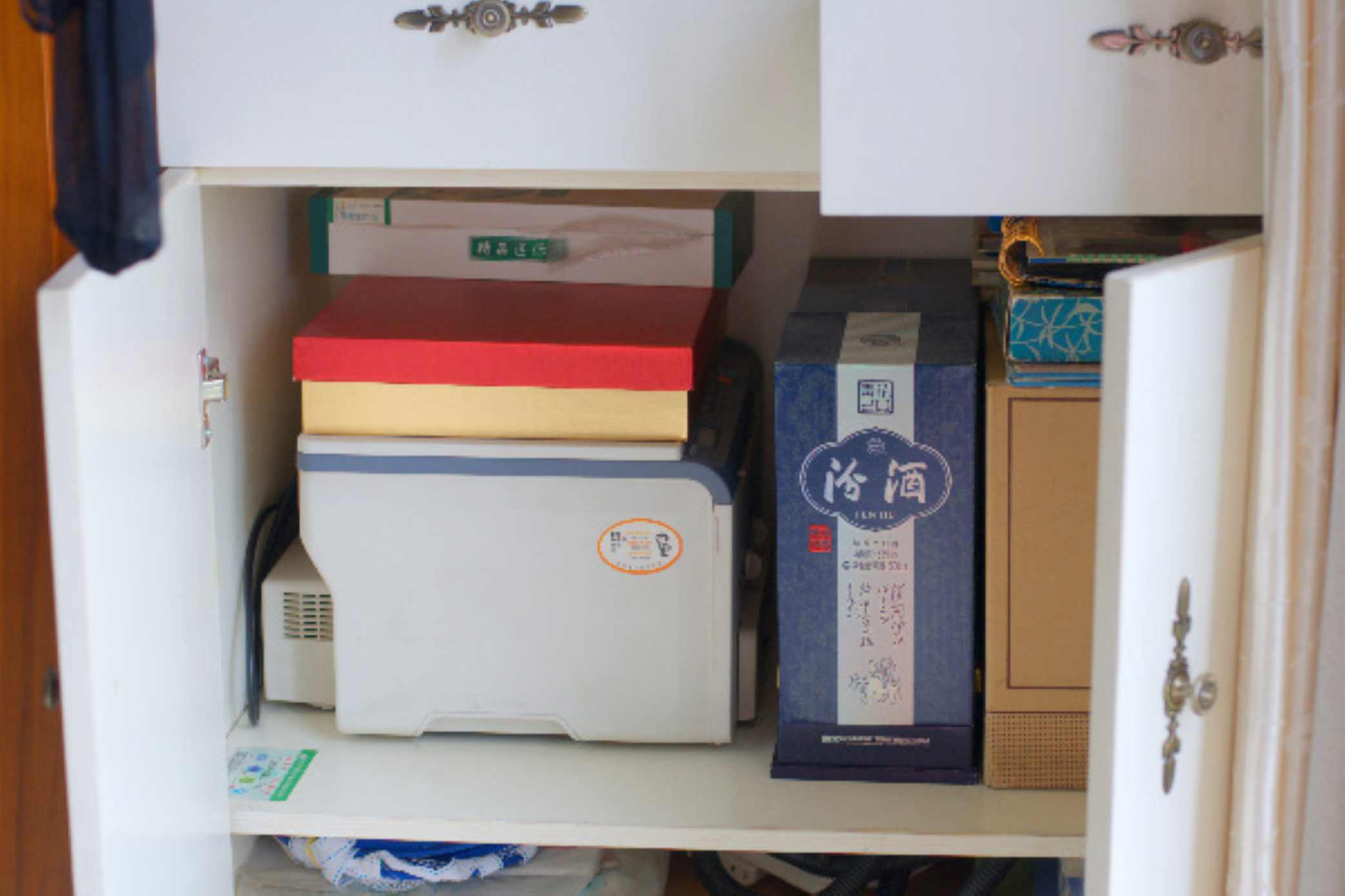}\vspace{2pt}
			\includegraphics[width=2.5cm]{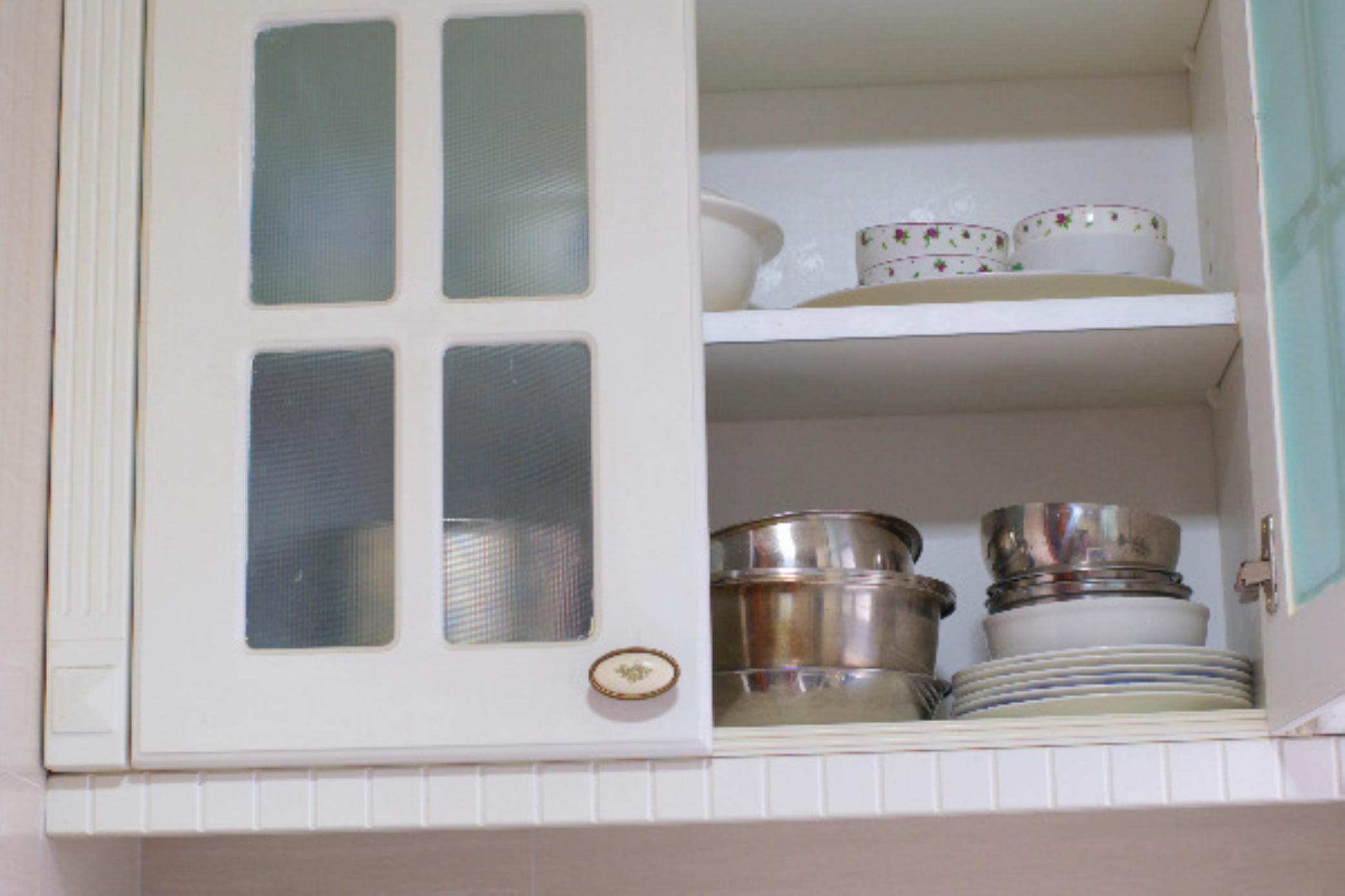}\vspace{2pt}
			\includegraphics[width=2.5cm]{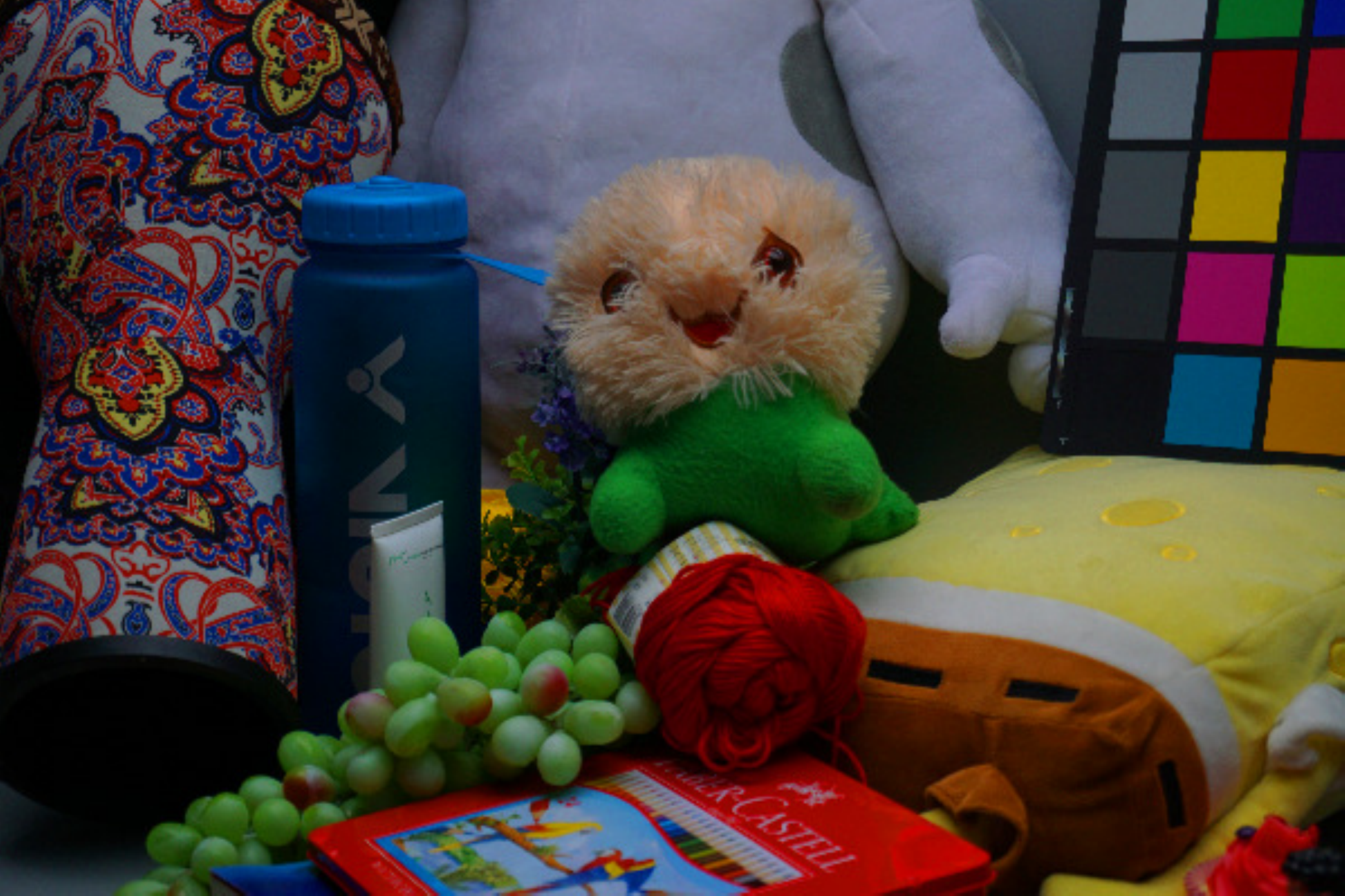}\vspace{2pt}
			\includegraphics[width=2.5cm]{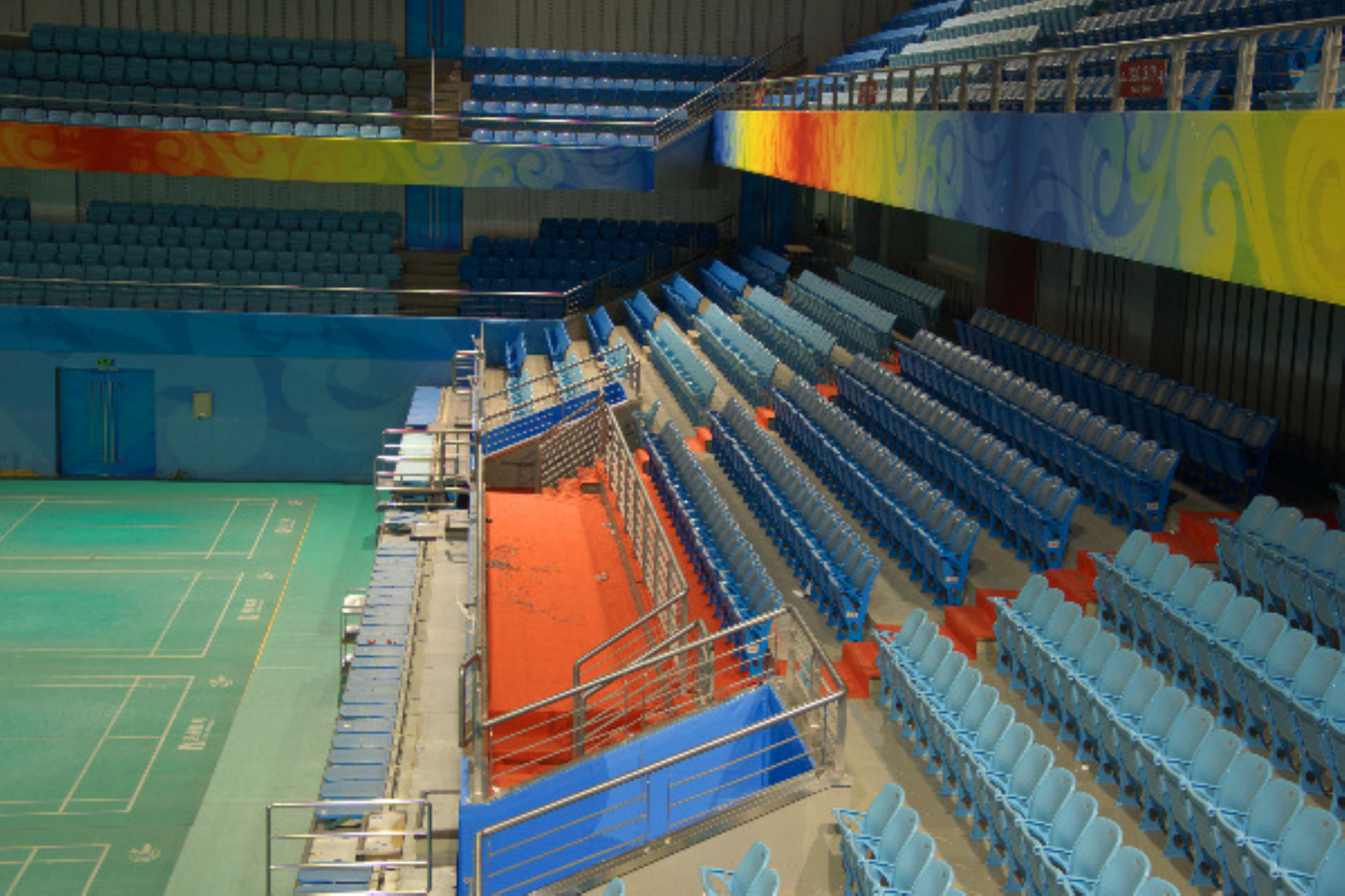}
		\end{minipage}
	}
	\caption{Visual comparison with other state-of-the-art methods on the LOL real-world validation dataset.}
	\label{lol}
\end{figure*}

\begin{figure*}
	\centering
	
	
	\subfigure[Input]{
		\begin{minipage}[b]{0.13\textwidth}
			\includegraphics[width=2.5cm]{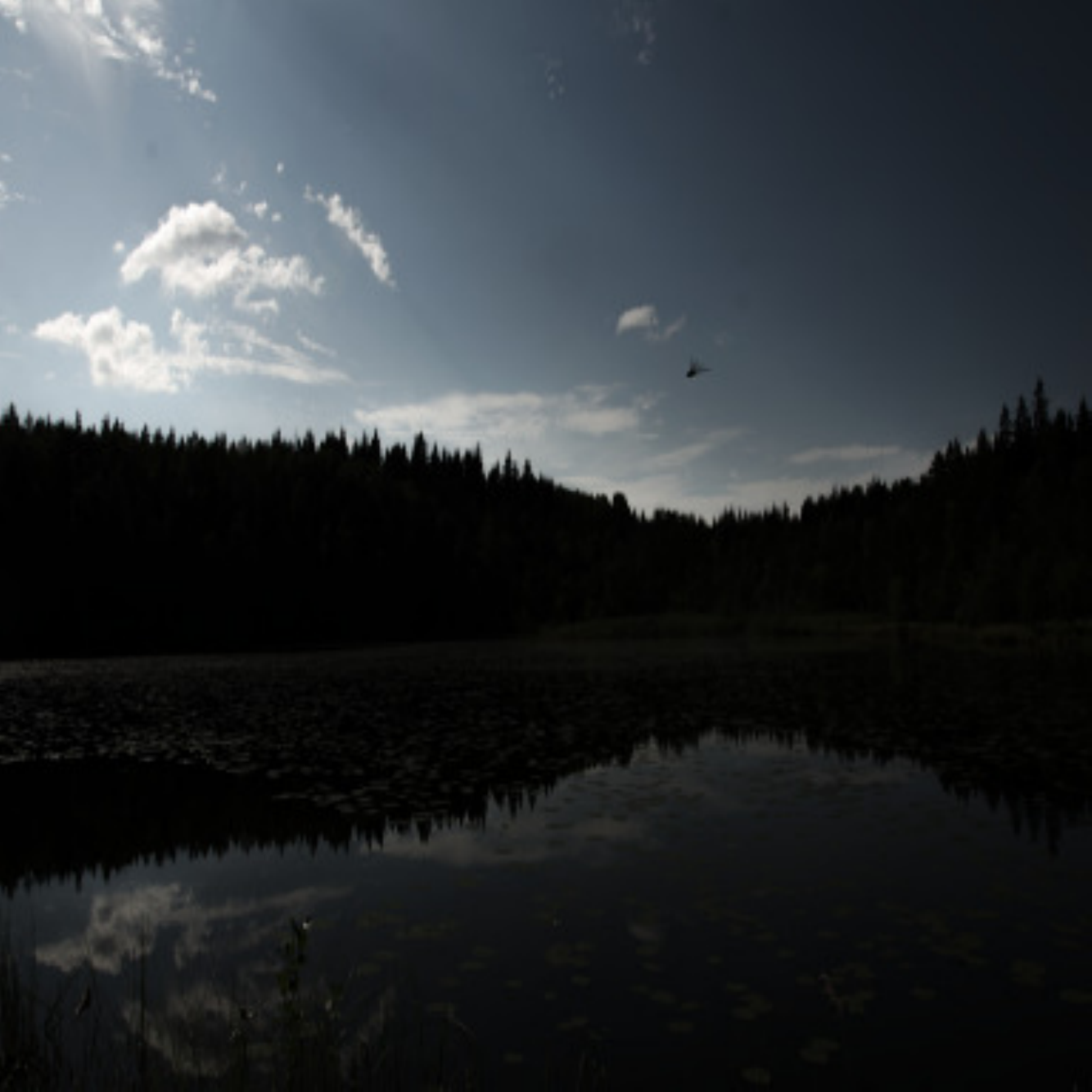}\vspace{2pt} \\
			\includegraphics[width=2.5cm]{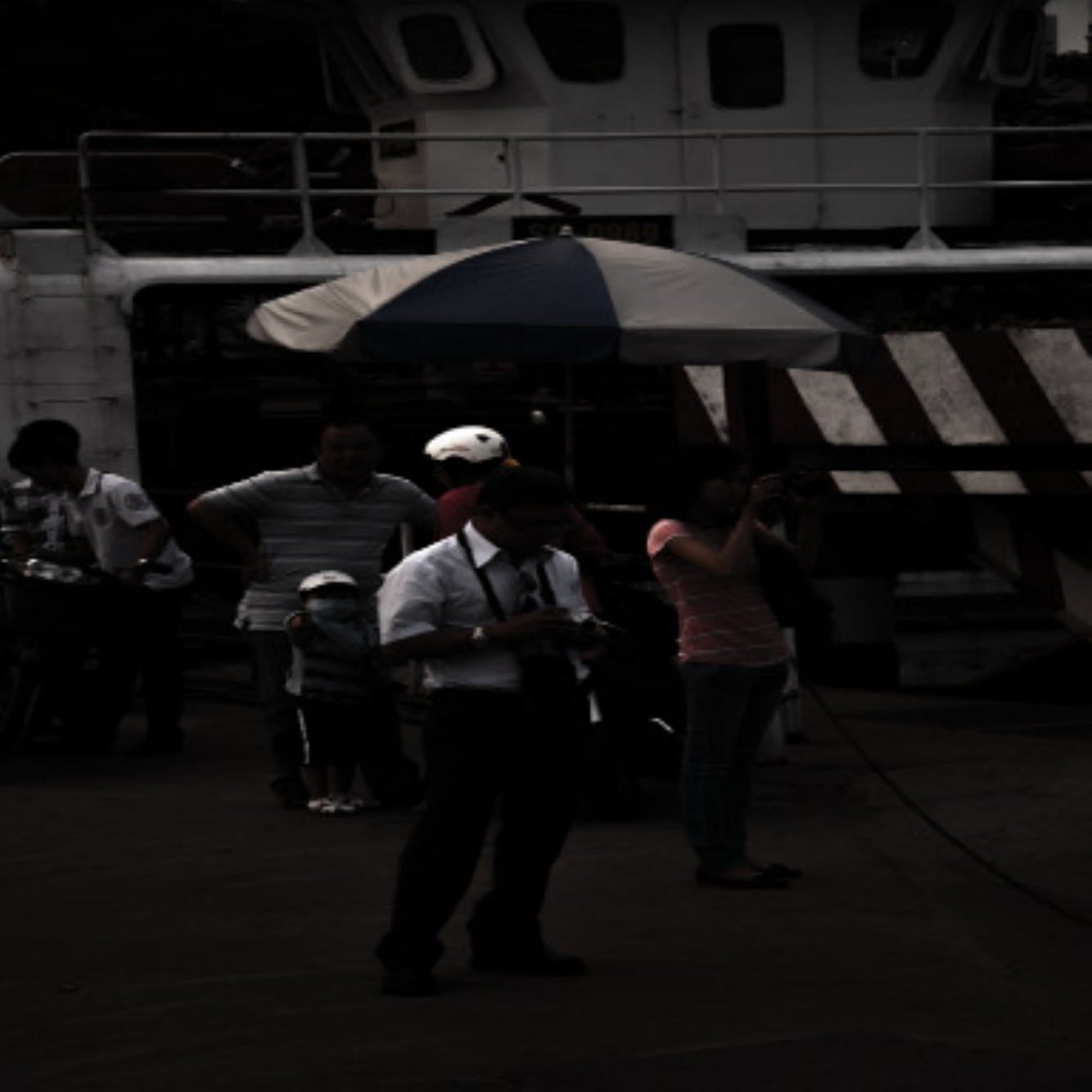}\vspace{2pt}
			\includegraphics[width=2.5cm]{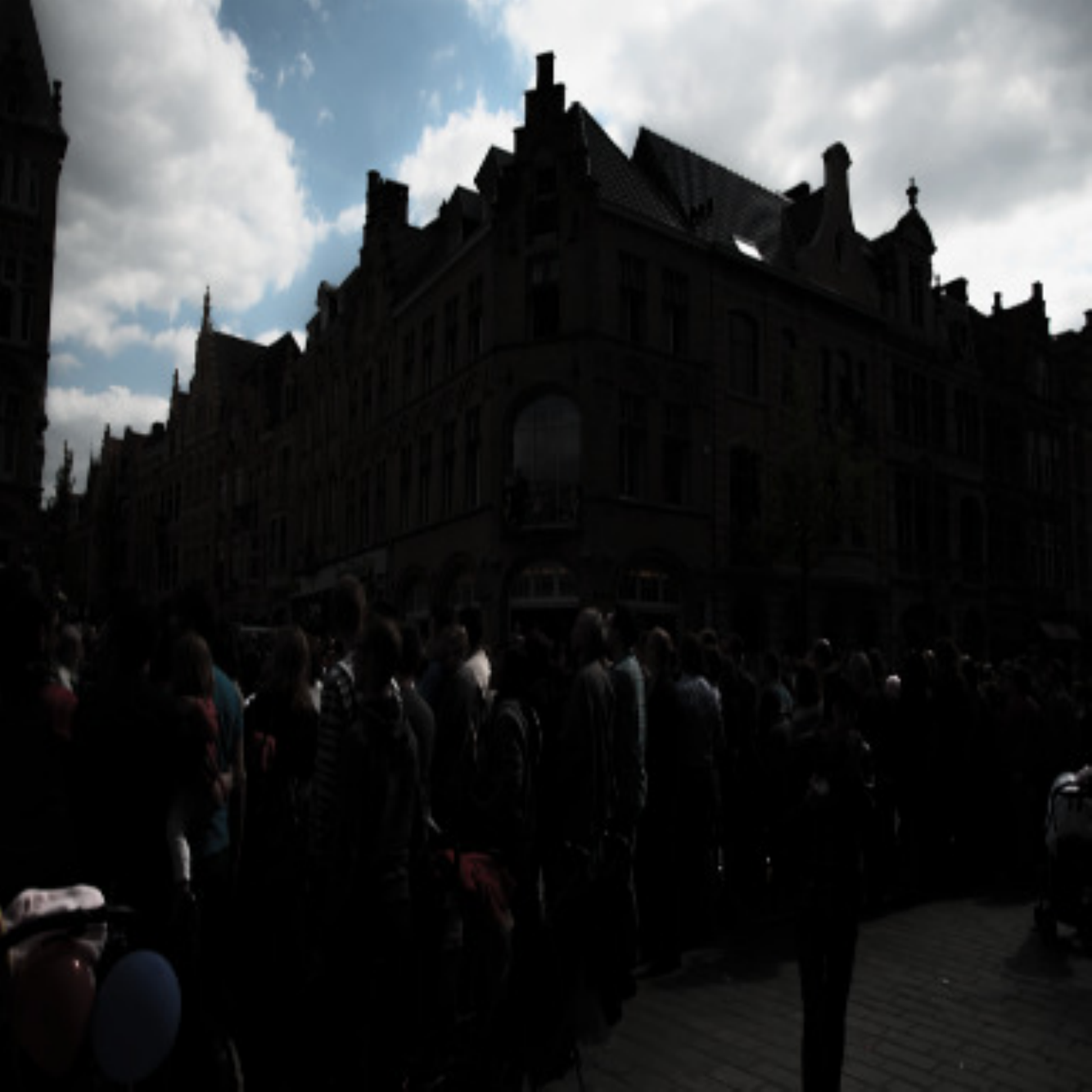}\vspace{2pt}
			\includegraphics[width=2.5cm]{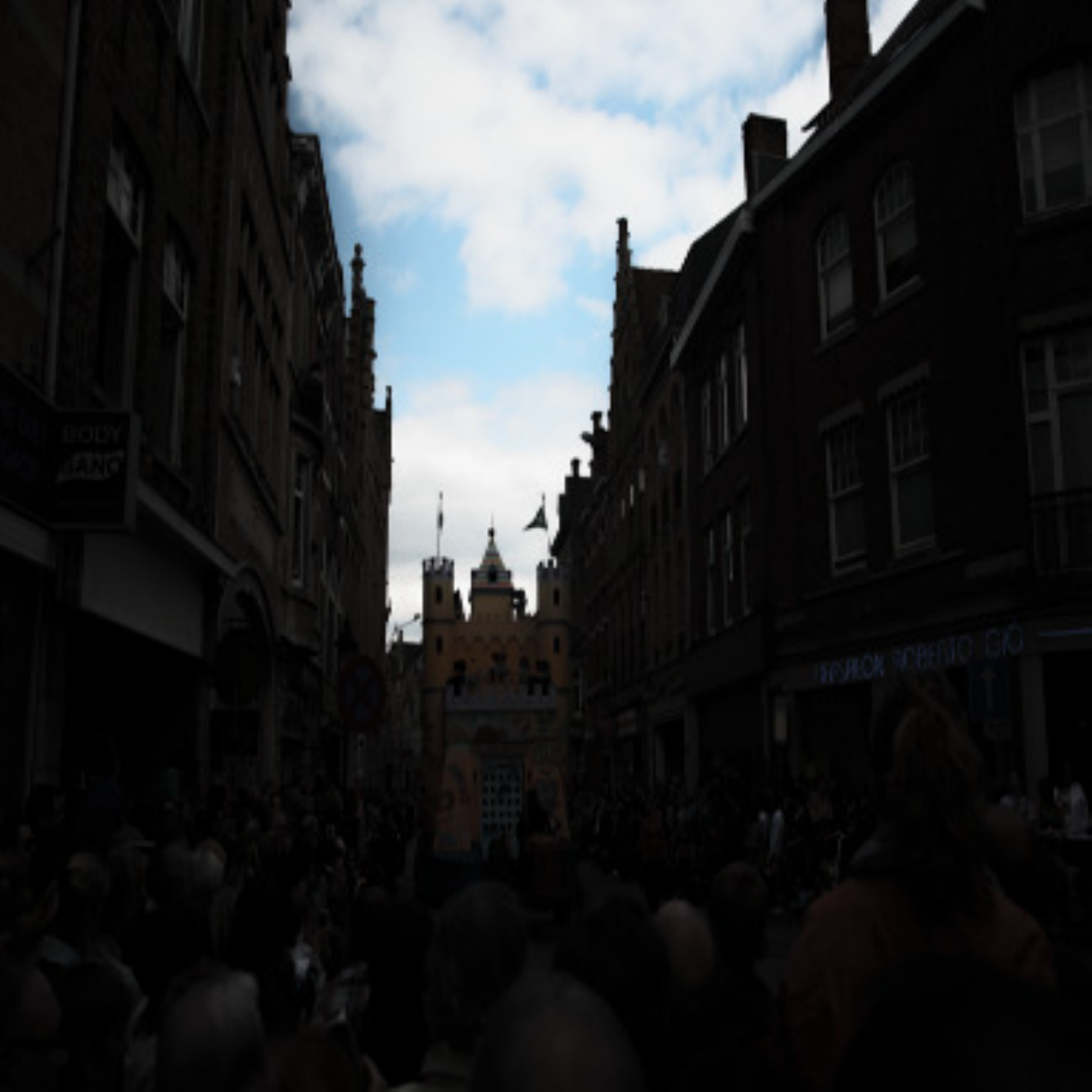}\vspace{2pt} 
		\end{minipage}
	}\hspace{-5pt}
	\subfigure[MBLLEN\cite{lv2018mbllen}]{
		\begin{minipage}[b]{0.13\textwidth}
			\includegraphics[width=2.5cm]{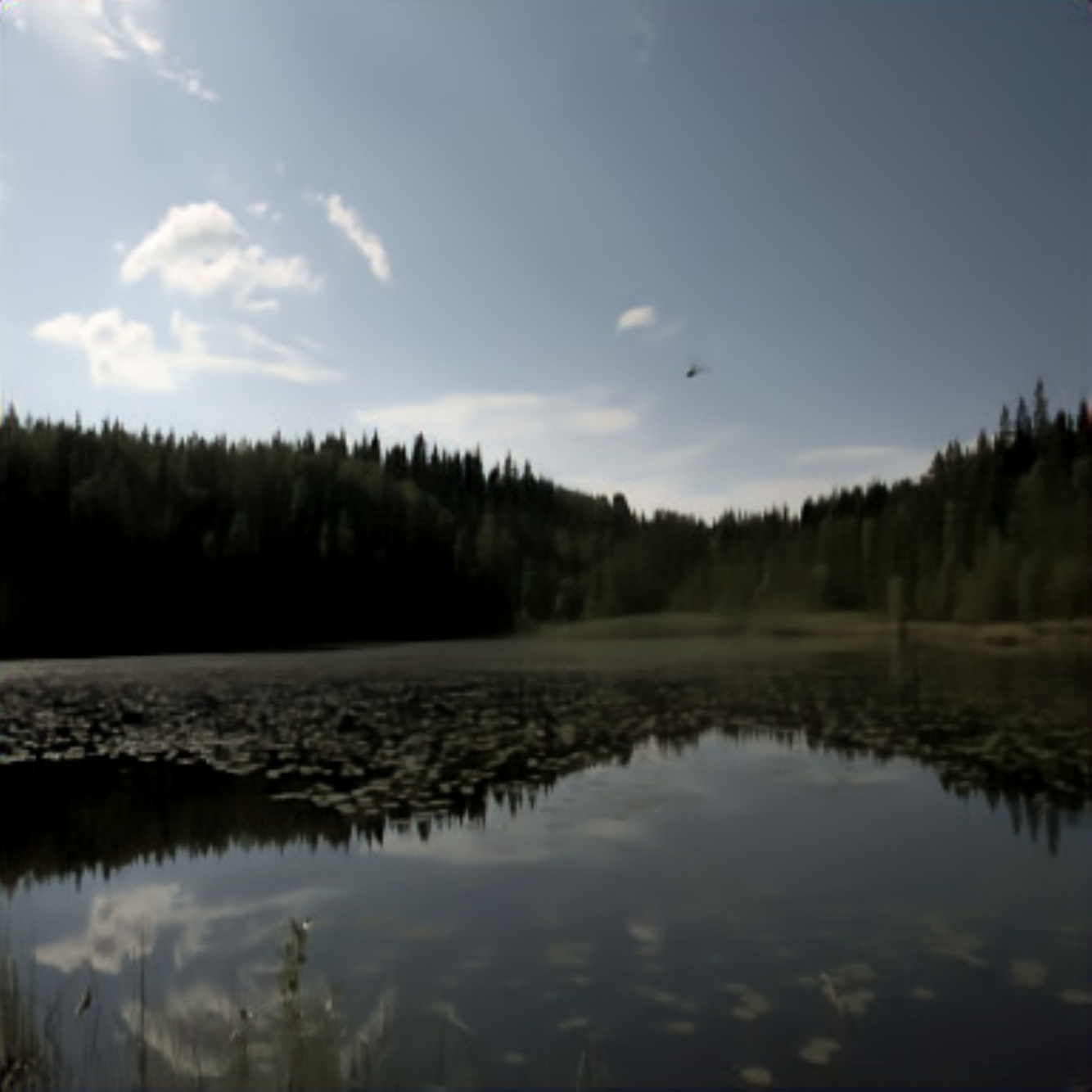}\vspace{2pt} \\
			\includegraphics[width=2.5cm]{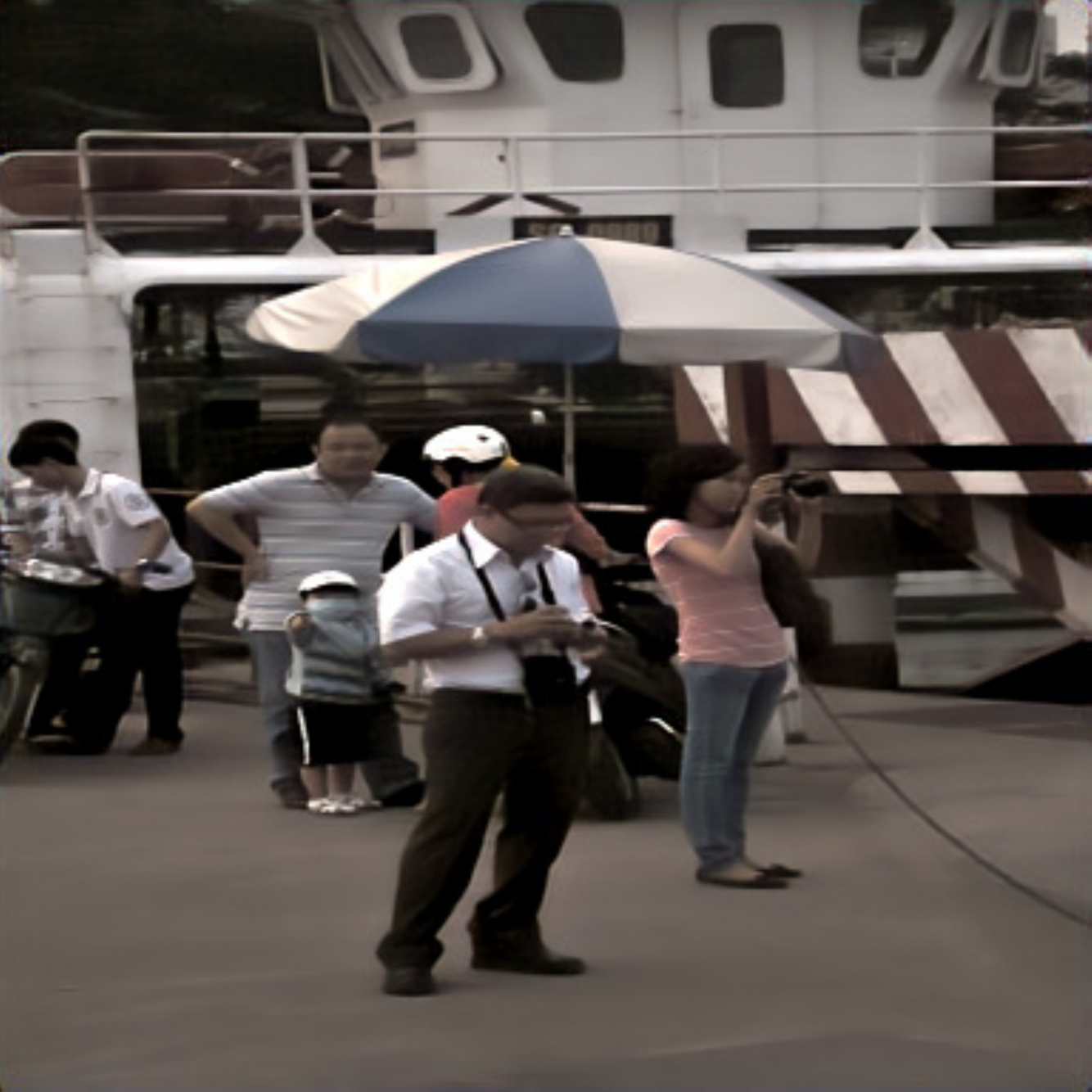}\vspace{2pt}
			\includegraphics[width=2.5cm]{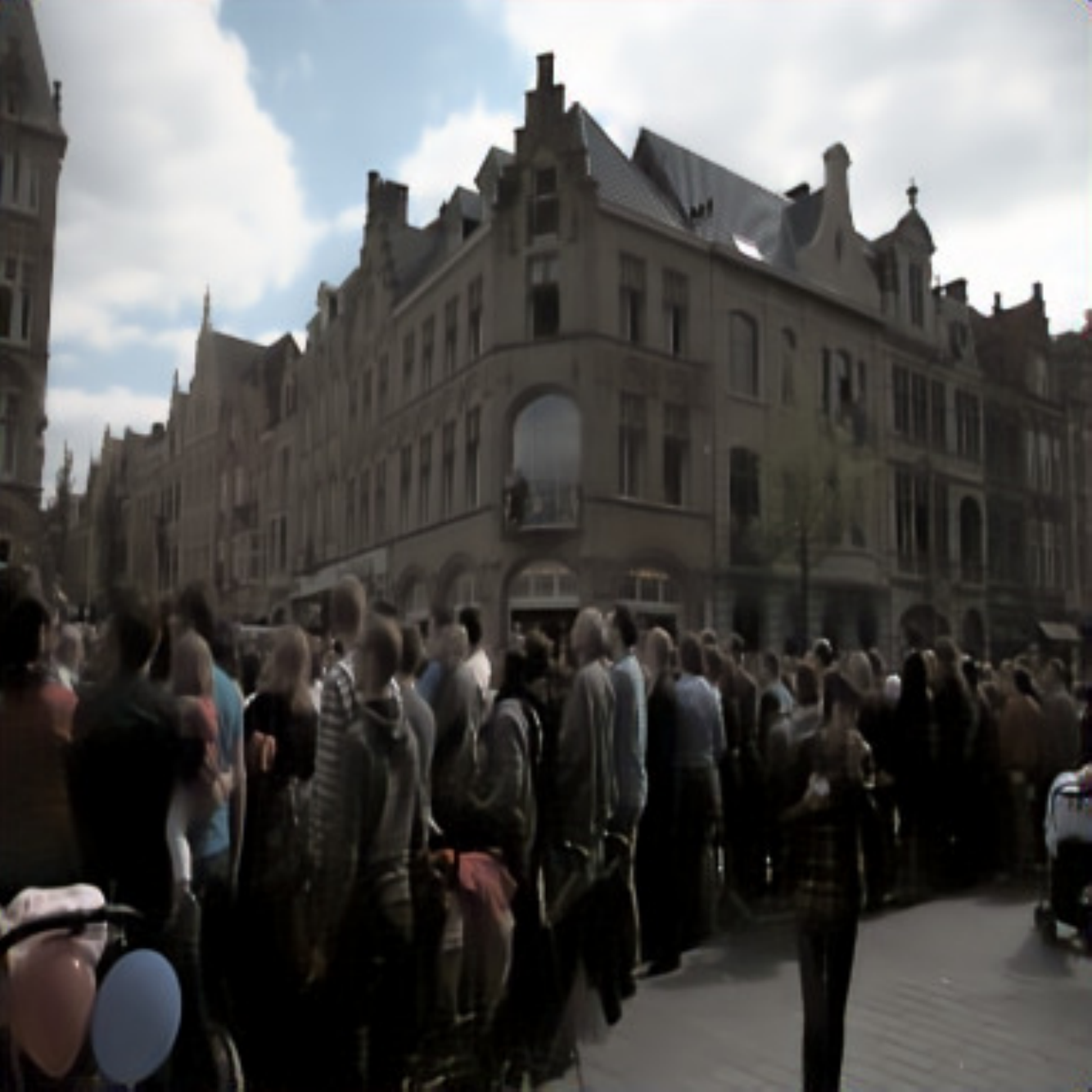}\vspace{2pt}
			\includegraphics[width=2.5cm]{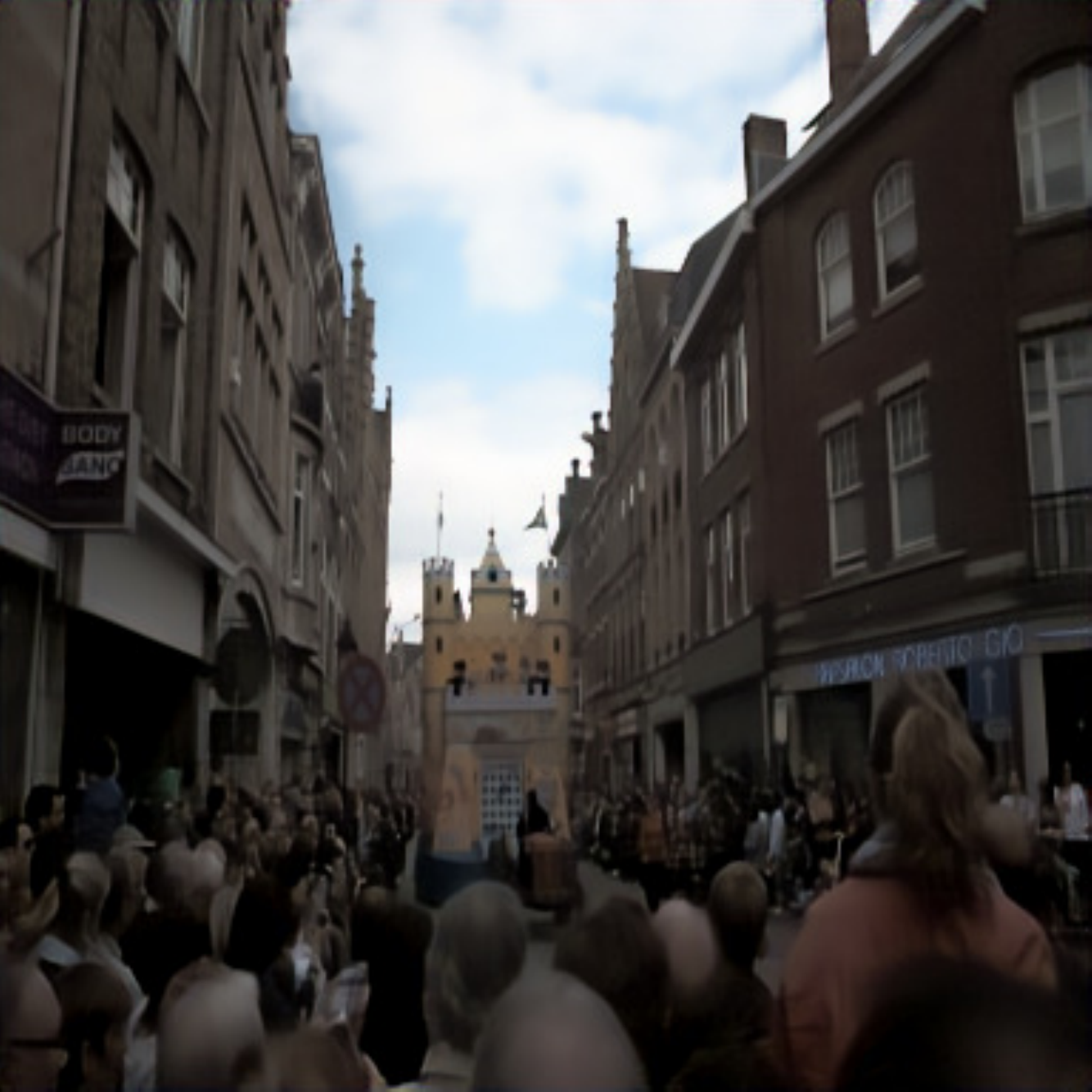}\vspace{2pt} 
		\end{minipage}
	}\hspace{-5pt}
	\subfigure[Zero-DCE\cite{guo2020zero}]{
		\begin{minipage}[b]{0.13\textwidth}
			\includegraphics[width=2.5cm]{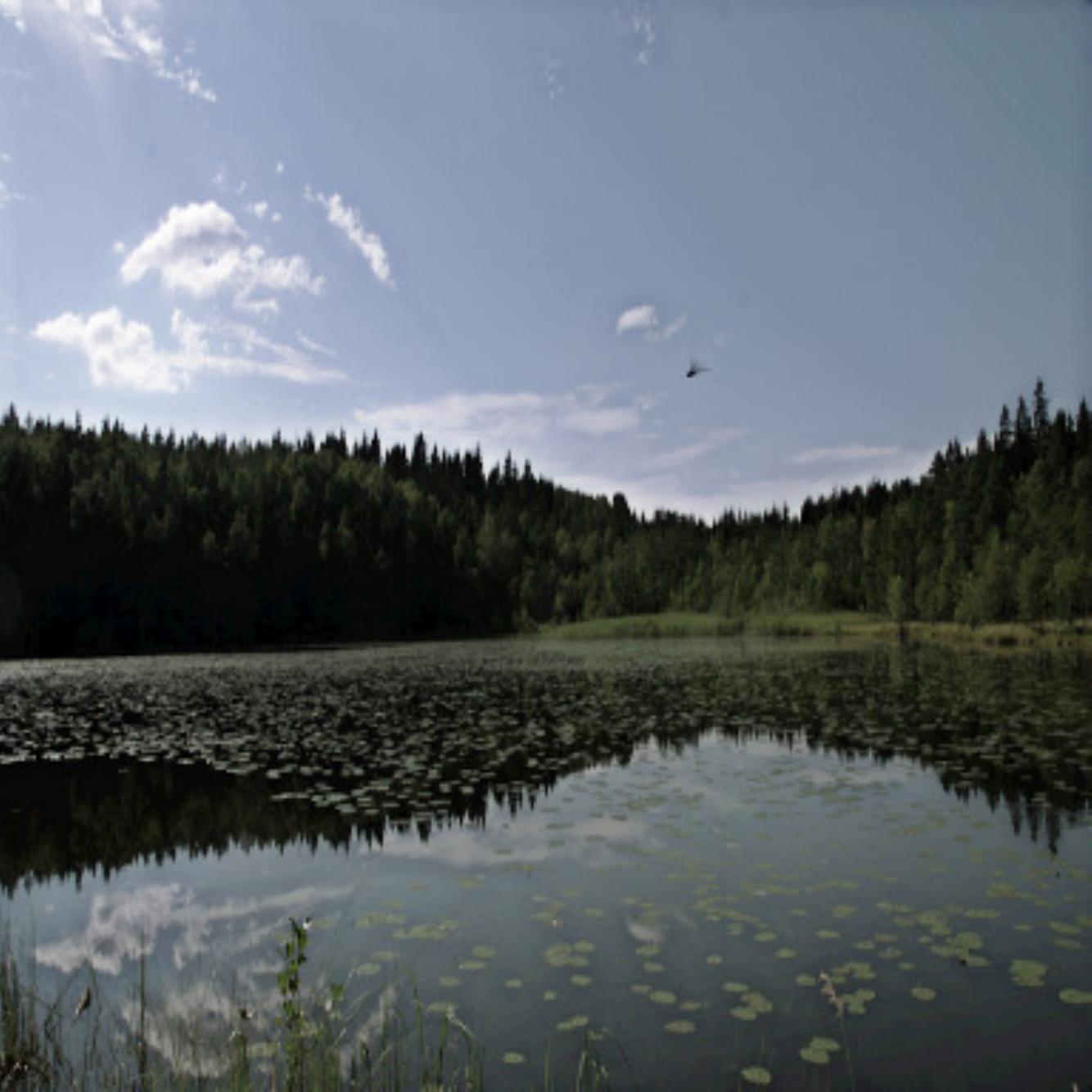}\vspace{2pt} \\
			\includegraphics[width=2.5cm]{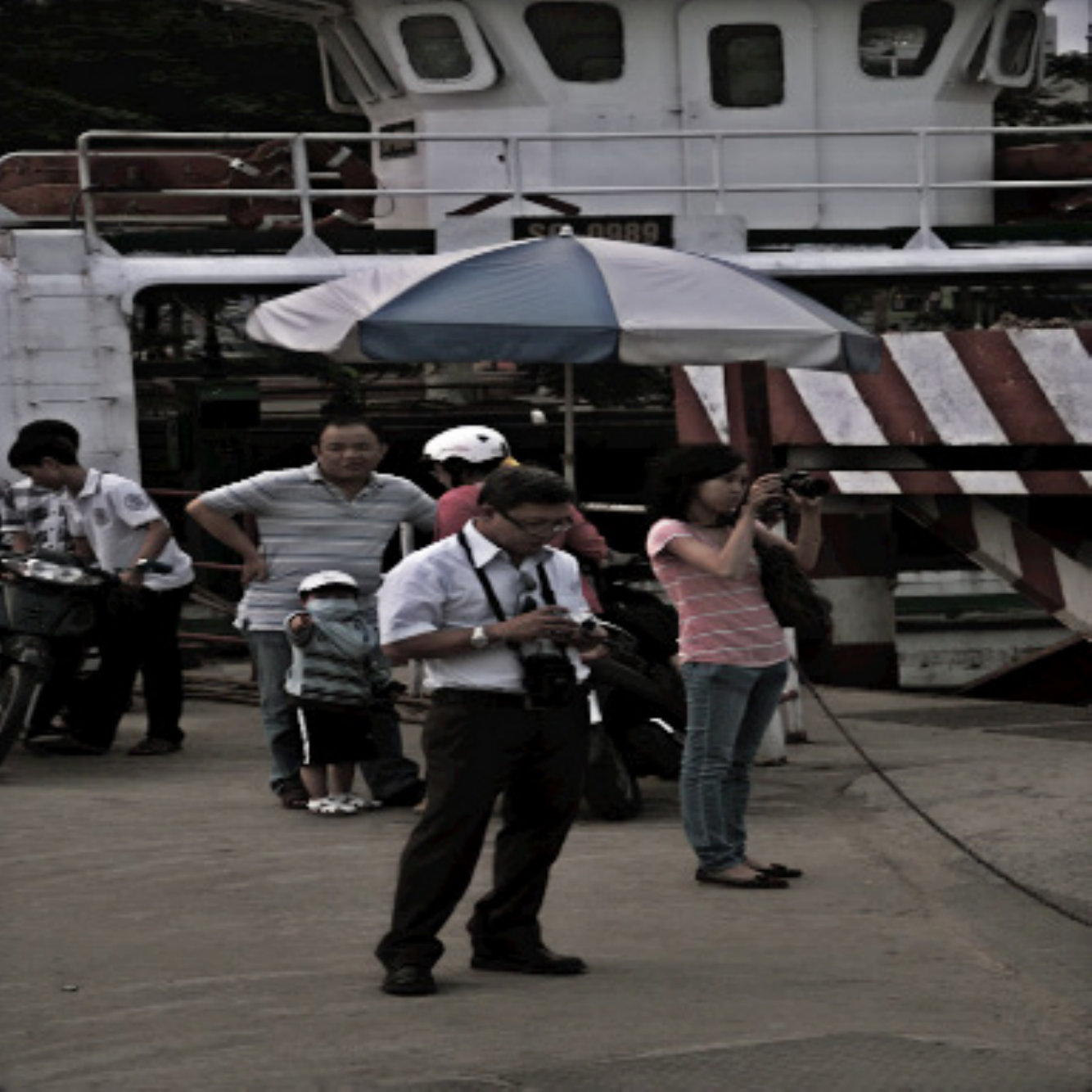}\vspace{2pt}
			\includegraphics[width=2.5cm]{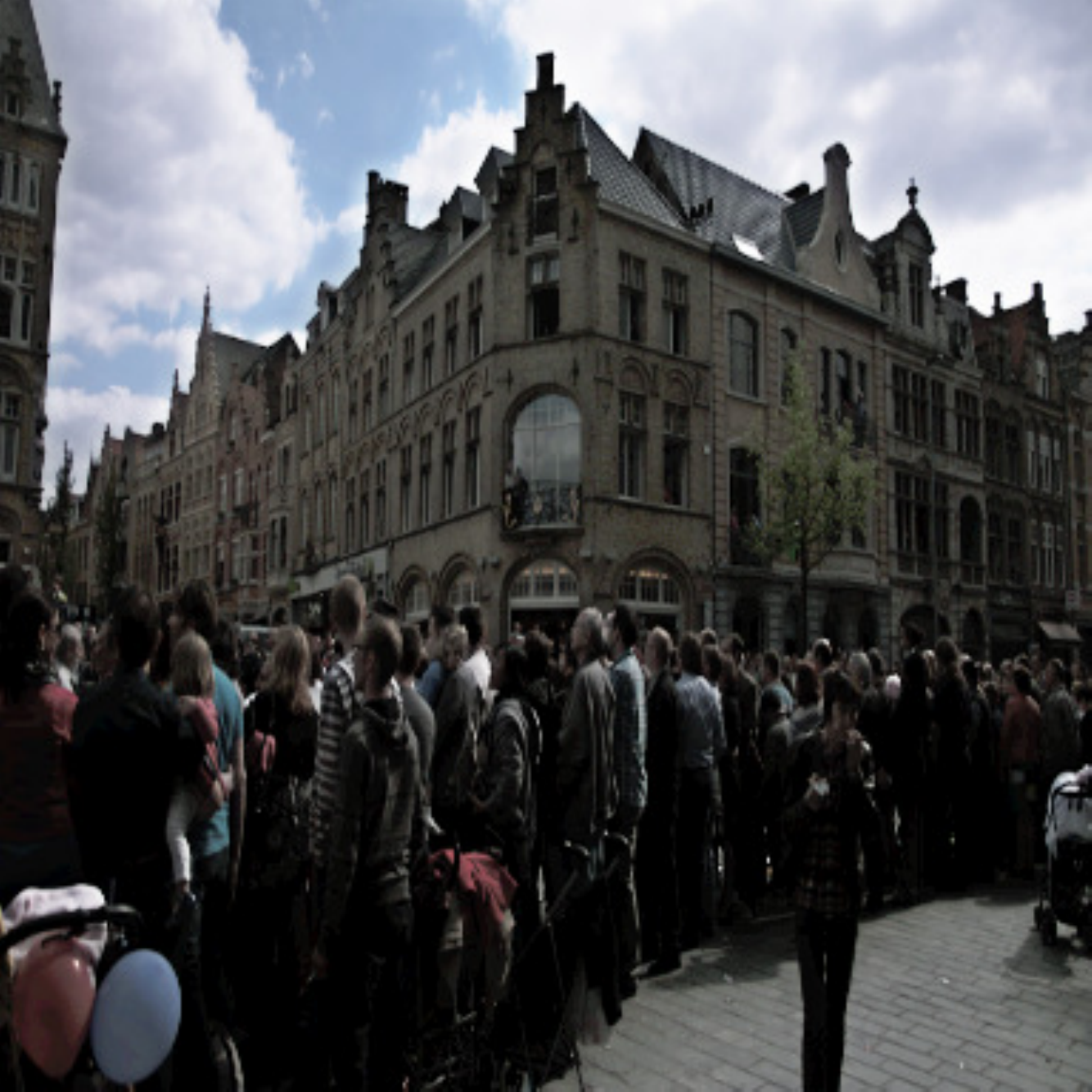}\vspace{2pt}
			\includegraphics[width=2.5cm]{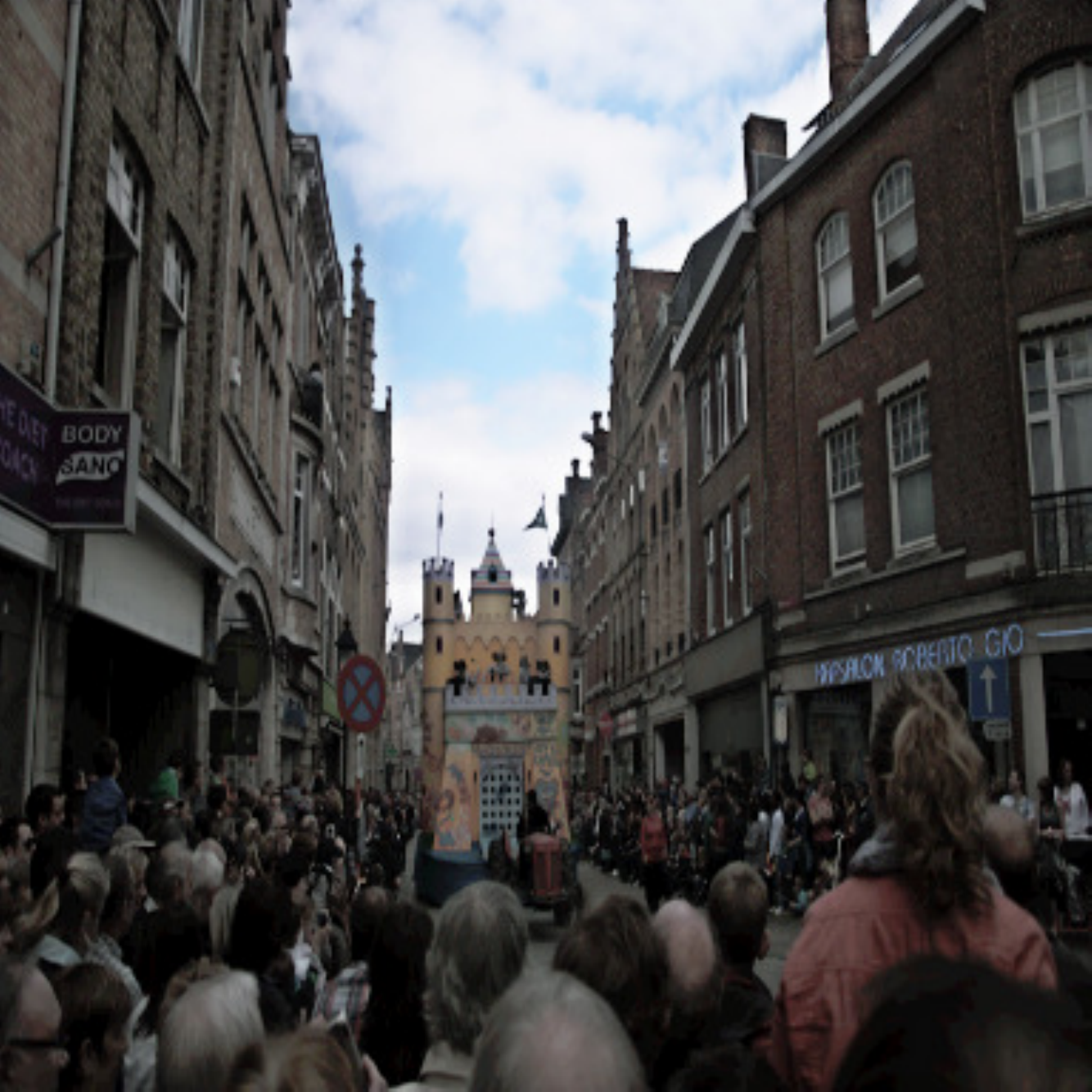}\vspace{2pt} 
		\end{minipage}
	}\hspace{-5pt}
	\subfigure[KinD++\cite{zhang2021beyond}]{
		\begin{minipage}[b]{0.13\textwidth}
			\includegraphics[width=2.5cm]{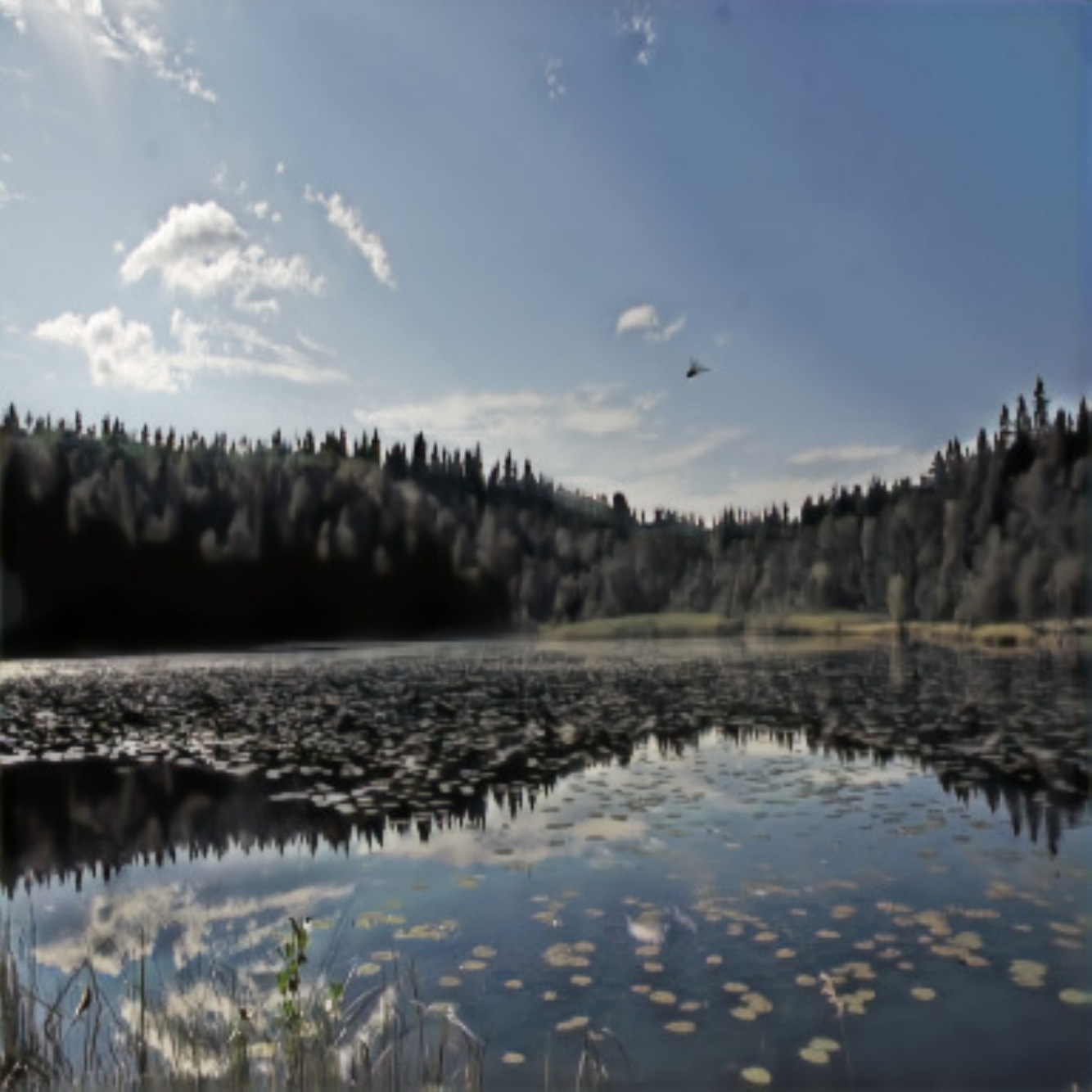}\vspace{2pt} \\
			\includegraphics[width=2.5cm]{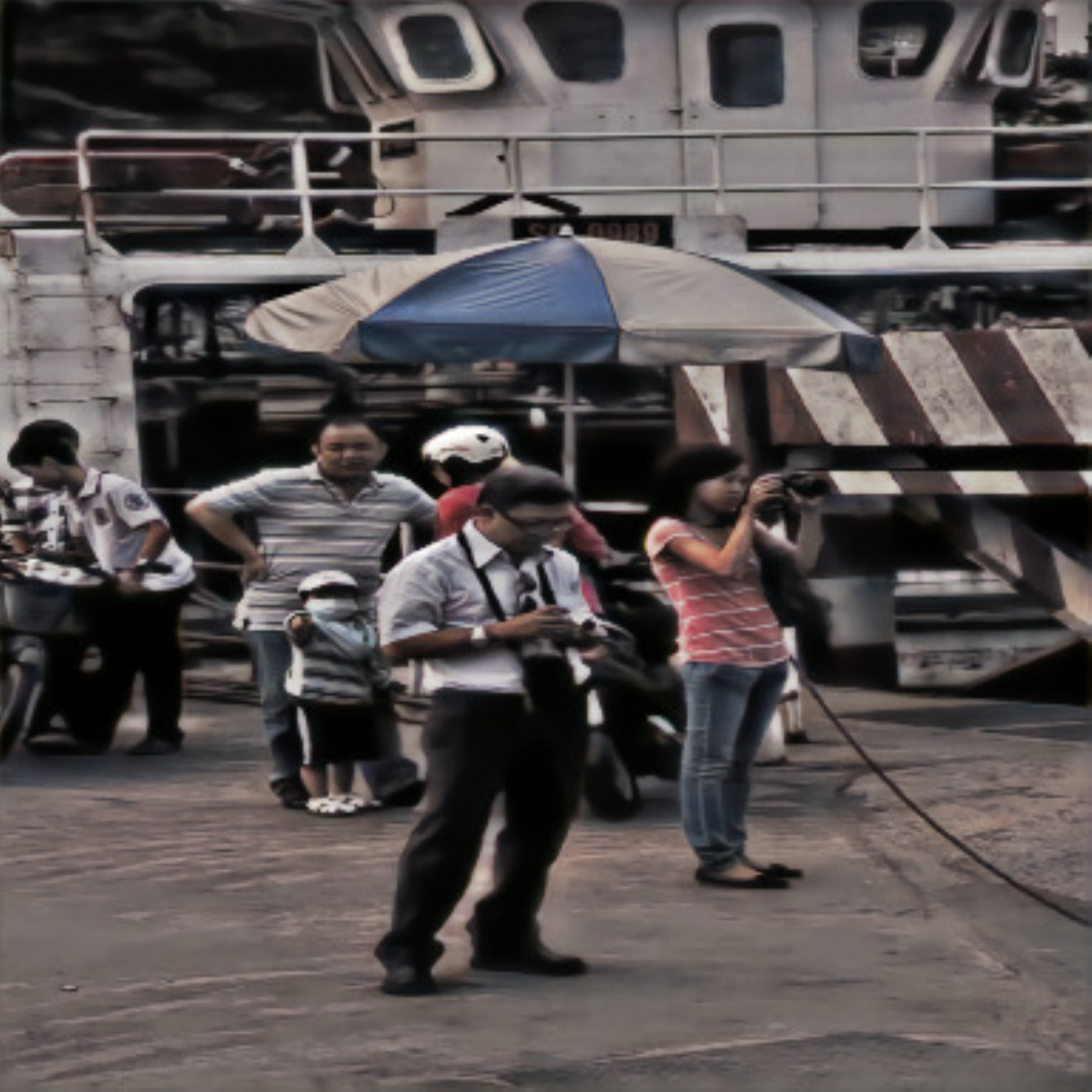}\vspace{2pt}
			\includegraphics[width=2.5cm]{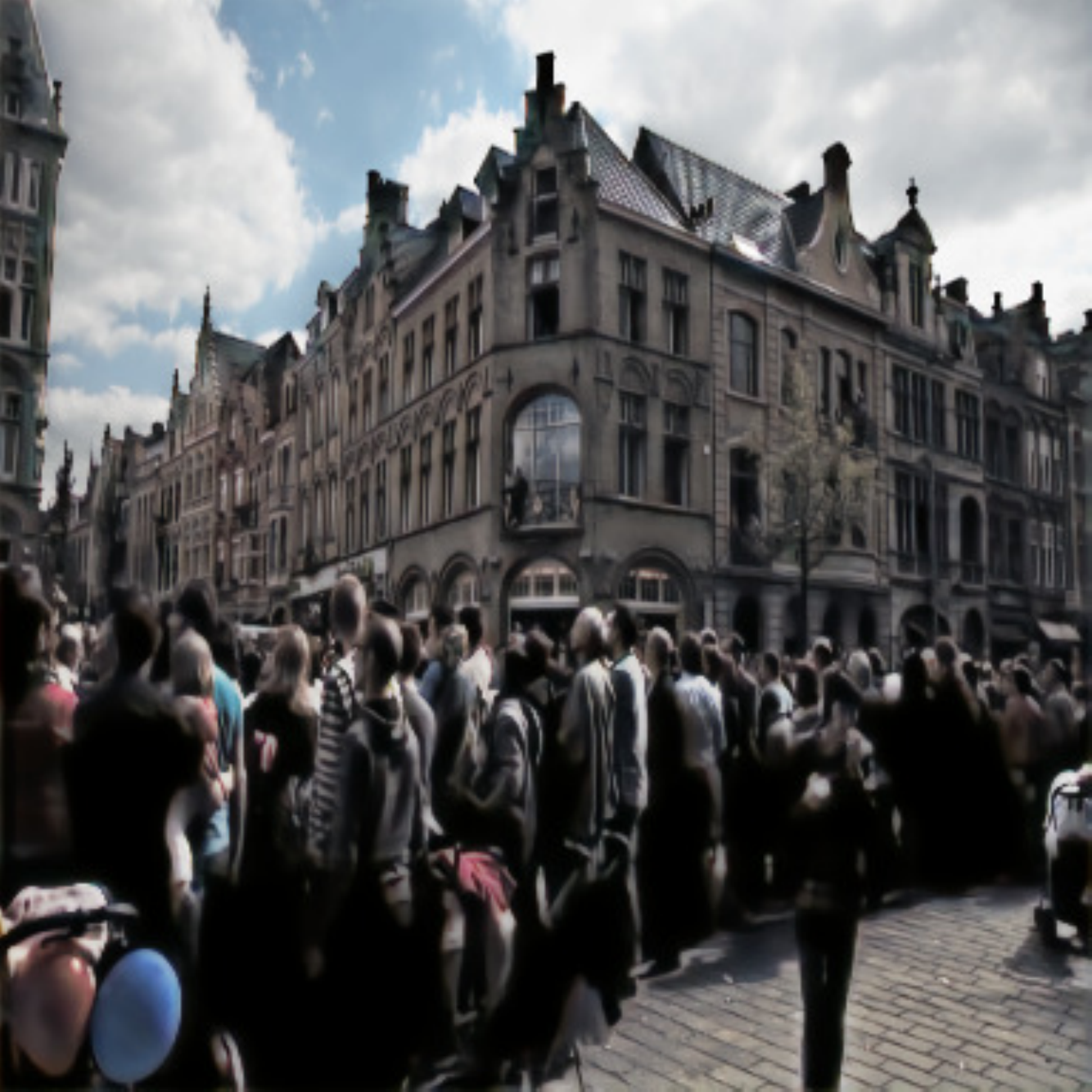}\vspace{2pt}
			\includegraphics[width=2.5cm]{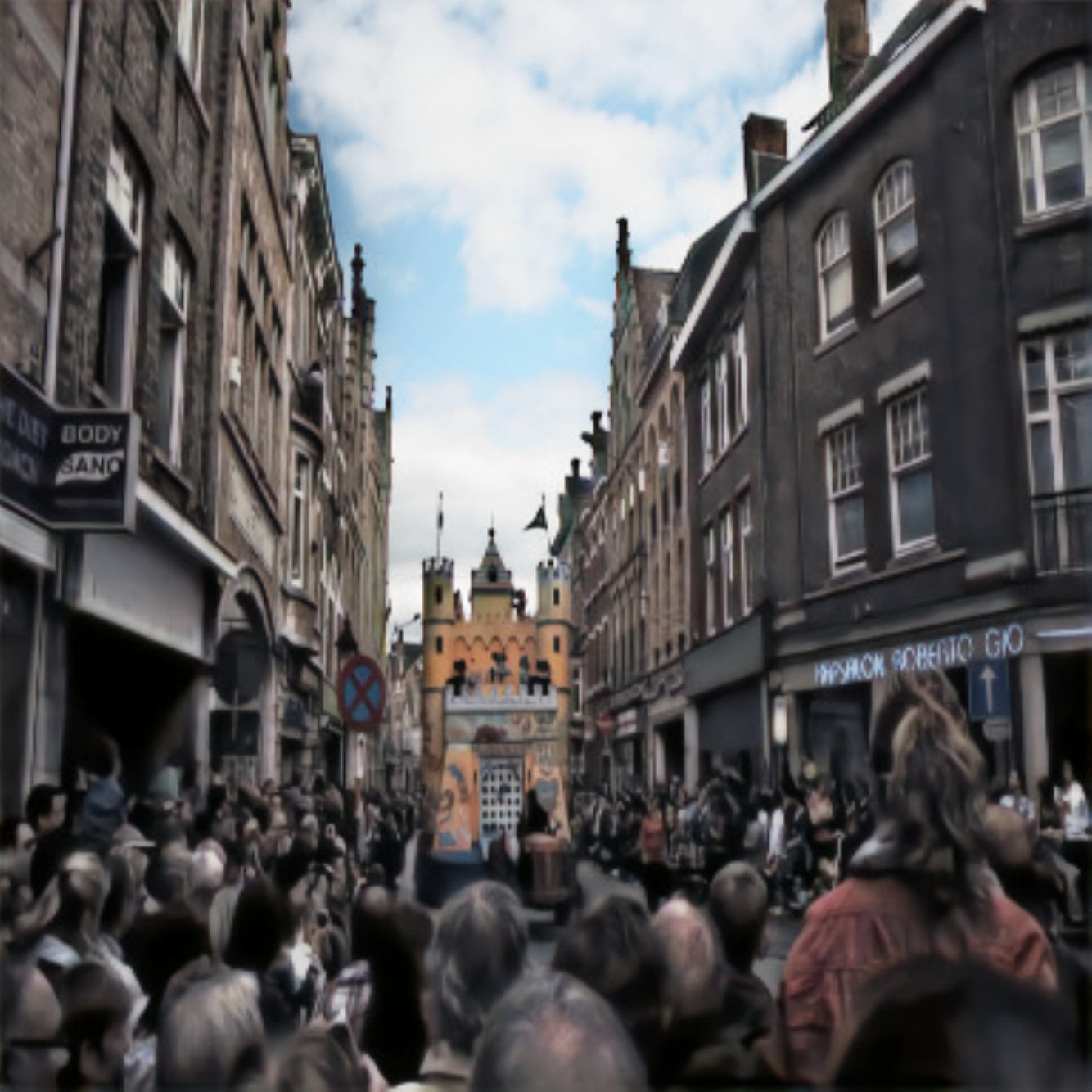}\vspace{2pt} 
		\end{minipage}
	}\hspace{-5pt}
	\subfigure[DA-DRN\cite{wei2021dadrn}]{
		\begin{minipage}[b]{0.13\textwidth}
			\includegraphics[width=2.5cm]{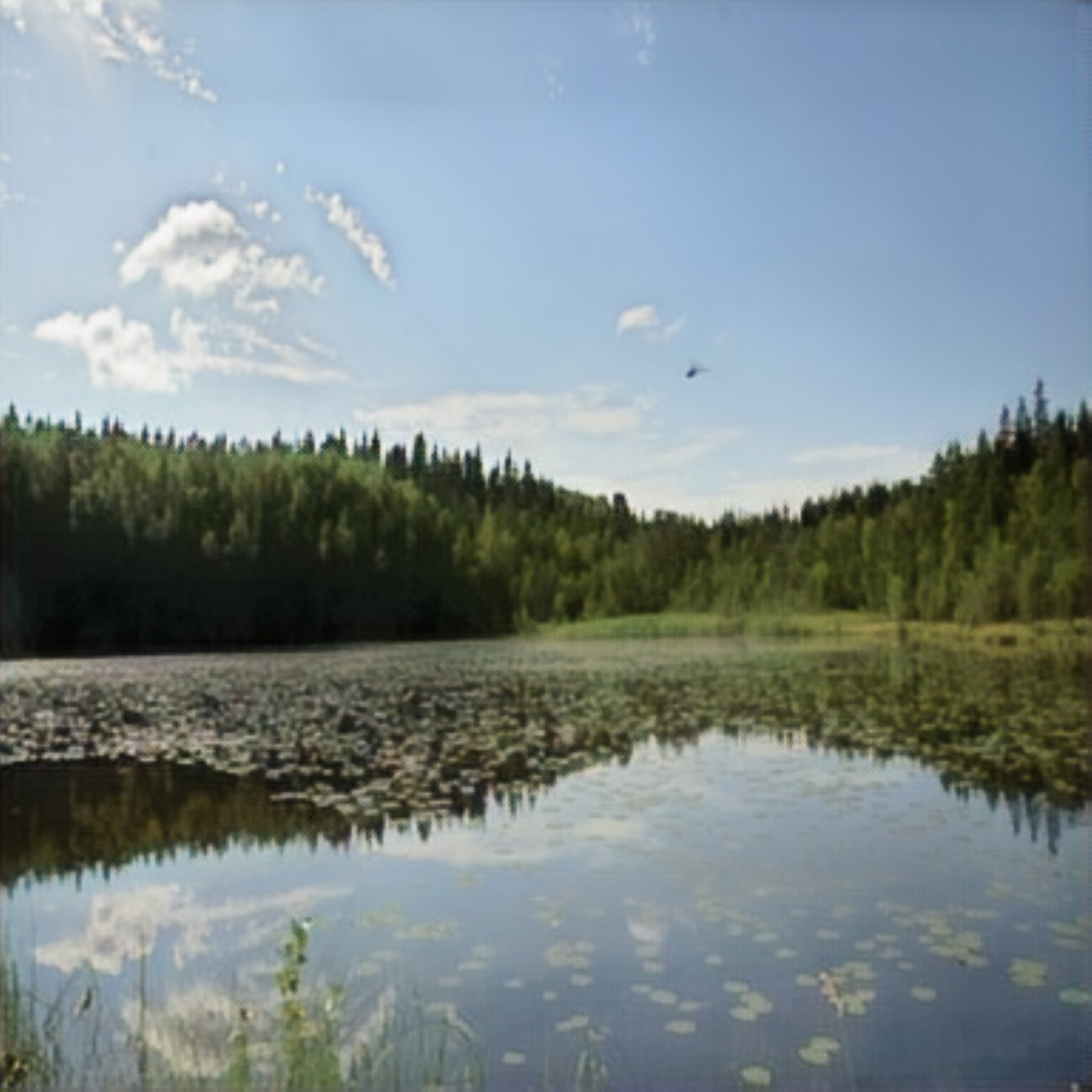}\vspace{2pt} \\
			\includegraphics[width=2.5cm]{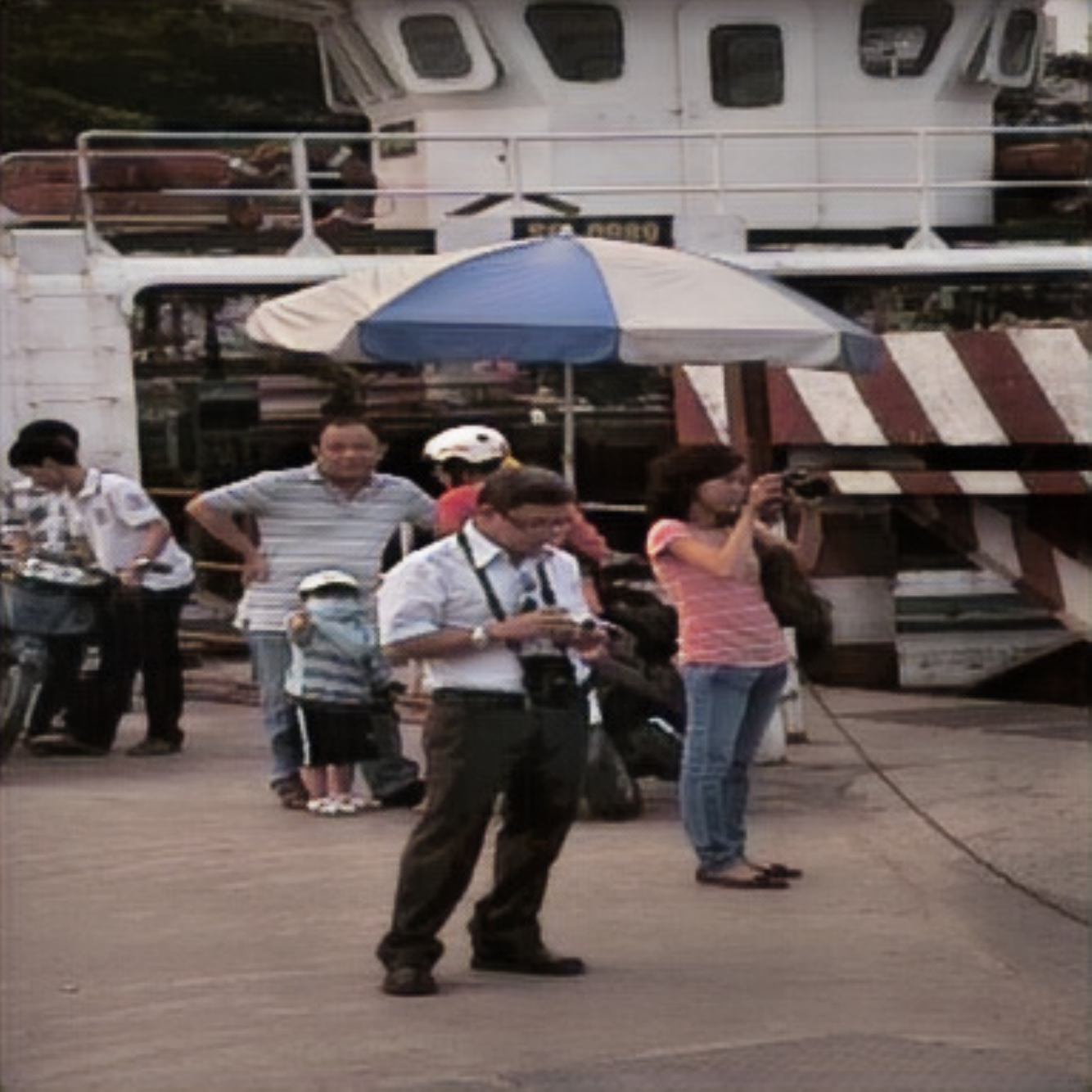}\vspace{2pt}
			\includegraphics[width=2.5cm]{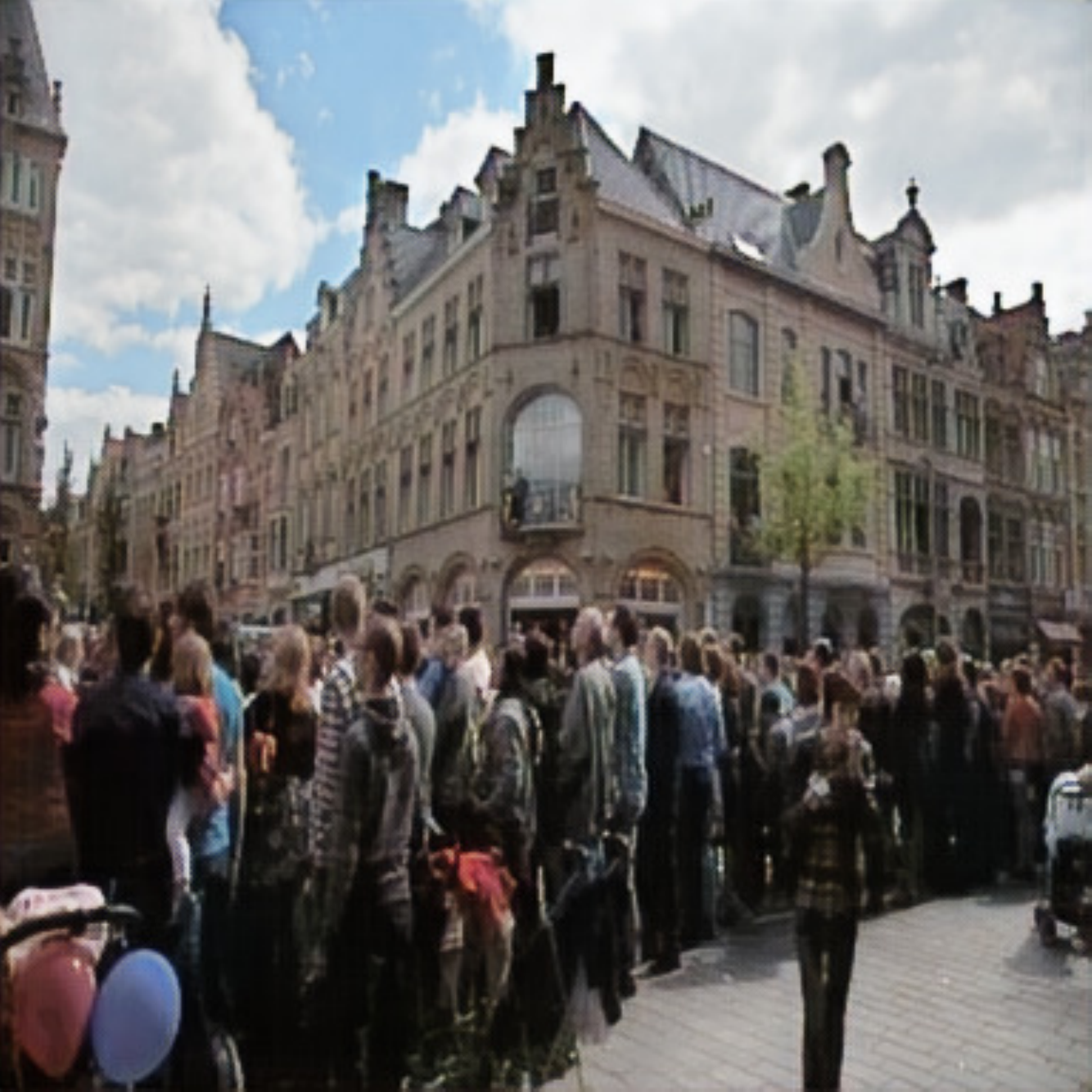}\vspace{2pt}
			\includegraphics[width=2.5cm]{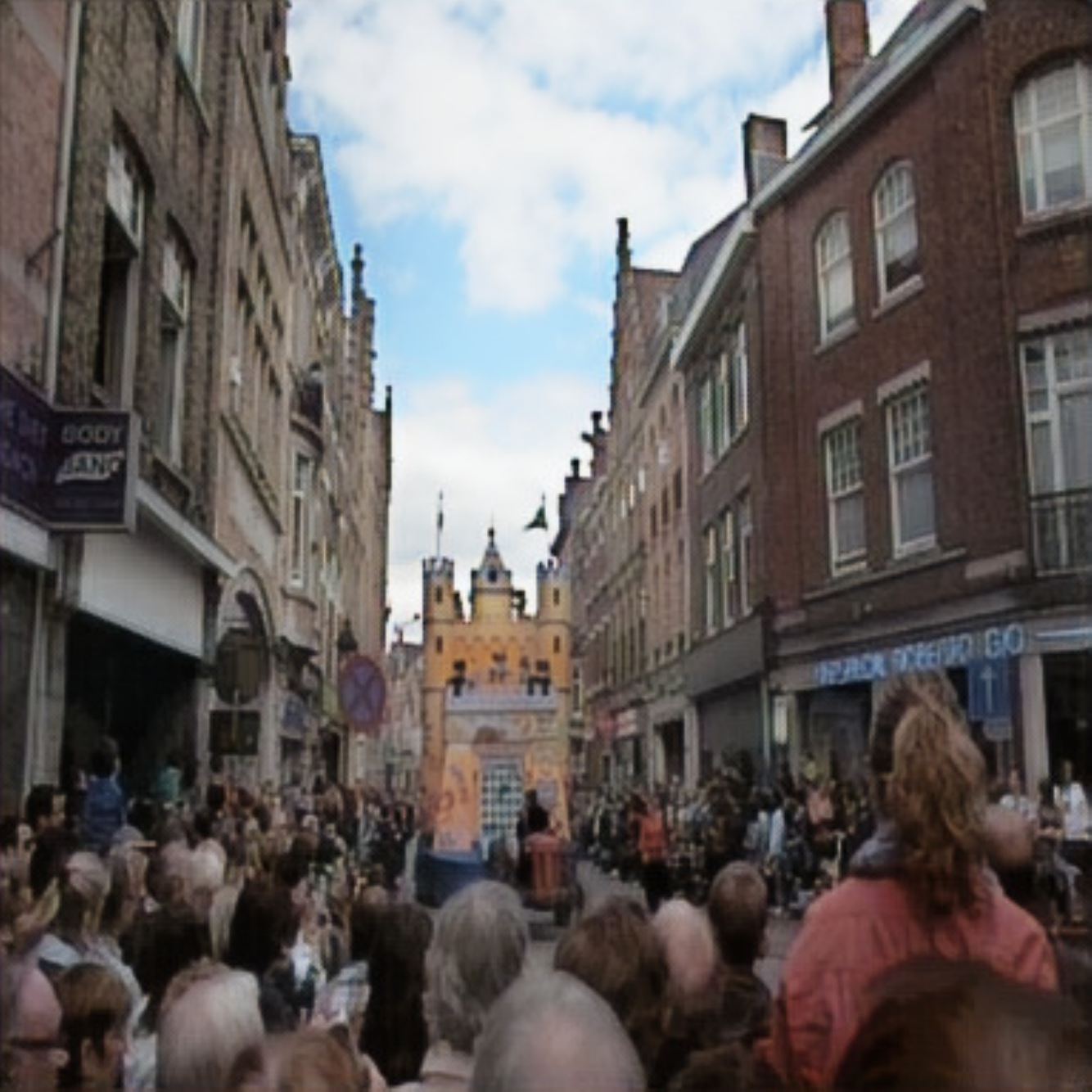}\vspace{2pt} 
		\end{minipage}
	}\hspace{-5pt}
	\subfigure[TSN-CA]{
		\begin{minipage}[b]{0.13\textwidth}
			\includegraphics[width=2.5cm]{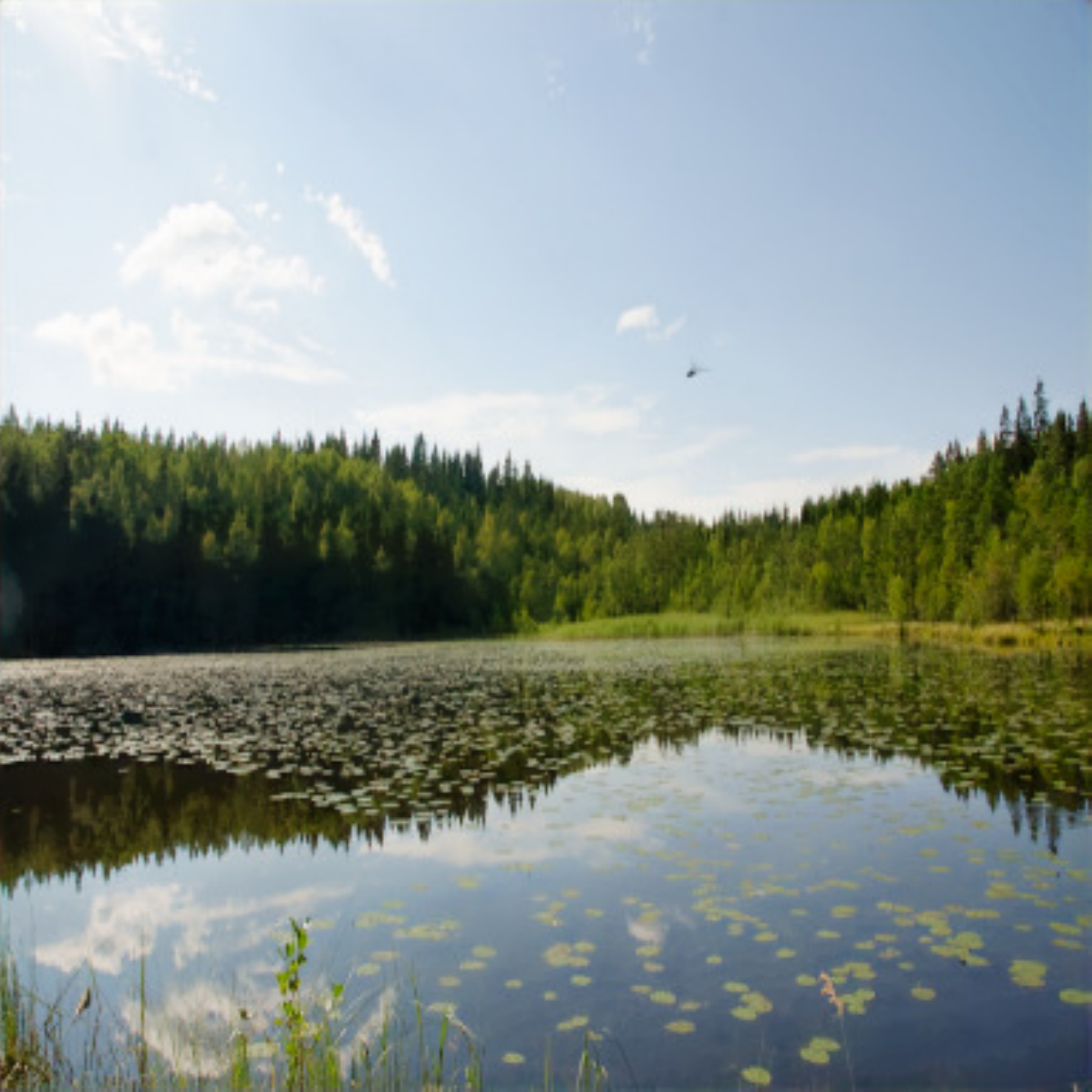}\vspace{2pt} \\
			\includegraphics[width=2.5cm]{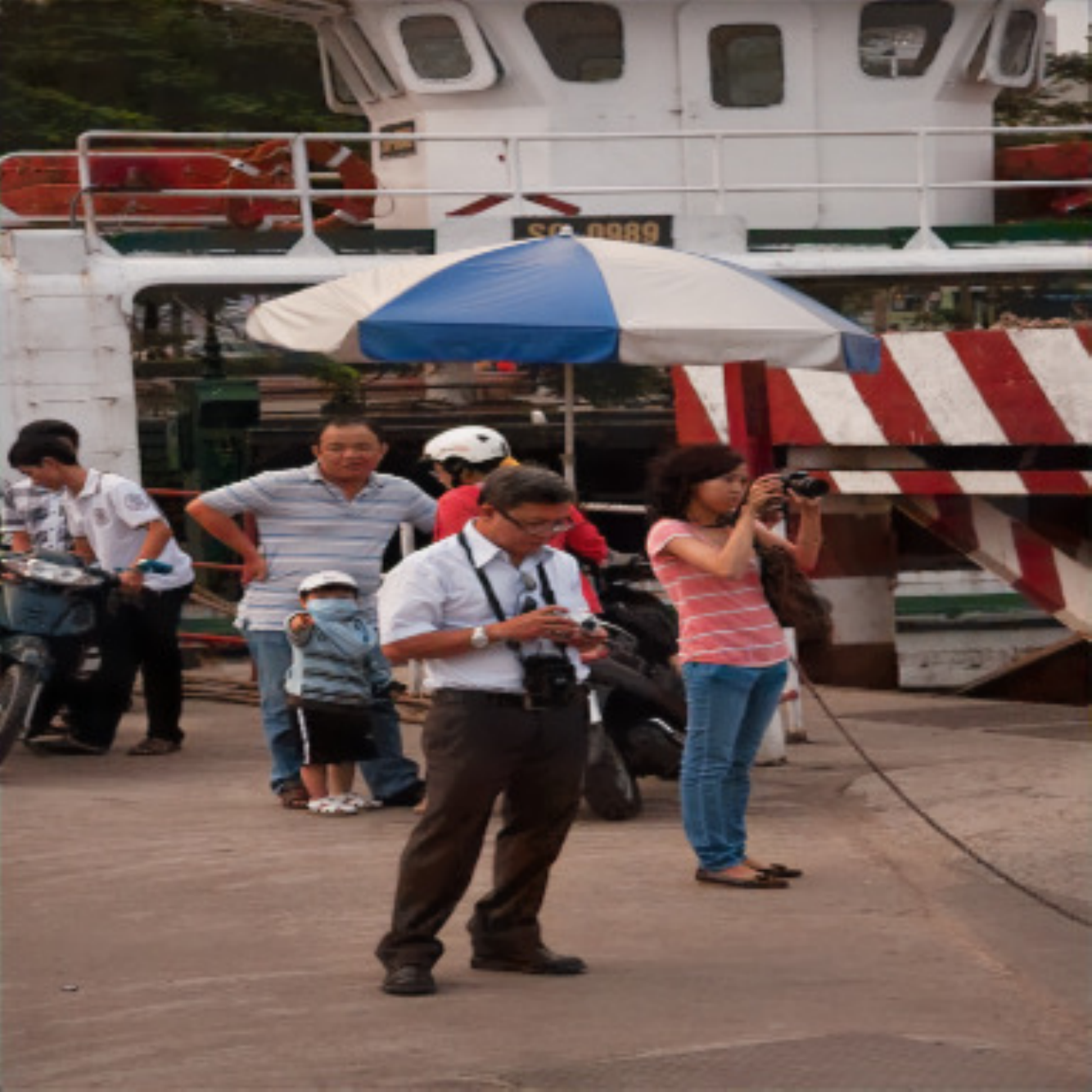}\vspace{2pt}
			\includegraphics[width=2.5cm]{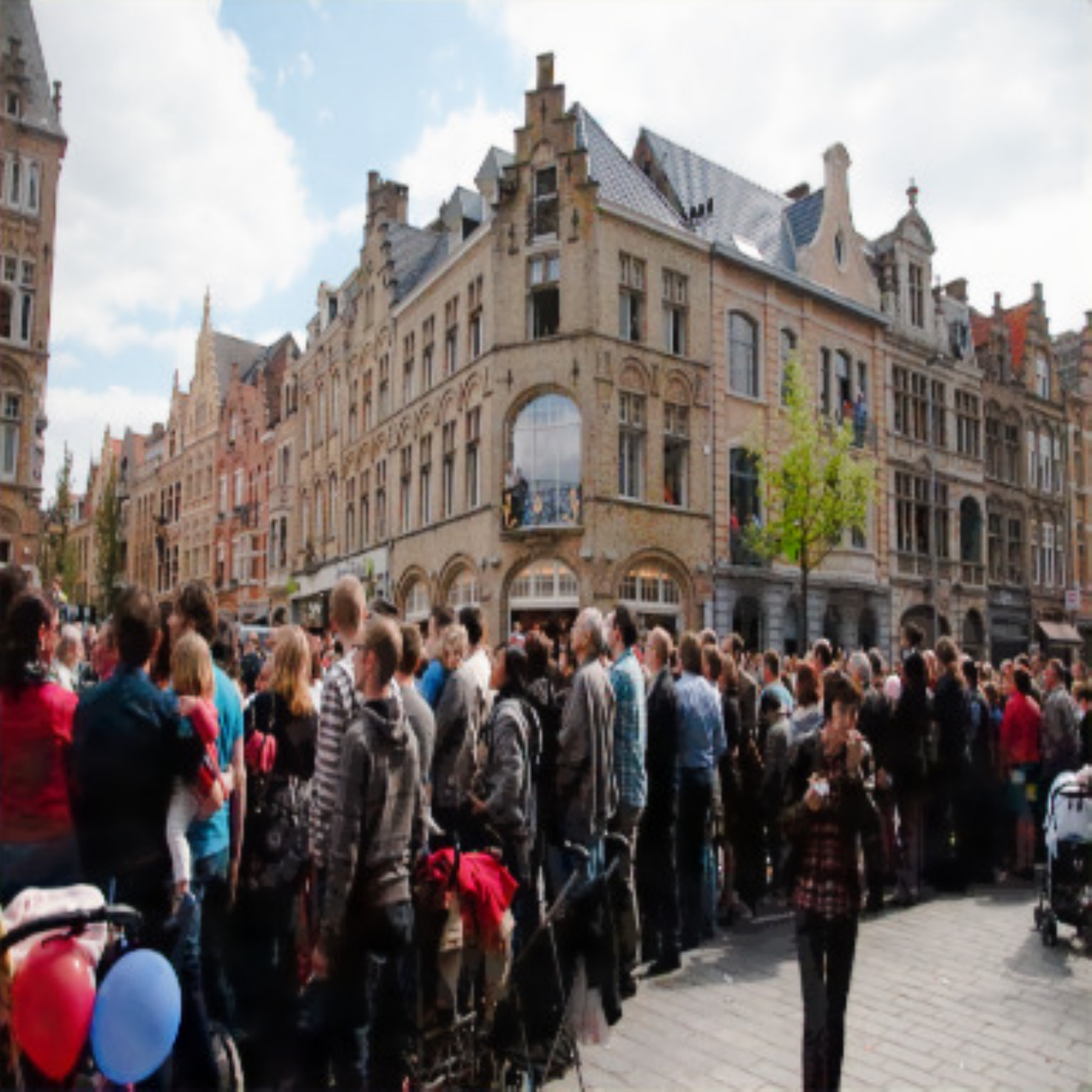}\vspace{2pt}
			\includegraphics[width=2.5cm]{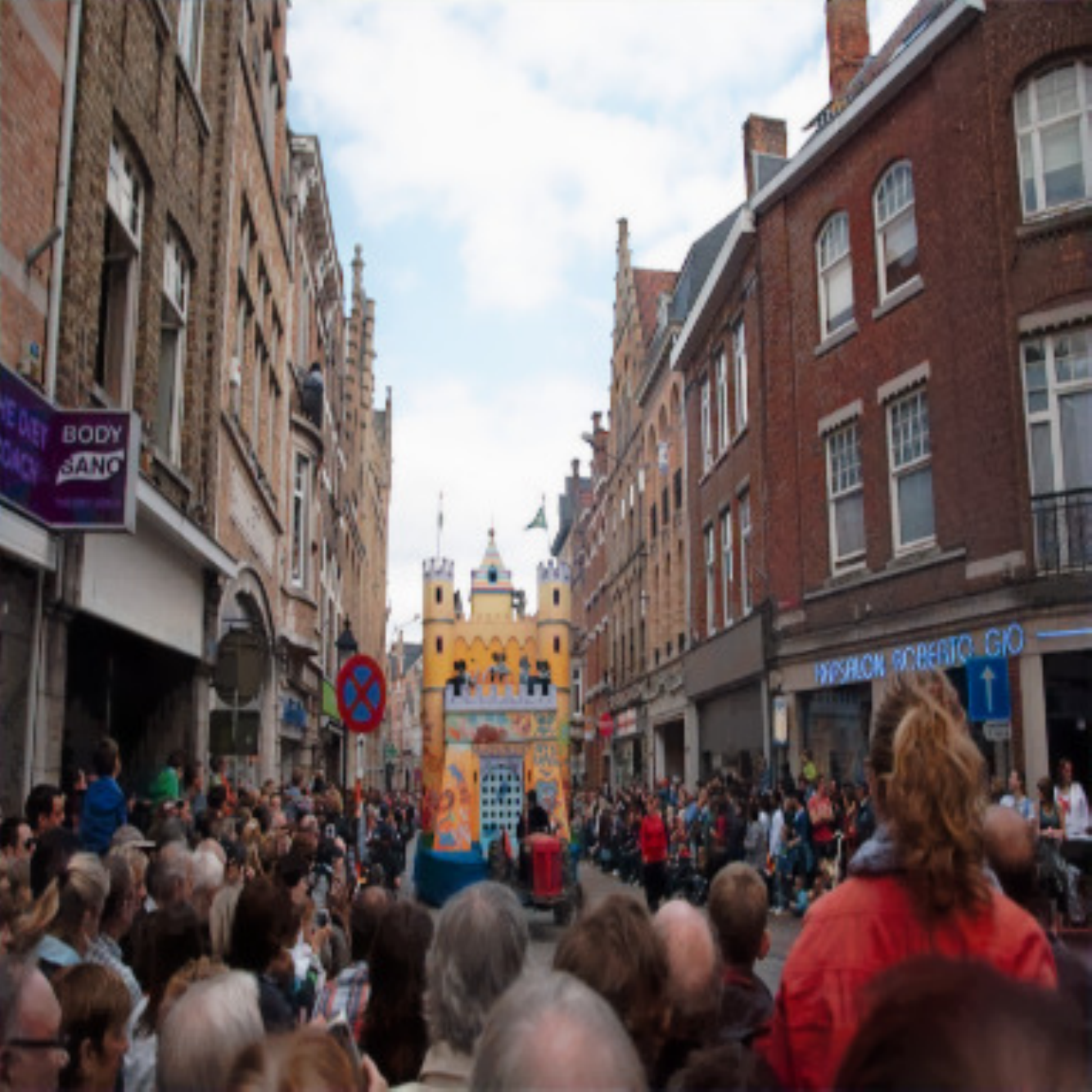}\vspace{2pt} 
		\end{minipage}
	}\hspace{-5pt}
	\subfigure[Ground-Truth]{
		\begin{minipage}[b]{0.13\textwidth}
			\includegraphics[width=2.5cm]{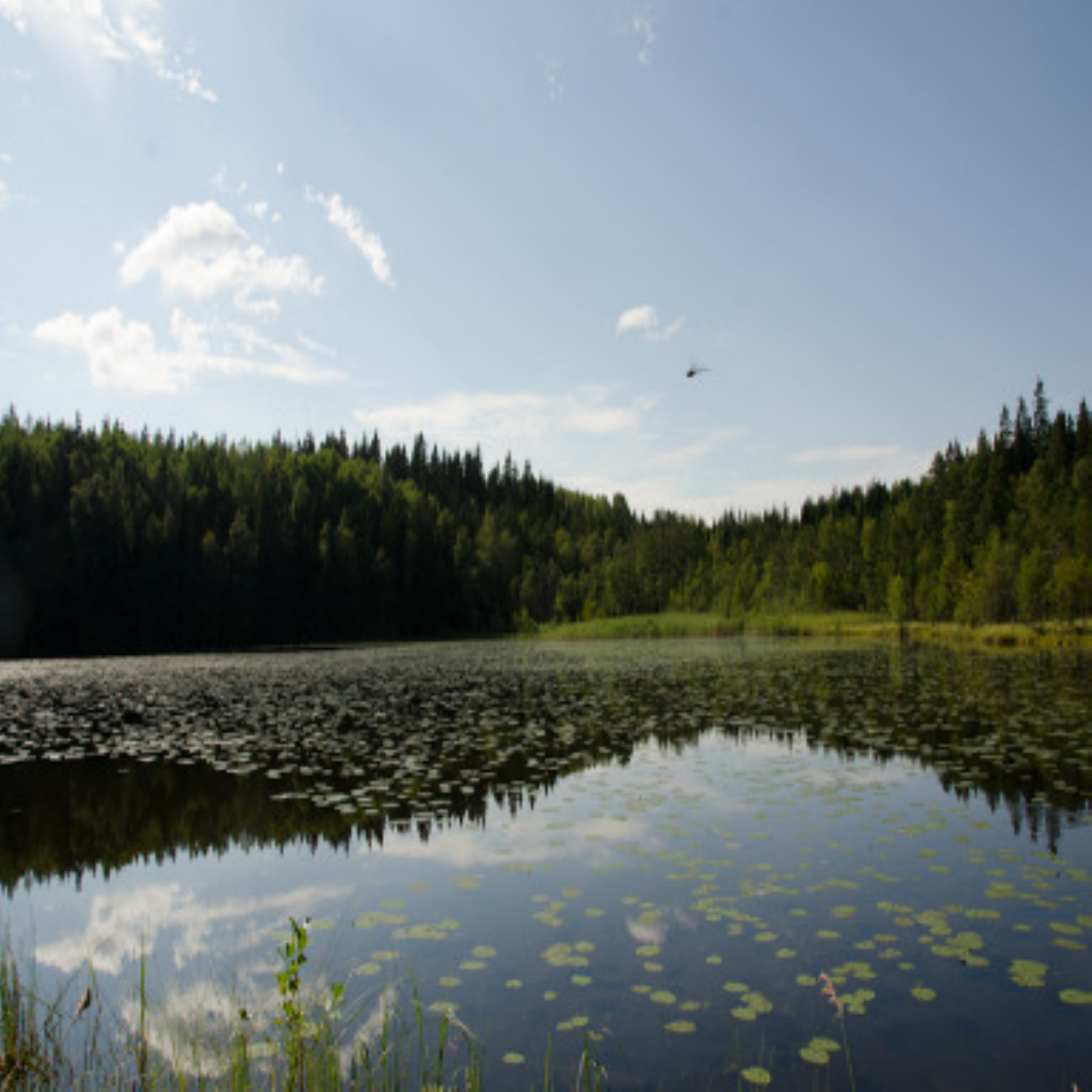}\vspace{2pt} \\
			\includegraphics[width=2.5cm]{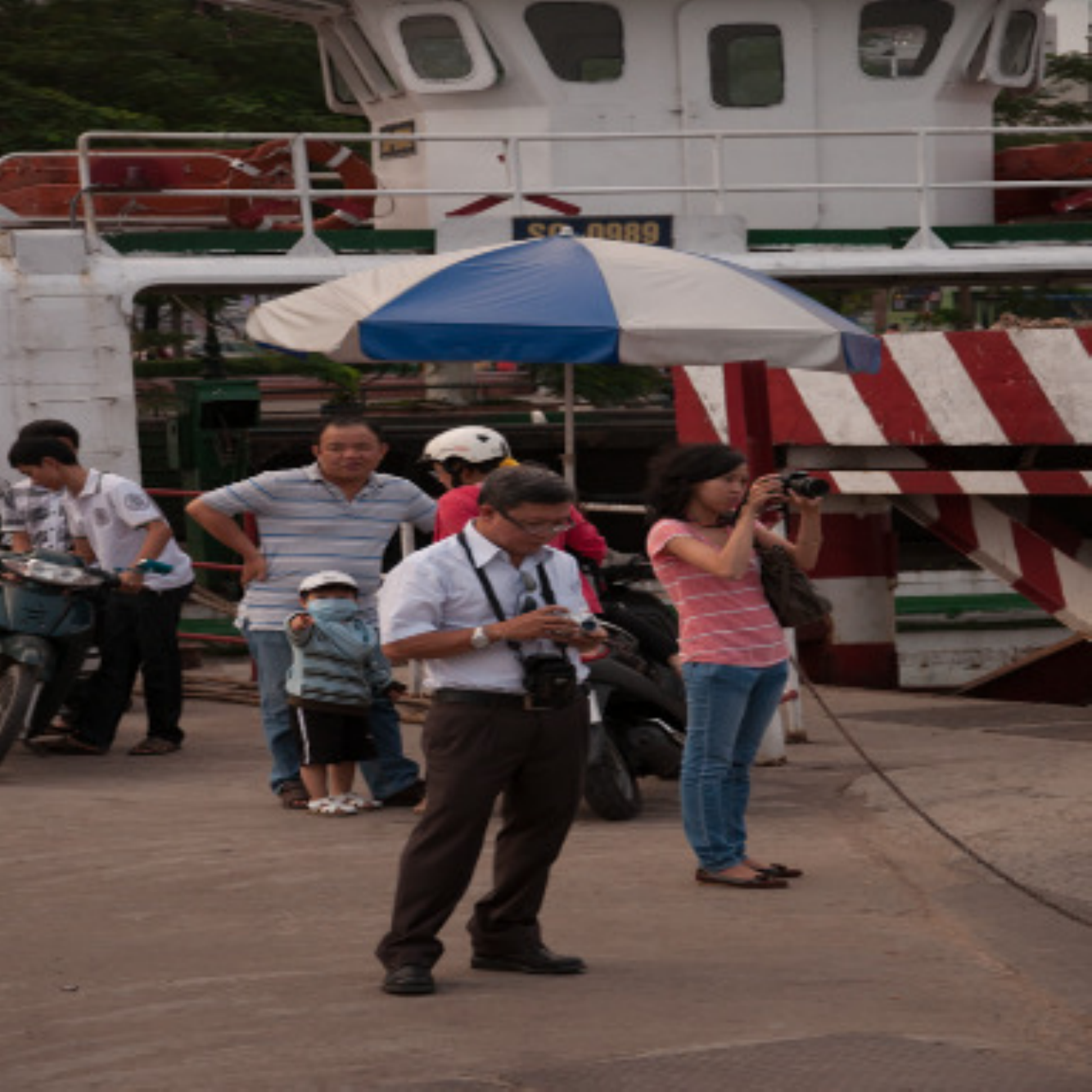}\vspace{2pt}
			\includegraphics[width=2.5cm]{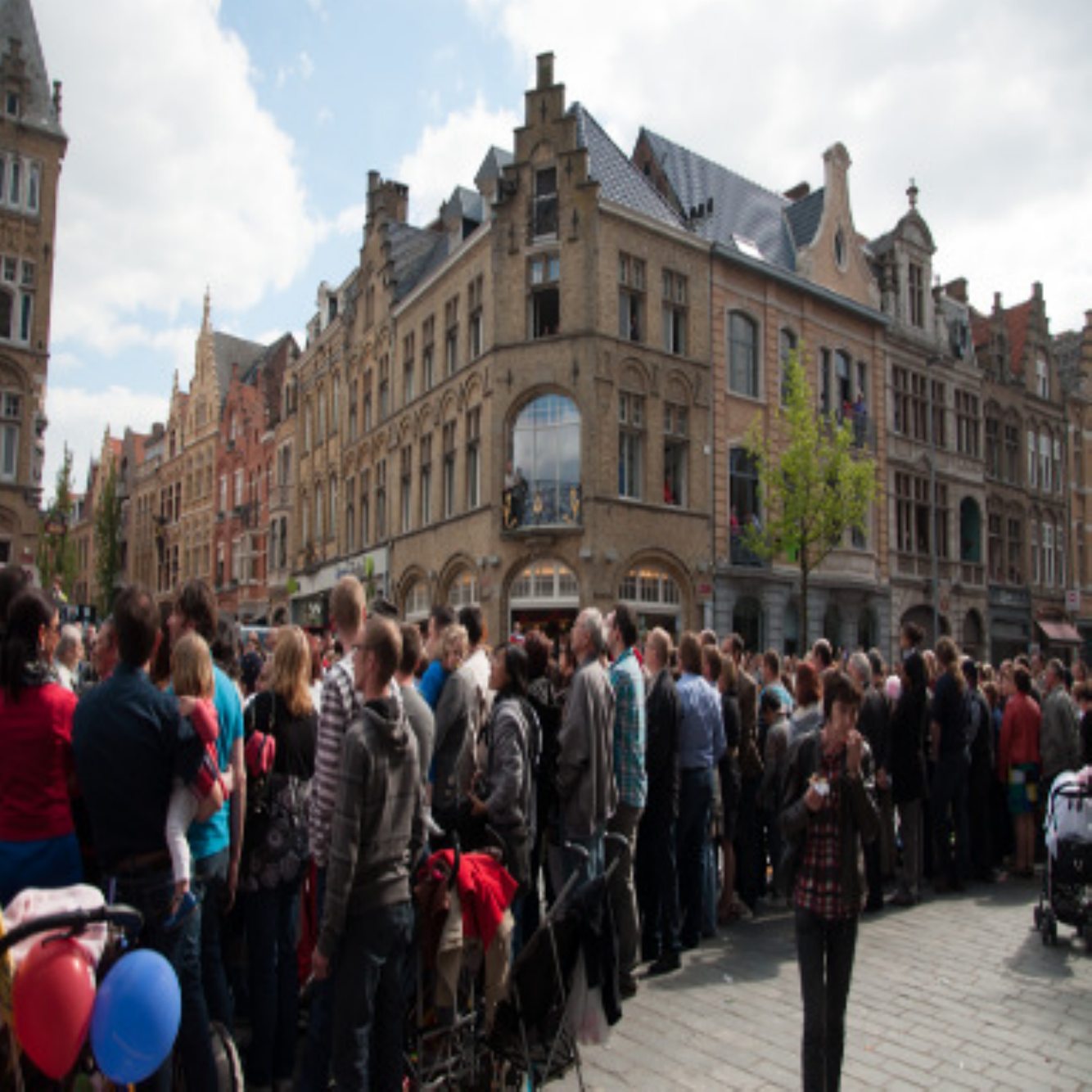}\vspace{2pt}
			\includegraphics[width=2.5cm]{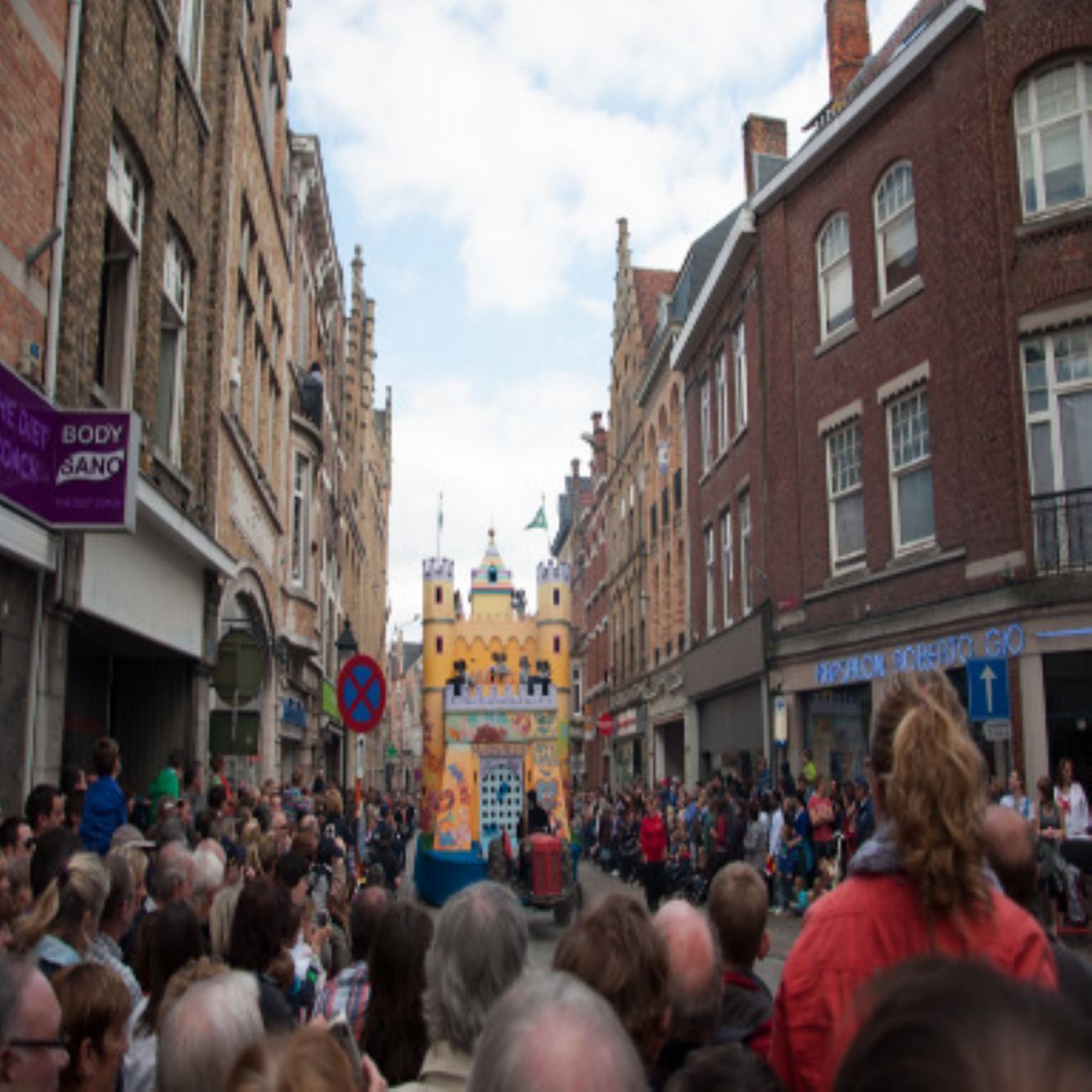}\vspace{2pt} 
		\end{minipage}
	}
	\caption{Visual comparison with other state-of-the-art methods on the LOL synthetic validation dataset.}
	\label{syn}
\end{figure*}

\begin{figure*}
	\centering
	
	
	\subfigure[Input]{
		\begin{minipage}[b]{0.13\textwidth}
			\includegraphics[width=2.5cm]{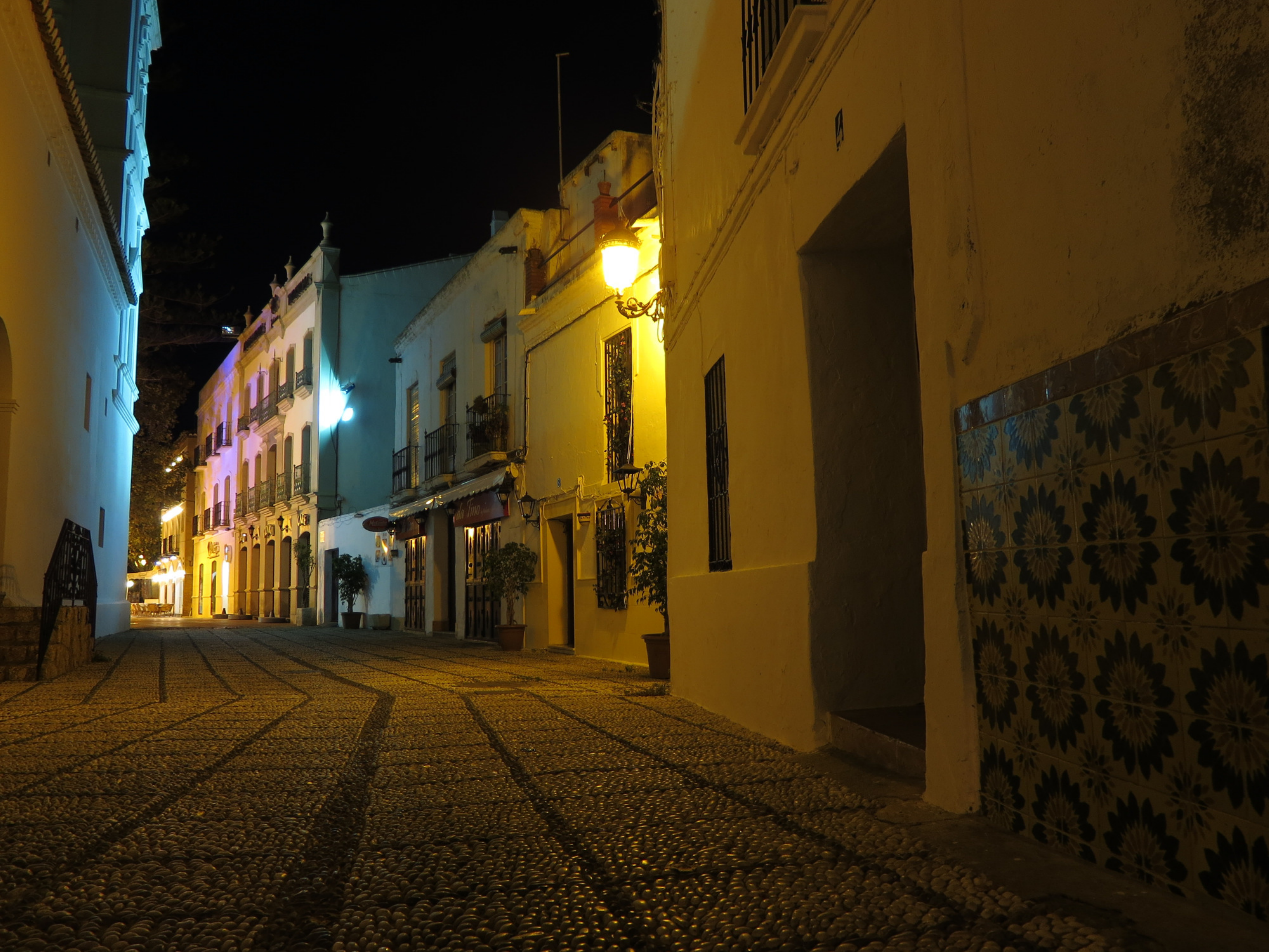}\vspace{2pt} \\
			\includegraphics[width=2.5cm]{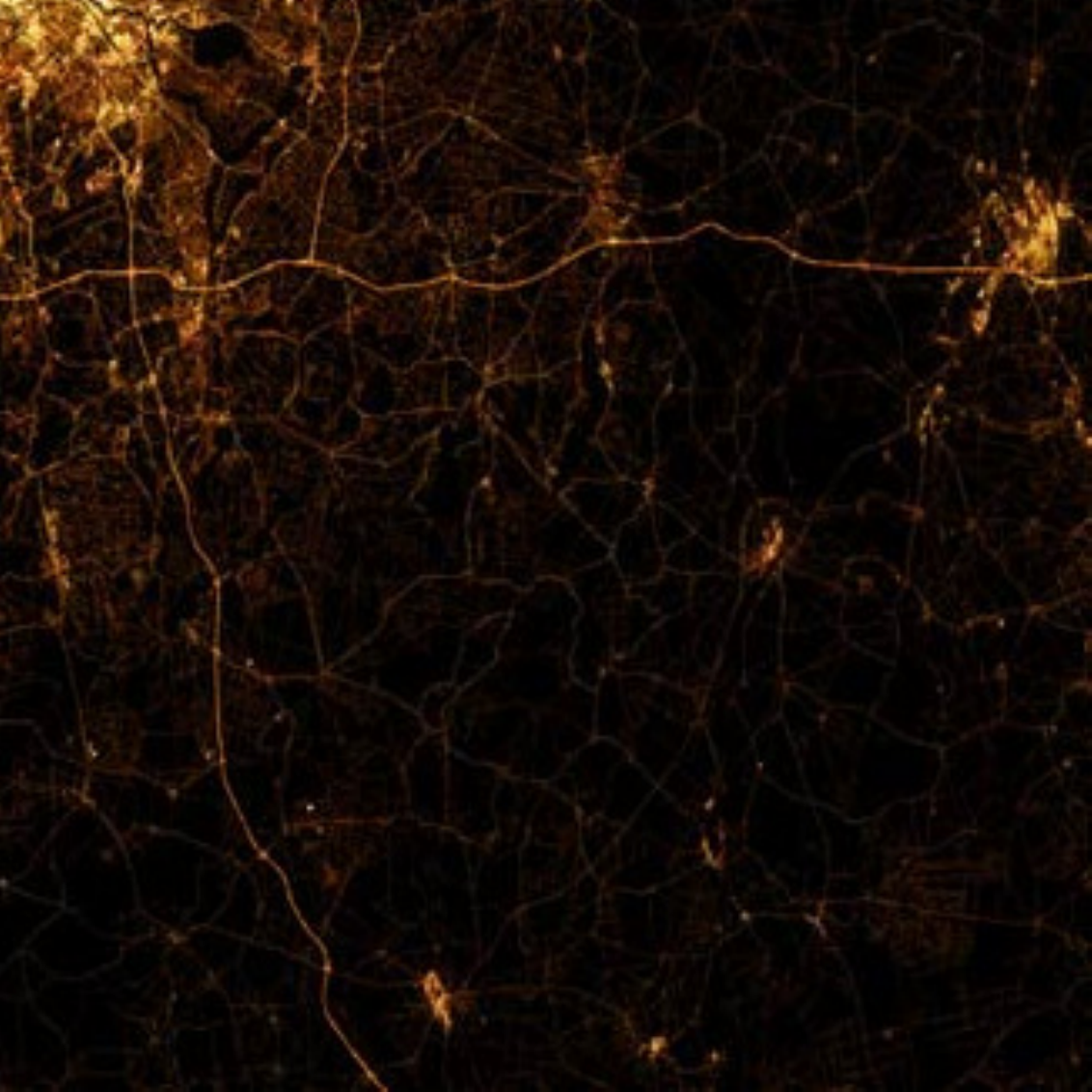}\vspace{2pt}
			\includegraphics[width=2.5cm]{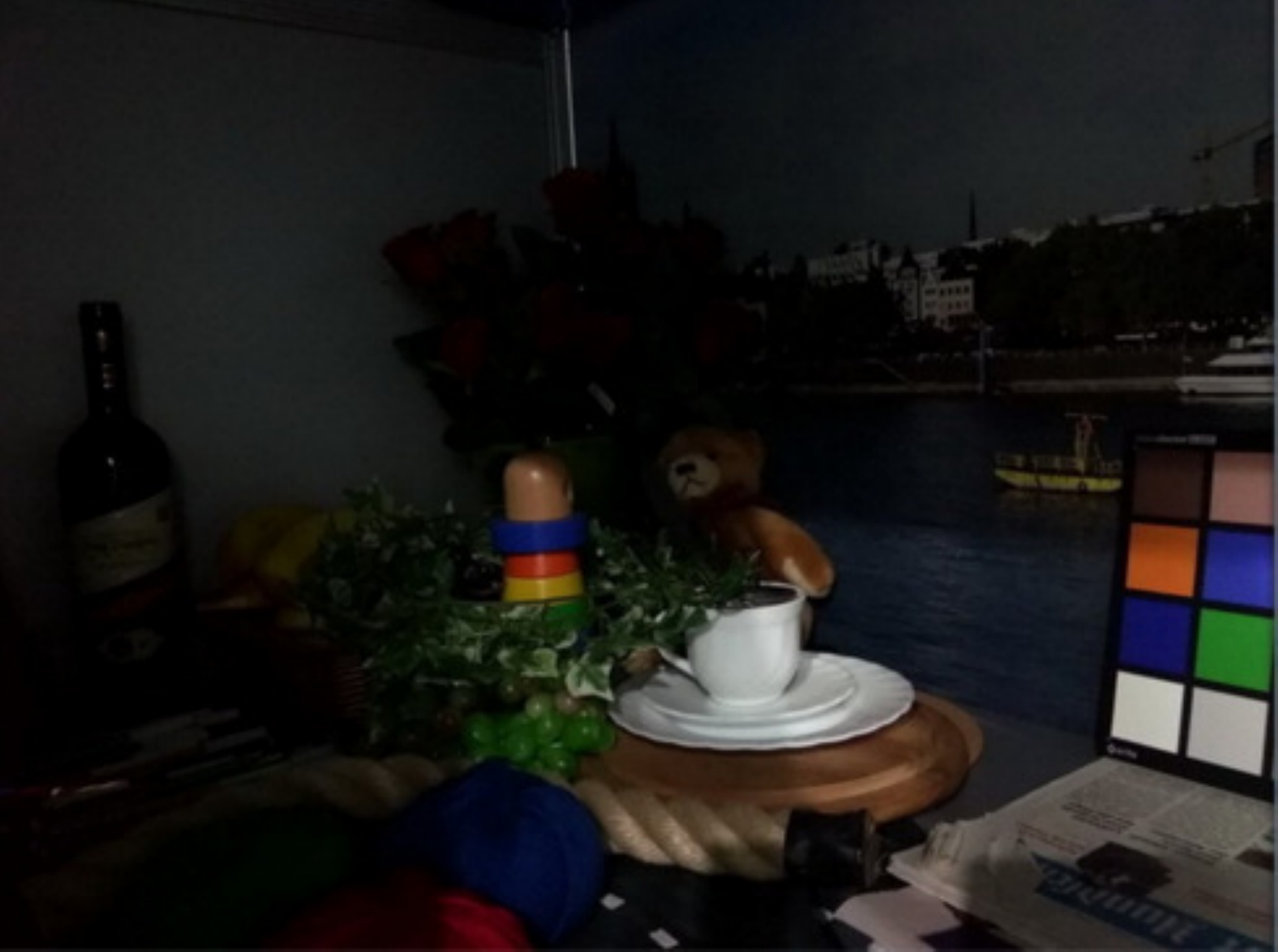}\vspace{2pt}
			\includegraphics[width=2.5cm]{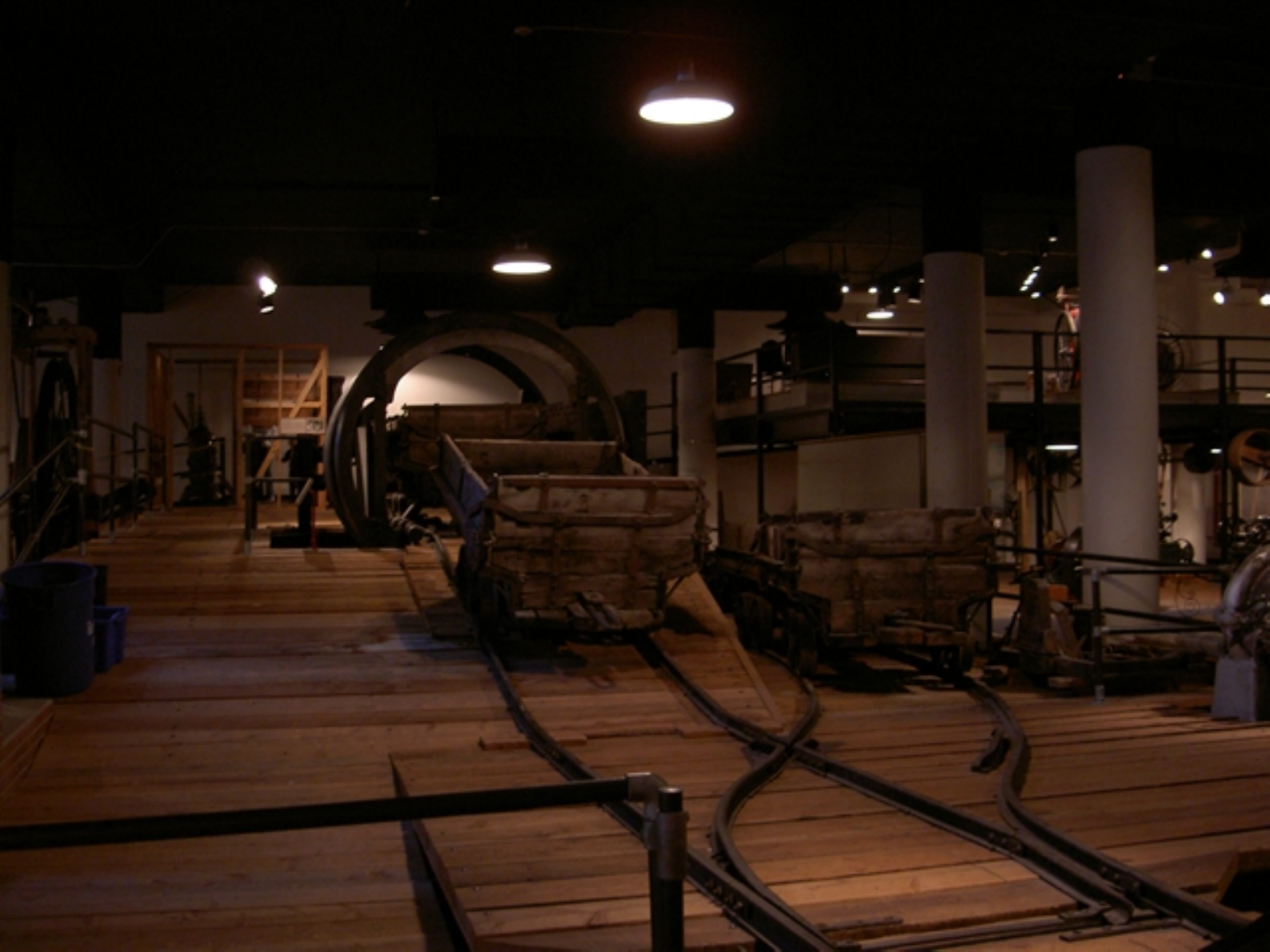}\vspace{2pt}
			\includegraphics[width=2.5cm]{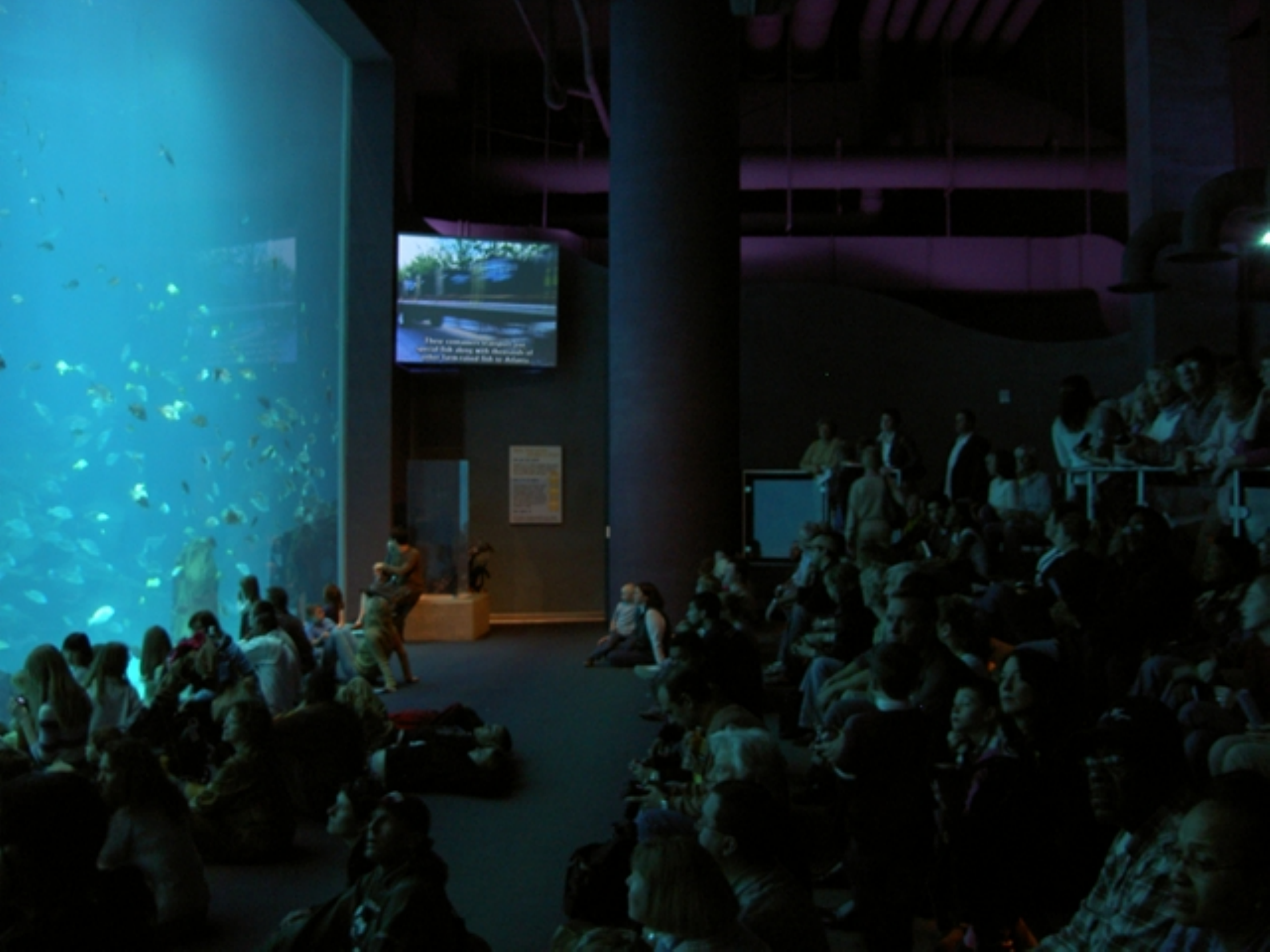}\vspace{2pt}
			\includegraphics[width=2.5cm]{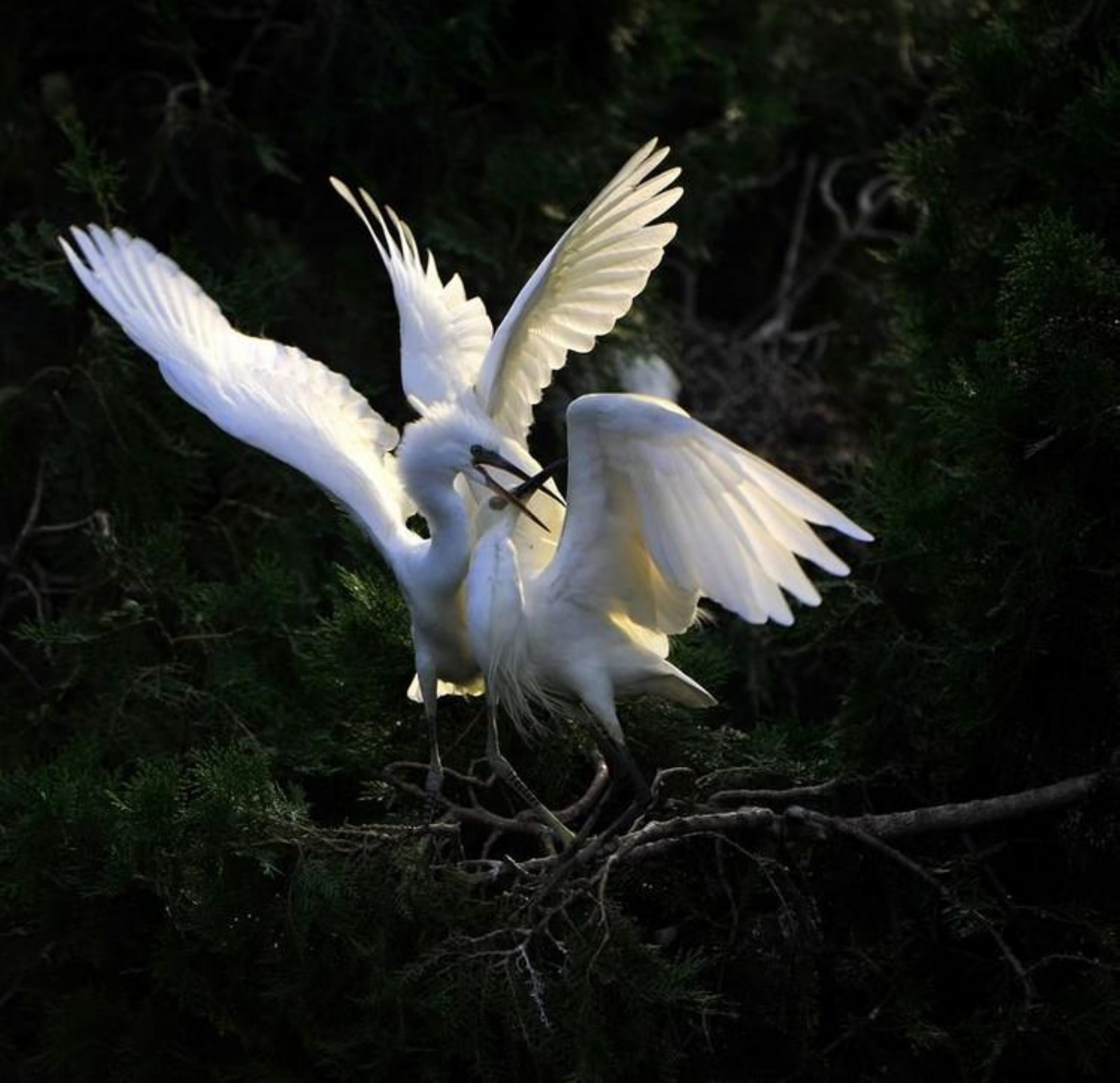}
		\end{minipage}
	}\hspace{-5pt}
	\subfigure[NPE\cite{wang2013naturalness}]{
		\begin{minipage}[b]{0.13\textwidth}
			\includegraphics[width=2.5cm]{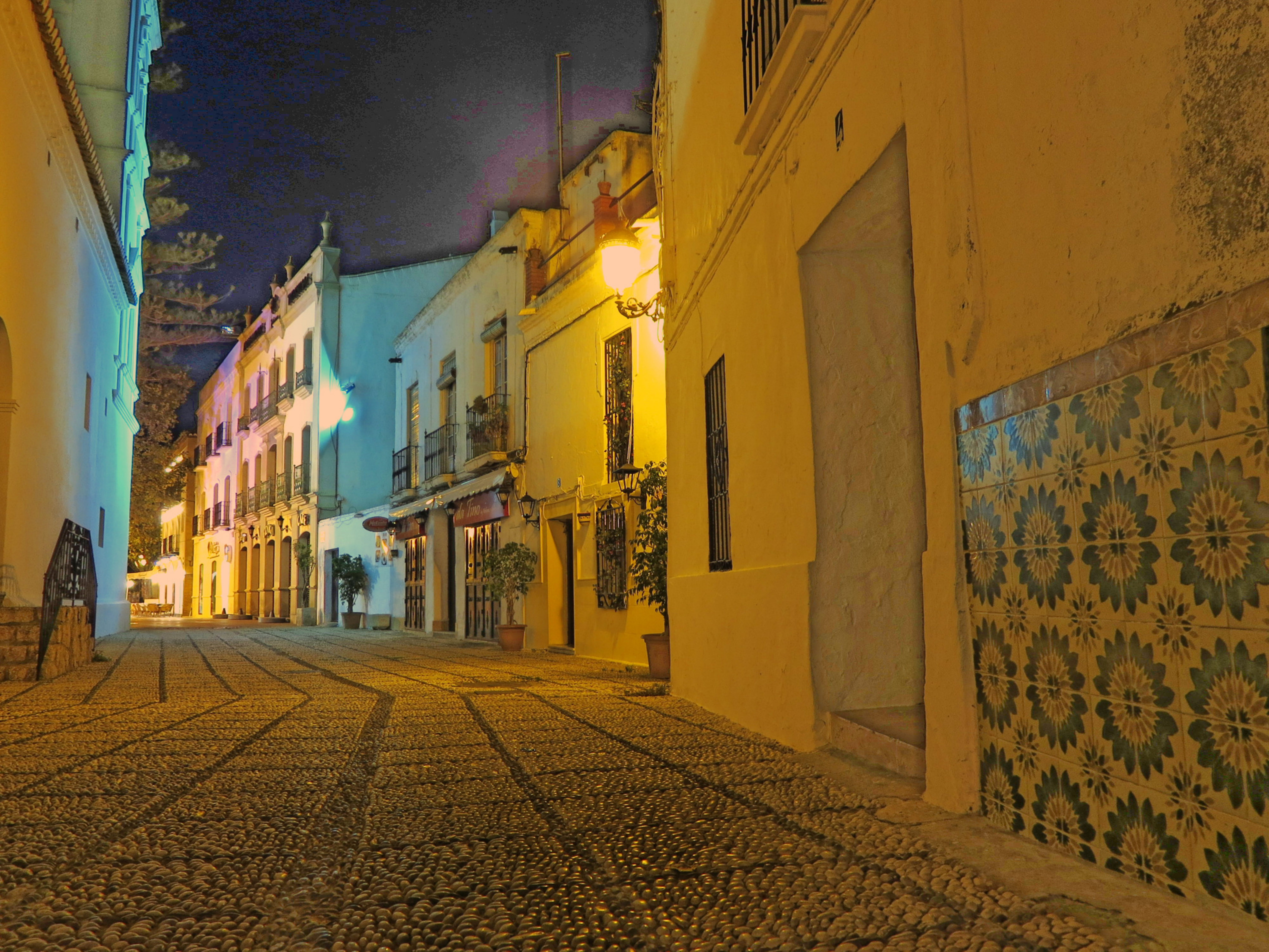}\vspace{2pt} \\
			\includegraphics[width=2.5cm]{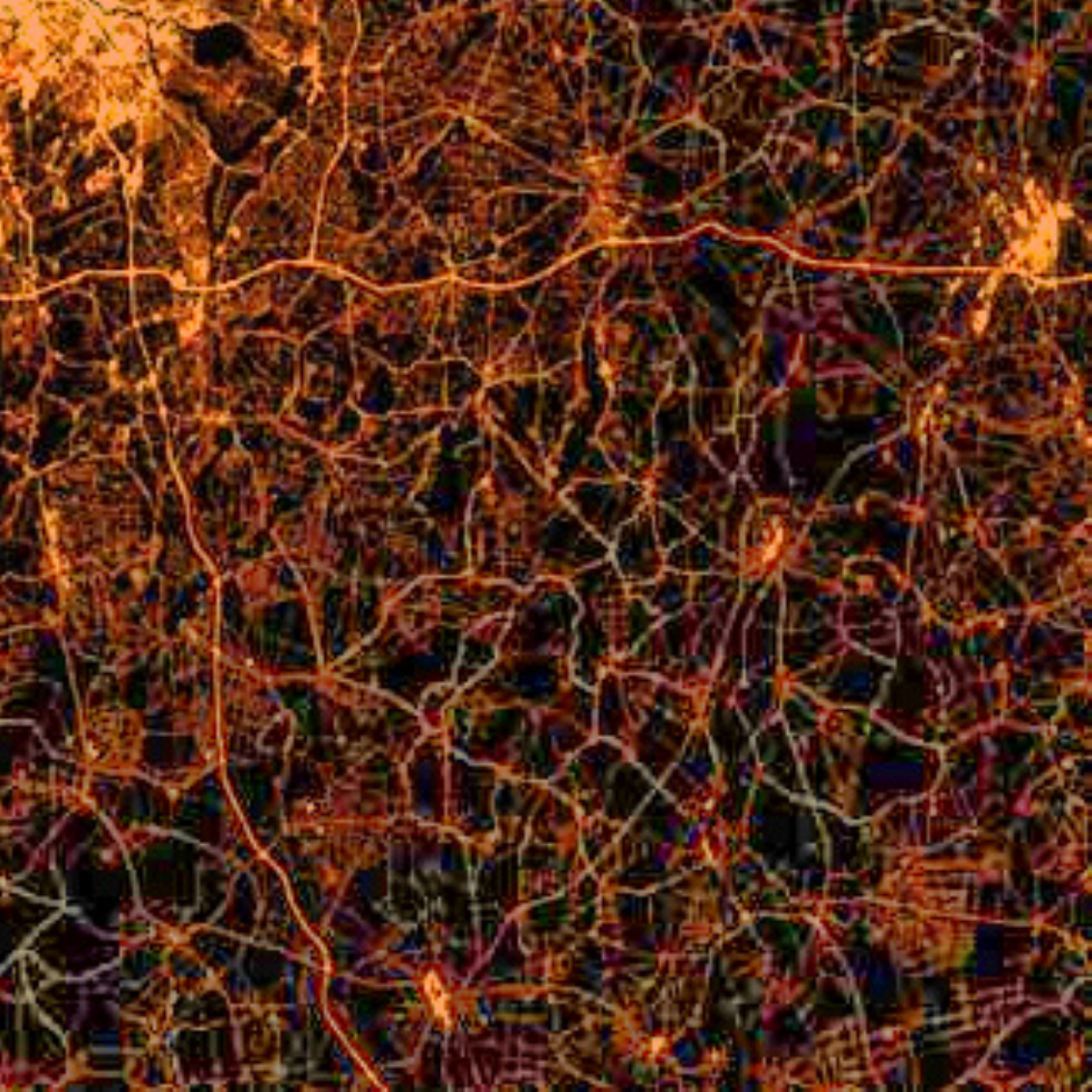}\vspace{2pt}
			\includegraphics[width=2.5cm]{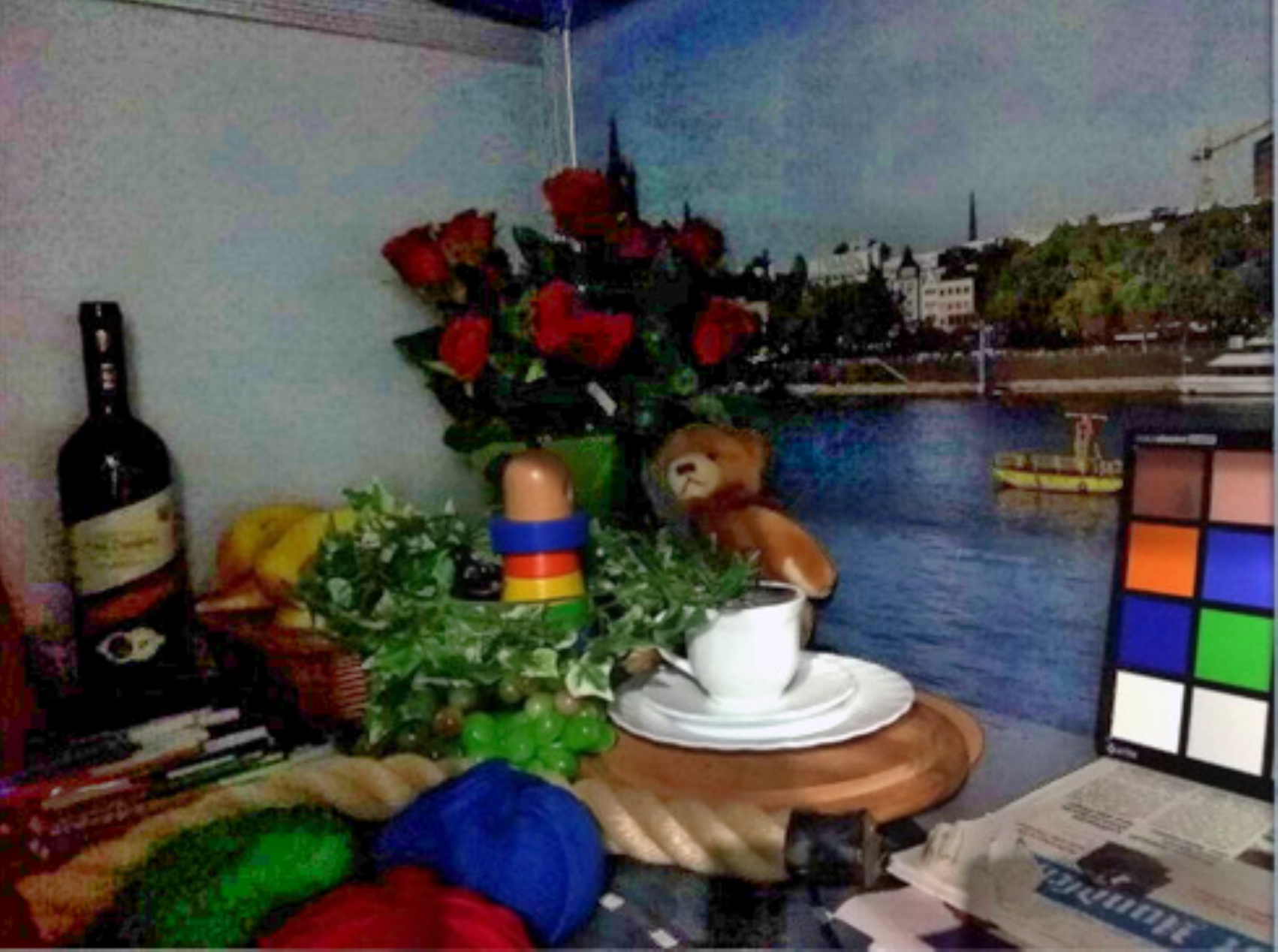}\vspace{2pt}
			\includegraphics[width=2.5cm]{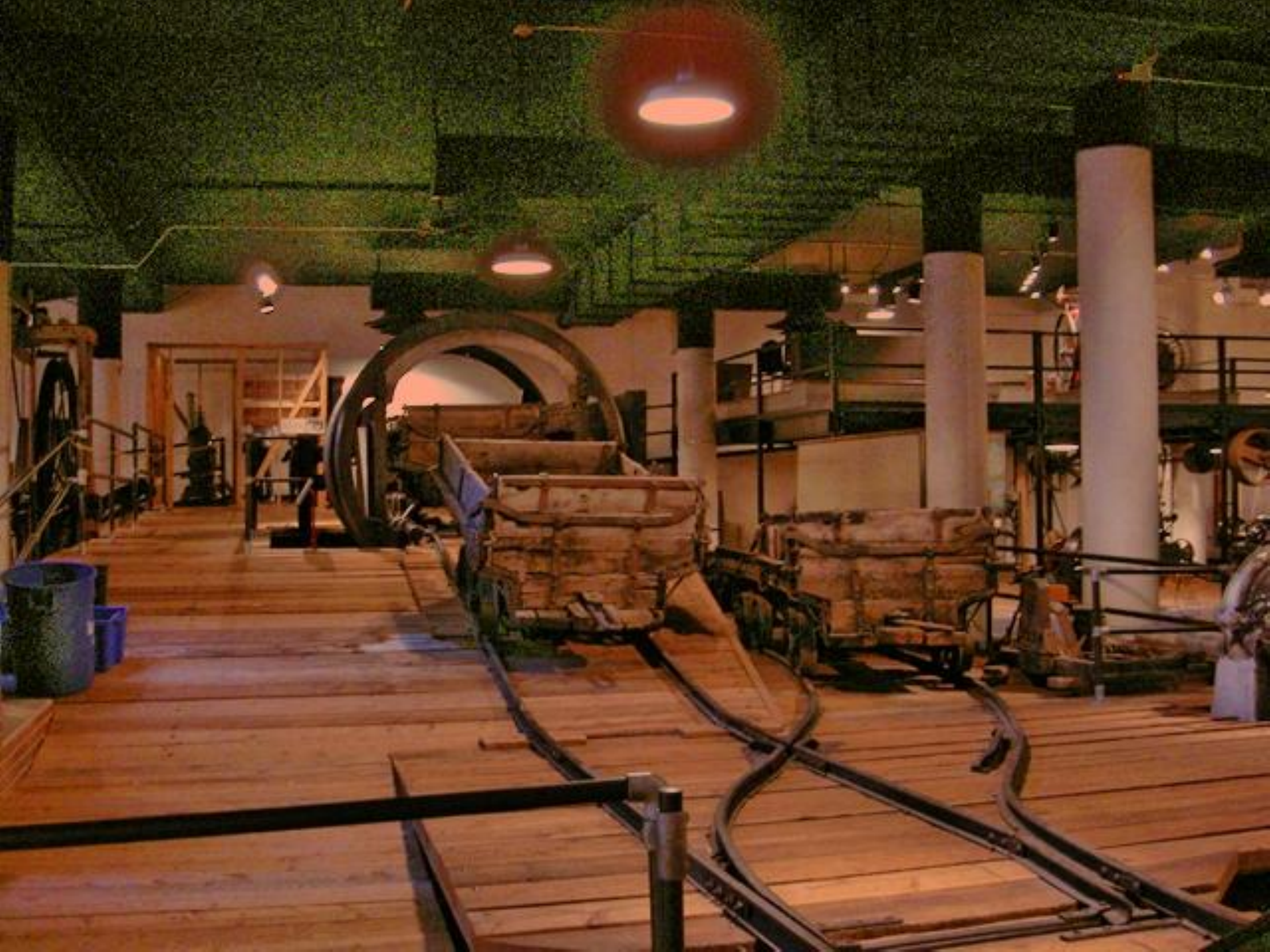}\vspace{2pt}
			\includegraphics[width=2.5cm]{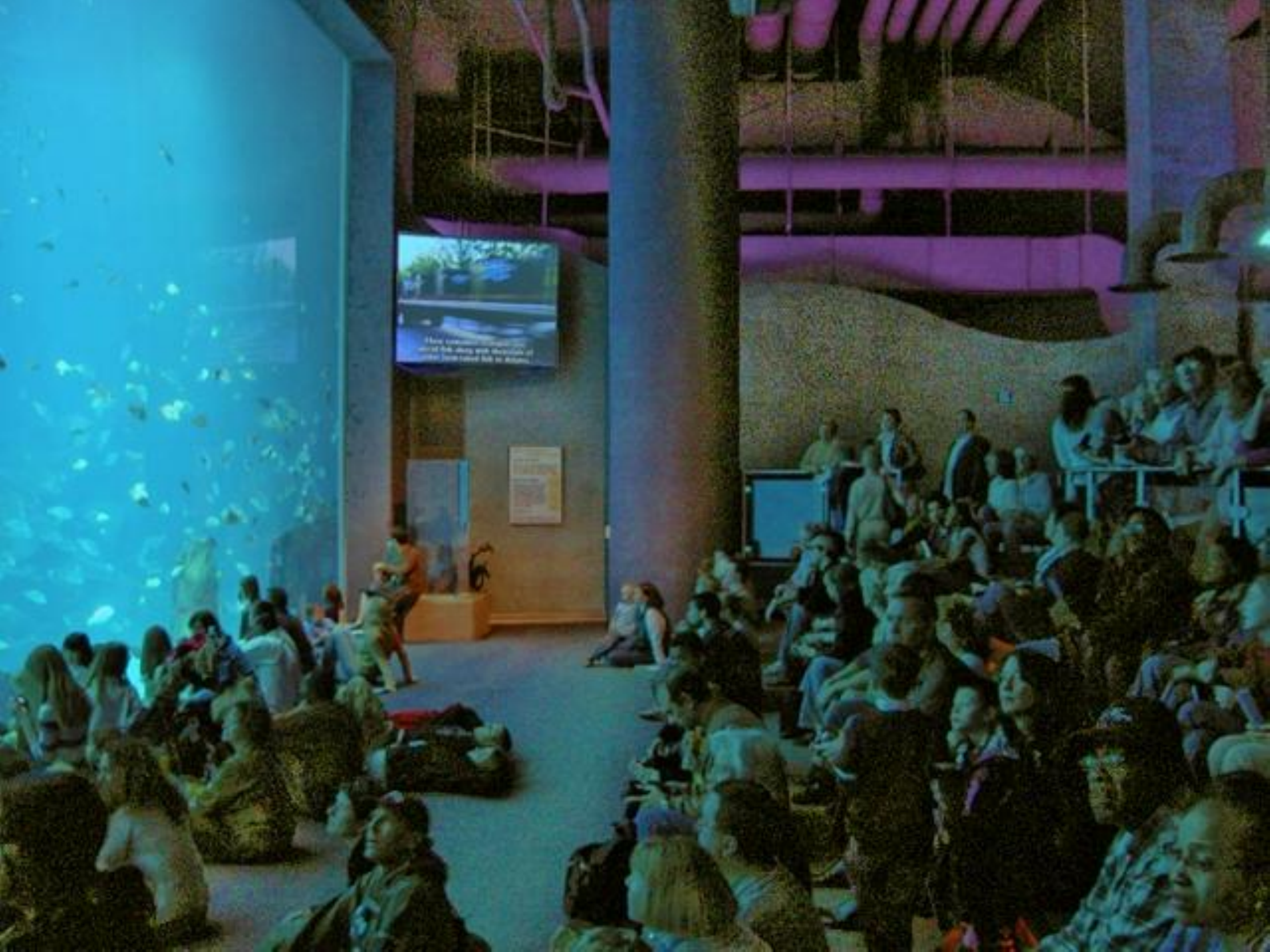}\vspace{2pt}
			\includegraphics[width=2.5cm]{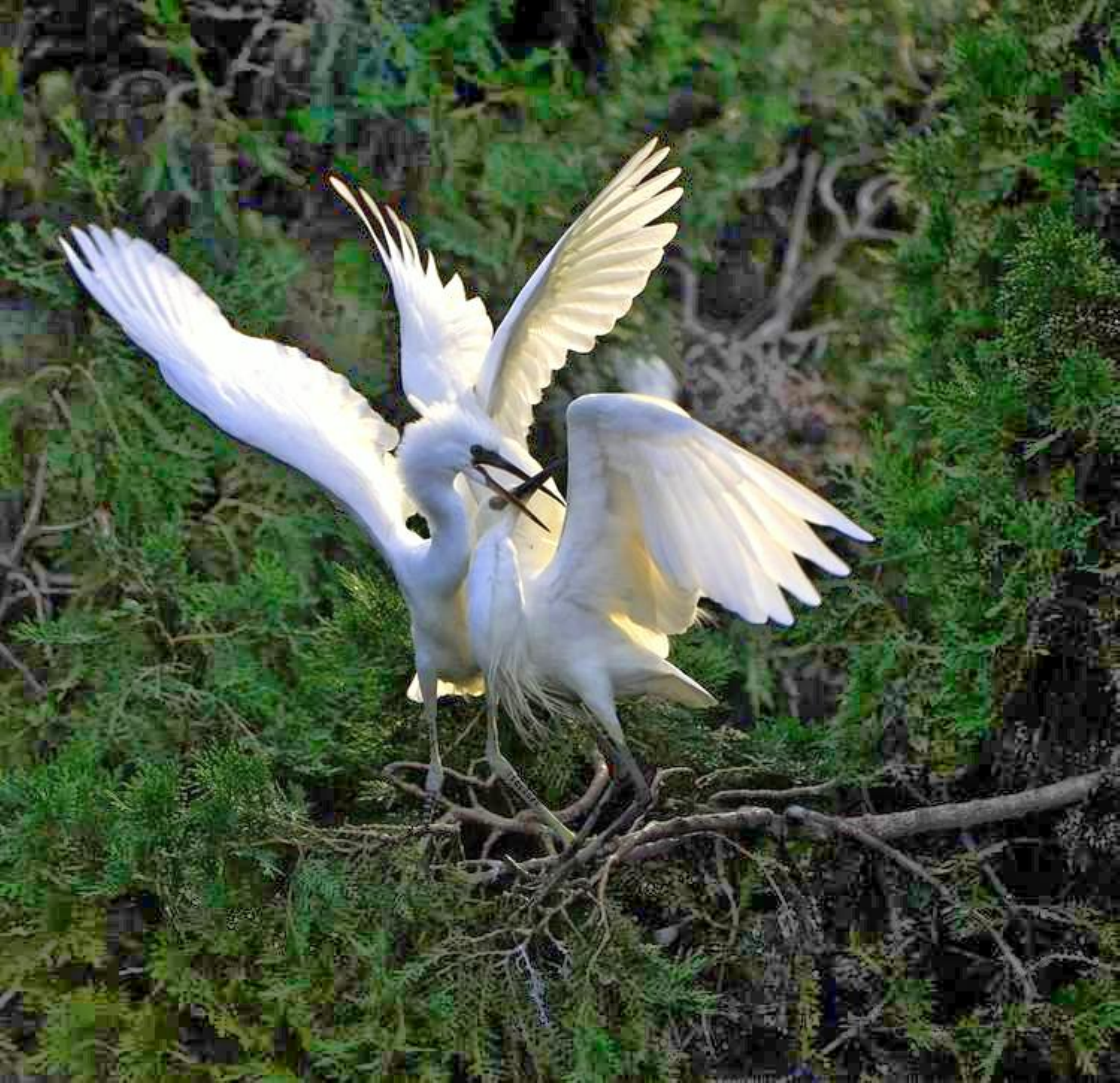}
		\end{minipage}
	}\hspace{-5pt}
	\subfigure[GLAD\cite{wang2018gladnet}]{
		\begin{minipage}[b]{0.13\textwidth}
			\includegraphics[width=2.5cm]{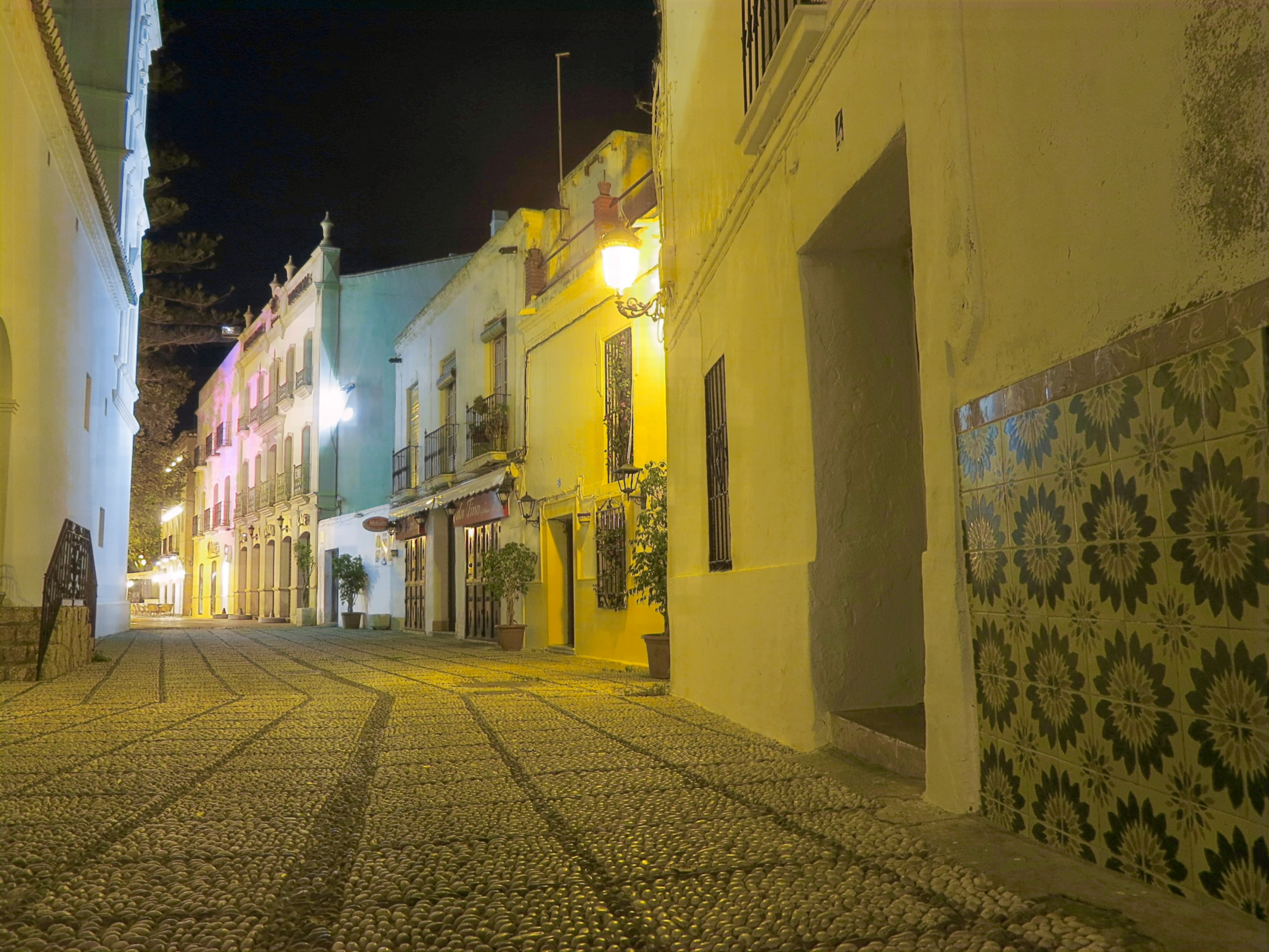}\vspace{2pt} \\
			\includegraphics[width=2.5cm]{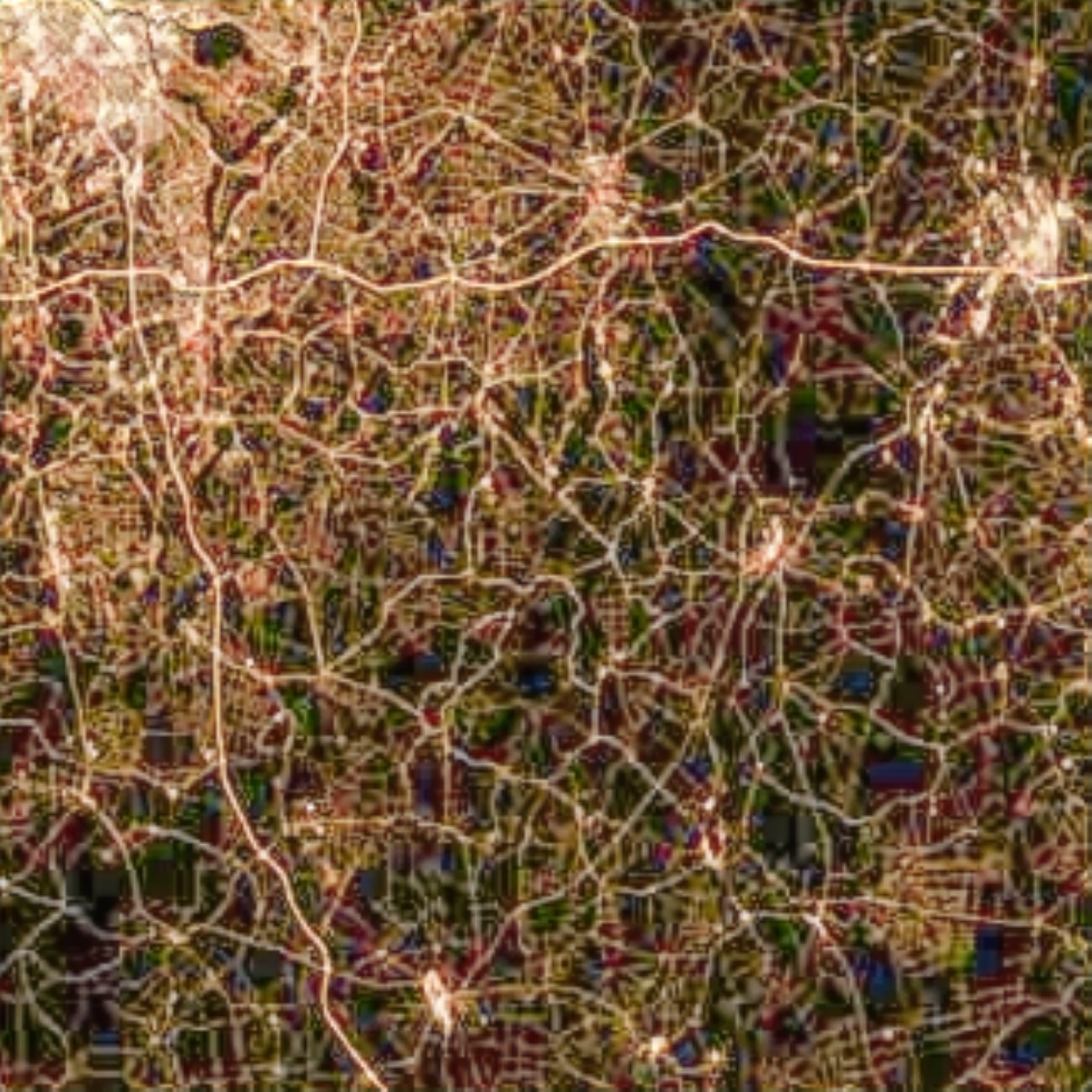}\vspace{2pt}
			\includegraphics[width=2.5cm]{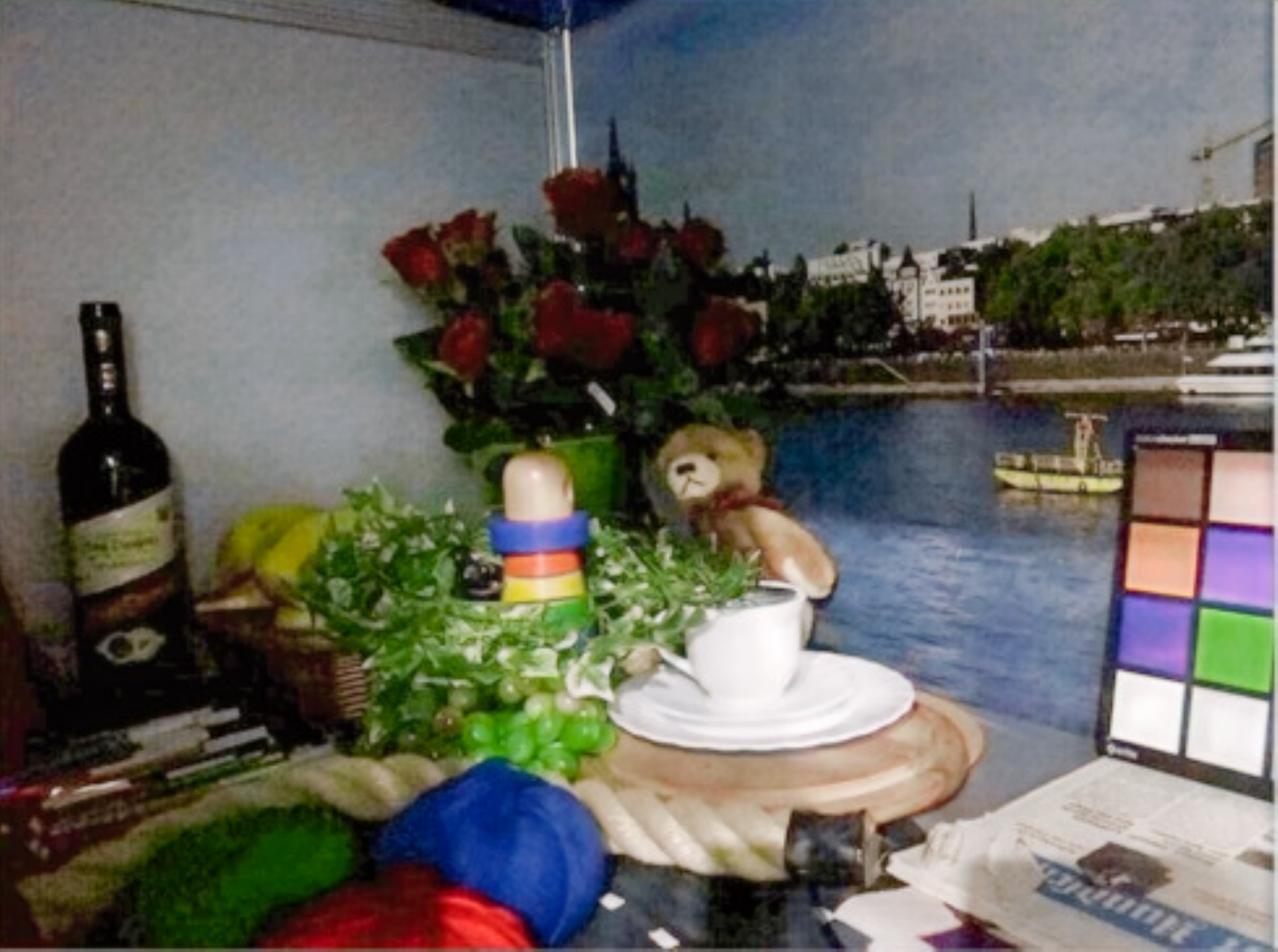}\vspace{2pt}
			\includegraphics[width=2.5cm]{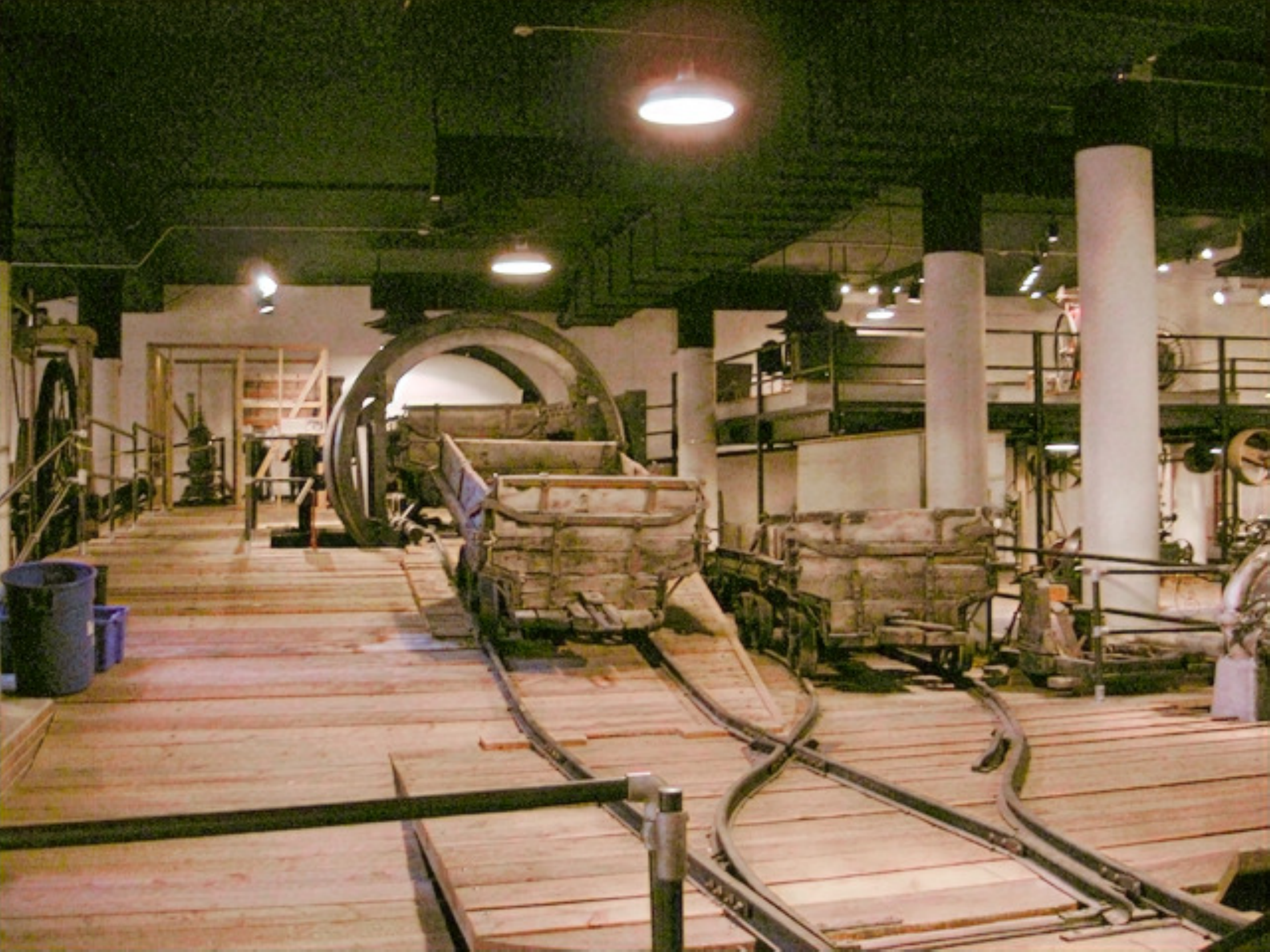}\vspace{2pt}
			\includegraphics[width=2.5cm]{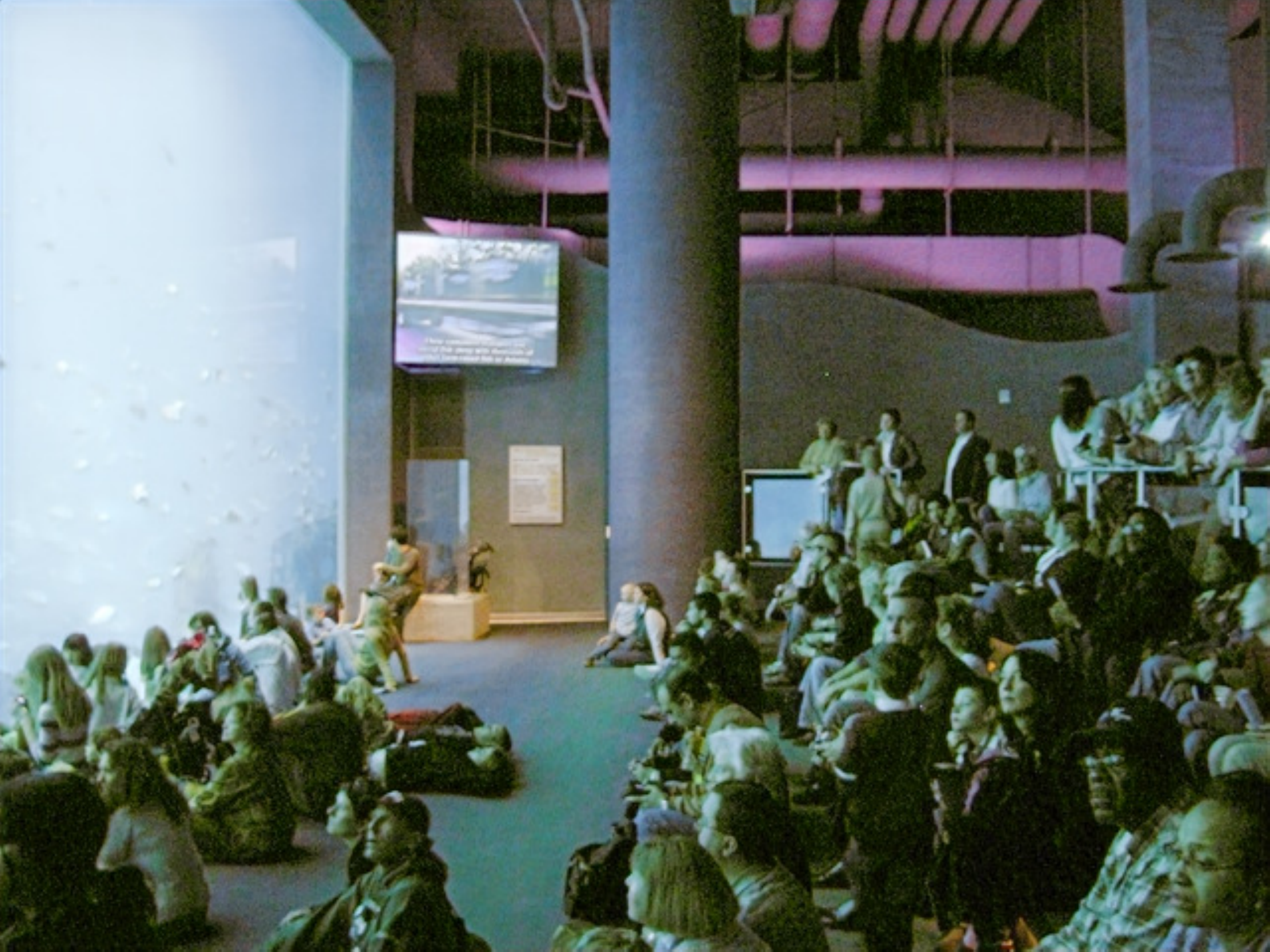}\vspace{2pt}
			\includegraphics[width=2.5cm]{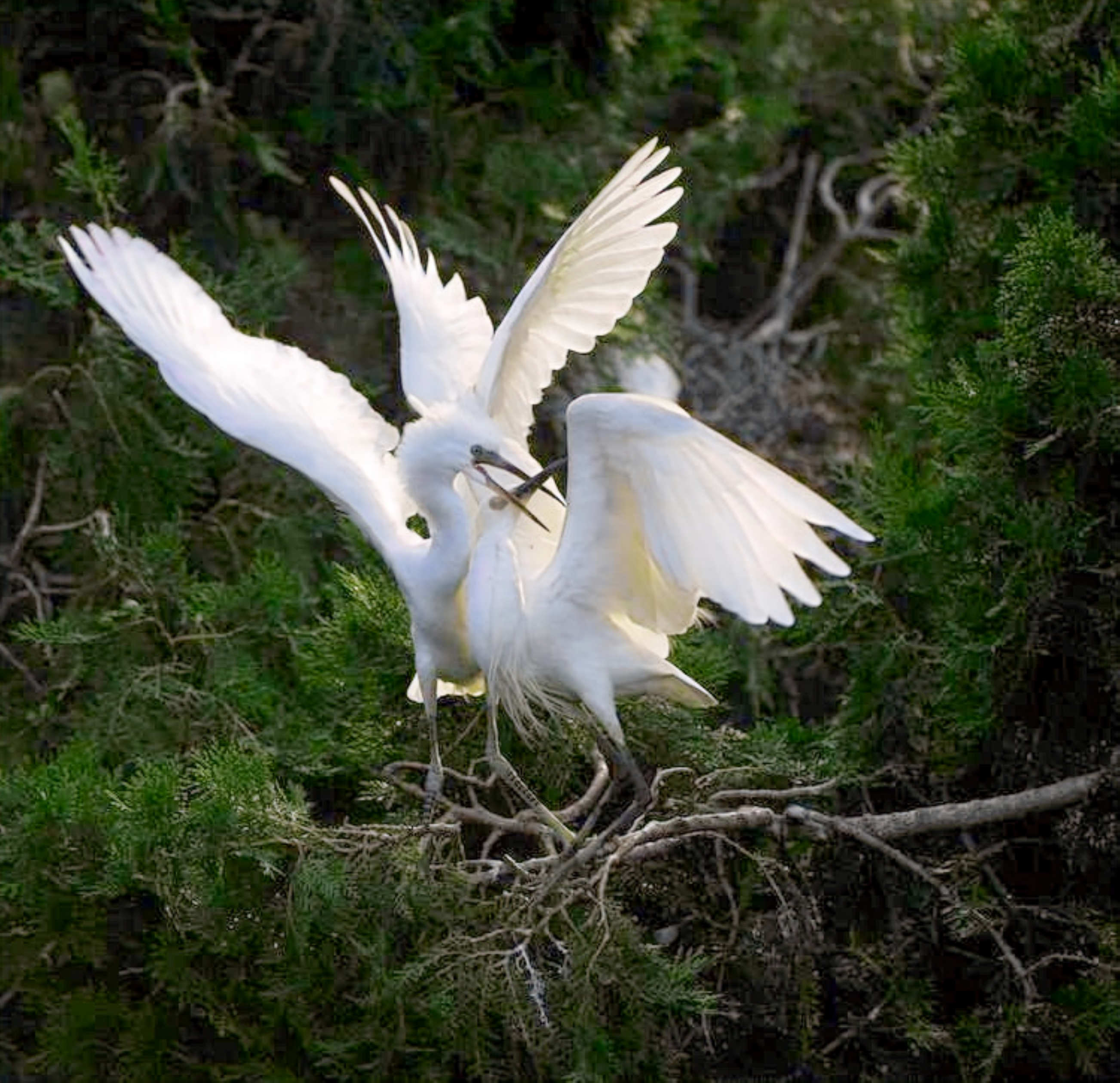}
		\end{minipage}
	}\hspace{-5pt}
		\subfigure[RetinexNet\cite{wei2018deep}]{
		\begin{minipage}[b]{0.13\textwidth}
			\includegraphics[width=2.5cm]{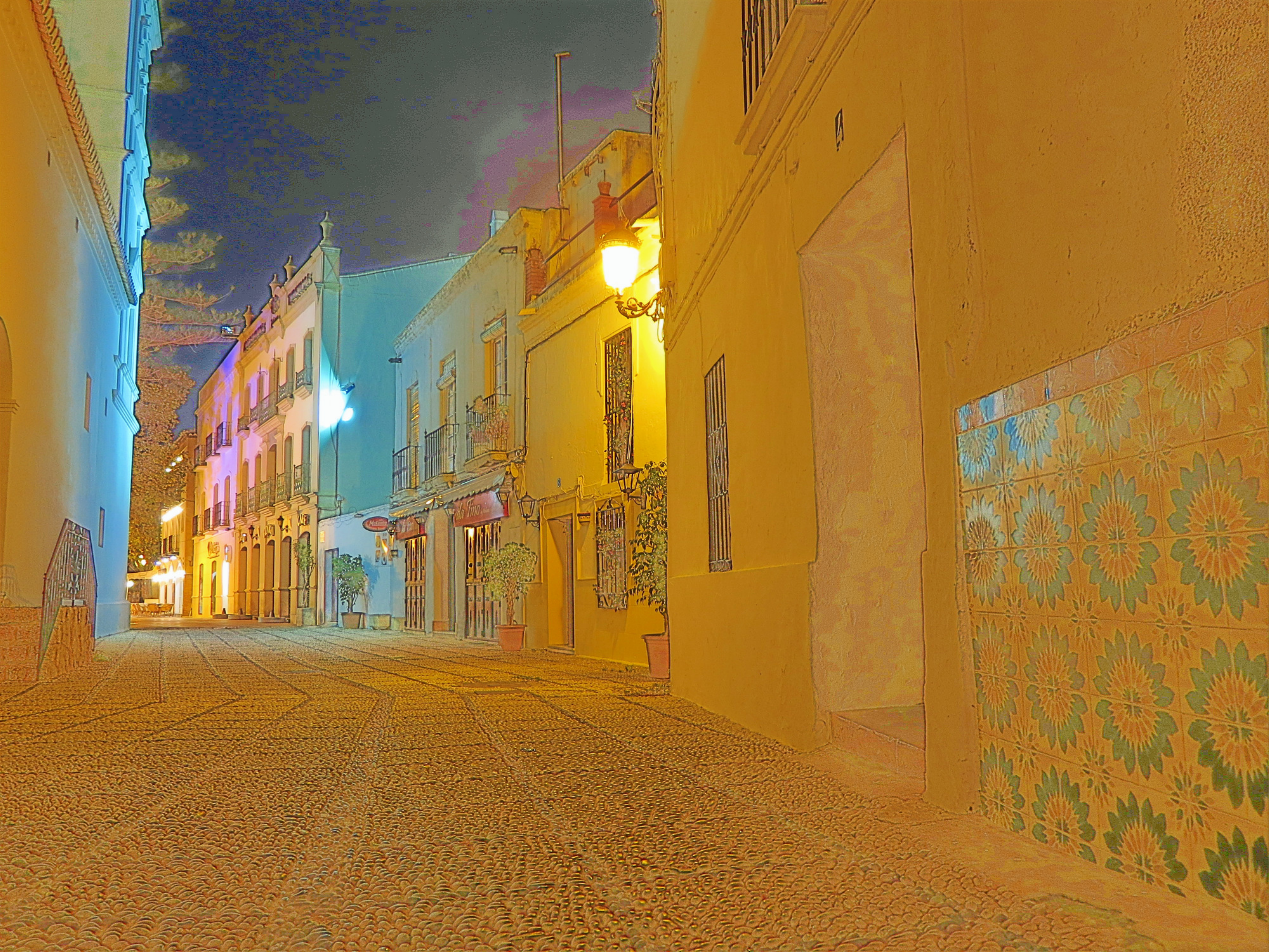}\vspace{2pt} \\
			\includegraphics[width=2.5cm]{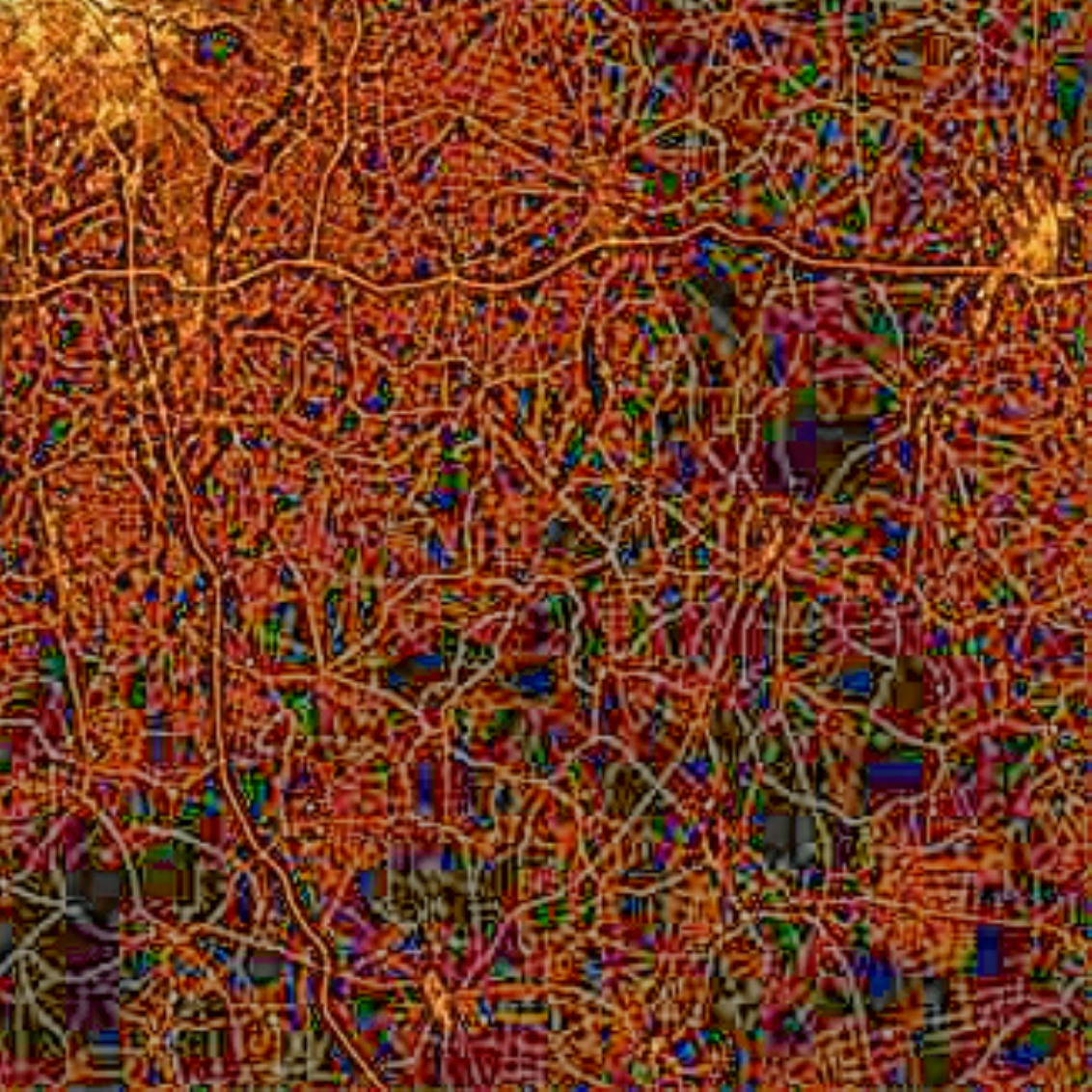}\vspace{2pt}
			\includegraphics[width=2.5cm]{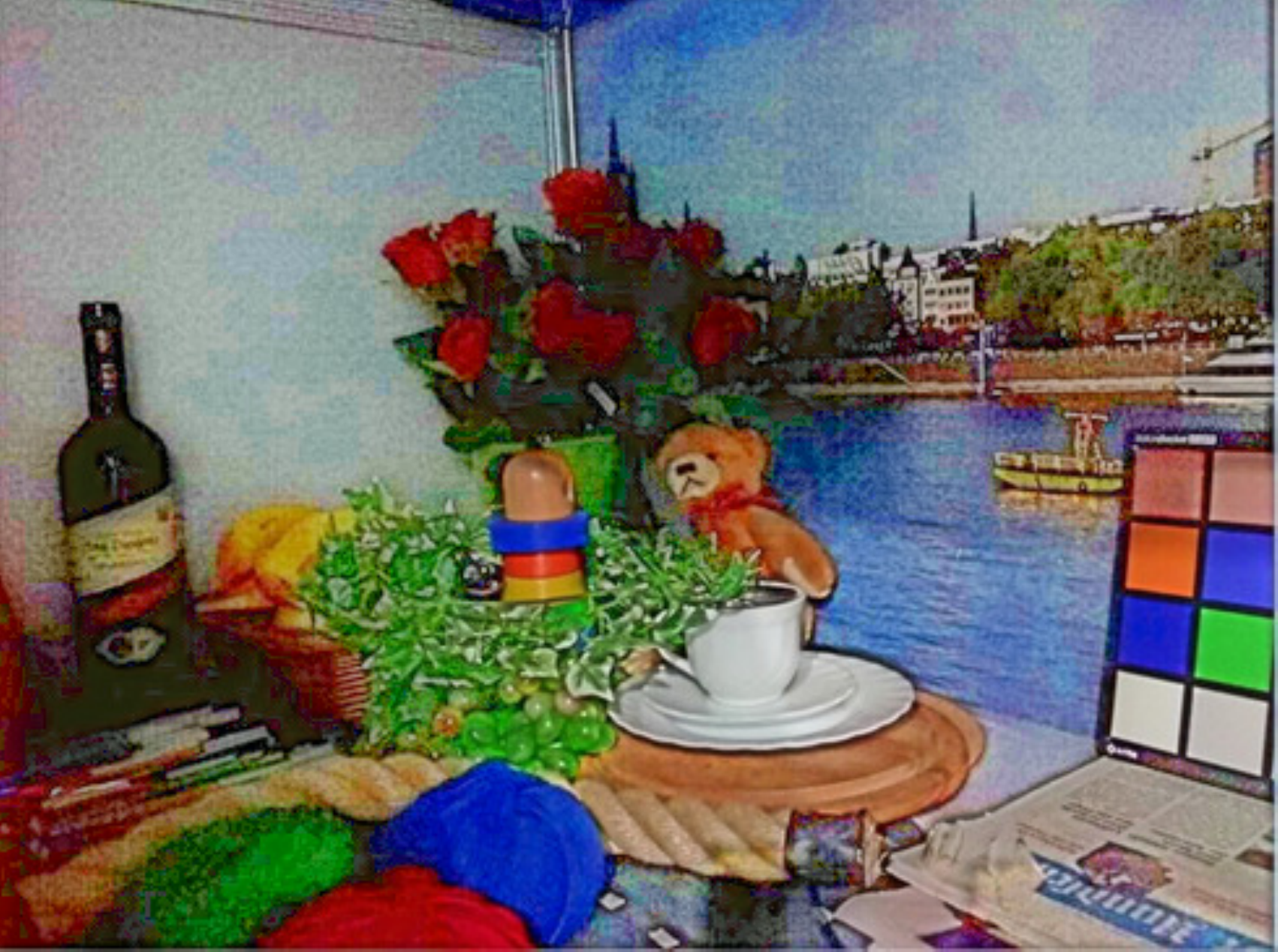}\vspace{2pt}
			\includegraphics[width=2.5cm]{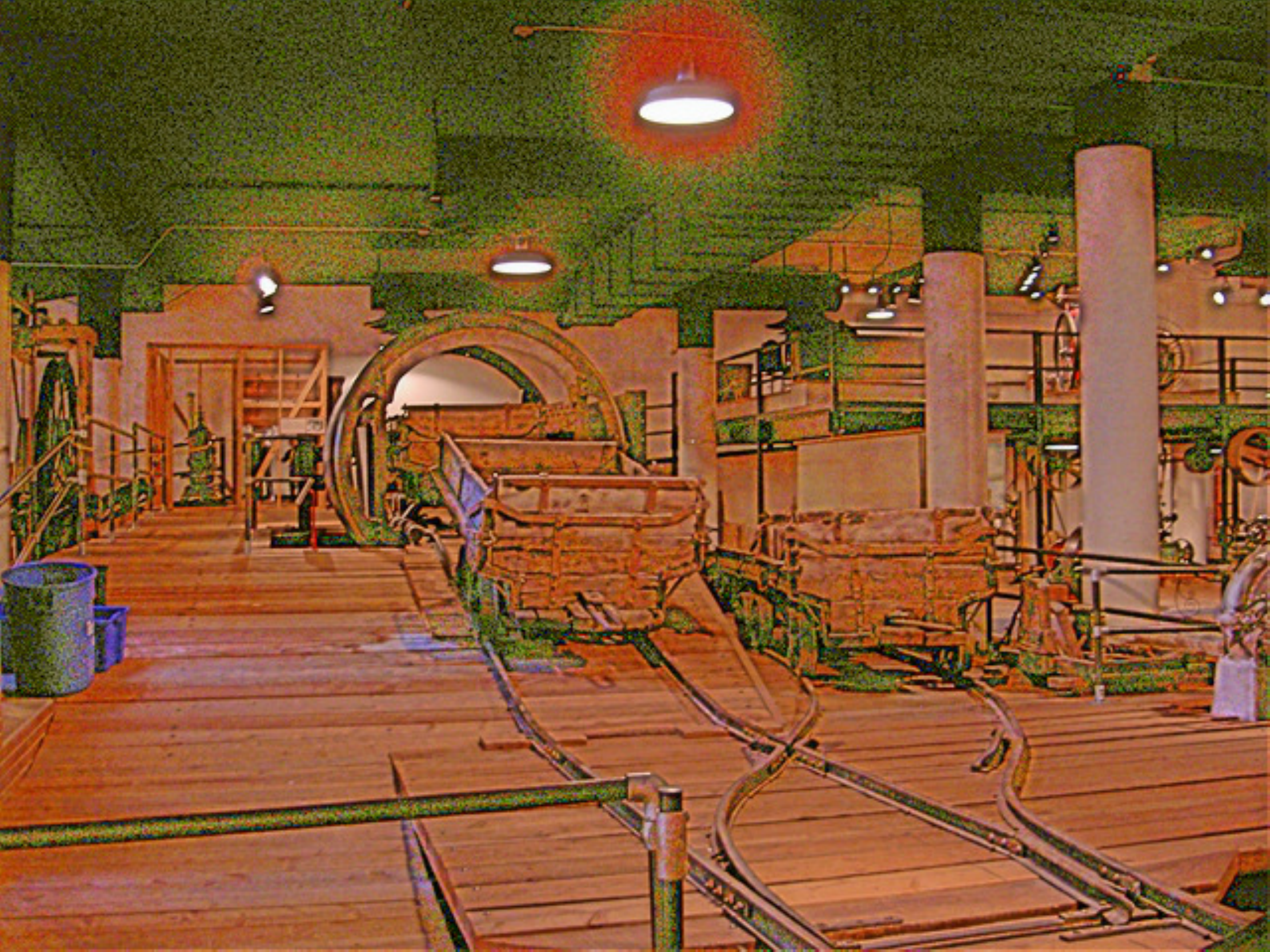}\vspace{2pt}
			\includegraphics[width=2.5cm]{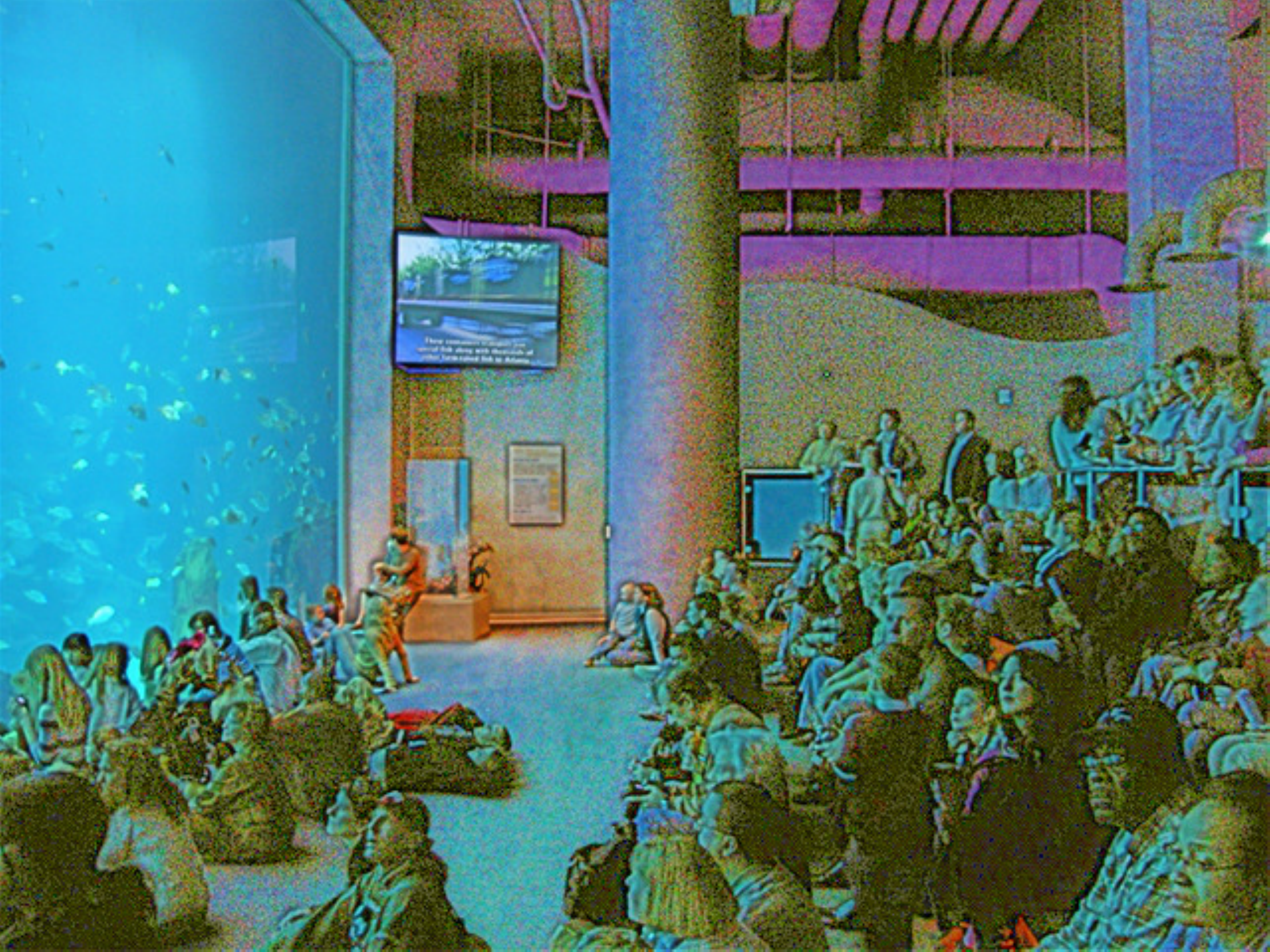}\vspace{2pt}
			\includegraphics[width=2.5cm]{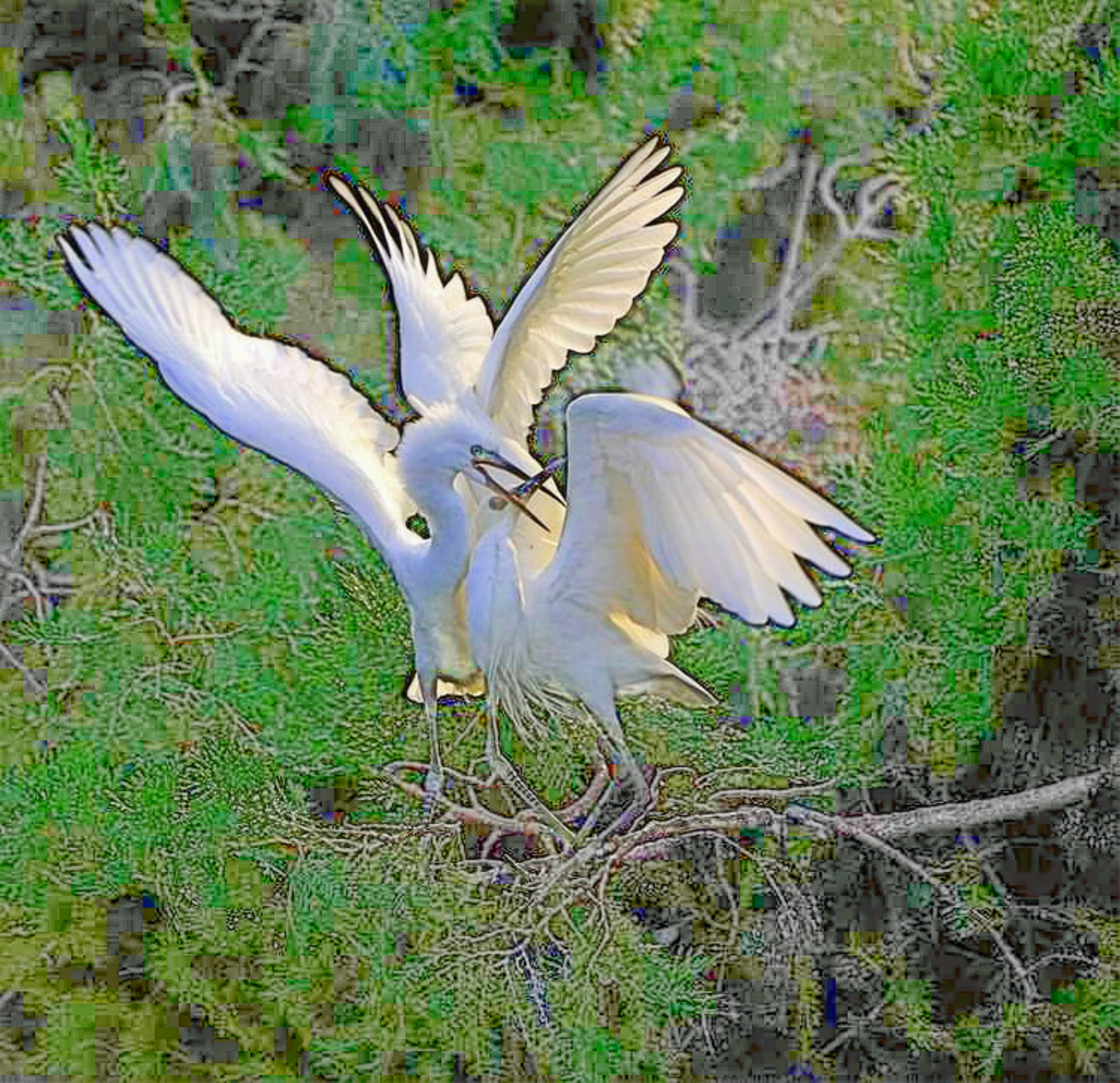}
		\end{minipage}
	}\hspace{-5pt}
	\subfigure[KinD\cite{zhang2019kindling}]{
		\begin{minipage}[b]{0.13\textwidth}
			\includegraphics[width=2.5cm]{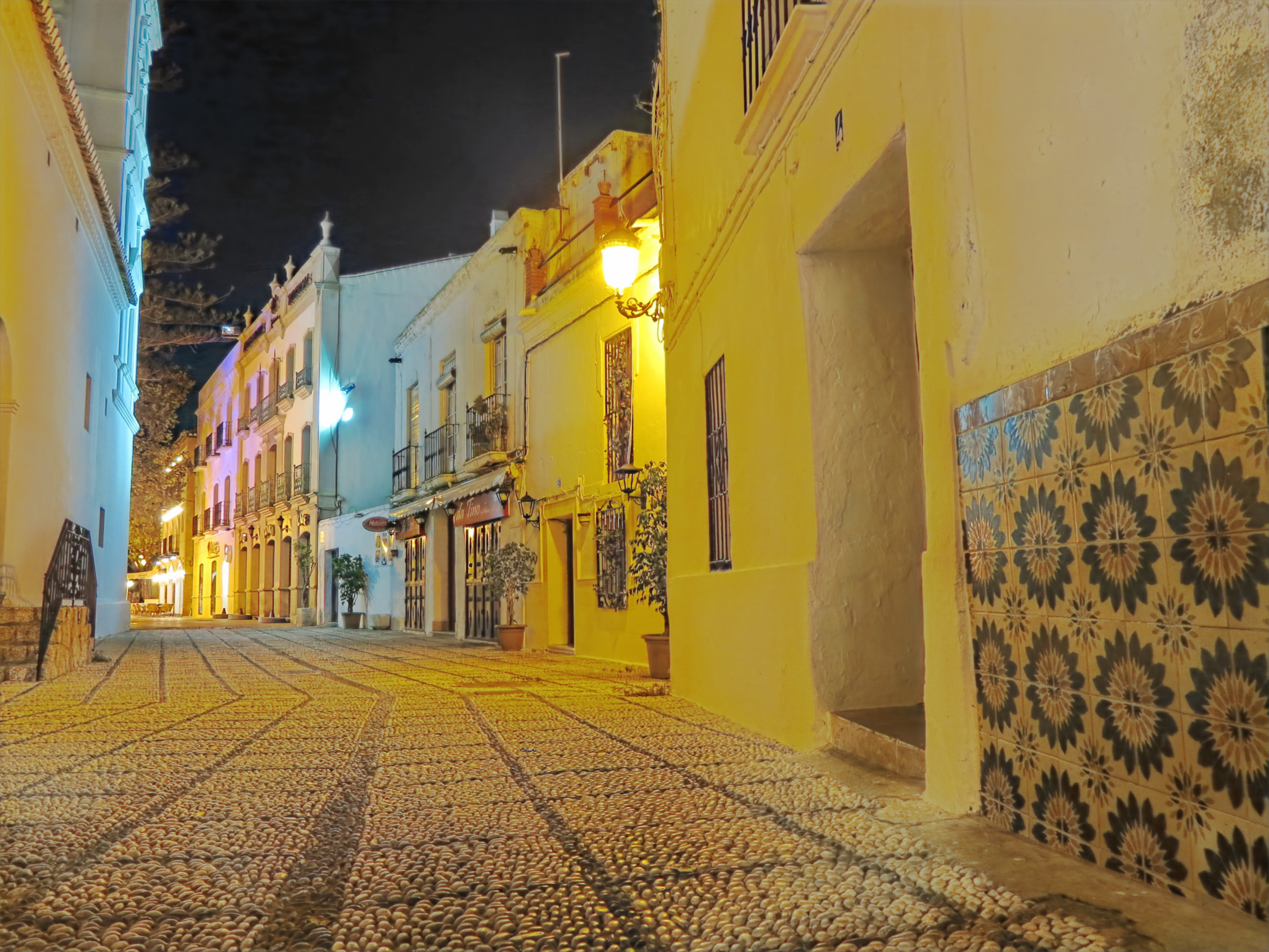}\vspace{2pt} \\
			\includegraphics[width=2.5cm]{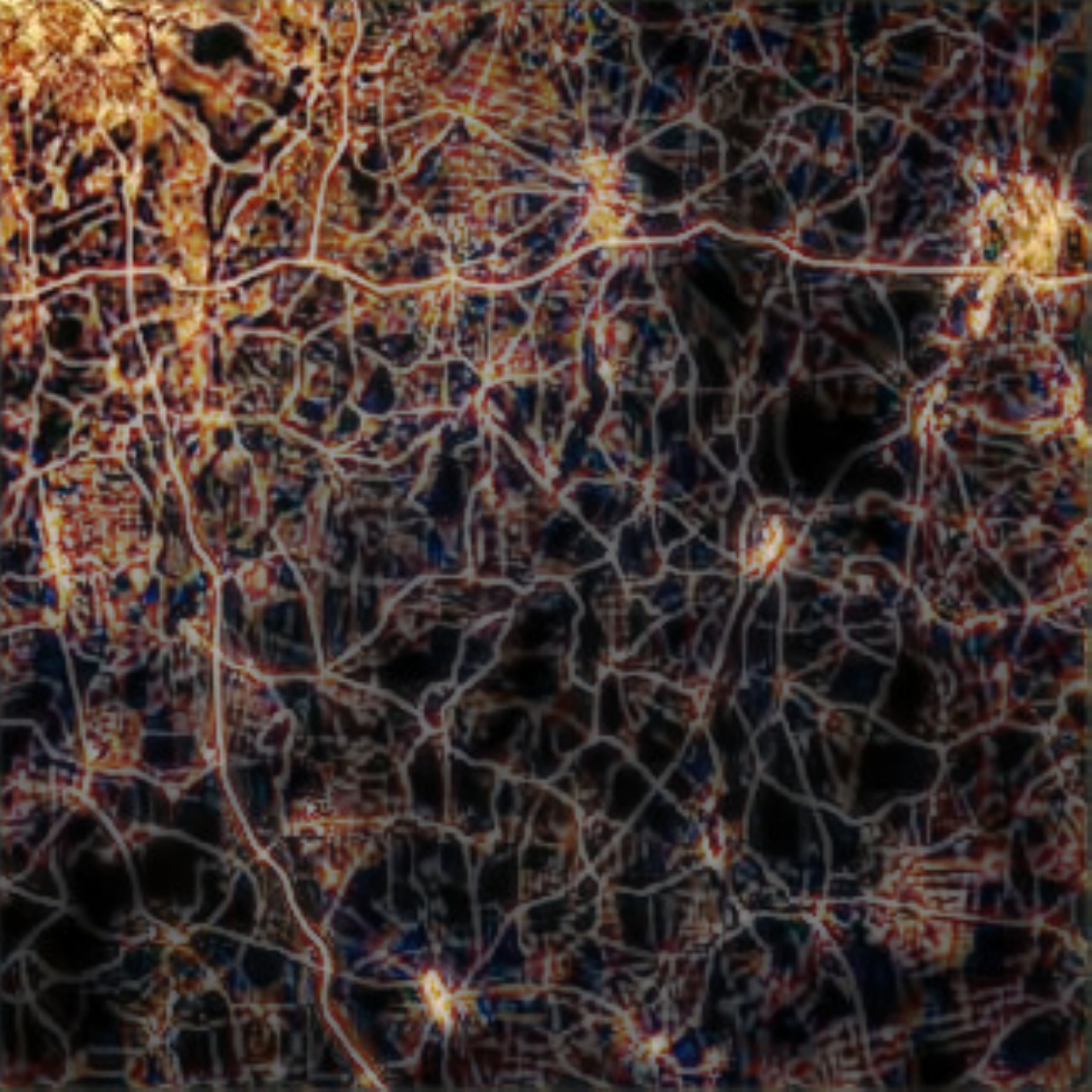}\vspace{2pt}
			\includegraphics[width=2.5cm]{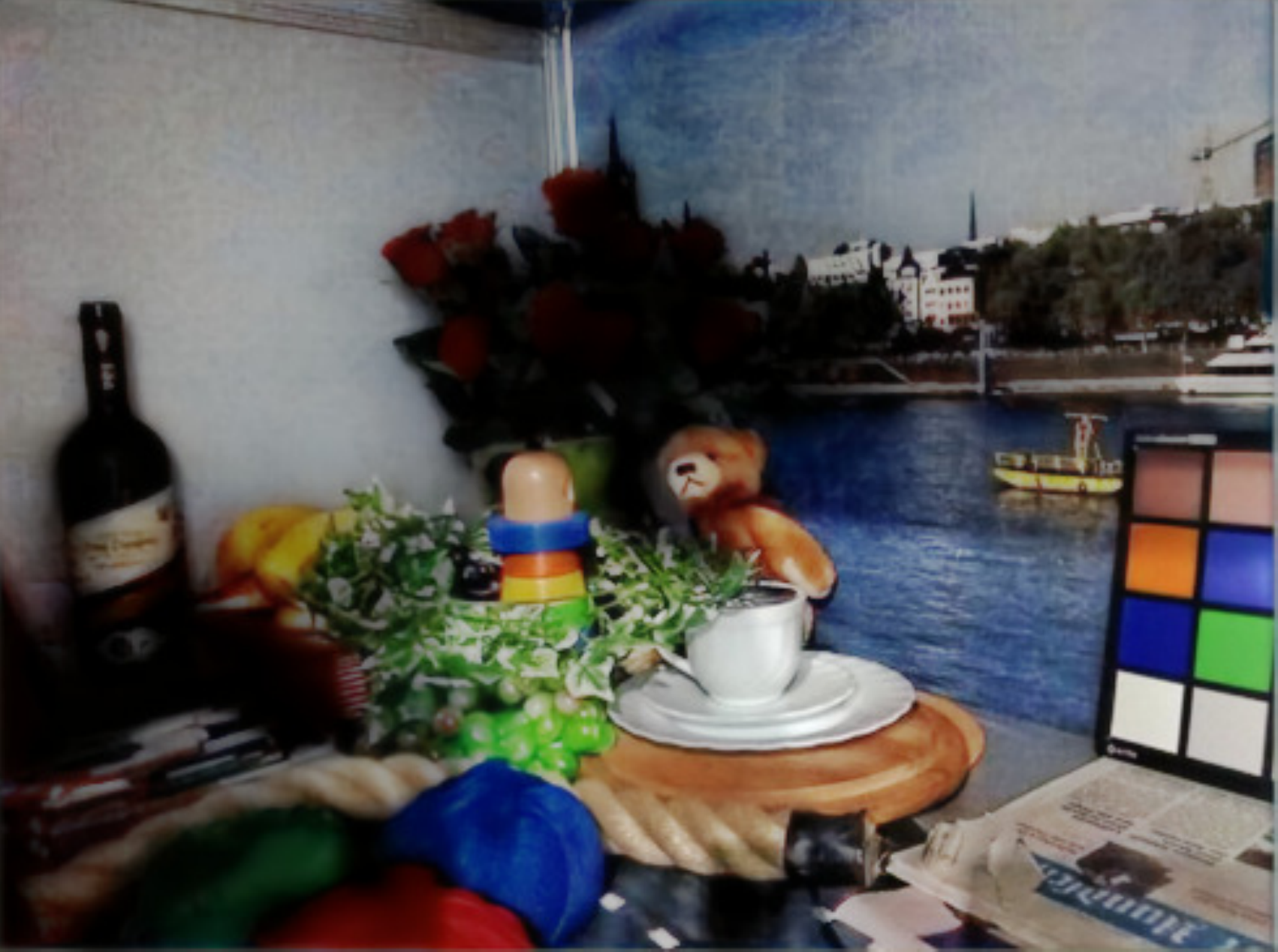}\vspace{2pt}
			\includegraphics[width=2.5cm]{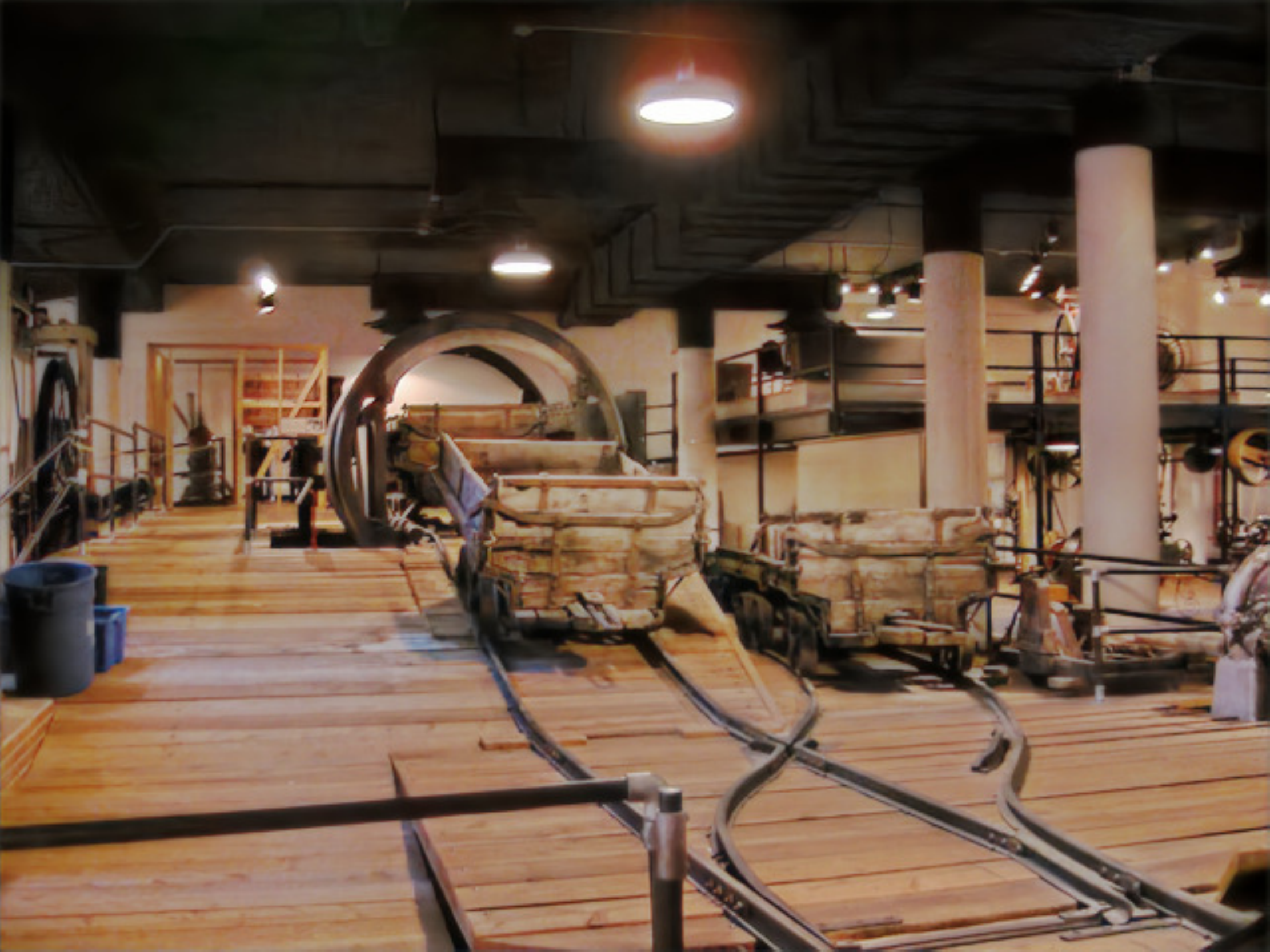}\vspace{2pt}
			\includegraphics[width=2.5cm]{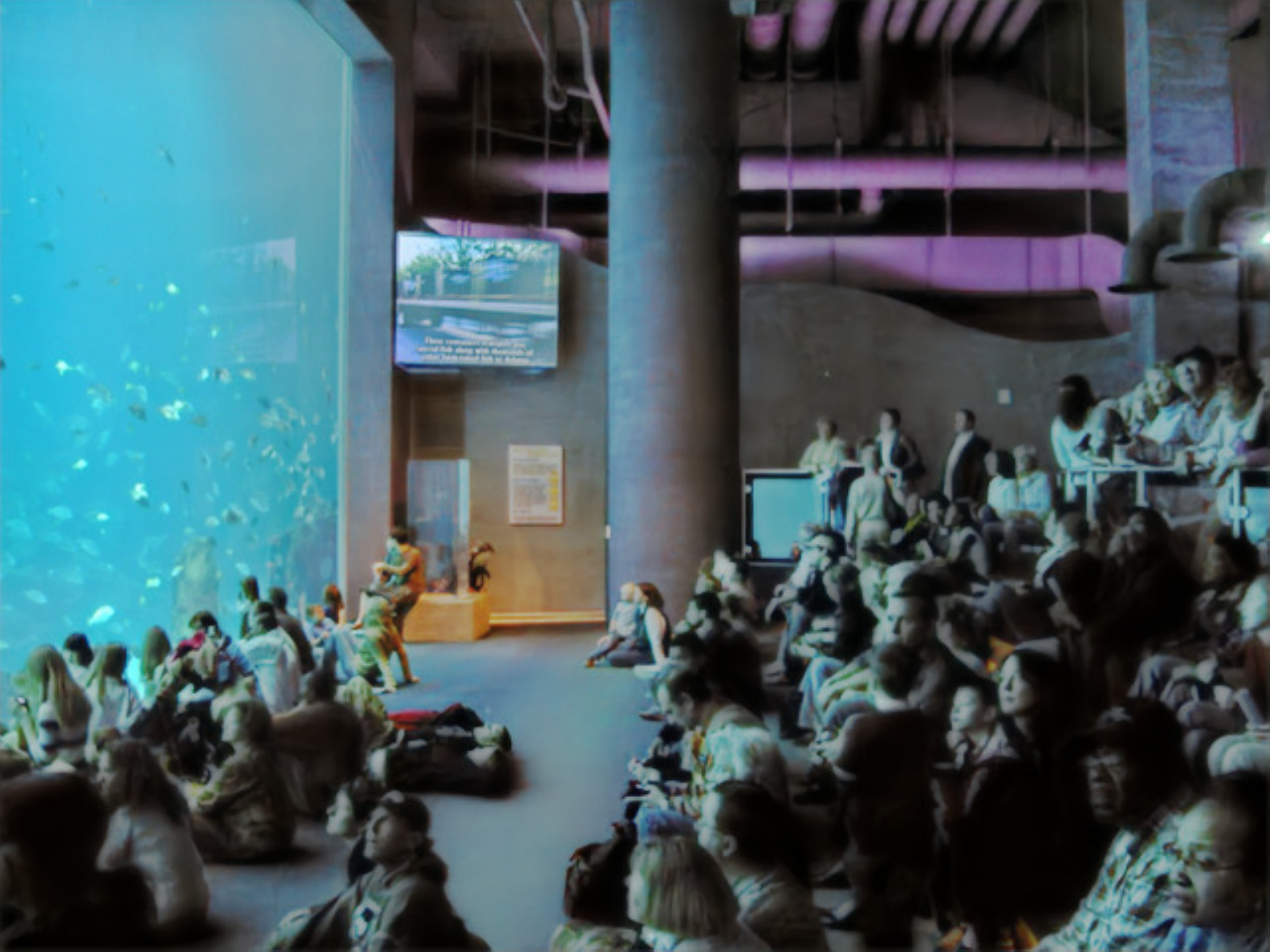}\vspace{2pt}
			\includegraphics[width=2.5cm]{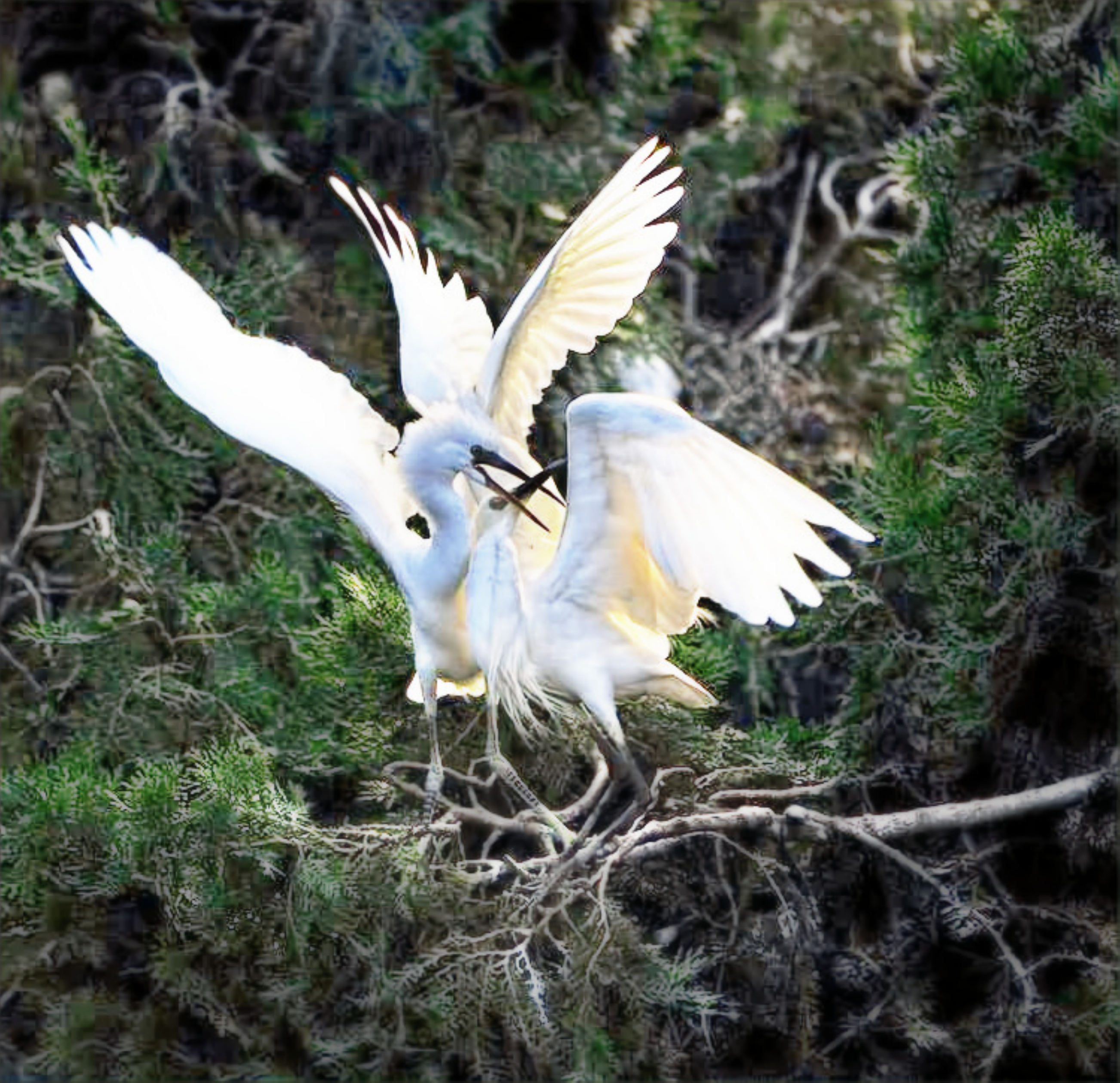}
		\end{minipage}
	}\hspace{-5pt}
	\subfigure[KinD++\cite{zhang2021beyond}]{
		\begin{minipage}[b]{0.13\textwidth}
			\includegraphics[width=2.5cm]{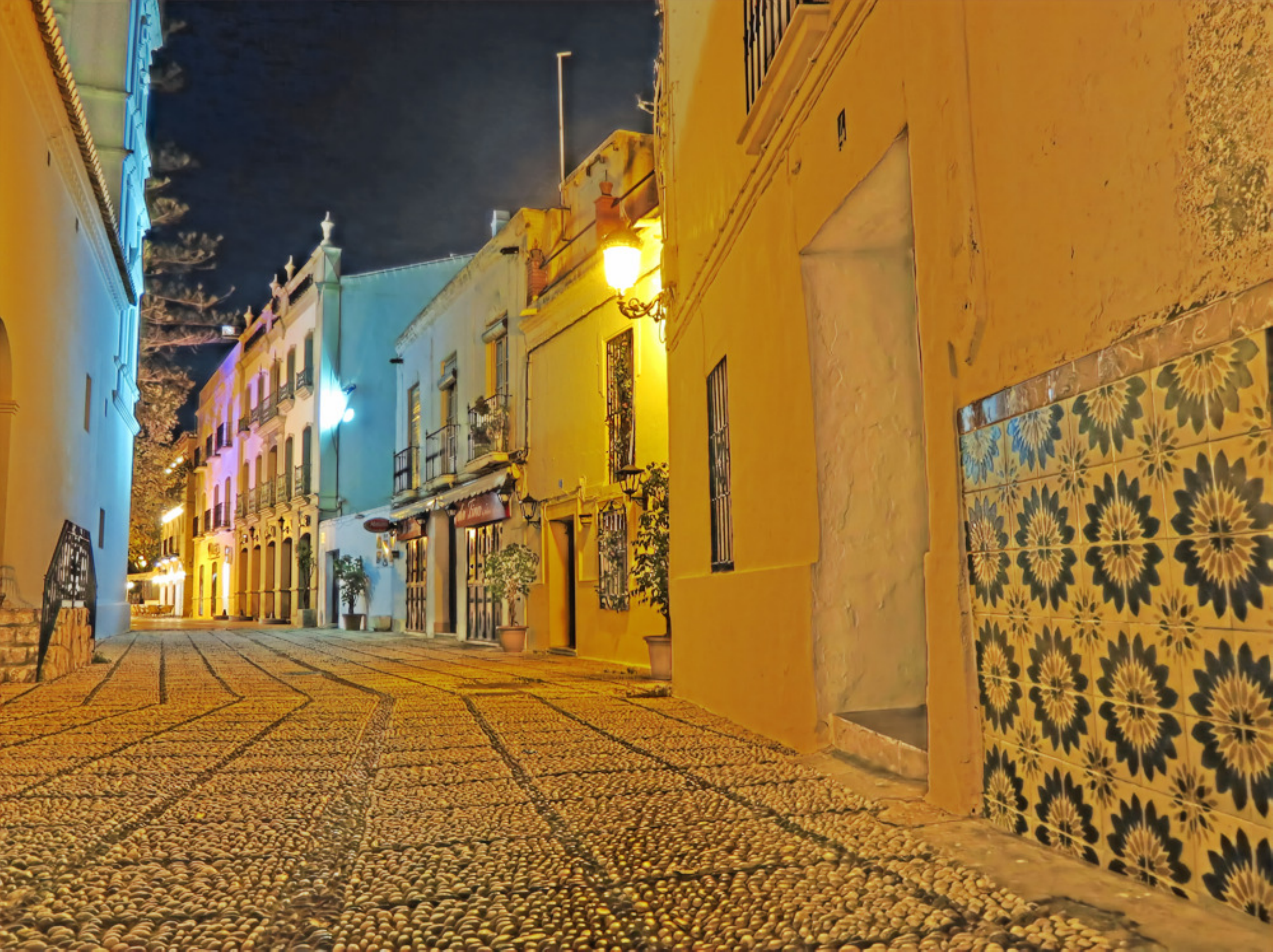}\vspace{2pt} \\
			\includegraphics[width=2.5cm]{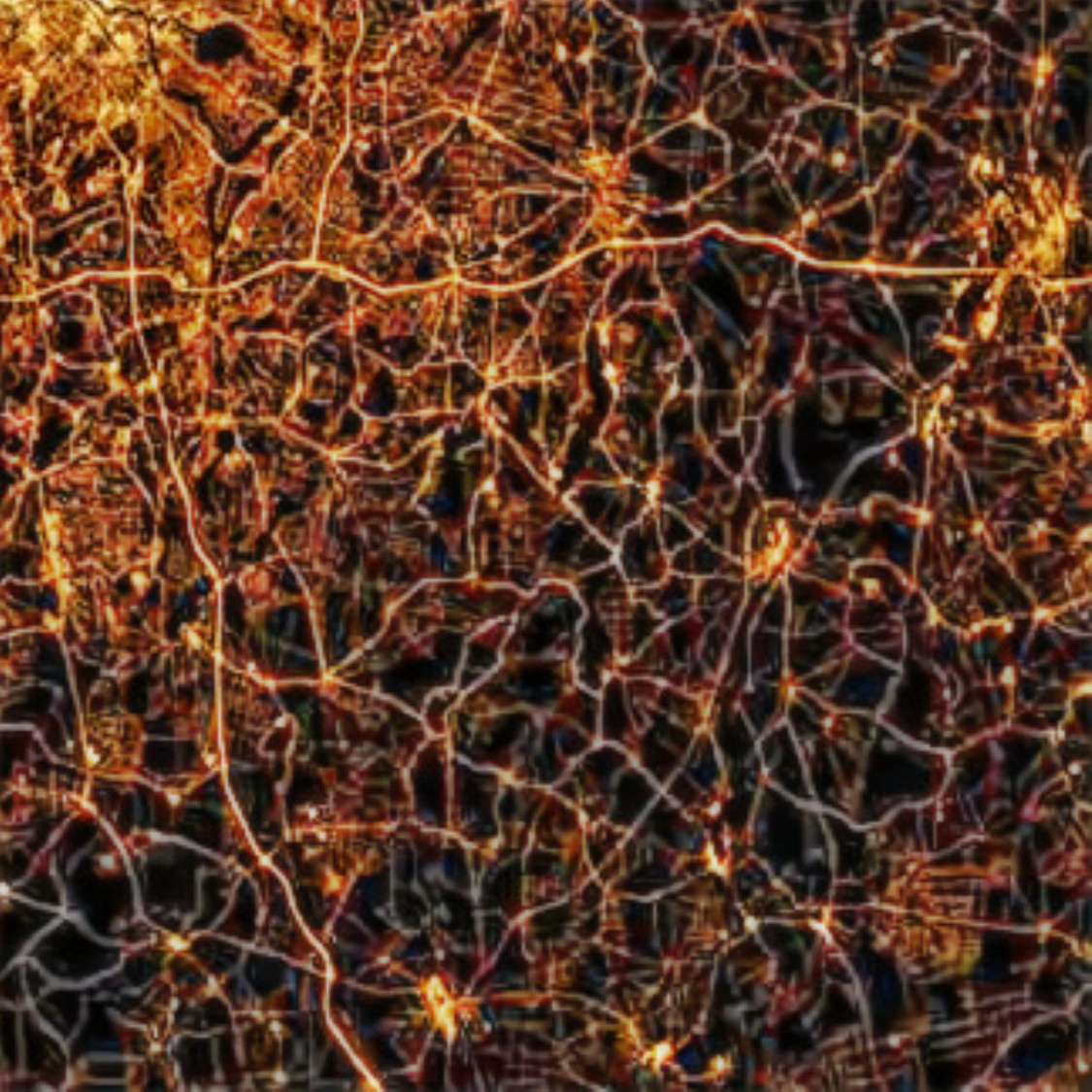}\vspace{2pt}
			\includegraphics[width=2.5cm]{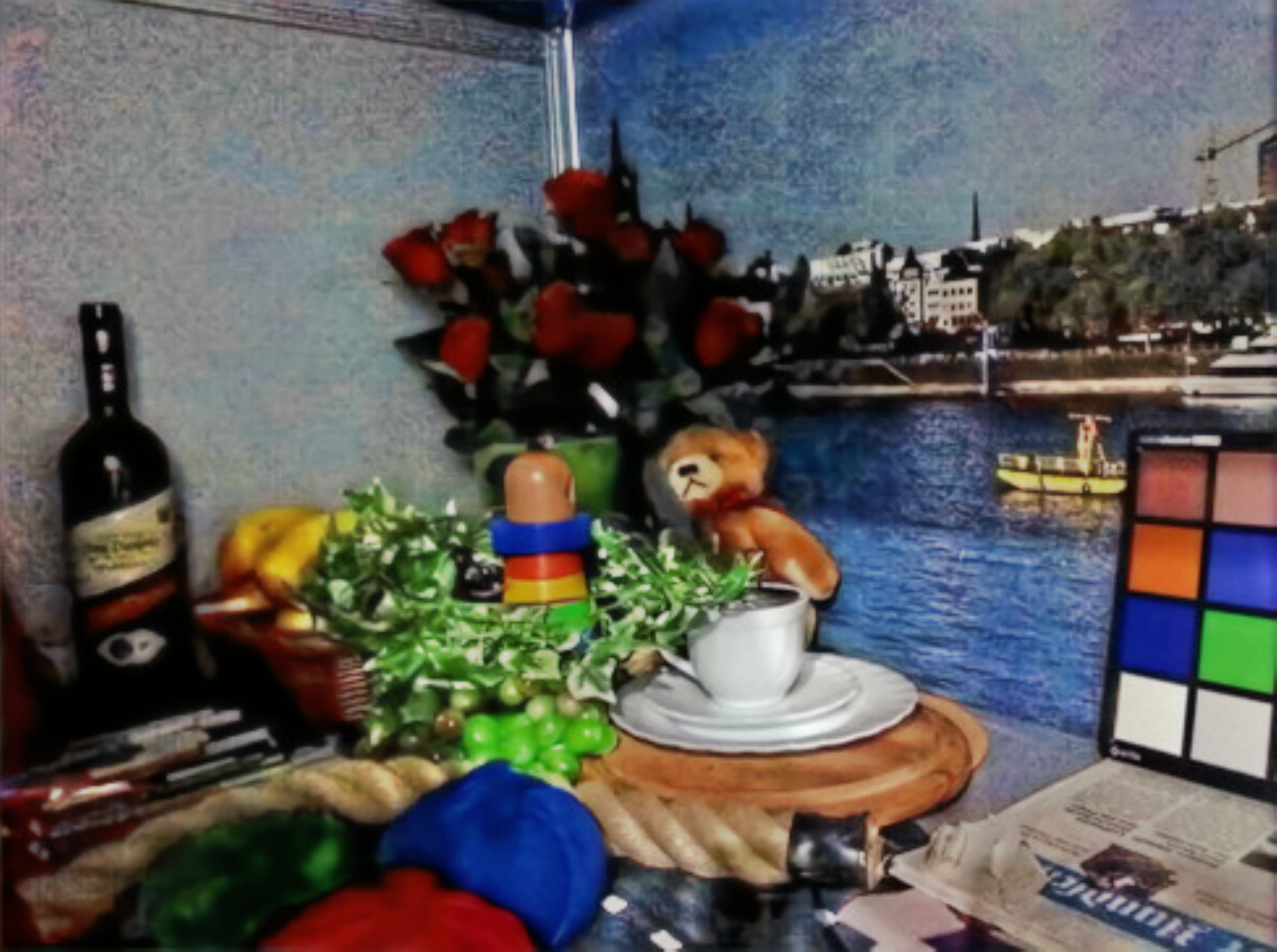}\vspace{2pt}
			\includegraphics[width=2.5cm]{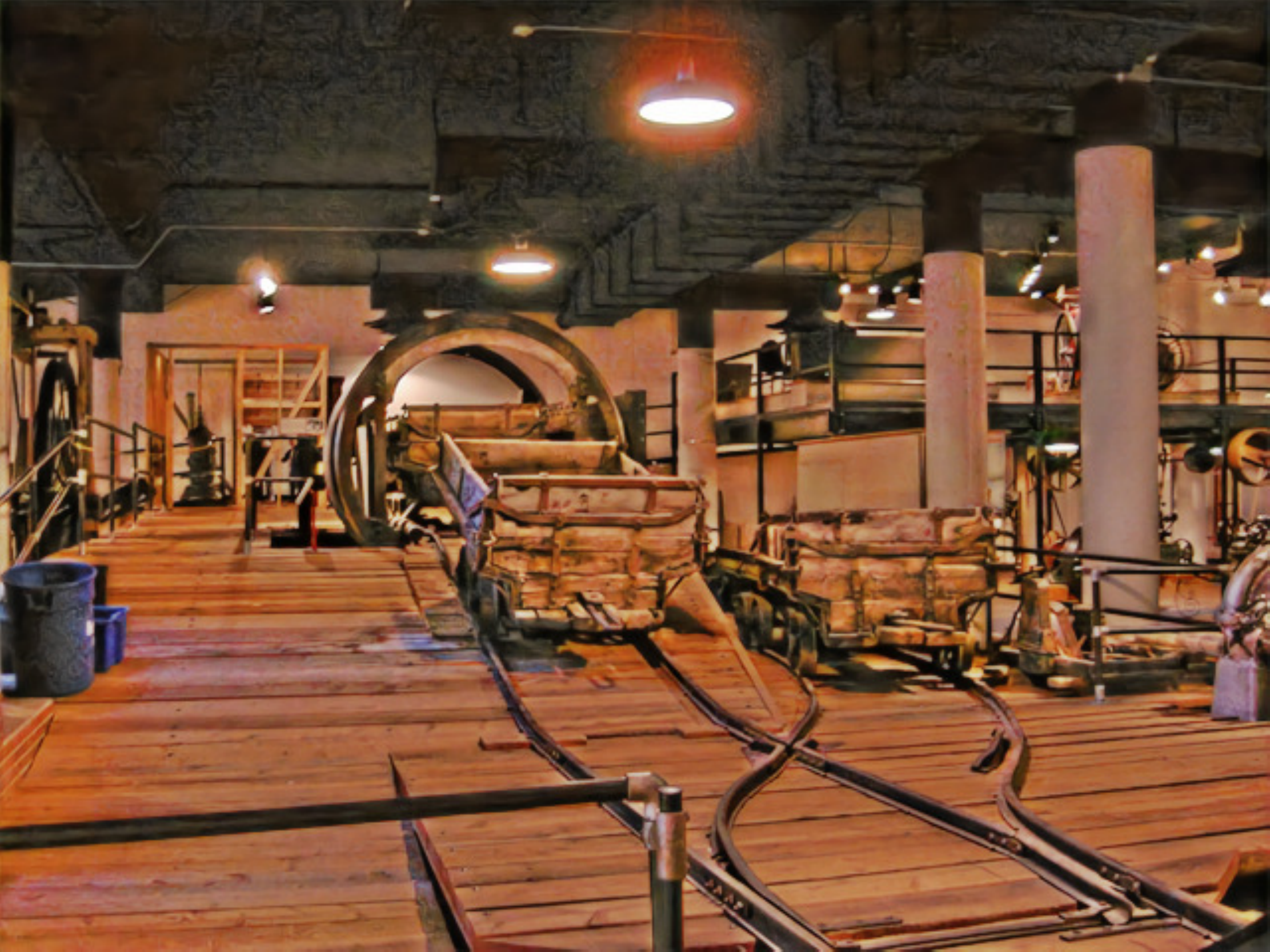}\vspace{2pt}
			\includegraphics[width=2.5cm]{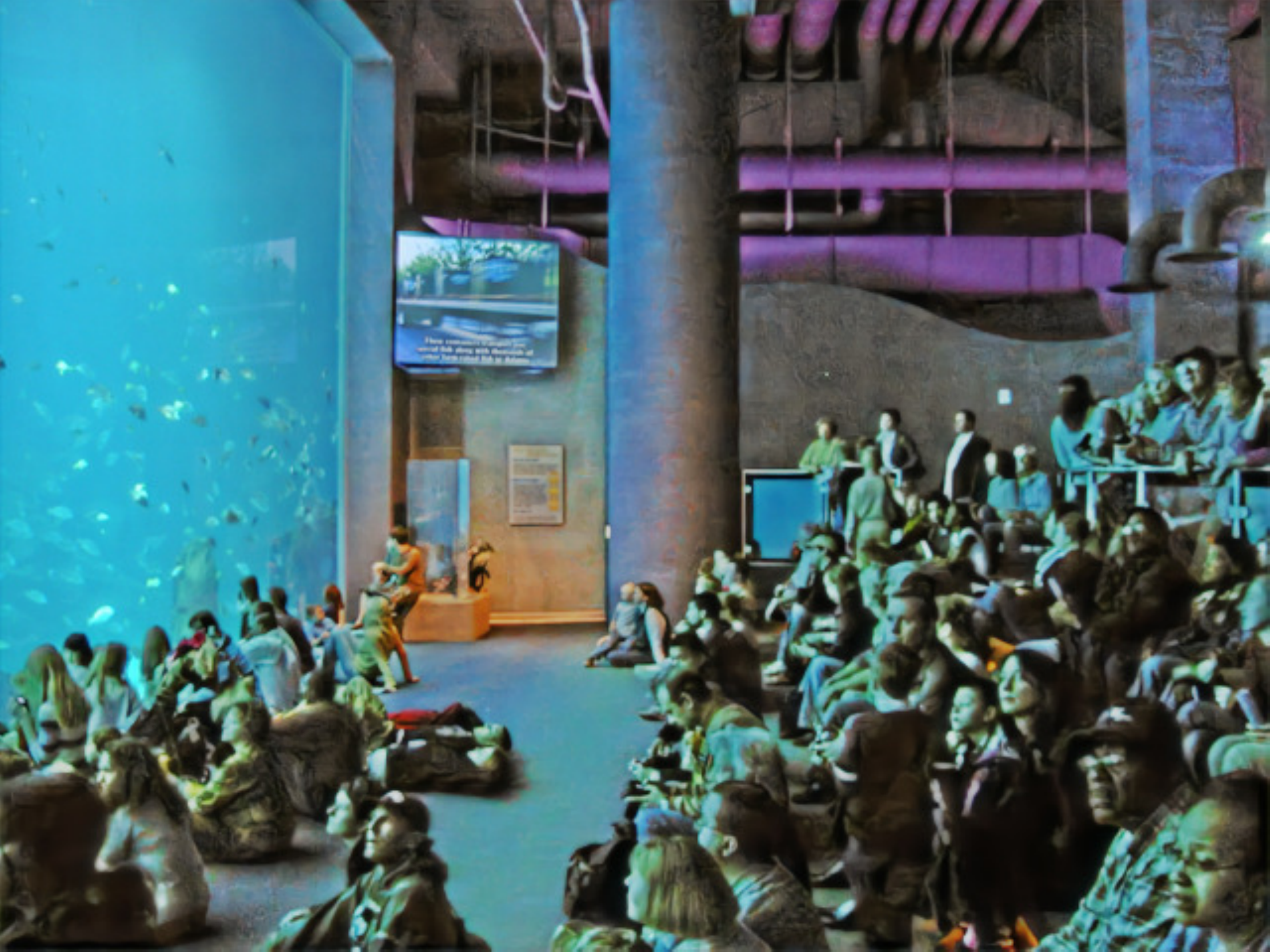}\vspace{2pt}
			\includegraphics[width=2.5cm]{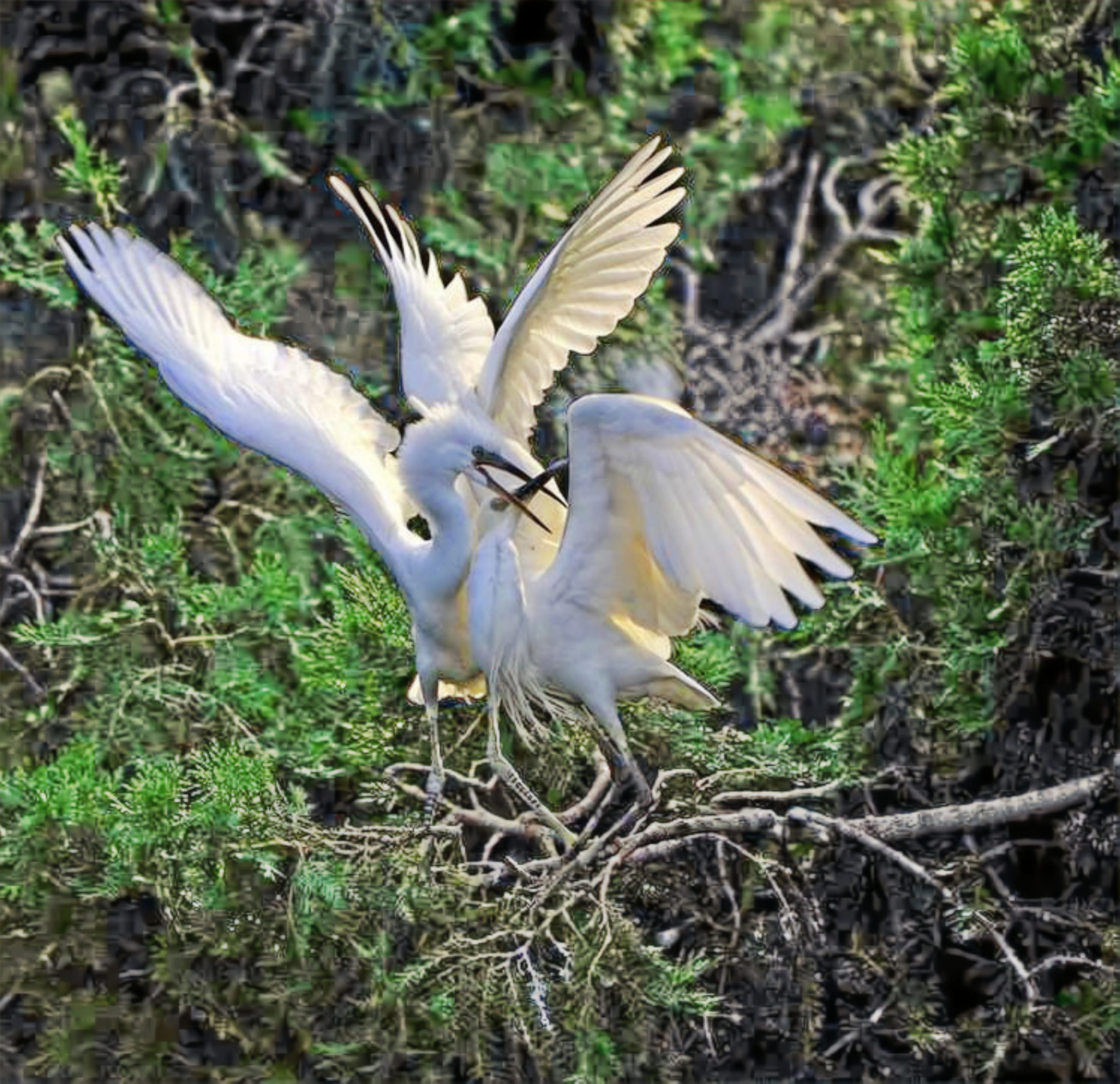}
		\end{minipage}
	}\hspace{-5pt}
	\subfigure[TSN-CA]{
		\begin{minipage}[b]{0.13\textwidth}
			\includegraphics[width=2.5cm]{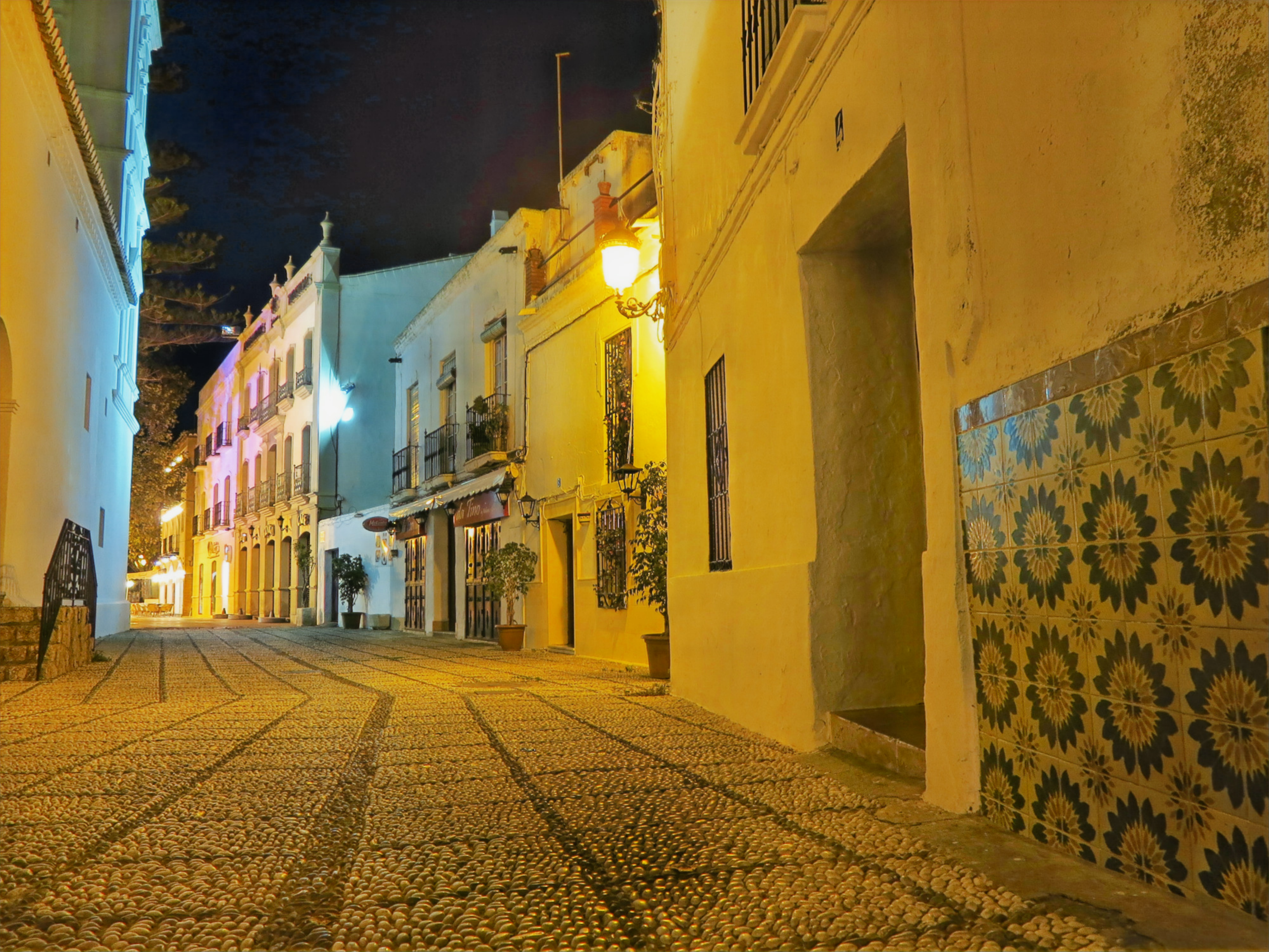}\vspace{2pt} \\
			\includegraphics[width=2.5cm]{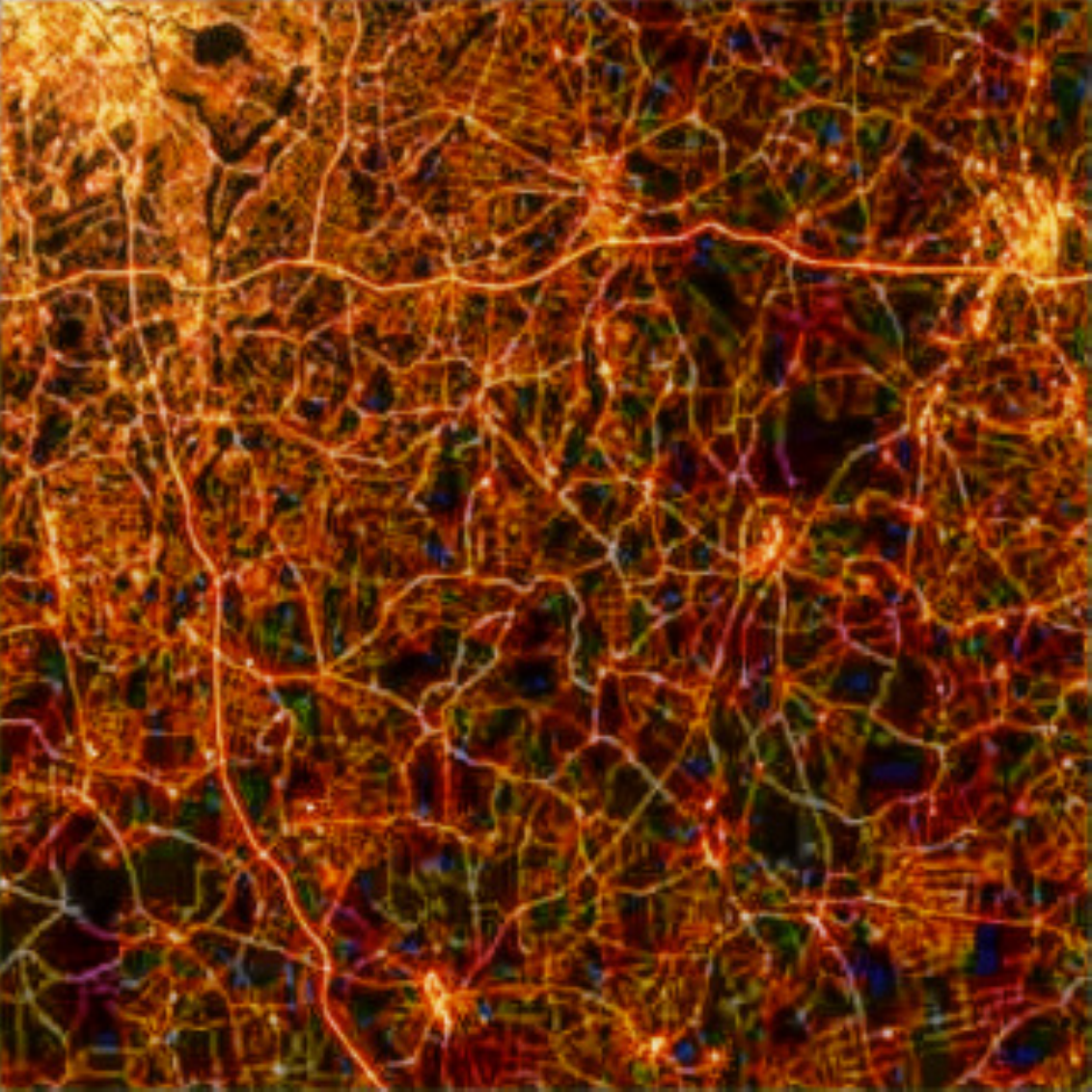}\vspace{2pt}
			\includegraphics[width=2.5cm]{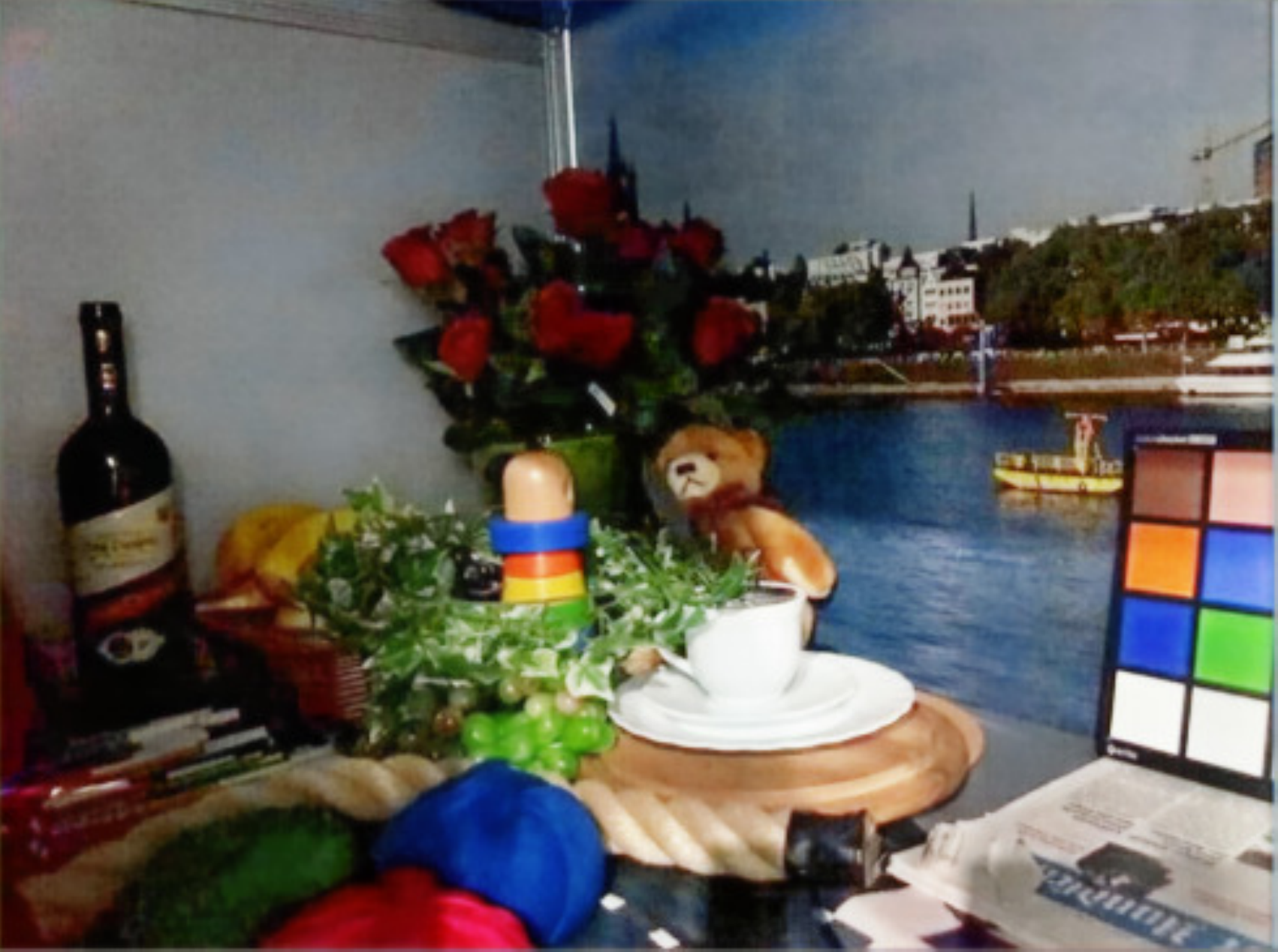}\vspace{2pt}
			\includegraphics[width=2.5cm]{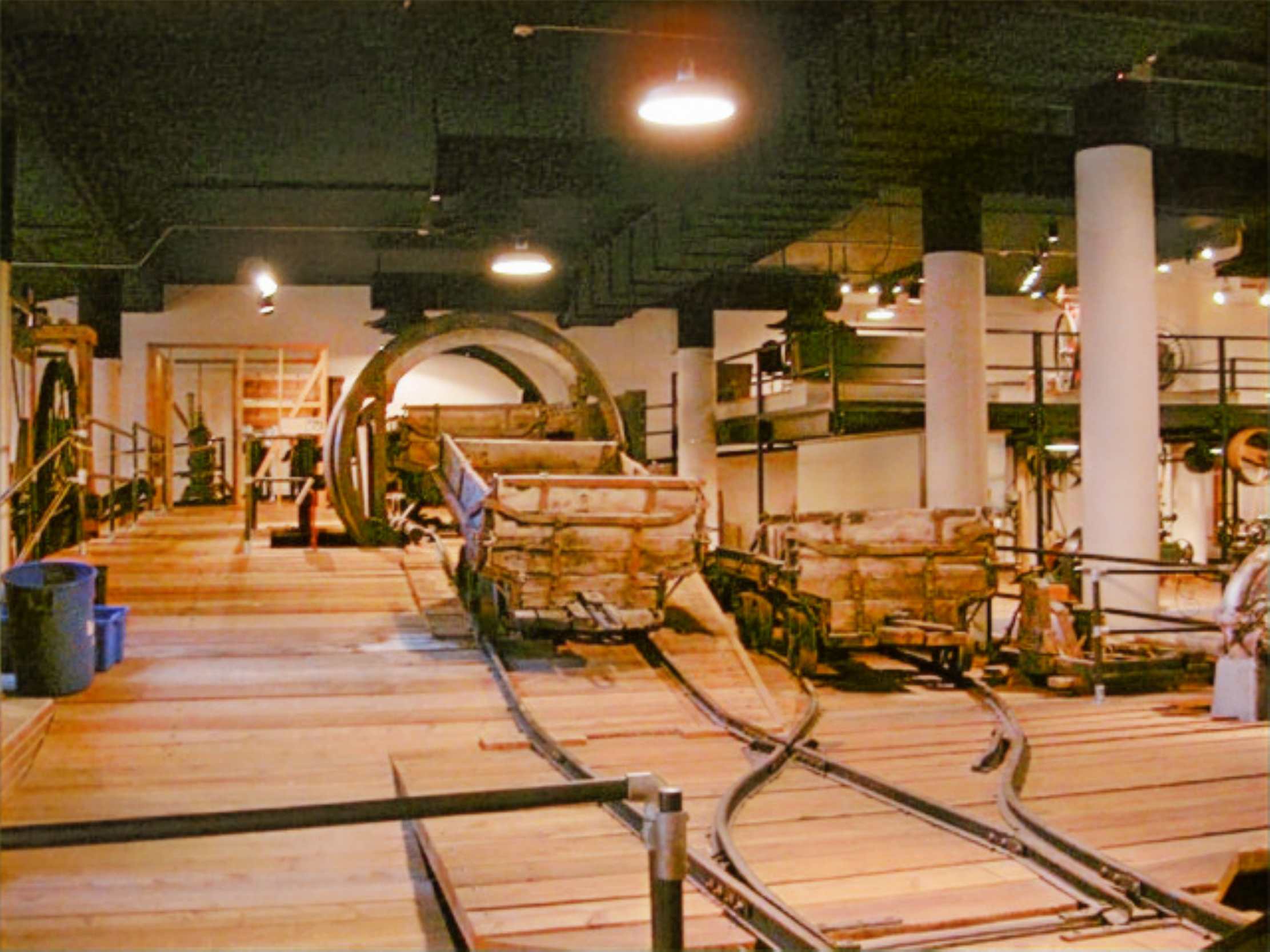}\vspace{2pt}
			\includegraphics[width=2.5cm]{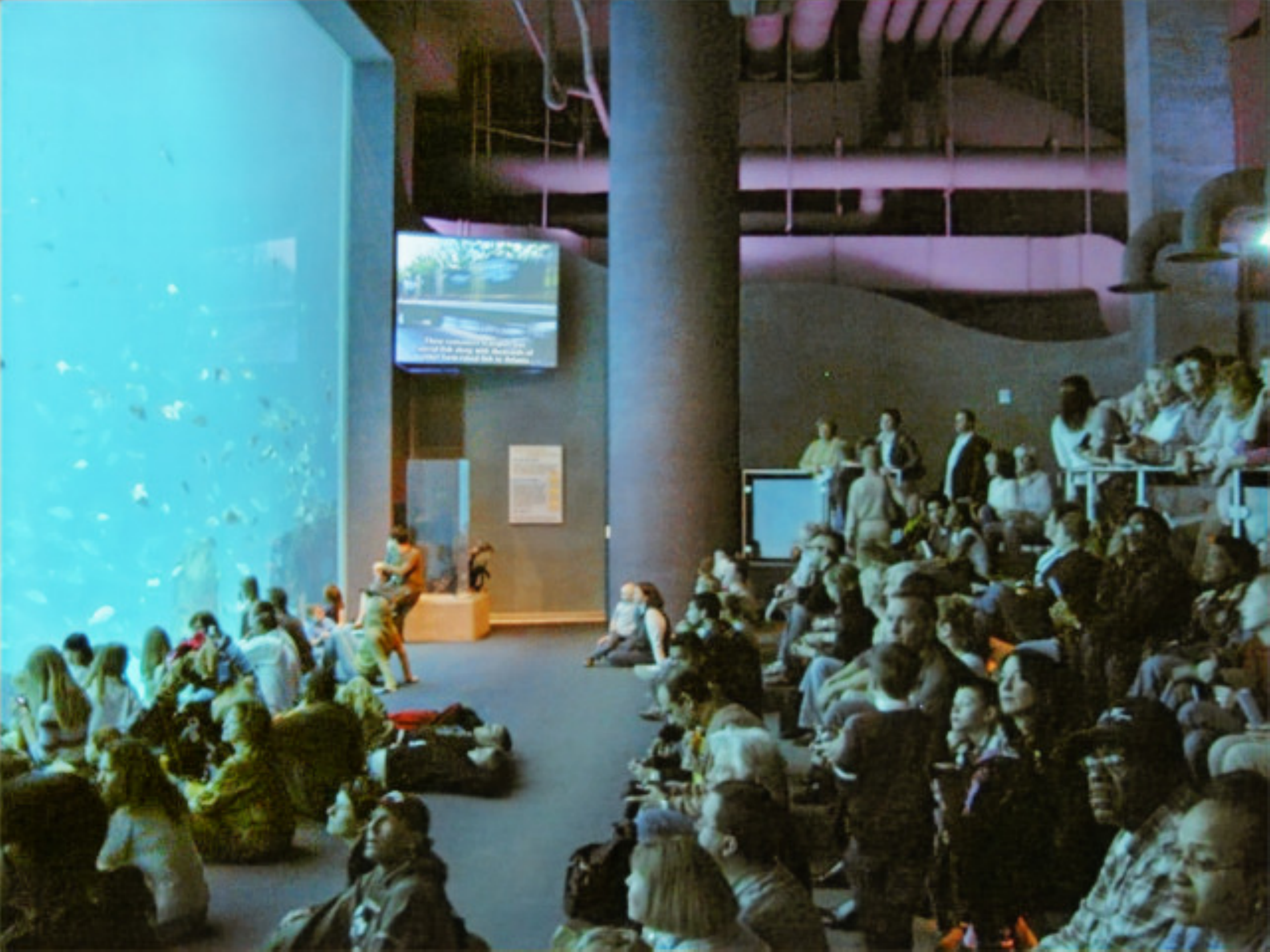}\vspace{2pt}
			\includegraphics[width=2.5cm]{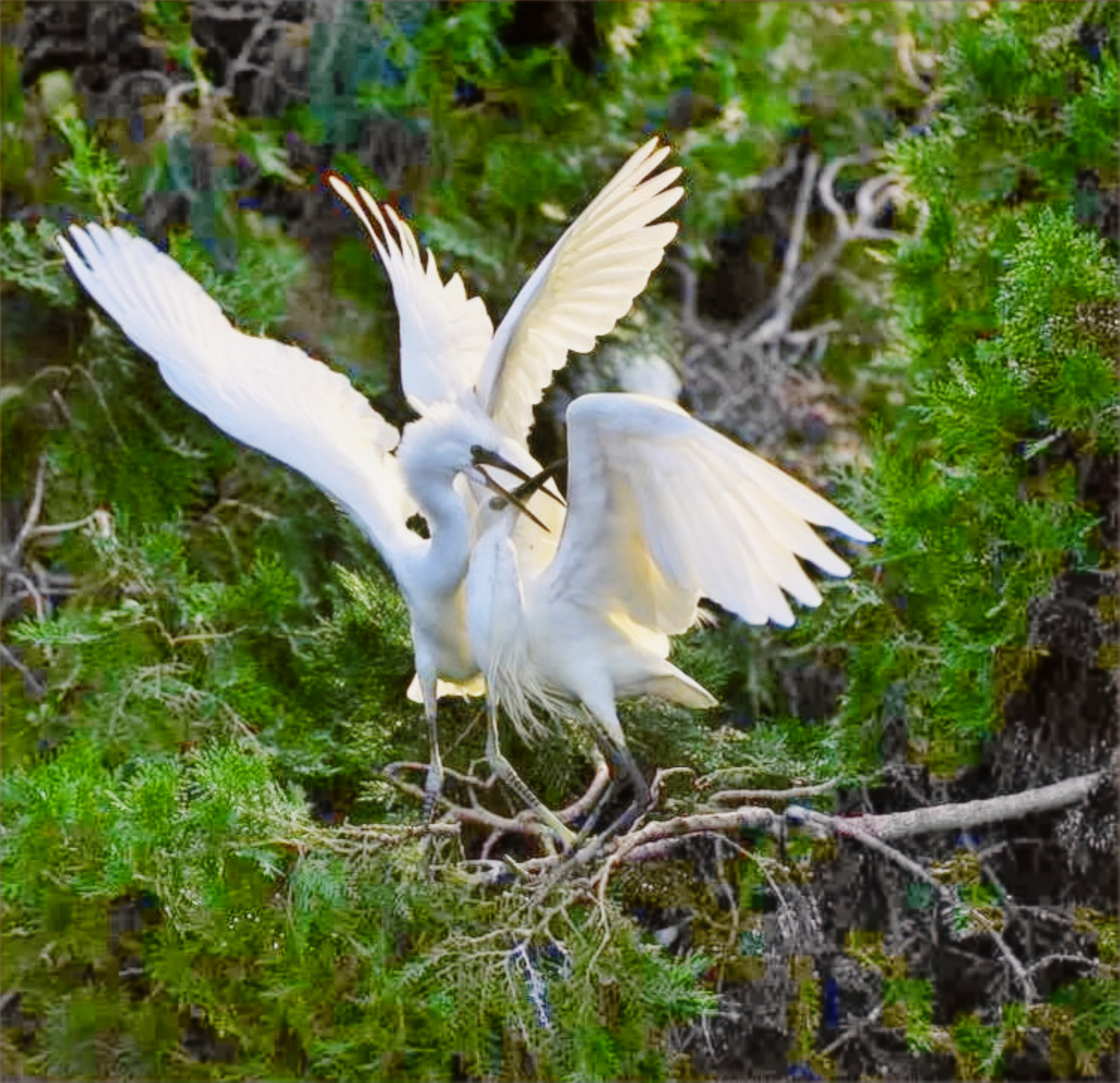}
		\end{minipage}
	}\hspace{-5pt}
	\caption{Visual comparison with other state-of-the-art methods on several frequently used datasets without Ground-Truth. The model of TSN-CA was trained on the LOL real-world dataset.}
	\label{test}
\end{figure*}

\subsection{Qualitative Comparison}
As shown in Fig.\ref{lol}, \ref{syn} and \ref{test}, in other enhancement methods, there is still a lot of noise and severe color distortion, as well as lots of unpleasant shadow blocks and halo artifacts. By contrast, our method achieves good effect of noise removal, color distortion correction, in addition, with combining channel attention (CA) mechanism with the skip connection of U-Net and embedded SE Module into the skip connection, shadow blocks and halo artifacts can be eliminated very well without introducing too much extra computation cost.\\
Our results are even better than Ground-Truth in terms of visual effects and more consistent with human visual perception.

\section*{Conclusion}
In this paper, we propose a two-stage network for low-light image enhancement and restoration. In stage one, we firstly transform the low-light image from RGB space to HSV space, train the network to enhance the brightness of the V channel, and leverage the information of H and S to help the V channel reconstruct the details information during the enhancement process. In stage two, we combined the brightness-enhanced and detail-preserved V channel with the original degraded H and S channels and converted them from HSV space back to RGB space. And then we train a U-Net to restore the enhanced but degraded images. We introduce channel attention mechanism to help the restoration network remove noise, restore details better as well as eliminate shadow blocks and halo artifacts.

\bibliographystyle{IEEEtran}
\bibliography{ref}

\end{document}